\documentclass[a4paper,12pt,twoside,openright,titlepage,oldfontcommands]{memoir}
 
\usepackage[utf8]{inputenc}

\usepackage[T1]{fontenc}
\usepackage{lmodern}

\usepackage[francais,english]{babel}
\usepackage{amsmath}
\usepackage{amsfonts}
\usepackage{amssymb}
\usepackage{float}

\usepackage{xspace}
\usepackage{graphicx}
\usepackage{amsthm}

\usepackage{dsfont}
\usepackage{mathrsfs}

\usepackage{framed}

\usepackage{sidecap}
\sidecaptionvpos{figure}{c}

\usepackage{enumitem}

\usepackage{pifont}

\usepackage{hhline}

\usepackage{multirow}

\usepackage[inner=2.9cm,outer=2.4cm,top=3cm,bottom=3cm]{geometry}
\usepackage{subfiles}

\usepackage{xr}

\usepackage[usenames,dvipsnames]{xcolor}

\definecolor{Cblue}{HTML}{045FB4}
\definecolor{Cred}{HTML}{DF0101}

\definecolor{ViolUPS}{HTML}{63003C}

\usepackage{hyperref}
\hypersetup{
    colorlinks=true, 
    linktoc=all,     
    linkcolor=Cred,
    citecolor=Cblue,
    urlcolor=Cblue 
}

\usepackage{tikz}

\usepackage{pdfpages}
\newcounter{ArtCount}

\newcommand{\AddArticle}[2]{%
  \refstepcounter{ArtCount}%
  \chapter*{Article \arabic{ArtCount}}%
  \label{#1}%
  \addcontentsline{toc}{chapter}{Article~\arabic{ArtCount}: #2}%
}

\newcommand{\ArtRef}[1]{\hyperref[#1]{Article~\ref*{#1}}}

\usepackage{etoolbox}

\newbool{IncArt}
\booltrue{IncArt} 

\newbool{IncPage}

\usepackage{cite}

\makepagestyle{mystyle}
\makeevenfoot{mystyle}{}{\textbf{--~\thepage~--}}{}
\makeoddfoot{mystyle}{}{\textbf{--~\thepage~--}}{}
\nouppercaseheads
\makeevenhead{mystyle}{\leftmark}{}{}
\makeoddhead{mystyle}{}{}{\rightmark}
\makeheadrule{mystyle}{\textwidth}{\normalrulethickness}

\makepagestyle{chapter}
\makeevenfoot{chapter}{}{\textbf{--~\thepage~--}}{}
\makeoddfoot{chapter}{}{\textbf{--~\thepage~--}}{}

\makepagestyle{part}

\makechapterstyle{mychap}{%
  \chapterstyle{default}
  \renewcommand*{\chapnamefont}{\normalfont\Large\sffamily}
  
  \renewcommand*{\printchaptername}{%
    \chapnamefont\centering Chapter}

}

\chapterstyle{mychap}

\newcommand\Chap[1]{%
  \chapter*{#1}%
  \addcontentsline{toc}{chapter}{#1}%
  \markboth{#1}{#1}
}

\setsecnumdepth{subsection}

\graphicspath{{Figures/}{../Figures/}}

\renewcommand{\leq}{\leqslant}
\renewcommand{\geq}{\geqslant}

\def\I{\mathrm{i}}

\newcommand{\nn}{\nonumber}
\newcommand{\Prob}{{\rm Prob}}

\DeclareMathOperator{\Ai}{Ai}
\DeclareMathOperator{\Bi}{Bi}
\DeclareMathOperator{\Li}{Li}
\DeclareMathOperator{\Tr}{Tr}

\DeclareMathOperator{\J}{J}
\DeclareMathOperator{\Det}{Det}

\DeclareMathOperator{\erfc}{erfc}
\DeclareMathOperator{\erf}{erf}
\newcommand{\moy}[1]{\ensuremath{\langle #1 \rangle}}

\newcommand{\be}{\begin{equation}}
\newcommand{\ee}{\end{equation}}
\newcommand{\var}[1]{{\rm Var}\left(#1\right)}

\begin{document}

\frontmatter
\begin{titlingpage}

  \includepdf[pages={1}]{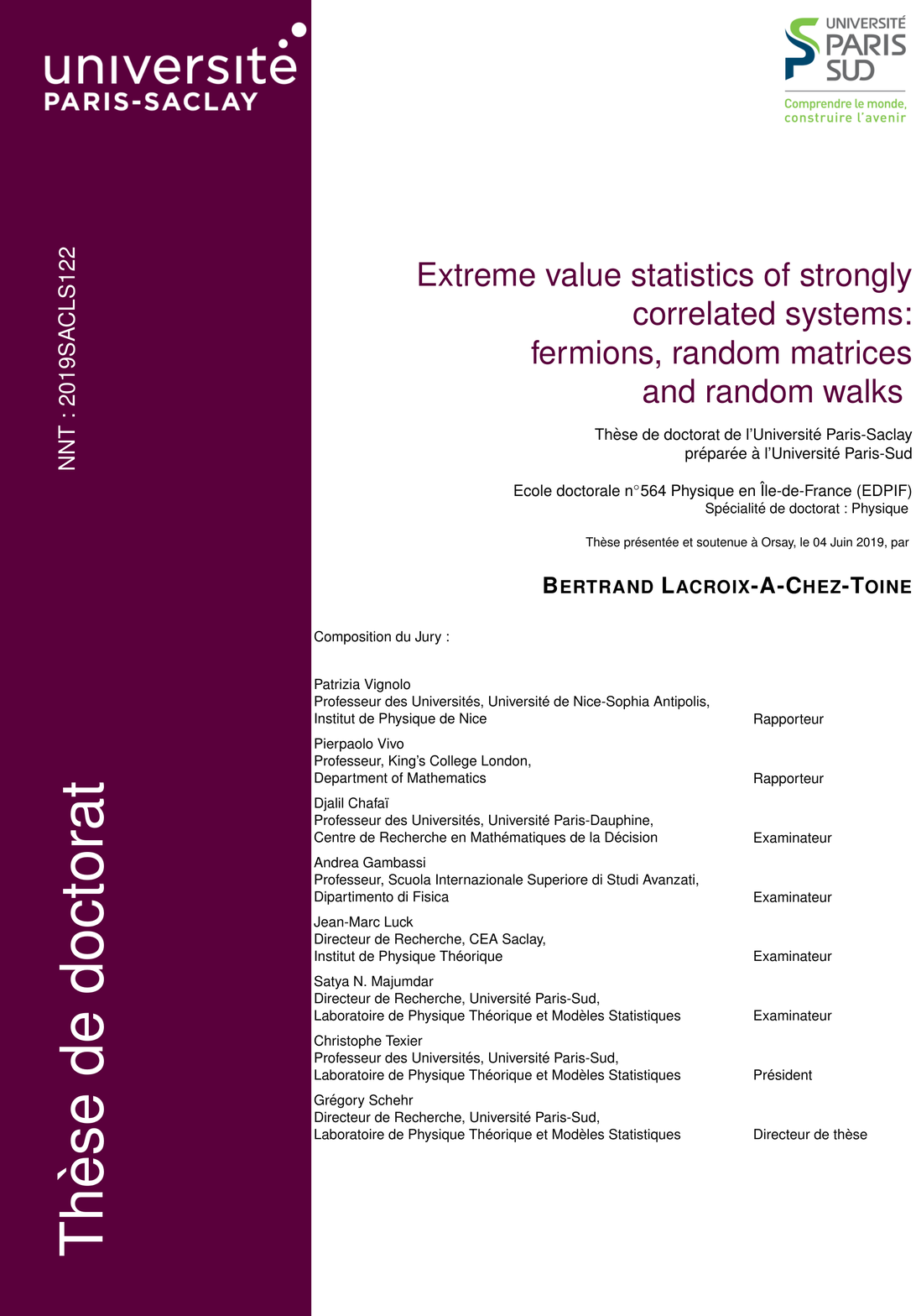}

\end{titlingpage}

\clearpage
\begingroup
  \pagestyle{empty}
\endgroup

\Chap{Remerciements}
Je tiens tout d'abord à remercier mon directeur de thèse Grégory Schehr. En duo avec Satya Majumdar, qui a été plus qu'un collaborateur mais véritablement un co-directeur, ils m'ont enseigné plus que je n'aurais pu l'espérer scientifiquement et sur le mode de fonctionnement de la recherche. Il a été par ailleurs toujours agréable de les côtoyer tant au laboratoire que dans un cadre plus informel lors de conférences. Je remercie l'ensemble de mes collaborateurs d'avoir partagé leur savoir et Pierre Le Doussal en particulier. Merci à Dio pour ses suggestions toujours intéressantes et son travail de relecture.

Je remercie particulièrement Patrizia Vignolo et Pierpaolo Vivo d'avoir accepté la tâche complexe de rapporteurs. Merci également Djalil Chafaï, Andrea Gambassi, Jean-Marc Luck et Christophe Texier de partager leur expertise en prenant part à mon jury de thèse. 

Je remercie l'ensemble des membres du LPTMS de m'avoir accompagné durant ces trois années. Un grand merci en particulier à Claudine Le Vaou et Karolina Kolodziej pour leur support technique toujours très efficace et au directeur du laboratoire Emmanuel Trizac. Merci également à Christophe Texier de m'avoir accordé sa confiance pour participer à ses enseignements et d'avoir pris le temps de répondre à mes nombreuses questions. Je suis également très reconnaissant à Aurélien Grabsch qui a toujours su m'aider lorsque je venais le perturber dans son travail. Je pense également à mes camarades doctorants Inés, Samuel, Mathieu, Thibault $\times 2$, Lucas $\times 2$, Ivan, Hao, Nina, Nadia, Sebastian, Hugo grâce auxquels les journées au laboratoire et en dehors ont toujours été sympathiques.

J'ai également eu la chance de partager de très bons moments en dehors du travail avec mes anciens colocataires Vincent et Romain dont la joie de vivre est communicative. Je pense également à Michaël, Richard, Frédéric, Félix, Maxime, Mathieu et Rémy qui m'ont offert des distractions bienvenues au quotidien de la recherche. Une pensée aux physiciens Hugues, Marion, Louis, Pascal qui m'ont accompagné depuis nos premières heures à Cachan et comprendront je l'espère certains passages de ce manuscrit. Un grand merci également aux amis de toujours Hugo, Sylvain, Simon, Antonin, Maxime, Bazil, Anaïs, Yoann et Anta que je retrouve toujours avec beaucoup de plaisir. Je remercie mes parents qui m'ont toujours encouragé à poursuivre ma voie ainsi que mes frères. Finalement, je tiens à remercier Doriane qui a partagé mon quotidien depuis de nombreuses années désormais et a toujours été là pour moi. Elle m'a toujours encouragé à être la meilleure version de moi-même.



\clearforchapter

{\hypersetup{linkcolor=black}
  \setcounter{tocdepth}{2}
  \tableofcontents*
}







\Chap{R\'esum\'e en fran\c cais}

Une question cruciale dans de nombreux contextes, allant de la m\'et\'eorologie (ouragans, canicule, inondations) \`a la g\'eologie (tremblements de terre), en passant par la finance (crash boursier) et la physique est celle de pr\'evoir la fr\'equence des \'ev\'enements extr\^emes. Ces \'{e}v\'{e}nements, bien qu'atypiques, ont g\'{e}n\'{e}ralement un impact d\'{e}sastreux. Pour mieux se pr\'{e}parer \`{a} des effets aussi catastrophiques, il est essentiel d'avoir une compr\'{e}hension plus approfondie de ces ph\'{e}nom\`{e}nes.  Les statistiques d'extr\^{e}mes ont \'{e}t\'{e} \'{e}tudi\'{e}es en d\'{e}tail depuis de nombreuses ann\'{e}es, soit depuis les travaux fondateurs de Gumbel \cite{gumbel2012statistics} qui a identifi\'{e} les trois classes d'universalit\'{e} dans le cas des variables al\'{e}atoires ind\'{e}pendantes et \`{a} distribution identique (i.i.d.) : (I) Gumbel, (II) Fr\'echet et (III) Weibull. De nombreuses \'{e}tudes ont \'{e}t\'{e} men\'{e}es depuis lors pour obtenir des r\'{e}sultats concernant les valeurs extr\^{e}mes dans le cas de variables al\'{e}atoires non r\'{e}parties de fa\c{c}on identique ou corr\'{e}l\'{e}es (voir par exemple \cite{krapivsky2000traveling, tracy1994level, bertin2006generalized, carpentier2001glass, majumdar2014extreme, bouchaud1997universality, majumdar2005airy}) mais on ne dispose pas de th\'{e}orie g\'{e}n\'{e}rale et chaque cas doit \^{e}tre examin\'{e} s\'{e}par\'{e}ment.

Dans le contexte de la physique, les applications de la th\'{e}orie des valeurs extr\^{e}mes sont nombreuses. Ainsi, on peut citer l'\'{e}tude des syst\`{e}mes d\'{e}sordonn\'{e}s \cite{derrida1981random, bouchaud1997universality, dean2001extreme, le2003exact, schawe2018ground} comme les verres (de spin) o\`{u} la physique \`{a} basse temp\'{e}rature est domin\'{e}e par les propri\'{e}t\'{e}s de l'\'{e}tat fondamental, c.-\`{a}-d. l'\'{e}tat de plus basse \'{e}nergie (voir aussi \cite{biroli2007extreme} pour une revue sur le sujet). Les statistiques des valeurs extr\^{e}mes ont \'{e}galement fait l'objet d'\'{e}tudes approfondies dans le contexte des processus de croissance stochastique dans la classe d'universalit\'{e} $(1+1)d$ Kardar-Parisi-Zhang \cite{kardar1986dynamic, johansson2000shape, prahofer2000universal, sasamoto2010one, calabrese2010free, amir2011probability, baik2012joint, baik1999distribution} et le probl\`{e}me associ\'{e} de polym\`{e}res dirig\'{e}s (voir \cite{majumdar2007course, TAKEUCHI201877} pour des pr\'{e}sentations p\'{e}dagogiques).
Un point d\'{e}cisif a \'{e}t\'{e} de relier cette classe de probl\`{e}mes \`{a} la th\'{e}orie des matrices al\'{e}atoires (RMT), o\`{u} Tracy et Widom ont obtenu des r\'{e}sultats exacts \cite{tracy1994level}. La distribution de Tracy-Widom est devenue d\`{e}s lors une pierre angulaire des statistiques de valeurs extr\^{e}mes pour les syst\`{e}mes corr\'{e}l\'{e}s \cite{majumdar2014top}. Elle a \'{e}t\'{e} observ\'{e}e exp\'{e}rimentalement dans la croissance de cristaux liquides n\'{e}matiques \cite{takeuchi2010universal,takeuchi2011growing} en relation directe avec les mod\`{e}les de croissance stochastique mais aussi dans un contexte tr\`{e}s diff\'{e}rent dans des exp\'{e}riences de fibres optiques coupl\'{e}es \cite{fridman2012measuring}. Toutefois, dans la plupart des probl\`{e}mes physiquement pertinents, l'obtention des statistiques d'extr\^{e}mes reste un probl\`{e}me ouvert.  L'un des principaux objectifs de cette th\`{e}se est d'\'{e}largir la connaissance des statistiques d'extr\^{e}mes pour les syst\`{e}mes corr\'{e}l\'{e}s en explorant des mod\`{e}les pour lesquels ces statistiques peuvent \^{e}tre obtenues exactement. 
%
%
%

Dans cette th\`{e}se, nous consid\'{e}rons trois grandes classes de mod\`{e}les, avec des applications physiques directes, o\`{u} les statistiques d'extr\^{e}mes peuvent \^{e}tre obtenues exactement : (i) les fermions sans interaction, (ii) les matrices al\'{e}atoires et (iii) les marches al\'{e}atoires.

\subsubsection{Fermions sans interaction}

Consid\'{e}rons d'abord certaines propri\'{e}t\'{e}s physiques des gaz de fermions pi\'{e}g\'{e}s sans spin (ou polaris\'{e}s) et sans interaction. M\^{e}me en l'absence d'interaction, le principe d'exclusion de Pauli introduit de fortes corr\'{e}lations quantiques dans le syst\`{e}me car deux fermions ne peuvent occuper le m\^{e}me \'{e}tat quantique. Les fluctuations thermiques \'{e}liminent ces corr\'{e}lations \`{a} haute temp\'{e}rature et les corr\'{e}lations quantiques sont donc plus importantes \`{a} basse temp\'{e}rature. Les statistiques quantiques engendrent des propri\'{e}t\'{e}s spatiales non triviales pour le syst\`{e}me de fermions. Les exp\'{e}riences sur les atomes froids constituent la plateforme id\'{e}ale pour \'{e}tudier ces corr\'{e}lations car les r\'{e}cents progr\`{e}s exp\'{e}rimentaux permettent un contr\^{o}le sans pr\'{e}c\'{e}dent des param\`{e}tres (voir \cite{bloch2008many, giorgini2008theory} pour des revues).
Dans ces syst\`{e}mes, les interactions entre atomes peuvent \^{e}tre ajust\'{e}es via la r\'{e}sonance de Feshbach \cite{regal2003tuning} et en particulier, le r\'{e}gime sans int{e}raction est accessible. Cela  permet de sonder et d'isoler les effets purement quantiques \'{e}mergents du principe de Pauli. Ces exp\'{e}riences sont maintenant utilis\'{e}es comme simulateurs quantiques pour les syst\`{e}mes de mati\`{e}re condens\'{e}e, o\`{u} l'on peut r\'{e}gler les param\`{e}tres du Hamiltonien de mani\`{e}re contr\^{o}l\'{e}e \cite{bloch2012quantum}. Pour les gaz de Fermi, le d\'{e}veloppement r\'{e}cent de microscopes \`{a} gaz de Fermi \cite{haller2015single, cheuk2015quantum, parsons2015site} permet de sonder les positions des particules individuelles comme observ\'{e} sur la Fig. \ref{Fig_qu_gas_micro} (voir \cite{kuhr2016quantum, ott2016single} pour des revues r\'{e}centes). Ce type d'imagerie pourrait \'{e}galement permettre de tester les propri\'{e}t\'{e}s dynamiques et hors \'{e}quilibre des gaz quantiques qui ont g\'{e}n\'{e}r\'{e} beaucoup d'int\'{e}r\^{e}t th\'{e}orique ces derni\`{e}res ann\'{e}es \cite{calabrese2006time, calabrese2007quantum, vignolo2001one, krapivsky2019return, krapivsky2018quantum, eisler2013full, perfetto2017ballistic}. 

Pour mener ce type d'exp\'{e}riences, le gaz quantique doit \^{e}tre confin\'{e} par un {\it potentiel de pi\'{e}geage}. Ce potentiel cr\'{e}e toujours un bord fini au-del\`{a} duquel la densit\'{e} de particules dans le gaz est essentiellement nulle. Bien que des techniques standard telles que l'approximation de la densit\'{e} locale \cite{castin2006basic} aient \'{e}t\'{e} mises au point pour d\'{e}crire l'essentiel du gaz, elles ne permettent pas de d\'{e}crire les statistiques spatiales pr\`{e}s du bord \cite{kohn1998edge, vignolo2000exact}.

Compte tenu des installations exp\'{e}rimentales d\'{e}crites ci-dessus, il est crucial de bien comprendre la physique du gaz de Fermi froid \`{a} proximit\'{e} du bord. Cette question a \'{e}t\'{e} abord\'{e}e pour les fermions sans interaction dans une r\'{e}cente s\'{e}rie d'articles \cite{dean2015universal, dean2016noninteracting, dean2015finite} utilisant le cadre des processus d\'{e}terminantaux \cite{hough2006determinantal, johansson2005random}. En particulier, beaucoup de progr\`{e}s ont r\'{e}sult\'{e} d'une connexion directe entre l'\'{e}tat fondamental d'un syst\`{e}me unidimensionnel de fermions \`{a} temp\'{e}rature nulle et confin\'{e} par un potentiel harmonique et l'ensemble Gaussien unitaire (GUE). La probabilit\'{e} jointe des positions dans l'\'{e}tat fondamental  de ce gaz de Fermi peut \^{e}tre calcul\'{e}e exactement
\be
|\Psi_0(x_1,\cdots,x_N)|^2=\frac{1}{Z_N(\alpha)} \prod_{i<j}|x_i-x_j|^2 \prod_{i=1}^N e^{-\alpha^2 x_i^2 }\;,\;\;\alpha=\sqrt{\frac{m\omega}{\hbar}}\;.\label{jPDF_OH1d_intro_fr}
\ee
On reconna\^ it d'apr\`{e}s la correspondance exacte $x_i=\sqrt{N}\lambda_i/\alpha$ la c\'{e}l\`{e}bre distribution jointe des valeurs propres du GUE. Cet ensemble matriciel est un ensemble invariant de la th\'{e}orie des matrices al\'{e}atoires, o\`{u} les matrices hermitiennes sont construites avec des entr\'{e}es complexes gaussiennes  ind\'{e}pendantes. Le r\'{e}sultat dans Eq. \eqref{jPDF_OH1d_intro_fr} n'est valable que pour un potentiel harmonique, mais il est exp\'{e}rimentalement pertinent de r\'ealiser diff\'{e}rentes formes pour le potentiel de confinement \cite{mukherjee2017homogeneous, hueck2018two}. En r\'{e}alit\'{e}, il a \'{e}t\'{e} d\'{e}montr\'{e} que les r\'{e}sultats obtenus pour les corr\'{e}lations au bord de la densit\'{e} s'\'{e}tendent \`{a} tout potentiel qui varie lentement, par exemple $V(x)\sim |x|^p$ avec $p>0$.
 Ces potentiels produisent une variation continue et r\'{e}guli\`{e}re de la densit\'{e} pr\`{e}s du bord. Dans la th\'{e}orie des matrices al\'{e}atoires, cette classe d'universalit\'{e} pour le comportement au bord est g\'{e}n\'{e}ralement qualifi\'{e}e de ``soft edge''. Cependant, il existe des ensembles de matrice al\'{e}atoires, par exemple les ensembles Wishart ou Jacobi, o\`{u} la densit\'{e} pr\'{e}sente des bords avec des divergences ou des discontinuit\'{e}s que l'on qualifie de ``hard edges''. Une question naturelle est donc de se demander si ces mod\`{e}les de matrices al\'{e}atoires ont des analogues dans les mod\`{e}les de fermions qui pr\'{e}sentent un comportement de ``hard edge'' similaire et s'il existe une classe d'universalit\'{e} associ\'{e}e. R\'{e}pondre \`{a} cette question constitue l'un des principaux objectifs de cette th\`{e}se. 

\subsubsection{Matrices al\'eatoires}

Nous consid\'{e}rons un probl\`{e}me connexe, celui des propri\'{e}t\'{e}s statistiques des matrices al\'{e}atoires \cite{mehta2004random, anderson2010introduction, tracy1993introduction, livan2018introduction, forrester2010log}. Depuis leur premi\`{e}re apparition dans la litt\'{e}rature statistique \cite{wishart1928generalised}, les matrices al\'{e}atoires ont \'{e}t\'{e} largement utilis\'{e} en math\'{e}matiques, t\'{e}l\'{e}communications, \'{e}cologie ou finance. En physique, elles ont d'abord \'{e}t\'{e} introduites par Wigner \cite{wigner1951statistical} pour d\'{e}crire l'espacement entre les niveaux d'\'{e}nergie dans les noyaux mais a \'{e}t\'{e} utilis\'{e} depuis en physique statistique afin de d\'{e}crire des marcheurs al\'{e}atoires malveillants (sans intersection) \cite{Forrester1989,tracy2007nonintersecting, NAGAO200329, nadal2009nonintersecting, PhysRevLett.101.150601, FORRESTER2011500}, des mod\`{e}les de plasmas \cite{chafai2014note, forrester1998exact, cunden2016large, cunden2017universality}, en physique m\'esoscopique \cite{Jayannavar1989, beenakker1997random,Grabsch2017_2 , PhysRevLett.101.216809, PhysRevB.81.104202, PhysRevLett.82.4220, PhysRevLett.78.4737} ou en chromodynamique quantique \cite{wadia1980n,gross1993possible} (voir \cite{majumdar2014top, biroli2007extreme} pour une revue concernant les applications physiques de la RMT). Les mod\`{e}les de matrices al\'{e}atoires offrent un cadre tr\`{e}s utile pour analyser les statistiques de valeurs extr\^{e}mes : alors que leurs valeurs propres sont fortement corr\'{e}l\'{e}es, leurs statistiques d'extr\^{e}mes peuvent \^{e}tre obtenues exactement et ont \'{e}t\'{e} \'{e}tudi\'{e}es en d\'{e}tail \cite{tracy1994fredholm, tracy1994level, tracy1994level, tracy1994level2, duenez2010lowest, dumitriu2008distributions, rider2003limit}. Dans cette th\`{e}se, nous explorons la connexion entre les fermions et les matrices al\'{e}atoires \cite{dean2015finite,dean2015universal,dean2018wigner,eisler2013universality,grabsch2018fluctuations,le2016exact,marino2016number, eisler2013full, calabrese2015random} et consid\'{e}rons principalement l'ensemble unitaire de Jacobi (JUE) et l'ensemble complexe de Ginibre. A partir de ces connexions, nous r\'{e}solvons plusieurs questions ouvertes pour les grandes d\'{e}viations des statistiques de comptage (FCS) et des statistiques d'extr\^{e}mes dans ces ensembles.

\subsubsection{Marches al\'eatoires}

Le dernier syst\`{e}me que nous examinerons est celui des marches al\'{e}atoires et de leur \'{e}quivalent continu, les mouvements Browniens. Ce mod\`{e}le fondamental de la m\'{e}canique statistique a \'{e}t\'{e} introduit pour la premi\`{e}re fois il y a plus d'un si\`{e}cle par Louis Bachelier dans le contexte de la finance \cite{bachelier1900theorie} (voir aussi \cite{Pearson1905} pour la premi\`{e}re apparition du nom {\it random walk} et \cite{RevModPhys.15.1} pour une revue). Dans ce mod\`{e}le, les positions prises par le marcheur al\'{e}atoire forment un ensemble de variables al\'{e}atoires fortement corr\'{e}l\'{e}es et constituent donc un laboratoire pr\'{e}cieux pour tester les effets des fortes corr\'{e}lations.
En particulier, de nombreux r\'{e}sultats ont \'{e}t\'{e} obtenus pour les statistiques du maximum global \cite{feller1968introduction, Mounaix_2018, comtet2005precise, majumdar2010universal}. Les statistiques d'ordre, c'est-\`{a}-dire les statistiques des maxima ordonn\'{e}s (deuxi\`{e}me, troisi\`{e}me, etc.) ont \'{e}galement \'{e}t\'{e} examin\'{e}es en d\'{e}tail pour une marche al\'{e}atoire r\'{e}guli\`{e}re \cite{wendel1960order, port1962elementary, feller1968introduction, schehr2012universal, schehr2014exact, dassios1996sample, 10.2307/2959757} et pour les mouvements browniens ramifi\'{e}es \cite{Brunet_2009, Derrida2011, ramola, ramola2}. Les statistiques d'ordre font partie plus g\'en\'eralement de la {\it th\'{e}orie de la fluctuation} qui a \'{e}t\'{e} largement \'{e}tudi\'{e}e par la communaut\'{e} math\'{e}matique \cite{wendel1960order, port1962elementary, feller1968introduction,pitman2018guide,revuz2013continuous }.  Une vaste litt\'{e}rature a \'{e}galement \'{e}merg\'{e} sur le sujet connexe des records pour ces marches al\'{e}atoires \cite{majumdar_ziff, PhysRevE.86.011119, Sabhapandit_2011, majumdar2012record, schehr2014exact, berkowitz_records_noise, wergen_borgner_krug, Godr_che_2017, Godr_che_2014}. Dans ces deux cas, le probl\`{e}me devient plus simple dans la limite de grand $n$ pour laquelle le processus converge (pour une distribution de sauts \`{a} variance finie) vers le mouvement brownien, pour lequel un grand nombre de r\'{e}sultats ont \'{e}t\'{e} obtenus \cite{yor1995distribution, perret2013near, ramola, dassios1995distribution}. Il existe cependant moins de r\'{e}sultats pour les statistiques d'\'{e}carts entre positions maximales cons\'{e}cutives (gaps) pour les marches al\'{e}atoires \cite{Brunet_2009, Derrida2011,  ramola, battilana2017gap,schehr2012universal}, qui n\'{e}cessitent de prendre en compte le caract\`{e}re intrins\`{e}quement discret du processus. Ces probl\`{e}mes ne peuvent \^{e}tre r\'{e}solus en utilisant la convergence vers le mouvement brownien. L'un des objectifs de cette th\`{e}se est donc d'obtenir de nouveaux r\'{e}sultats pour ce probl\`{e}me int\'{e}ressant.


\section*{Panorama de la th\`ese et r\'esultats principaux}

Nous pr\'esentons ici un panorama rapide de la th\`ese et r\'esumons les r\'esultats principaux. Ces r\'esultats sont encadr\'es dans le texte ($\boxed{x}$)  et les r\'esultats non publi\'es sont encadr\'es deux fois ($\boxed{\frame{\boxed{x}}}$).

Bien que nous consid\'erons des mod\`eles sp\'ecifiques, ces r\'esultats sont souvents valides de mani\`ere plus g\'en\'erale, comme c'est souvent le cas en physique statistique.

\subsection*{Premi\`ere partie: Description spatiale de gaz de Fermi sans interaction}

La partie \ref{Part:Fer} de cette th\`ese est d\'edi\'ee \`a l'\'etude des fermions sans interaction et de leur connexion avec les valeurs propres de matrices al\'eatoires.

Dans le chapitre \ref{chap:fermions_intro}, nous introduisons le cadre th\'eorique permettant de d\'ecrire les propri\'et\'es spatiales des fermions et passons en revue les r\'esultats concernant des potentiels \`a variations lentes.

Dans le chapitre \ref{ch: ferm_hard_edge}, nous \'etendons la description des statistiques spatiales de fermions proches du bord au cas de potentiels de type {\it hard edge}. Nous d\'emontrons une correspondance exacte entre l'\'etat fondamentale d'un mod\`ele de fermions dans un potentiel de bo\^ite et l'ensemble unitaire de Jacobi.
Nous obtenons de mani\`ere exacte le noyau de correlation de ce processus determinantal et d\'emontrons que ce r\'esultat s'\'etend \`a toute une nouvelle classe d'universalit\'e dite de potentiels de hard edge. Nous \'etendons ce r\'esultat dans deux directions: pour des mod\`eles en plus grande dimension et \`a temperature non-nulle. Nous appliquons ces r\'esultats en calculant la distribution de la position du fermion le plus \'eloign\'e du centre du pi\`ege. Nous obtenons en particulier l'\'emergence d'un r\'egime de d\'eviation interm\'ediaire, connectant les fluctuations typiques aux grandes d\'eviations. Ce r\'egime n'appara\^it pas dans les ensembles invariants standard comme le GUE.

L'\'etude de ces syst\`emes de fermions sans interaction dans des potentiels type hard edge a conduit \`a la publication de deux articles:\\

\ref{Art:fermions_lett} {\it Statistics of fermions in a d-dimensional box near a hard wall}

 B. Lacroix-A-Chez-Toine, P. Le Doussal, S.~N. Majumdar, G. Schehr,

 Europhys. Lett. {\bf 120} (1), 10006 (2018).\\

\ref{Art:ferm_long} {\it Non-interacting fermions in hard-edge potentials},

 B. Lacroix-A-Chez-Toine, P. Le Doussal, S.~N. Majumdar, G. Schehr,

 J. Stat. Mech {\bf 12}, 123103 (2018).\\
 
Dans le chapitre \ref{ch: rot_trap}, nous d\'eveloppons une correspondance exacte entre l'\'etat fondamental d'un syst\`eme de fermions en rotation et l'ensemble complexe de Ginibre. Nous calculons de mani\`ere exacte les statistiques de comptage et l'entropie d'intrication pour ce syst\'eme et ce pour toute valeure finie du nombre $N$ de fermions. Il existe des mod\`eles de plasmas confin\'es par un potentiel harmonique en dimension deux, dans lequel les particules interagissent via la r\'epulsion coulombienne, correspondant exactement \`a ce probl\`eme. Nous montrons que nos r\'esultats sont universels et ne d\'ependent pas des d\'etails du potentiel de confinement dans ce mod\`ele de plasma. Ce faisant, nous r\'esolvons deux probl\`emes de matching entre fluctuations typiques et grandes d\'eviations en r\'evelant la pr\'esence d'un r\'egime de {\it d\'eviation interm\'ediaire} pour (i) les fluctuations de la particule la plus \'eloign\'ee du centre du pi\`ege et (ii) les statistiques de comptage. 

L'\'etude de ces fermions sans interaction dans des pi\`eges tournants, leur connexion avec l'ensemble de complexe de Ginibre et l'extension \`a des plasmas plus g\'en\'eraux a conduit \`a la publication de deux articles:\\

\ref{Art:rot} {\it Rotating trapped fermions in two dimensions and the complex Ginibre ensemble: Exact results for the entanglement entropy and number variance},\\
B. Lacroix-A-Chez-Toine, S.~N. Majumdar, G. Schehr,\\
Phys. Rev. A {\bf 99} (2), 021602 (2019).\\

\ref{Art:r_max} {\it Extremes of $2d$ Coulomb gas: universal intermediate deviation regime}\\
B. Lacroix-A-Chez-Toine, A. Grabsch, S.~N. Majumdar, G. Schehr,\\
J. Stat. Mech {\bf 1}, 013203 (2018).\\

Nous avons \'egalement soumis un autre article sur le sujet:\\

 \ref{Art:FCSGin} {\it Intermediate deviation regime for the full eigenvalue statistics
in the complex Ginibre ensemble}\\
B. Lacroix-A-Chez-Toine, J.~A.  Monroy Garz\'on, C.~S. Hidalgo Calva, I. P\'erez Castillo, A.  Kundu, S.~N. Majumdar, G. Schehr\\
arXiv preprint, arXiv: \textbf{1904.01813}, (2019).\\

\subsection*{Deuxi\`eme partie: statistiques des gaps de marches al\'eatoires}

La partie \ref{Part:gaps} de cette th\`ese est consacr\'ee \`a l'\'etude des statistiques d'extr\^eme, d'ordre et de gaps des marches al\'eatoires.

Dans le chapitre \ref{intro_iid}, nous passons en revue quelques r\'esultats pour les statistiques d'ordre et de gaps de variables al\'eatoires i.i.d.

Dans le chapitre \ref{ch:gaps_intro}, nous passons en revue les statistiques d'extr\^emes du mouvement brownien et des marches al\'eatoires.

Dans le chapitre \ref{ch:maxk}, nous consid\'erons les statistiques d'ordre des marches al\'eatoires. Nous obtenons une formule exacte pour le temps d'atteinte du $k^{\rm ieme}$ maximum de la marche. Nous passons en revue quelques propri\'et\'es de la distribution de la valeur prise par le $k^{\rm ieme}$ maximum  d'une marche al\'eatoire avec une distribution de saut \`a variance finie. Nous introduisons et calculons la densit\'e des maxima tremp\'es et recuits \`a la fois pour une variance finie et pour des vols de L\'evy. Ces r\'esultats ne sont pas encore publi\'es.
 
Dans le chapitre \ref{ch:gapk}, nous consid\'erons les statistiques des gaps de marches al\'eatoires. Nous montrons comment obtenir exactement la distribution de probabilit\'e des gaps pour la marche al\'eatoire avec une distribution de sauts donn\'ee par la loi de Laplace. Nous soutenons \`a partir de simulations num\'eriques que ce r\'esultat est universel dans la limite de grands $n$ pour toute distribution de saut avec variance finie.	

Nous avons r\'ecemment soumis un article sur les statistiques de gaps pour les marches al\'eatoires:\\

\ref{Art:gap} {\it Gap statistics close to the quantile of a random walk},\\
B. Lacroix-A-Chez-Toine, S.~N. Majumdar, G. Schehr,\\
arXiv preprint,  arXiv: \textbf{1812.08543}, (2018).\\

Mentionnons enfin un article r\'ecemment soumis et largement d\'econnect\'e des statistiques d'extr\^eme pour \'evaluation par les pairs:\\

\ref{Art:BMcoi} {\it Distribution of Brownian coincidences}\\
A. Krajenbrink, B. Lacroix-A-Chez-Toine, P. Le Doussal,\\
 arXiv preprint,  arXiv: \textbf{1903.06511}, (2019).\\

Dans cet article, nous calculons la distribution du temps de co\"incidence ${\cal T}_N(t)$, c'est-\`a-dire le temps local total de toutes les co\"incidences par paires de $N$ marcheurs Browniens ind\'ependants. Nous montrons que ce probl\`eme est li\'e (i) au probl\`eme de Lieb-Liniger de $N$ bosons en interaction de c\oe urs durs \cite{PhysRev.130.1605} et (ii) les moments de la fonction de partition canonique de polym\`eres dirig\'es (en relation directe avec l'\'equation de Kardar-Parisi-Zhang \cite{kardar1986dynamic}). Nous obtenons la distribution exacte de ${\cal T}_N(t)$ pour $N=2,3$ pour plusieurs conditions initiales et finales et les comportements asymptotiques de la distribution pour toute valeur de $N$.

 \clearpage
 \begingroup
   \pagestyle{empty}
   \null
   \newpage
   \null
   \newpage
 \endgroup



\mainmatter

\counterwithout{table}{chapter}







\Chap{Introduction}

Predicting the occurrence of extreme events is a crucial issue in many contexts, ranging from meteorology (hurricanes, heatwave, floods), geology (earthquakes), finance (stock market crash), all the way to physics. These events, although atypical, usually have a disastrous impact. 
To better prepare for such dramatic effects, it is vital to have a deeper understanding of these phenomena.  Extreme value statistics have been studied in detail for a number of years since the seminal work of Gumbel \cite{gumbel2012statistics} who identified the three universality classes in the case of independent and identically distributed (i.i.d) random variables: (I) Gumbel, (II) Fr\'echet and (III) Weibull. Many studies have been conducted ever since to obtain extreme value results in the case of either non-identically distributed or correlated random variables (see e.g. \cite{krapivsky2000traveling, tracy1994level, bertin2006generalized, carpentier2001glass, majumdar2014extreme, bouchaud1997universality, majumdar2005airy}) but there is no general theory and each case needs to be studied separately. 

In the context of physics, the applications of extreme value theory are numerous. A seminal example is the study of disordered systems \cite{derrida1981random, bouchaud1997universality, dean2001extreme, le2003exact, schawe2018ground} such as (spin) glasses where the physics at low temperature is dominated by the ground state properties, i.e. the state with lowest energy (see also \cite{biroli2007extreme} for a review). Extreme value statistics have also been extensively studied in the context of growth processes in the $(1+1)d$ Kardar-Parisi-Zhang universality class \cite{kardar1986dynamic, johansson2000shape, prahofer2000universal, sasamoto2010one, calabrese2010free, amir2011probability, baik2012joint, baik1999distribution} and the related problem of directed polymer (see \cite{majumdar2007course, TAKEUCHI201877} for pedagogical introductions). The turning point has been to connect this class of problems to random matrix theory (RMT), where Tracy and Widom have obtained exact results \cite{tracy1994level}. The Tracy-Widom distribution has become ever-since a cornerstone of extreme value statistics for correlated systems \cite{majumdar2014top}. It has been observed experimentally in the growth of nematic liquid crystals \cite{takeuchi2010universal,takeuchi2011growing} in direct relation to this class of models but also in a very different context in coupled optical fibres experiments \cite{fridman2012measuring}. However, in most physically relevant problems, obtaining the extreme value statistics remains an open problem. One of the main goal of this thesis is to enlarge the knowledge of extreme value statistics for correlated systems by exploring models for which these statistics can be obtained  exactly. 
%
%
%

In this thesis, we consider three large classes of models, with direct physical applications, where the extreme value statistics can be obtained exactly: (i) non-interacting fermions, (ii) random matrices and (iii) random walks.

\subsubsection{Non-interacting fermions}

\begin{figure}
\centering
\includegraphics[width=0.5\textwidth]{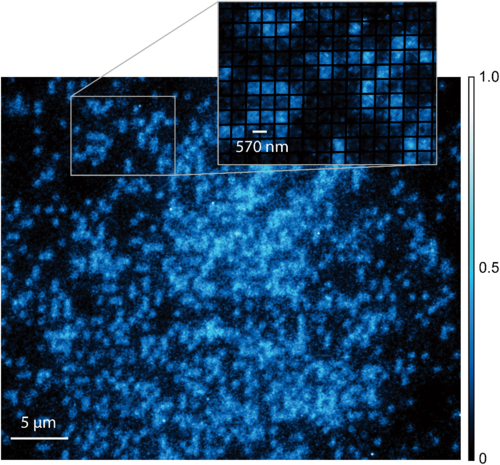}
\caption{Fluorescence image of $^{6}$Li atoms in a single layer of a cubic lattice, figure from Parsons et al. \cite{parsons2015site}.}\label{Fig_qu_gas_micro}
\end{figure}

We first discuss some physical properties of trapped gas of spin-less (or spin-polarised) non-interacting fermions. Even in the absence of interaction, the Pauli exclusion principle introduces strong quantum correlations in the system as two fermions cannot occupy the same quantum state. Of course, thermal fluctuations wash out these correlations at higher temperature and the quantum correlations are therefore more prominent at low temperature. The quantum statistics yields non-trivial spatial properties for the system of fermions. Cold atom experiments are the ideal platform to study these correlations as the recent experimental progress allows an unprecedented control over the parameters (see \cite{bloch2008many, giorgini2008theory} for reviews). In these systems, the interactions between atoms can be tuned via the Feshbach resonance \cite{regal2003tuning} and in particular, the non-interacting regime is reachable, allowing to probe and single-out the purely quantum effects emerging form the Pauli exclusion principle. These experiments are now used as quantum simulators for condensed-matter systems, where one can tune the parameters of the Hamiltonian in a controlled manner \cite{bloch2012quantum}. For Fermi gases the recent development of {\it Fermi gas microscope} \cite{haller2015single, cheuk2015quantum, parsons2015site} allows to probe the positions of single particles as observed in Fig. \ref{Fig_qu_gas_micro} (see \cite{kuhr2016quantum, ott2016single} for recent reviews). This type of imaging could also allow to test dynamical and non-equilibrium properties of quantum gases which has generated a lot of theoretical interest over the years \cite{calabrese2006time, calabrese2007quantum, vignolo2001one, krapivsky2019return, krapivsky2018quantum, eisler2013full, perfetto2017ballistic}. 

To conduct these types of experiments, the quantum gas needs to be confined by a {\it trapping potential}. This confining potential always creates a finite {\it edge} beyond which the density of particles in the gas is essentially zero. Although standard techniques such as local density approximation \cite{castin2006basic} were developed to describe the gas in the bulk of the density, they are not able to capture the spatial statistics close to the edge \cite{kohn1998edge, vignolo2000exact}.
In view of the aforementioned experimental set-ups, it becomes crucial to understand properly the physics of the cold Fermi gas close to the edge. This issue was tackled in a recent series of paper \cite{dean2015universal, dean2016noninteracting, dean2015finite} for non-interacting fermions, using the framework of determinantal point processes \cite{hough2006determinantal, johansson2005random}. In particular, a lot of progress ensued from a direct connection between the ground state of a one-dimensional system of fermions at zero temperature confined by a harmonic potentials and the Gaussian Unitary Ensemble (GUE). The ground state joint probability of the positions of this Fermi gas can be computed exactly
\be
|\Psi_0(x_1,\cdots,x_N)|^2=\frac{1}{Z_N(\alpha)} \prod_{i<j}|x_i-x_j|^2 \prod_{i=1}^N e^{-\alpha^2 x_i^2 }\;,\;\;\alpha=\sqrt{\frac{m\omega}{\hbar}}\;.\label{jPDF_OH1d_intro}
\ee
One recognises under the exact mapping $x_i=\sqrt{N}\lambda_i/\alpha$ the famous joint distribution of the eigenvalues of the GUE. This matrix ensemble is an invariant ensemble of random matrix theory, where Hermitian matrices are built with complex Gaussian independent entries. The result in Eq. \eqref{jPDF_OH1d_intro} only holds for a harmonic potential, but it is experimentally relevant to design different shapes for the confining potential \cite{mukherjee2017homogeneous, hueck2018two}. In fact, it was shown that the results obtained for the correlations at the edge of the density extend for any smoothly varying potential, e.g. $V(x)\sim |x|^p$ with $p>0$. These potentials yield a smooth variation of the density close to the edge. In random matrix theory, this universality class for the edge behaviour is usually referred to as {\it soft edge}. However, there are many examples in RMT, e.g. the Wishart or Jacobi ensembles, where the density has {\it hard edges} where it vanishes abruptly due to the presence of effective hard walls. It is therefore natural to ask if these RMT models have any counterparts in models of fermions, which present a similar hard edge behaviour and if there is a universality class associated to this different behaviour. This is one of the main purposes of this thesis.


%
%
%
%
%

\subsubsection{Random matrices}

A related problem that we consider is the statistical properties of random matrices \cite{mehta2004random, anderson2010introduction, tracy1993introduction, livan2018introduction, forrester2010log}.
%
%
%
Since its first appearance in the statistical literature \cite{wishart1928generalised}, RMT has been used extensively in mathematics, telecommunication, ecology or finance. In physics, it was first introduced by Wigner \cite{wigner1951statistical} to describe the level spacing between energies of nuclei but has been used since in statistical physics in the context of vicious (non-intersecting) random walkers \cite{Forrester1989,tracy2007nonintersecting, NAGAO200329, nadal2009nonintersecting, PhysRevLett.101.150601, FORRESTER2011500}, one component plasma \cite{chafai2014note, forrester1998exact, cunden2016large, cunden2017universality}, mesoscopic physics \cite{Jayannavar1989, beenakker1997random,Grabsch2017_2 , PhysRevLett.101.216809, PhysRevB.81.104202, PhysRevLett.82.4220, PhysRevLett.78.4737} or quantum chromodynamics \cite{wadia1980n,gross1993possible} (see \cite{majumdar2014top, biroli2007extreme} for reviews on the physical applications of RMT).  The models of random matrices offer a very useful setting to analyse extreme value statistics: while their eigenvalues are strongly correlated, their extreme value statistics can be obtained exactly and have been studied in detail \cite{tracy1994fredholm, tracy1994level, tracy1994level2, duenez2010lowest, dumitriu2008distributions, rider2003limit}. In this thesis, we explore the connection between fermions and random matrices \cite{dean2015finite,dean2015universal,dean2018wigner,eisler2013universality,grabsch2018fluctuations,le2016exact,marino2016number, eisler2013full, calabrese2015random} and mainly consider the Jacobi Unitary Ensemble (JUE) and complex Ginibre Ensemble. From these connections, we solve several open questions for the large deviations of the full counting statistics (FCS) and extreme value statistics in these ensembles.


\subsubsection{Random walks}

The last system that we will consider is random walks and its associated continuous counterpart Brownian motions. This seminal model of statistical mechanics was introduced for the first time more than a century ago by Louis Bachelier in the context of finance \cite{bachelier1900theorie} (see also \cite{Pearson1905} for the first appearance of the name {\it random walk} and \cite{RevModPhys.15.1} for a review). In this model, the positions taken by the random walker form a strongly correlated set of random variables and is therefore a useful laboratory to test the effects of strong correlations.
In particular, many results were obtained for the statistics of the global maximum \cite{feller1968introduction, Mounaix_2018, comtet2005precise, majumdar2010universal}. The order statistics i.e. the statistics of the ordered maxima (second, third, etc) were also considered in detail both for a regular random walk \cite{wendel1960order, port1962elementary, feller1968introduction, schehr2012universal, schehr2014exact, dassios1996sample, 10.2307/2959757} and for branching Brownian motions \cite{Brunet_2009, Derrida2011, ramola, ramola2}. Order statistics is part of the general {\it fluctuation theory} which has been extensively studied in the mathematics community \cite{wendel1960order, port1962elementary, feller1968introduction,pitman2018guide,revuz2013continuous }. A large literature has also emerged on the related topic of records for these random walks \cite{majumdar_ziff, PhysRevE.86.011119, Sabhapandit_2011, majumdar2012record, schehr2014exact, berkowitz_records_noise, wergen_borgner_krug, Godr_che_2017, Godr_che_2014}. In both cases, the problem is simplified in the large $n$ limit and for finite variance jump distribution by using the convergence of this process towards Brownian motion, for which a number of results have been obtained \cite{yor1995distribution, perret2013near, ramola, dassios1995distribution}. There exists however fewer results for the gap statistics of random walks \cite{Brunet_2009, Derrida2011,  ramola, battilana2017gap,schehr2012universal}, which is inherently linked to the {\it discrete} nature of the process. These problems cannot be solved using the convergence to Brownian motion. A goal of this thesis is therefore to obtain new results for this interesting and versatile problem.


\section*{Overview of the thesis and main results}

We present a quick overview of the thesis and summary of the main results. These main results are framed ($\boxed{x}$) in the text and the unpublished results are doubly framed ($\boxed{\frame{\boxed{x}}}$).

While some of the considered models are quite specific, many of their properties 
exhibit universality, as in many instances in statistical mechanics, and hold in a more general context. 

\subsection*{First part: Spatial description of non-interacting fermions}

Part \ref{Part:Fer} of this thesis is devoted to the study of non-interacting fermions and their connections to eigenvalues of random matrices.

In chapter \ref{chap:fermions_intro}, we introduce the framework to describe the spatial properties of fermions and review the results for {\it smooth} confining potentials.

In chapter \ref{ch: ferm_hard_edge}, we extend the description of the edge statistics of non-interacting fermions to {\it hard edge} potentials. We show an exact mapping between the ground state of fermions trapped in a one-dimensional hard box potential and the Jacobi Unitary Ensemble of random matrices. We obtain exact results for the correlation kernel associated to this determinantal point process and show that these results extend to a new class of hard edge potentials. We extend these results in two directions: in higher dimension and at finite temperature. We apply these results to compute the fluctuations of the position of the fermion the farthest away from the centre of the trapping potential. In particular, we obtain the emergence of an {\it intermediate deviation regime} connecting the typical fluctuations to the large deviations, which does not appear for standard invariant ensembles as the GUE.

The study of these non-interacting fermions in hard edges led to the publication of two articles:\\

\ref{Art:fermions_lett} {\it Statistics of fermions in a d-dimensional box near a hard wall}

 B. Lacroix-A-Chez-Toine, P. Le Doussal, S.~N. Majumdar, G. Schehr,

 Europhys. Lett. {\bf 120} (1), 10006 (2018).\\

\ref{Art:ferm_long} {\it Non-interacting fermions in hard-edge potentials},

 B. Lacroix-A-Chez-Toine, P. Le Doussal, S.~N. Majumdar, G. Schehr,

 J. Stat. Mech {\bf 12}, 123103 (2018).\\


In chapter \ref{ch: rot_trap}, we unveil an exact mapping between the ground state of a model of non-interacting fermions in rotation and the complex Ginibre ensemble. We compute {\it exactly} for this system the full counting statistics and entanglement entropy for any finite number $N$ of fermions. This problem is mapped to a specific case of the two-dimensional one component plasma, where charged particles (interacting via the long-ranged $2d$ Coulomb logarithmic repulsion) are confined by a harmonic potential. Extending to more generic potentials, we show the universality of the results for the full counting statistics in the plasma model. Revealing the emergence of {\it intermediate deviation regimes} for (i) the fluctuations of the particle the farthest away from the centre of the trap and (ii) the full counting statistics, we solve two puzzles of matching between the typical fluctuations and the large deviations.  

The study of these non-interacting fermions in rotating traps, their connection to the complex Ginibre ensemble and their extension to general one component plasma led to the publication of two articles:\\

\ref{Art:rot} {\it Rotating trapped fermions in two dimensions and the complex Ginibre ensemble: Exact results for the entanglement entropy and number variance},\\
B. Lacroix-A-Chez-Toine, S.~N. Majumdar, G. Schehr,\\
Phys. Rev. A {\bf 99} (2), 021602 (2019).\\

\ref{Art:r_max} {\it Extremes of $2d$ Coulomb gas: universal intermediate deviation regime}\\
B. Lacroix-A-Chez-Toine, A. Grabsch, S.~N. Majumdar, G. Schehr,\\
J. Stat. Mech {\bf 1}, 013203 (2018).\\

We also recently submitted for peer-review another article on this subject:\\

 \ref{Art:FCSGin} {\it Intermediate deviation regime for the full eigenvalue statistics
in the complex Ginibre ensemble}\\
B. Lacroix-A-Chez-Toine, J.~A.  Monroy Garz\'on, C.~S. Hidalgo Calva, I. P\'erez Castillo, A.  Kundu, S.~N. Majumdar, G. Schehr\\
arXiv preprint, arXiv: \textbf{1904.01813}, (2019).\\


\subsection*{Second part: Statistics of the gaps of random walks}

Part \ref{Part:gaps} of this thesis is devoted to the study of extreme value, order and gap statistics of random walks.

In chapter \ref{intro_iid}, we review a few results for the order and gap statistics of i.i.d. random variables.

In chapter \ref{ch:gaps_intro}, we review the extreme value statistics of Brownian motion and random walks.

In chapter \ref{ch:maxk}, we consider the order statistics of random walks. We obtain an exact formula for the time to reach the $k^{\rm th}$ maximum of the walk. We review some properties of the distribution of the value taken by the $k^{\rm th}$ maximum of a random walk with finite variance jump distribution. We introduce and compute the quenched and annealed density of maxima both for finite variance and L\'evy flights. These results are still unpublished.
 
In chapter \ref{ch:gapk}, we consider the gap statistics of random walks. We show how to obtain exactly the probability distribution function of the gaps for the random walk with Laplace distribution of jumps. We argue from numerical simulations that this result is universal in the large $n$ limit for any jump distribution with finite variance.	

We recently submitted for peer-review an article on the gap statistics of random walks:\\

\ref{Art:gap} {\it Gap statistics close to the quantile of a random walk},\\
B. Lacroix-A-Chez-Toine, S.~N. Majumdar, G. Schehr,\\
arXiv preprint,  arXiv: \textbf{1812.08543}, (2018).\\

Note finally that we recently submitted another article for peer-review, which is largely disconnected to the extreme value statistics:\\

\ref{Art:BMcoi} {\it Distribution of Brownian coincidences}\\
A. Krajenbrink, B. Lacroix-A-Chez-Toine, P. Le Doussal,\\
 arXiv preprint,  arXiv: \textbf{1903.06511}, (2019).\\

In this article, we compute the distribution of coincidence time ${\cal T}_N(t)$, i.e. the total local time of all pairwise coincidences, of $N$ independent Brownian walkers. We show that this problem is related to (i) the Lieb-Liniger problem of $N$ hard-core interacting bosons \cite{PhysRev.130.1605} and (ii) the moments of the canonical partition function of directed polymers (in direct relation to the Kardar-Parisi-Zhang equation \cite{kardar1986dynamic}). We obtain the exact distribution of ${\cal T}_N(t)$ for $N=2,3$ for several initial and final conditions and the asymptotic behaviours of the distribution for any values of $N$.

\label{intro}

\counterwithin{table}{chapter}



\numberwithin{equation}{chapter}


\part{Spatial description of non-interacting fermions}
\label{Part:Fer}

\chapter{Introduction to non-interacting fermions}\label{chap:fermions_intro}

In this chapter, we consider the system formed by $N$ non-interacting, spin-less (or spin-polarised), identical fermions of mass $m$. As discussed in the introduction, the current experiments for cold Fermi gases allow to probe the positions of individual particles in the Fermi gas \cite{cheuk2015quantum,haller2015single,parsons2015site}. A prior requirement in order to conduct sophisticated measurements is to have a precise spatial description of the gas. We will now develop the main mathematical framework that will be useful in the following. As in experiments, the number of fermions is usually $N\sim 10^{2-3}$, we will mainly focus on the limit of large $N$.
The fermions are embedded in a $d$-dimensional space and we denote by $({\bf x}_i,{\bf p}_i)$ the positions and impulsions of the fermions. The Hamiltonian of the system reads
\be
\hat{\cal H}_N=\sum_{i=1}^N \hat{H}_i=\sum_{i=1}^N  H(\hat{\bf x}_i,\hat{\bf p}_i)\;\;{\rm with}\;\; H(\hat{\bf x},\hat{\bf p})=\frac{{\hat {\bf p}}^2}{2m}+V(\hat {\bf x})\;.
\ee
Note that we will pay particular attention in this chapter to the harmonic trapping potential $V({\bf x})=\frac{1}{2}m \omega^2 {\bf x}^2$. 
The single particle eigenstates $|{\bf k}\rangle$ satisfy
\be
\hat H|{\bf k}\rangle=\epsilon_{\bf k}|{\bf k}\rangle\;,\;\;{\rm and}\;\;\phi_{\bf k}({\bf x})=\langle{\bf x}|{\bf k}\rangle,
\ee
where $\epsilon_{\bf k}$ are their associated energy and $\phi_{\bf k}({\bf x})$ their associated wave-function in position space. These wave-functions form an orthonormal basis of the Hilbert space
\be\label{ortho}
\int d^d {\bf x}\, \phi_{\bf k}({\bf x})\overline{\phi}_{{\bf k}'}({\bf x})=\delta_{{\bf k},{\bf k}'}\;,
\ee
where we use the notation~~$\bar{}$~~for complex conjugation. For a single fermion in a given state $|{\bf k}\rangle$, the position ${\bf x}$ is a random variable and its quantum fluctuations are governed by the probability distribution function (PDF) $|\phi_{\bf k}({\bf x})|^2$.
The many-body state has energy $E_N=\sum_{i=1}^N \epsilon_{{\bf k}_i}$. The Pauli exclusion principle imposes that all occupied states must be different. At zero temperature the energy is minimum, which imposes to occupy only the $N$ lowest energy states $\epsilon_0\leq \epsilon_1\leq \cdots\leq \epsilon_{N-1}$ while the higher energy states remain empty. The energy $\epsilon_{N-1}=\epsilon_F=\mu$ of the highest energy occupied state, also called the Fermi energy, coincides at zero temperature with the chemical potential, i.e. the energy added to the system by adding a fermion. 
At finite temperature $T=1/(k_B \beta)$, the excited states are occupied with a non-zero probability, inducing thermal fluctuations on top of the quantum fluctuations. The relevant temperature scale is given for this many-body problem by the Fermi temperature $T_F=\epsilon_F/k_B$.
For these systems of non-interacting fermions, the temperature acts as a control parameter for the correlations, which emerge only from the Pauli exclusion principle. 

In the regime of high temperature $\beta \epsilon_F\ll 1$ many excited states are occupied, such that the occupation number $n_{\bf k}$ of each state $|\bf{k}\rangle$ is small. 
%
%
%
%
%
%
%
The Pauli exclusion principle becomes irrelevant in this limit, and the classical and quantum description of the system do coincide. The positions of the fermions are then described in the classical framework of statistical mechanics as i.i.d. random variables with individual PDF given by the Gibbs weight
\be
p_{\beta}({\bf x})=\frac{e^{-\beta V({\bf x})}}{z(\beta)}\;.
\ee 
The extreme value statistics associated to the positions of the fermions can be obtained using the standard theory of i.i.d. random variables and in particular, in the large $N$ limit, they fall into one of the three universality classes (Gumbel, Fr\'echet or Weibull).

Furthermore, in the low temperature regime $\beta \epsilon_F\gg 1$, which we will mainly focus on, only a few states are excited. The quantum fluctuations dominate in this regime and the Pauli exclusion principle must be taken into account. The positions of the fermions are strongly correlated and one expects that the associated extreme value statistics will therefore be non-trivial. The systems of fermions therefore offer an interesting laboratory to test the effects of correlations on extreme value statistics.

This chapter is organised as follows: in Section \ref{sec_1d_OH_GUE}, we explain the framework of determinantal point process that is associated to models of fermions on the case of a one-dimensional system at zero temperature. We show how it allows to retrieve the results of local density approximation in the bulk of the density and to describe precisely the statistics close to the edge. In Section \ref{sec_d_soft} we extend this description to higher dimension still at zero temperature. Finally in Section \ref{finite_T_smooth}, we review the extension of this framework to models of fermions at finite temperature.

We will now explain the framework of determinantal point process that is associated to non-interacting fermions starting by the case of dimension one and zero temperature, which is already non-trivial.



\section{One-dimensional system at zero temperature} \label{sec_1d_OH_GUE}

\begin{figure}
\centering
\includegraphics[width=0.6\textwidth]{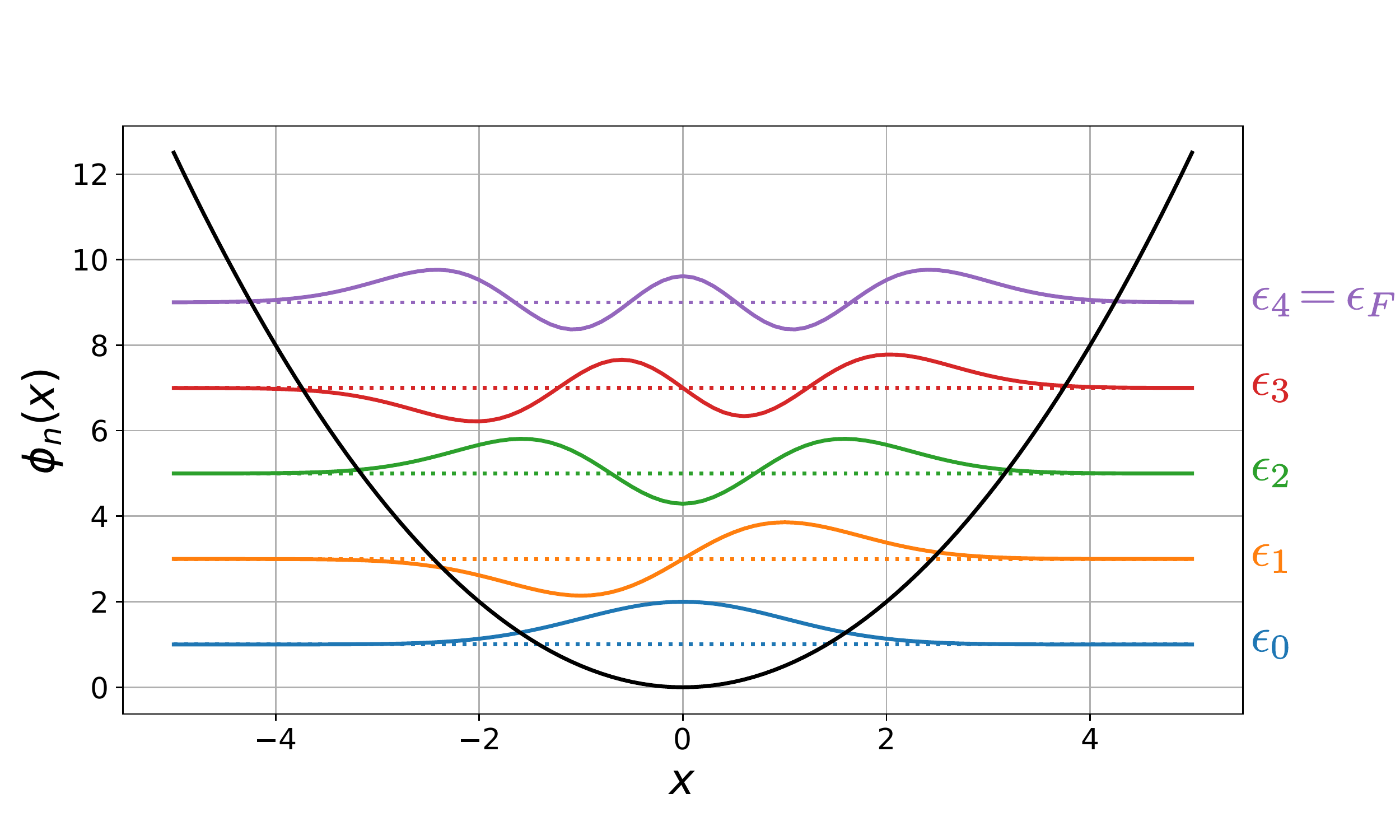}
\caption{Single-particle energies and wave-functions associated to a system of $N=5$ fermions at zero temperature in a harmonic trapping potential $V(x)=\frac{1}{2}m\omega^2 x^2$.}\label{Fig_phi_epsilon}
\end{figure}

Let us first consider a one-dimensional system where the effects of quantum correlations are the strongest. In this case, the quantum numbers are indexed by a single quantum number $n$. At zero temperature, only the $N$ lowest energy states are occupied with $\epsilon_0< \epsilon_1< \cdots< \epsilon_{N-1}=\epsilon_F$  (see Fig. \ref{Fig_phi_epsilon} for an example of filling). There are no degeneracies in this one-dimensional case. The many body wave function is given by the Slater determinant
\be
\Psi_0(x_1,\cdots,x_N)=\frac{1}{\sqrt{N!}}\det_{1\leq i,j\leq N} \phi_{{j-1}}({ x}_i)\;.
\ee

The joint quantum probability of the positions $x_1,\cdots, x_N$ is obtained by computing the modulus squared of this many body wave function
\be
\left|\Psi_0({ x}_1,\cdots,{x}_N)\right|^2=\frac{1}{N!}\det_{1\leq i,j\leq N} \phi_{{j-1}}({ x}_i)\det_{1\leq n,m\leq N} \overline{\phi}_{{n-1}}({ x}_m)\;.\label{psi_phi}
\ee
This joint probability can be re-written using the identity $\det(A)\det(B)=\det(AB)$ as
\be
\left|\Psi_0({ x}_1,\cdots,{x}_N)\right|^2=\frac{1}{N!}\det_{1\leq i,j\leq N}K_N(x_i,x_j)\;,
\ee
where the function $K_N(x,y)$ is called the correlation kernel and reads
\be
K_N(x,y)=\sum_{k=0}^{N-1} \overline{\phi}_k(x)\phi_k(y)\;.\label{K_N_gen}
\ee
It is straightforward to prove the reproducibility of this kernel, i.e.
\be
\int_{-\infty}^{\infty} dy\, K_N(x,y)K_N(y,z)=\sum_{k,l=0}^{N-1} \overline{\phi}_k(x){\phi}_l(z)\int_{-\infty}^{\infty}dy\, \overline{\phi}_k(y){\phi}_l(y)=K_N(x,z)\;,
\ee
where the orthonormality of the wave functions in Eq. \eqref{ortho} was used in the last step. 
This property implies that the positions of the fermions form a {\it determinantal point process} with kernel $K_N(x,y)$ \cite{johansson2005random, hough2006determinantal} and are therefore strongly correlated (the probability that any two positions are identical is zero).
In particular, the $p$ points correlation function can be obtained using the determinantal framework for any $1\leq p\leq N$ \cite{johansson2005random ,akemann2011oxford}
\begin{align}
R_p({ x}_1,\cdots,{ x}_p)&=\frac{N!}{(N-p)!}\sum_{i_1\neq i_2\neq \cdots\neq i_p}^N \moy{\prod_{l=1}^p\delta({ x}_k-{ x}_{i_k})}\nn\\
&=\frac{N!}{(N-p)!}\int dx_{p+1}\cdots\int dx_{N} |\Psi_0(x_1,\cdots,x_N)|^2\nn\\
&=\det_{1\leq i,j\leq p}\left(K_N({ x}_i,{ x}_j)\right)\;.\label{fermions_p_point}
\end{align}
This formula can also be obtained by writing the Hamiltonian in the second quantification as a quadratic operator of creation and annihilation operators and applying Wick's theorem \cite{mahan2013many}. In particular, this formula in Eq. \eqref{fermions_p_point} allows to recover the average density (normalised to $N$) of the Fermi gas
\be
R_1({\bf x})=N\rho_N({\bf x})=K_N({\bf x},{\bf x})=\sum_{n=0}^{N-1}\left|\phi_{{\bf k}_n}({\bf x})\right|^2\;.
\ee
As for the two point correlation function it is obtained explicitly as
\begin{align}
R_2({\bf x},{\bf y})&=\begin{vmatrix} K_N({\bf x},{\bf x}) & K_N({\bf x},{\bf y}) \\ K_N({\bf y},{\bf x}) & K_N({\bf y},{\bf y}) \end{vmatrix}\nn\\
&=\sum_{n_1,n_2=0}^{N-1}\left(\left|\phi_{{\bf k}_{n_1}}({\bf x})\right|^2 \left|\phi_{{\bf k}_{n_2}}({\bf y})\right|^2-\overline{\phi_{{\bf k}_{n_1}}}({\bf x})\phi_{{\bf k}_{n_1}}({\bf y})\overline{\phi_{{\bf k}_{n_2}}}({\bf y})\phi_{{\bf k}_{n_2}}({\bf x})\right)
\end{align}

The connection with fermions was actually part of the original motivations to consider determinantal point processes \cite{macchi1975coincidence}. However, only recently did the physics community really use this connection 
\cite{eisler2013universality, marino2014phase, dean2015finite, dean2015universal, calabrese2015random, dean2016noninteracting}. 
Note that this framework is not restricted to the position representation of the Hilbert space and one could work instead in the momentum representation. The $p$ point momenta correlation functions would be similar to Eq. \eqref{fermions_p_point} upon changing ${ x}\to { p}$ and
\be
\phi_{{ k}}({ x})\to \psi_{{ k}}({ p})=\int \frac{d{ x}}{\hbar}\,e^{ -\frac{\I p x}{\hbar}}\phi_{{ k}}({ x})\;.
\ee
We finally mention that for the harmonic oscillator, i.e. $V(x)=\frac{1}{2}m\omega^2 x^2$, $x$ and $p$ play symmetric roles. This is not the case in general and some studies have recently considered in details the determinantal point process associated to the momenta \cite{le2018multicritical}.

We will now see how this determinantal framework allows us to compute the extreme value statistics of the process.

\subsubsection{Statistics of the maximum}

Although the positions $x_i$'s of fermions form a determinantal point process -- and are therefore strongly correlated -- this system constitutes one of the rare examples where the extreme value statistics can be obtained explicitly. 
We consider the Cumulative Distribution Function (CDF) $\Prob\left[x_{\max}\leq x\right]$ of the position of the rightmost fermion $\displaystyle x_{\max}=\max_{1\leq i\leq N}x_i$,
%
%
%
%
which corresponds to the probability that there are no fermions in the interval $[x,\infty)$. Using the determinantal structure, this probability can be expressed as a Fredholm determinant \cite{johansson2005random, akemann2011oxford} of the kernel $K_N$, (see also \cite{gohberg2012traces} for precisions on Fredholm determinants)
\be\label{x_max_Det}
\Prob\left[x_{\max}\leq x\right]=\Det\left[\mathbb{I}-P_{[x,\infty)}K_N P_{[x,\infty)}\right]\;.
\ee

An alternative representation of this CDF can be obtained by integrating the joint PDF of the positions in Eq. \eqref{psi_phi} for all the variables ${ x}_i$'s over the interval $(-\infty,x]$. This yields
\be
\Prob\left[x_{\max}\leq x\right]=\frac{1}{N!}\int_{-\infty}^x d { x}_1\cdots\int_{-\infty}^x d { x}_N\, \det_{1\leq i,j\leq N} \phi_{{j-1}}({x}_i)\det_{1\leq n,m\leq N} \overline{\phi}_{{n-1}}({ x}_m)\;.
\ee 
The $N$-fold integral of product of two $N\times N$ determinants can be simplified using the Cauchy-Binet-Andr{\'e}ief formula \cite{andreief1883note}
\begin{align}
&\frac{1}{N!}\int d^d {\bf x}_1\cdots \int d^d {\bf x}_N \prod_{k=1}^N h({\bf x}_k)\det_{1\leq i,j\leq N} \chi_{{j}}({\bf x}_i)\det_{1\leq n,m\leq N}{\psi}_{{n}}({\bf  x}_m)\nn\\
&=\det_{1\leq i,j\leq x}\left(\int d^d{\bf x}\, h({\bf x})\chi_{i}({\bf x})\psi_j({\bf x})\right)\;,\label{Cauchy_Binet}
\end{align}
valid for general dimension $d$. Using this formula for $d=1$, we obtain an alternative representation of the CDF
\be\label{x_max_det}
\Prob\left[x_{\max}\leq x\right]=\det_{0\leq i,j\leq N-1}\left(\int_{-\infty}^x d y\, \phi_{{i}}({y})\overline{\phi}_{{j}}(y) \right)=\det_{0\leq i,j\leq N-1}\left(\delta_{i,j}-\int_{x}^{\infty} d y\, \phi_{{i}}({y})\overline{\phi}_{{j}}(y) \right)\;.
\ee 
Note that the two representations in Eq. \eqref{x_max_Det} and \eqref{x_max_det} are very similar, though different. In particular, they are expressed as a product over the eigenvalues of (i) a finite $N\times N$ matrix for \eqref{x_max_det} or (ii) an integral operator for \eqref{x_max_Det}. One can actually show that the non-zero eigenvalues of the projected kernel $P_I K_N P_I$ do coincide with the eigenvalues of the overlap matrix  $\mathbb{A}_{ij}=\int_I d^d {\bf x} \phi_{{\bf k}_i}({\bf x })\overline{\phi}_{{\bf k}_j}({\bf x})$ (see \cite{calabrese2011entanglement, calabrese2011entanglement2} for the case $d=1$). 

The behaviour of the CDF $\Prob\left[x_{\max}\leq x\right]$ is obtained in the large $N$ limit solely from the asymptotic behaviour of the correlation kernel using Eq. \eqref{x_max_Det} . 
%
%
In fact, the correlation kernel controls all the quantum fluctuations of the Fermi gas and we now analyse in details its large $N$ behaviour. 
%
%
%
%
%
%
%
%
%
%
%
%
%

\subsection{Local density approximation}\label{sec_LDA}

\begin{figure}
\centering
\includegraphics[width=0.6\textwidth]{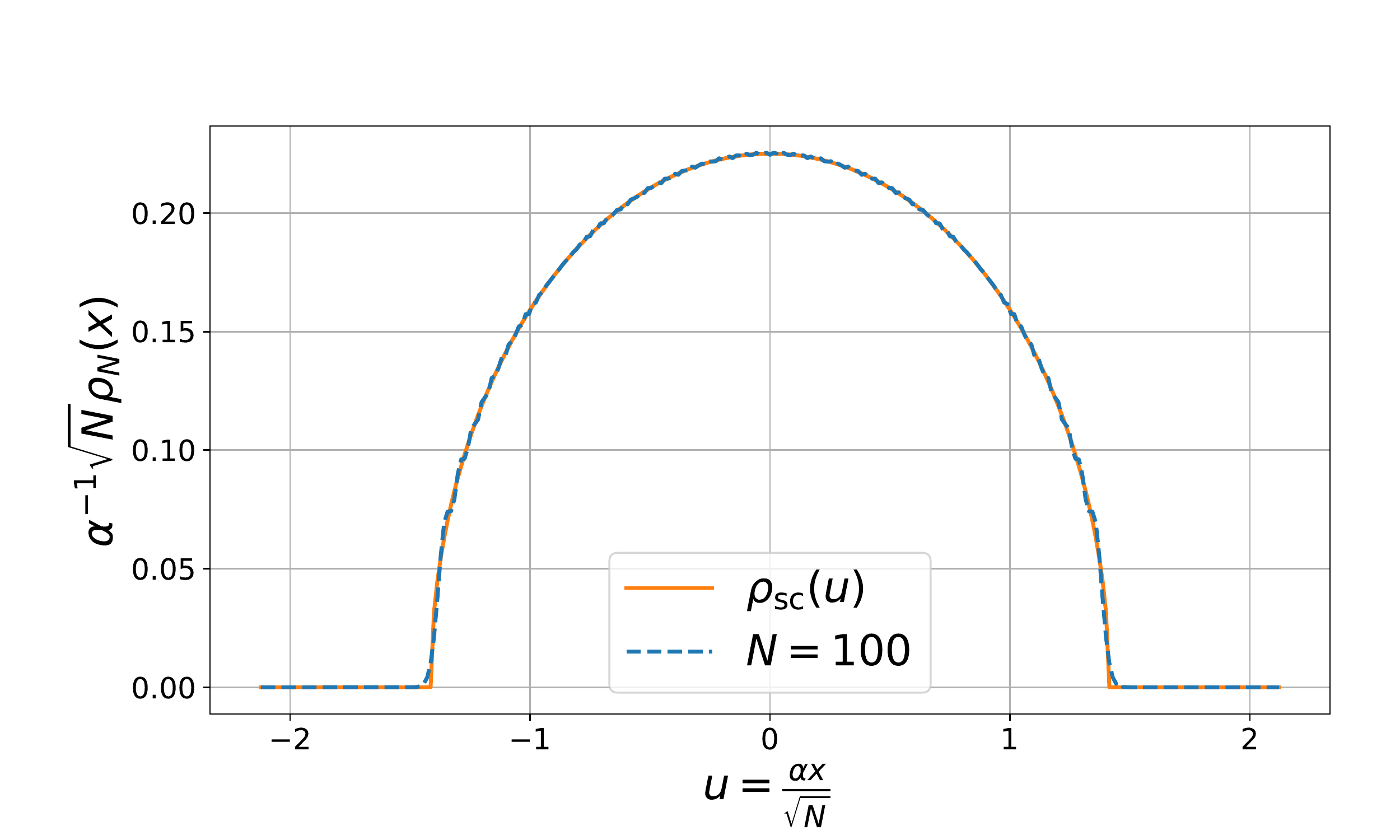}
\caption{Comparison between the bulk density obtained from LDA (orange) and the exact density profile for $N=100$ fermions in a one-dimensional harmonic potential $V(x)=\frac{1}{2}m\omega^2 x^2$ at zero temperature plotted as a function of the rescaled position $u=\alpha x/\sqrt{N}$ where $\alpha=\sqrt{m\omega/\hbar}$.}\label{Fig_bulk_LDA}
\end{figure}

In the bulk of the system, and in the large $N$ limit, there is a semi-classical approximation allowing us to obtain the correlation kernel: the local density (or Thomas-Fermi) approximation (LDA). There are several ways to derive this approximation \cite{castin2006basic}. We present here an elegant and useful manner by introducing the Wigner function $W_N({ x},{ p})$ of the $N$ fermions system, which defines a pseudo-probability (this function can become negative) in the phase space $({ x},{ p})$ \cite{wigner1997quantum, case2008wigner}. It is defined as
\be
W_N({ x},{ p})=\frac{N}{2\pi\hbar}\int dy dx_2\cdots dx_N e^{\frac{\I p y}{\hbar}}\overline{\Psi_0}\left(x+\frac{y}{2},x_2,\cdots,x_N\right)\Psi_0\left(x-\frac{y}{2},x_2,\cdots,x_N\right)\;.
\ee
In particular, the marginals of $W_N({ x},{ p})$ are the mean densities $\rho_N({ x})$ and $\hat{\rho}_N({ p})$ respectively in position and momentum space 
\begin{align}
\rho_N({x})=\frac{1}{N}\sum_{i=1}^N \moy{\delta({ x}-{ x}_i)}=\frac{1}{N}\int d{ p}\, W_N({ x},{ p})\;,\label{dens_pos_Wigner}\\
\hat{\rho}_N({ p})=\frac{1}{N}\sum_{i=1}^N \moy{\delta({ p}-{ p}_i)}=\frac{1}{N}\int d { x}\, W_N({ x},{ p})\;.
\end{align}
Note that we defined the densities such that $\int dx\,\rho_N(x)=1$. Interestingly, this Wigner function can be expressed as the Weyl transform of the correlation kernel \cite{ dean2018wigner}
\be
W_N({ x},{ p})=\int \frac{d{ y}}{2\pi \hbar} e^{-\frac{\I p y}{\hbar}}K_N\left({ x}+\frac{{ y}}{2},{ x}-\frac{{ y}}{2}\right)\;.
\ee 
In the limit $N\to \infty$, one expects to recover the classical limit $\hbar\to 0$, where the Wigner function becomes uniform over a finite region of space \cite{castin2006basic, dean2018wigner, bartel1985extended}
\be
W_N({ x},{ p})\approx \frac{\Theta[\mu-H({ x},{ p})]}{2\pi \hbar}=\frac{1}{2\pi \hbar}\Theta\left[\mu-\frac{{ p}^2}{2m}-V({ x})\right]\;,\label{Wigner_1d_t0}
\ee
where $\mu\approx\epsilon_{N}$ is the Fermi energy of the system and $\Theta(x)$ is the Heaviside step-function. The average density in position space $\rho_N(x)$ is obtained by integrating over the momentum $p$ as
\be\label{LDA_1d}
\rho_N(x)=\int \frac{dp}{2\pi \hbar}\Theta\left[\mu-\frac{{ p}^2}{2m}-V({ x})\right]=\frac{1}{N\pi \hbar}\sqrt{2m\left[\mu-V(x)\right]}\;.
\ee
For a smooth potential (e.g. of the type $V(x)\sim |x|^p$), one therefore expects this density to have finite edges $r_{\rm e}$ such that $V(r_{\rm e})=\mu$ where it vanishes as a square-root $\rho_N(x)\sim \sqrt{x-r_{\rm e}}$. Furthermore, taking the inverse Weyl transform of Eq. \eqref{Wigner_1d_t0} we obtain the correlation kernel in the local density approximation
\be
K_N({ x},{ y})\approx \int \frac{d { p}}{2\pi \hbar}e^{-\frac{\I p (x-y)}{\hbar}}\Theta\left[\mu-\frac{{ p}^2}{2m}-V\left(u\right)\right]
=\frac{1}{\ell_N(u)} K_{\rm b}^{1}\left(\frac{x-y}{\ell_N(u)}\right)\;,\label{k_scal_lda_bulk}
\ee
where $u=(x+y)/2$ is the centre of mass of $x$ and $y$, the scaling function $K_{\rm b}^{1}(r)$ reads
\be\label{LDA_1d_ker}
K_{\rm b}^{1}(r)=K_{\sin}(r)=\frac{\sin(r)}{\pi r}\;,\;\;{\rm and}\;\;\ell_N(u)=[\pi N \rho_N(u)]^{-1}\;.
\ee
Note that this approximation is called the ``local density approximation'' (LDA) because one assumes that on the small scale $\ell_N({u})=\left[N\pi\rho_N(u)\right]^{-1}$, defined from the point-wise density, the Fermi gas can be considered as a free, translation invariant gas (the correlation kernel rescaled by $\ell_N(u)$ in Eq. \eqref{k_scal_lda_bulk} is explicitly translationally invariant). 
This approximation is quite accurate to describe the bulk properties of the Fermi gas in the large $N$ limit \cite{castin2006basic, butts1997trapped} (see also Fig. \ref{Fig_bulk_LDA}). 
We already mentioned that the density in Eq. \eqref{LDA_1d} has finite edges at positions $r_{\rm e}$ such that $V(r_{\rm e})=\mu$ where the scale $\ell_N({u})$ diverges and the LDA description thus breaks down near these points $r_{\rm e}$. One therefore needs new tools to describe the statistics close to these edges \cite{kohn1998edge, vignolo2000exact}. 

We will now review the case of the harmonically confined Fermi gas, where this problem can be solved exactly using a mapping to random matrix theory.

\subsection{One-dimensional harmonic potential and the Gaussian Unitary Ensemble}

Let us consider a specific example to understand how the tools of random matrix theory (RMT) allow us to describe the edge statistics.
The single particle Hamiltonian associated to the harmonic potential reads
\be
\hat{H}=\frac{{\hat p}^2}{2m}+\frac{1}{2}m\omega^2 {\hat x}^2\;,
\ee
with single particle energies and wave functions labelled by an integer $n\in \mathbb{N}$,
\be
\epsilon_n=\hbar \omega\left(n+\frac{1}{2}\right)\;,\;\;{\rm and}\;\;\phi_n(x)=\frac{\alpha}{\pi^{1/4} 2^{n/2}\sqrt{n!}}e^{-\frac{\alpha^2 x^2}{2}}H_n(\alpha x)\;,\;\;\alpha=\sqrt{\frac{m\omega}{\hbar}}\;.
\ee
In this expression $H_n(x)=e^{x^2}(-\partial_x)^n e^{-x^2}$ is the Hermite polynomial of degree $n$. The Fermi energy is in this case $\mu=\hbar\omega(N-1/2)$.
The joint quantum probability of the $N$ positions of the fermions is given by the modulus squared of the Slater determinant
\be
|\Psi_0(x_1,\cdots,x_N)|^2=\frac{1}{N!}\det_{1\leq i,j\leq N} \phi_{{j-1}}({ x}_i)\det_{1\leq n,m\leq N} \overline{\phi}_{{n-1}}({ x}_m)\;,
\ee 
and can be computed exactly for this system. Indeed, we first remove column by column the Gaussian term from the determinants, yielding a product of Gaussian term over all the positions. 
Next, we use the Vandermonde identity allowing to obtain the determinant of any set of polynomials $\{p_0,\cdots p_{N-1}\}$ of ascending order as
\be\label{VdM}
\Delta(\Lambda)=\prod_{i<j}(\lambda_i-\lambda_j)=A_N(p)\det_{1\leq i,j\leq N} p_{j-1}(\lambda_i)\;,
\ee
where the weight $A_N(p)$ depends on the set of polynomials. This finally yields
\be
|\Psi_0(x_1,\cdots,x_N)|^2=\frac{1}{Z_N(\alpha)} \prod_{i<j}|x_i-x_j|^2 \prod_{i=1}^N e^{-\alpha^2 x_i^2 }\;,\label{jPDF_OH1d}
\ee 
where $Z_N(\alpha)$ is a constant ensuring the normalisation of the joint probability. Remarkably, this joint PDF appears in a very different context as the joint distribution of the eigenvalues of the Gaussian Unitary Ensemble (GUE), which we briefly review now.

\subsubsection{Gaussian Unitary Ensemble}

In this ensemble, the matrices have Hermitian symmetry and are filled with independent complex Gaussian entries
\begin{align}
m_{ii}&\sim {\cal N}\left(0,\frac{1}{\sqrt{2N}}\right)\nn\\
m_{ij}&\sim {\cal N}\left(0,\frac{1}{2\sqrt{N}}\right)+\I\; {\cal N}\left(0,\frac{1}{2\sqrt{N}}\right)\;,\;\;m_{ji}=\overline{m_{ij}}\;,\;\;i<j\;.
\end{align}
The probability weight $P(M)$ associated to a matrix $M$ in this ensemble therefore reads
\be
P(M)=\frac{1}{z_N}\prod_{i< j}e^{-2N m_{ij}m_{ji}}\prod_{i=1}^N e^{-N m_{ii}^2}=\frac{1}{z_N}e^{-N\left[2\sum_{i<j}m_{ij}m_{ji}+\sum_{i=1}^N m_{ii}\right]}=\frac{e^{-\frac{N}{2}\Tr[M^2]}}{z_N}\;.
\ee
Note that the name of this ensemble comes from the invariance of this measure by unitary transformation, which appears clearly in this expression. The joint probability of the eigenvalues is obtained after integration over the eigenvectors degrees of freedom (see Appendix \ref{RMT_app} for details) and reads
\be
P_{\rm joint}^{\rm GUE}(\lambda_1,\cdots,\lambda_N)=\frac{1}{Z_N^{\rm GUE}} \prod_{i<j}|\lambda_i-\lambda_j|^2 \prod_{i=1}^N e^{-N \lambda_i^2 }\;.\label{P_joint_GUE}
\ee 
Comparing Eq. \eqref{jPDF_OH1d} and Eq. \eqref{P_joint_GUE}, one realises that there is a one to one mapping between the joint PDF of the rescaled positions $\alpha x_i$'s of the fermions in this harmonic potential and the rescaled eigenvalues $\sqrt{N}\lambda_i$'s of a matrix belonging to the Gaussian Unitary Ensemble (GUE). Note that an alternative way to obtain this exact mapping is to compute the correlation kernel $K_N(x,y)$, which is identical for both of these determinantal point processes. The average density of eigenvalues in this ensemble converges in the large $N$ limit to the Wigner semi-circle \cite{wigner1993characteristic, wigner1958distribution} (see Fig. \ref{Fig_bulk_LDA})
\be\label{W_sc}
\rho_{\rm sc}(\lambda)=\frac{1}{\pi}\sqrt{2-\lambda^2}\;,\;\;-\sqrt{2}\leq \lambda\leq \sqrt{2}\;.
\ee   
Using the exact mapping, one can easily check that the density for the fermions coincides with the prediction from LDA in Eq. \eqref{LDA_1d}
\be
\rho_N(x)\approx \frac{\alpha}{\sqrt{N}}\rho_{\rm sc}\left(\frac{\alpha x}{\sqrt{N}}\right)\;,
\ee
where we used $\alpha=\sqrt{m\omega/\hbar}$. This density vanishes at the symmetric edges $\pm r_{\rm e}=\pm \alpha^{-1}\sqrt{2N}$. We now analyse the large $N$ limit for the correlation kernel associated to this determinantal point process.

\subsubsection{Bulk limit: Sine kernel}

\begin{figure}
\centering
\includegraphics[width=0.6\textwidth]{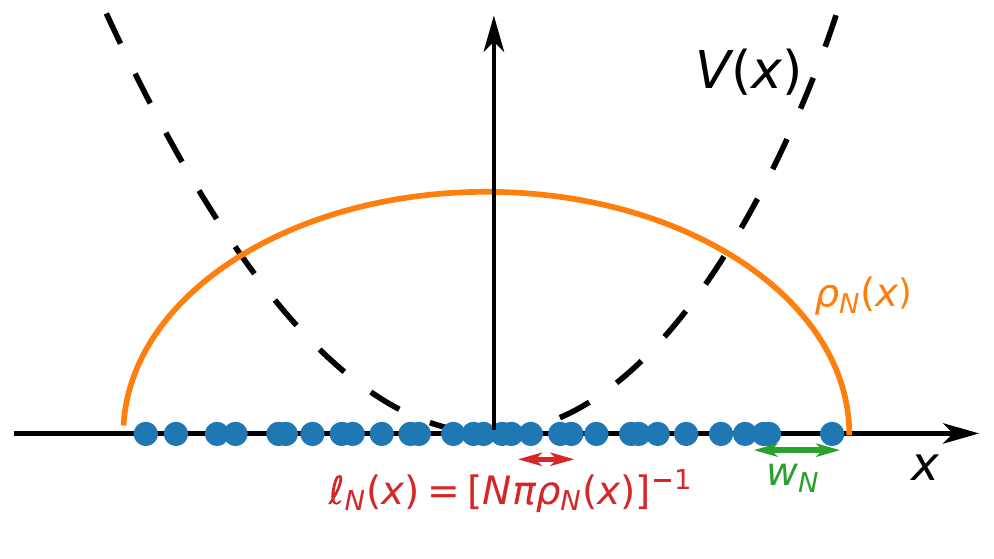}
\caption{Typical repartition of fermions in a trapping potential $V(x)$ (represented in dashed black). The density (in orange) has finite edges. As the density is smaller, the typical inter-particle distance at the edge $w_N$ is large $w_N\gg \ell_N(u)$ in comparison to the typical scale $\ell_N(u)$ in the bulk.}\label{Fig_sacles_b_e}
\end{figure}

In the large $N$ limit, the typical inter-particle distance $\ell_N(x)$ can be evaluated close to a point $x$ in the bulk of the density as (see also Fig. \ref{Fig_sacles_b_e})
\be
\int_{x}^{x+\ell_N(x)}\rho_{N}\left(x'\right)dx'\sim \frac{1}{N}\Rightarrow \ell_N(x)\sim \left[N \rho_{N}\left(x\right)\right]^{-1}\;.
\ee
Note that this scale corresponds exactly to the typical scale of LDA and diverges at the edge of the density. In the limit $N\to \infty$, it is well-known in the RMT literature \cite{mehta2004random,forrester2010log,akemann2011oxford} that the correlation kernel takes the scaling form
\be
K_N(x,y)\approx \rho_N(u) K_{ \sin}\left(\rho_N(u)(x-y)\right)\;,\;\;u=\frac{x+y}{2}
\ee
valid in the bulk, where the function $K_{ \sin}(r)$ is the sine kernel and reads
\be\label{k_sin}
K_{ \sin}(r)=\int_{-1}^{1}\frac{e^{i k r}}{2\pi}dk=\frac{\sin(r)}{\pi r}\;.
\ee
This result coincides exactly with the prediction from LDA in Eq. \eqref{LDA_1d_ker}, which provides a more rigorous proof of this bulk description. Before considering the edge properties of this system, let us mention a first application of this exact mapping that may prove to be experimentally relevant.

\subsubsection{Full counting statistics and entanglement entropy}\label{FCS_EE}

Using the exact mapping to GUE, it is possible to obtain the statistics of the number $N_L$ of fermions inside the interval $[-L,L]$. We refer to this observable as the {\it full counting statistics} (FCS). At zero temperature, and using the connection to random matrix, one can show a central limit theorem for the number of fermions inside this interval in the large $N$ limit \cite{costin1995gaussian,marino2014phase, marino2016number} and obtain the asymptotic behaviour of the number variance for $0<\zeta=\alpha L/\sqrt{N}<\sqrt{2}$,
\be
\var{N_L}\approx \frac{1}{\pi^2}\ln\left[N\zeta(2-\zeta^2)^{3/2}\right]\;.
\ee
Note that these results were recently extended to the case of finite temperature \cite{grabsch2018fluctuations}. For non-interacting Fermi gases, one can show that the FCS is directly related to another quantity: the bipartite entanglement entropy  \cite{klich2009quantum}. The R\'enyi bipartite entanglement entropy $S_q(N,L)$ of the domain ${\cal D}_L=[-L,L]$ is defined as
\be
S_q(N,L)=\frac{1}{1-q}\ln \Tr\left[\rho_{L}^q\right]\;,
\ee
where $\rho_L=\Tr_{\bar{\cal D}_L}[\rho]$ is the reduced density matrix, obtained from the full density matrix $\rho=|\Psi_0\rangle\langle \Psi_0|$ by tracing out the degrees of freedom of the complement $\bar{\cal D}_L=(-\infty,-L)\cup(L,\infty)$ of ${\cal D}_L$. This entanglement entropy allows to characterise in particular the critical and topological phases of matter \cite{amico2008entanglement, calabrese2004entanglement}. One can show that for a general system of non-interacting fermions, there exists an exact expression for the R\'enyi entanglement entropy as a series of the cumulants $\moy{N_{\cal D}^p}_c$ of order $p>2$ of the particles number \cite{klich2009quantum, song2011entanglement, song2012bipartite}
\be
S_q({\cal D})=\sum_{p=2}^{\infty}\eta_{q,p}\moy{N_{\cal D}}_c^p\;,\label{S_q_cum}
\ee
with in particular $\eta_{q,2}=\frac{\pi^2}{6q}(q+1)$. We discuss in further details this connection in Section \ref{FCS_EE_sec}. This relation was exploited in \cite{calabrese2015random} to obtain the entanglement entropy in the bulk of this one-dimensional Fermi gas at zero temperature. This remark closes this apart\'e and we consider now the edge properties of this system by first obtaining the correlation kernel.

\subsubsection{Edge limit: Airy kernel}

\begin{figure}
\centering
\includegraphics[width=0.6\textwidth]{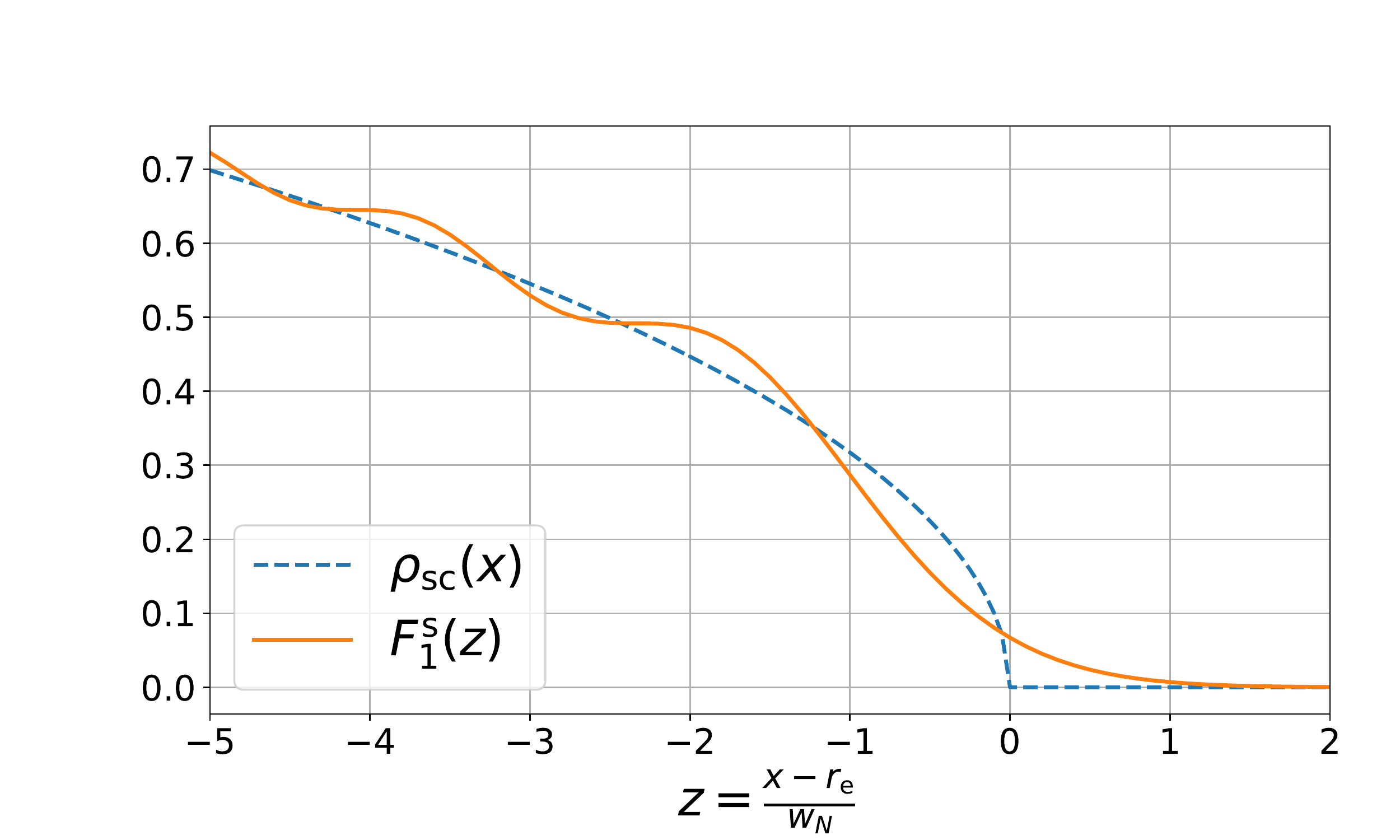}
\caption{Plot of the density profile $F_1^{\rm s}(z)$ close to the soft edge given in Eq. \eqref{F_1_soft} (in orange) and comparison with the prediction from local density approximation (in dashed blue).}\label{Fig_dens_soft_1d}
\end{figure}

The exact mapping with the GUE is especially convenient for
the description of the statistics at the edge of the density. As seen in section \ref{sec_LDA}, the description introduced by the LDA is no longer valid close to the edge. In particular, as the density becomes small close to the edge, the typical inter-particle distance $w_N$ is larger than in the bulk (c.f. Fig. \ref{Fig_sacles_b_e}). It can be evaluated by ensuring that there is $O(1)$ particles in the interval $[r_{\rm e}-w_N, r_{\rm e}]$ close to the edge, i.e.
\be
\int_{r_{\rm e}-w_N}^{r_{\rm e}}\rho_{N}\left(x'\right)dx'\approx \int_{0}^{w_N}\frac{\alpha}{\sqrt{N}}\sqrt{\frac{\alpha x'}{\sqrt{N}}} dx'\sim \frac{1}{N}\Rightarrow w_N\sim N^{-1/6}\;.
\ee
%
The edge properties of the spectrum have been extensively studied in RMT \cite{mehta2004random,forrester2010log,akemann2011oxford}. Using the exact mapping, one can then show that the kernel converges on a typical scale $w_N=(\alpha \sqrt{2}N^{1/6})^{-1}$ close to the edge $r_{\rm e}$ towards the Airy scaling form \cite{forrester2010log}
\be\label{K_airy_scal_oh}
K_{N}(x,y)\approx \frac{1}{w_N}K_{\Ai}\left(\frac{x-r_{\rm e}}{w_N},\frac{y-r_{\rm e}}{w_N}\right)\;,
\ee
where the function $K_{ \Ai}\left(u,v\right)$ is called the Airy kernel and reads
\be\label{k_airy}
K_{ \Ai}(u,v)=\int_{0}^{\infty}ds\Ai(s+u)\Ai(s+v)=\frac{\Ai(u)\Ai'(v)-\Ai(v)\Ai'(u)}{u-v}\;.
\ee
The Airy function $\Ai(u)=\lim_{x\to \infty}\frac{1}{\pi}\int_0^{x}dt \cos\left(\frac{t^3}{3}+u t\right)$ is the real solution of $f''(u)=u f(u)$ that vanishes as $u\to +\infty$. This scaling form allows a more precise description of the edge properties. In particular, the Airy kernel controls the density profile close to the soft edge
\be
\rho_N(x)=\frac{1}{N}K_N(x,x)\approx \frac{1}{N w_N}F_1^{\rm s}\left(\frac{x-r_{\rm e}}{w_N}\right)\;,\;\;{\rm with}\;\;F_1^{\rm s}(z)=K_{ \Ai}(z,z)={\Ai'}^2(z)-z\Ai^2(z)\;.\label{F_1_soft}
\ee
The asymptotic behaviours of this scaling function read
\be
F_1^{\rm s}(z)\approx \begin{cases}
\displaystyle \frac{\sqrt{|z|}}{\pi}&\;,\;\;z\to -\infty\;,\\
&\\
\displaystyle \frac{e^{-\frac{4}{3}z^{3/2}}}{8\pi z}&\;,\;\;z\to +\infty\;.
\end{cases}\label{F_1_s_as}
\ee
Note that the behaviour for $z\to -\infty$ matches smoothly with the square-root behaviour of the density obtained from LDA in Eq. \eqref{LDA_1d}. This density profile is plotted in Fig. \ref{Fig_dens_soft_1d} together with a comparison with the LDA description.

Even though the framework developed only extends to non-interacting Fermi gases, it was recently argued \cite{dean2016noninteracting, stephan2019free} that as the gas is very dilute close to the edge, the physics should not depend crucially on the interactions between particles, provided they are not too strong. These edge results describing a non-interacting Fermi gas should therefore extend to finite interaction. It is indeed well-known that the many-body ground state probability in Eq. \eqref{jPDF_OH1d} is recovered in the Lieb-Liniger model of bosons with contact repulsion, \cite{PhysRevA.63.033601,vignolo2002degenerate}
\be
{\cal H}_N=\sum_{i=1}^N \left(-\partial_{x_i}^2+x_i^2\right)+c\sum_{i< j}\delta(x_i-x_j)\;,
\ee
in the Tonks-Girardeau limit where $c\to \infty$ \cite{girardeau1960relationship, girardeau1965permutation}. The results at the edge should also extend to bosons with finite repulsive short range interactions.

One interesting counter-example where the strong interactions modify the behaviour at the edge is the Calogero-Moser-Sutherland model \cite{sutherland2004beautiful} described by the one-dimensional Hamiltonian for $N$ particles
\be
{\cal H}_N=\sum_{i=1}^N \left(-\partial_{x_i}^2+x_i^2\right)+\sum_{i< j}\frac{\beta(\beta-2)}{(x_i-x_j)^2}\;.
\ee
In this model the ground state joint probability of the positions can be obtained exactly \cite{ stephan2019free}
\be
\left|\Psi_0(x_1,\cdots,x_N)\right|^2=\frac{1}{Z_N^{\rm CMS}} \prod_{i<j}|x_i-x_j|^{\beta} \prod_{i=1}^N e^{-x_i^2 }\;.
\ee
After the rescaling $x_i=\sqrt{N\beta/2}\,\lambda_i$, this joint PDF matches exactly the PDF of eigenvalues in the Gaussian $\beta$ Ensemble, which is a natural extension of the GUE,
\be
P_{\rm joint}^{\rm G\beta E}(\lambda_1,\cdots,\lambda_N)=\frac{1}{Z_N^{\rm G\beta E}} \prod_{i<j}|\lambda_i-\lambda_j|^{\beta} \prod_{i=1}^N e^{-\frac{N\beta}{2} \lambda_i^2 }\;.\label{P_joint_GbE}
\ee
The values $\beta=1,2,4$ correspond to Dyson's famous three-fold  way \cite{dyson1962threefold} (see also Appendix \ref{RMT_app}). Note that the eigenvalues only form a determinantal point process for $\beta=2$, which prevents to generalise the framework developed in this chapter to this more general situation.

The Airy scaling form of the correlation kernel in Eq. \eqref{K_airy_scal_oh} controls the spatial statistics at the edge, and therefore the extreme value statistics of the Fermi gas. We now exploit this result and turn to the statistics of the rightmost fermion $\displaystyle x_{\max}=\max_{1\leq i\leq N}x_i$. 

\subsubsection{Extreme value statistics: Tracy-Widom distribution}
\begin{figure}
\centering
\includegraphics[width=0.6\textwidth]{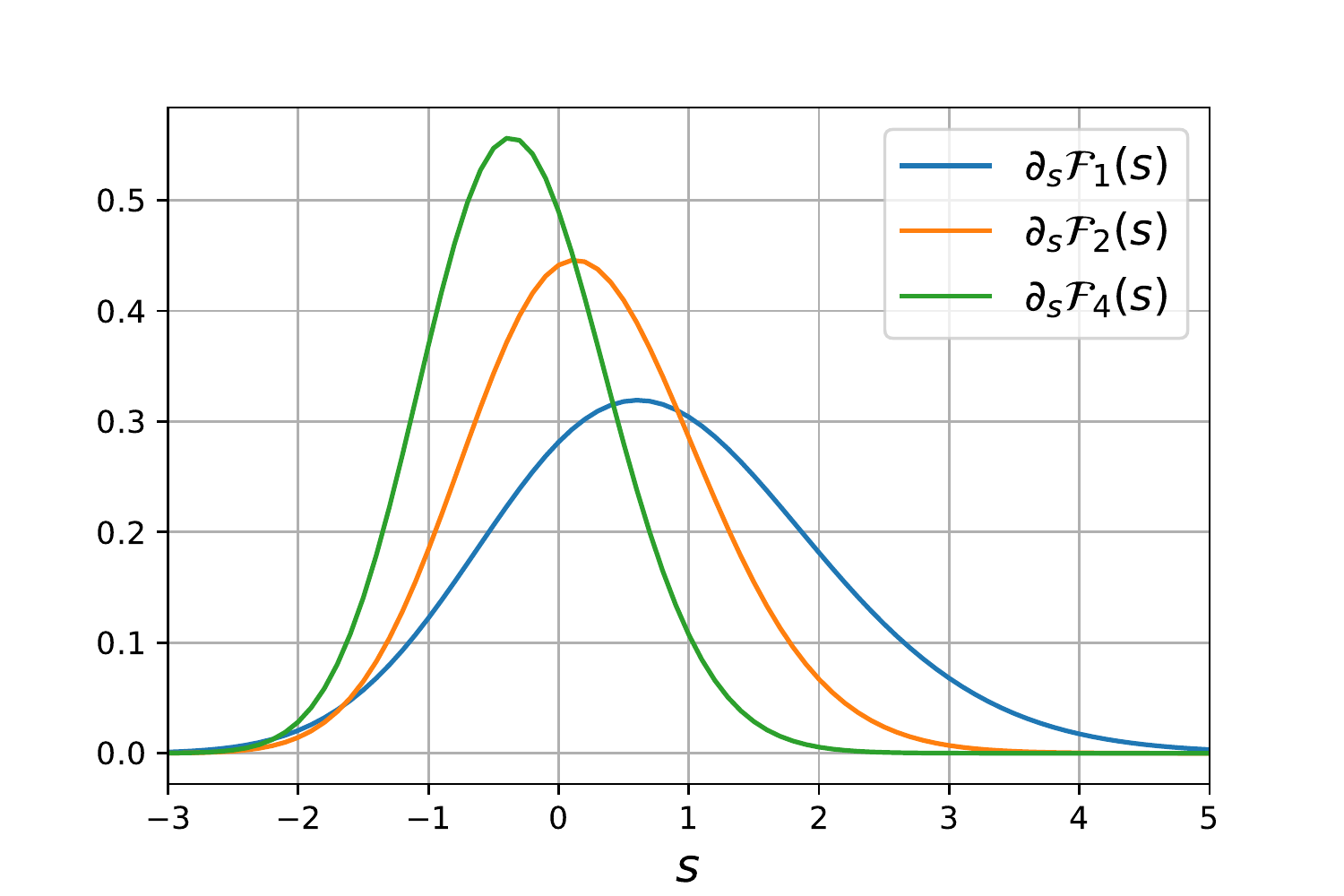}
\caption{Plot of the Tracy-Widom distribution ${\cal F}_{\beta}'(s)$ for $\beta=1,2,4$ respectively in blue, orange and green.}\label{Fig_TWb}
\end{figure}

%
The problem considered is completely symmetric, and in particular $\Prob\left[x_{\min}\geq -\lambda\right]=\Prob\left[x_{\max}\leq \lambda\right]$. We therefore only focus on the case of $x_{\max}$. As seen in Eq. \eqref{x_max_Det}, the CDF of the position $x_{\max}$ of the rightmost fermion (or the largest eigenvalue $\lambda_{\max}$ of the GUE) can be expressed as a Fredholm determinant. From the scaling form of the kernel at the edge in Eq. \eqref{K_airy_scal_oh}, one naturally obtains the scaling form for the CDF \cite{mehta2004random,forrester1993spectrum}
\be
\lim_{N\to \infty }\Prob\left[x_{\max}\leq \sqrt{2}+\frac{s}{\sqrt{2} N^{1/6}}\right]=\lim_{N\to \infty }\Prob\left[\lambda_{\max}\leq \sqrt{2N}+\frac{s}{\sqrt{2} N^{2/3}}\right]={\cal F}_2(s)\;
\ee
where the scaling function ${\cal F}_2(s)$ reads
\be
{\cal F}_2(s)=\Det\left[\mathbb{I}-P_{[s,\infty)}K_{\Ai}P_{[s,\infty)}\right]=\exp\left[-\sum_{p=1}^{\infty}\frac{1}{p}\Tr\left(P_{[s,\infty)}K_{\Ai}P_{[s,\infty)}\right)^p\right]
\ee
and where we recall that $K_{\Ai}(x,y)$ is given in Eq. \eqref{k_airy}, $P_{I}$ is the projector on the interval $I$ and $\mathbb{I}$ is the identity operator. A few basic definitions for Fredholm determinant are given in Appendix \ref{Fred_det_app}. We refer the interested reader to \cite{gohberg2012traces} for more precisions on Fredholm determinants.
In 1994, Tracy and Widom realised a tour de force by obtaining another expression for the scaling function ${\cal F}_2(s)$  -- which now holds the name of Tracy-Widom $\beta=2$ distribution -- of the rescaled variable $\chi_2=\sqrt{2}N^{2/3}(\lambda_{\max}-\sqrt{2})$, as \cite{tracy1994level}
\be\label{TW_2}
\lim_{N\to \infty}\Prob\left[\chi_2=\sqrt{2}N^{2/3}(\lambda_{\max}-\sqrt{2})\leq s\right]={\cal F}_2(s)=\exp\left(-\int_s^{\infty}(r-s)q^2(r)dr\right)\;,
\ee
where $q(r)$ is solution of the Painlev\'e II equation
\be
q''(r)=2q^3(r)+rq(r)\;,\;\;{\rm with}\;\;q(r)\sim\Ai(r)\;,\;\;{r\to +\infty}\;.
\ee
There exists an expression for general values of $\beta$ for the Tracy-Widom $\beta$ distribution ${\cal F}_{\beta}(s)$ corresponding to the typical fluctuations of $\lambda_{\max}$ in the Gaussian $\beta$ Ensemble \eqref{P_joint_GbE} in terms of the ground state of a stochastic Airy operator \cite{bloemendal2013limits}, but it is rather complex. Simple expressions only exists for $\beta=1,2,4$, corresponding to Dyson's three fold way (see \cite{dyson1962threefold} and Appendix \ref{RMT_app} for precisions on  Dyson's classification). Note however that an expression has recently been obtained for $\beta=6$ \cite{grava2016tracy}. We emphasise that it is one of the rare occurrence where the distribution of the maximum of strongly correlated random variables can be obtained explicitly. The tails of this distribution are given by
\be\label{TW_tails}
{\cal F}_\beta'(s)\approx\begin{cases}
\displaystyle \exp\left(-\frac{\beta}{24}|s|^3\right)&\;,\;\;s\to -\infty\;\\
&\\
\displaystyle \exp\left(-\frac{2\beta}{3}s^{\frac{3}{2}}\right)&\;,\;\;s\to +\infty\;,
\end{cases}
\ee
showing the strong asymmetry of the distribution as seen in Fig. \ref{Fig_TWb}. Note that the tail for $s\to \infty$ matches the tail of the density profile in Eq. \eqref{F_1_s_as}. The Tracy-Widom PDFs ${\cal F}'_{\beta}(s)$ are plotted in Fig. \ref{Fig_TWb} for $\beta=1,2,4$.
 This distribution turns out to be ubiquitous \cite{majumdar2014top}. It has for instance been observed experimentally in the growth of nematic liquid crystals \cite{takeuchi2010universal,takeuchi2011growing} in link with the well-known Kardar-Parisi-Zhang $(1+1)d$ universality class or in coupled optical fibres experiments  \cite{fridman2012measuring}. Using this explicit link with the position of the rightmost fermion, one could hope to measure this distribution in a cold atom experiment.

\subsubsection{Extreme value statistics: large deviations}\label{TW_3rd}

\begin{figure}
\centering
\includegraphics[width=0.55\textwidth]{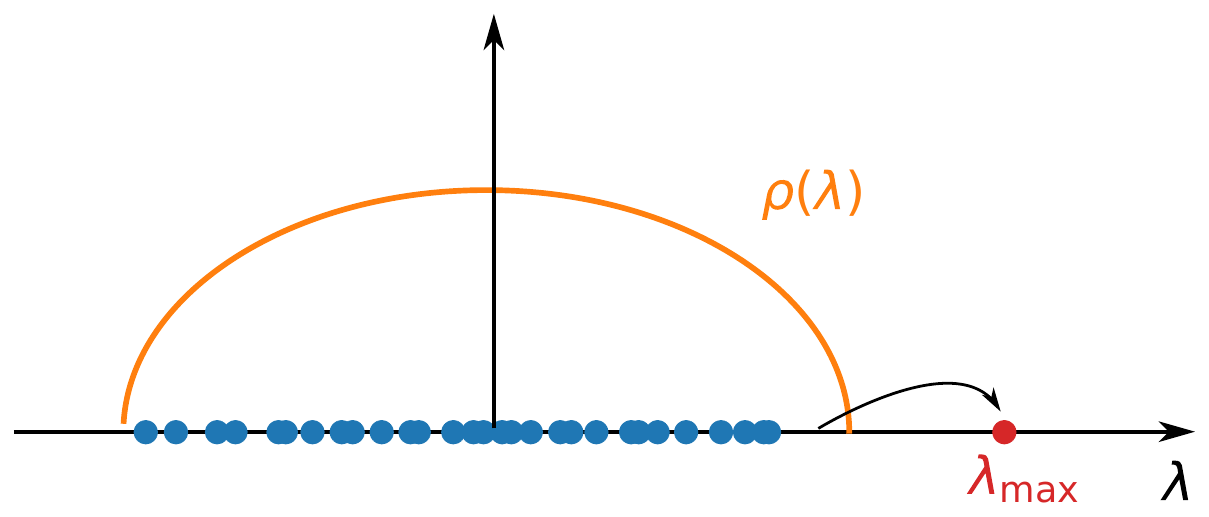}
\includegraphics[width=0.35\textwidth]{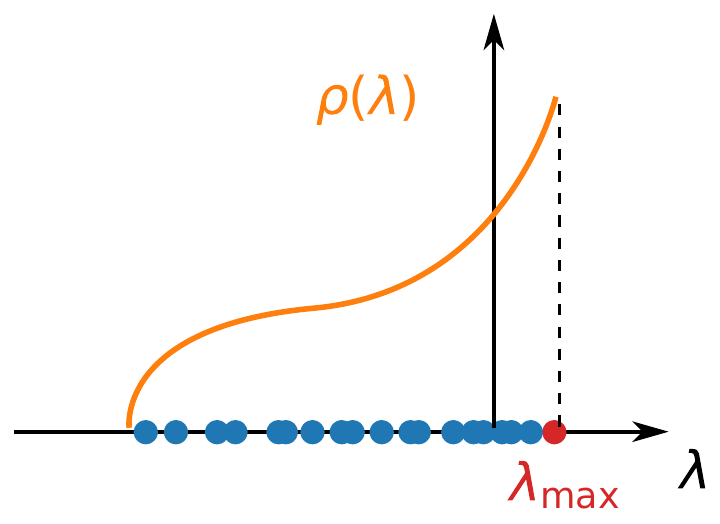}
\caption{Scheme of the equilibrium repartition of eigenvalues for the pulled (left) or pushed (right) log gas.}\label{Fig_pushed_pulled}
\end{figure}

The atypical fluctuations of $\lambda_{\max}$ (and therefore of $x_{\max}$) were found to follow large deviation principles, yielding the three different regimes of fluctuations \cite{arous2001aging,dean2006large, dean2008extreme,majumdar2009large}
\be\label{suma_GUE}
\partial_\lambda\Prob\left[\lambda_{\max}\leq \lambda\right]\approx\begin{cases}
\displaystyle \exp\left(-\beta N^2 \Phi_-^{{\rm G}\beta{\rm E}}(\lambda)\right)&\;,\;\;\sqrt{2}-\lambda=O(1)\\
&\\
\displaystyle \sqrt{2}N^{2/3}{\cal F}_{\beta}'\left(\sqrt{2}N^{2/3}(\lambda-\sqrt{2})\right)&\;,\;\;|\lambda-\sqrt{2}|=O(N^{-2/3})\\
&\\
\displaystyle \exp\left(-\beta N \Phi_+^{{\rm G}\beta{\rm E}}(\lambda)\right)&\;,\;\;\lambda-\sqrt{2}=O(1)\;,
\end{cases}
\ee
where the left large deviation function $\Phi_-^{{\rm G}\beta{\rm E}}(\lambda)$ was obtained for general values of $\beta$ \cite{dean2006large, dean2008extreme} while the right large deviation function $\Phi_+^{{\rm G}\beta{\rm E}}(\lambda)$ was obtained first $\beta=1$ \cite{arous2001aging} and then for general $\beta$ \cite{majumdar2009large}. These large deviation functions were obtained
using a Coulomb gas method (or rather log gas). To introduce this method, we rewrite the probability weight in Eq. \eqref{P_joint_GbE} as
\begin{align}
P_{\rm joint}(\lambda_1,\cdots,\lambda_N)&=\frac{\displaystyle e^{-\frac{\beta N^2}{2} E_N(\lambda_1,\cdots,\lambda_N)}}{Z_N} \;,\nn\\
E_N(\lambda_1,\cdots,\lambda_N)&=\frac{1}{N}\sum_{i=1}^N \lambda_i^2 -\frac{2}{N^2}\sum_{i<j}\ln|\lambda_i-\lambda_j|\;.\label{E_gas}
\end{align}
The joint probability is then reinterpreted as the Gibbs weight of a gas of ``charges'' (with logarithmic $2d$ Coulomb repulsion) confined by a one-dimensional potential $v(r)=r^2$ \cite{dyson1962statistical,dyson1962statistical2,dyson1962statistical3}. Introducing the empirical density of eigenvalues
\be
\hat \rho_N(\lambda)=\frac{1}{N}\sum_{i=1}^N \delta(\lambda-\lambda_i)\;,
\ee 
which is normalised to unity, allows us to rewrite the energy in Eq. \eqref{E_gas} as a functional of this distribution \cite{dyson1962statistical,dyson1962statistical2,dyson1962statistical3}
\begin{align}
E_N(\lambda_1,\cdots,\lambda_N)&=S[\hat\rho_N]+o(1)\;,\nn\\
S[\hat\rho_N]&=\int d\lambda \lambda^2\hat \rho_N(\lambda)-\int d\lambda  \int d\lambda'\hat \rho_N(\lambda)\rho_N(\lambda')\ln|\lambda-\lambda'|\;.\label{CG_herm}
\end{align}
In the large $N$ limit, we replace $\hat \rho_N(\lambda)\to \rho(\lambda)$ where $\rho(\lambda)$ is a continuous function. The PDF $\partial_\lambda\Prob\left[\lambda_{\max}\leq \lambda\right]$ is recast as a functional integral over the density $\rho(\lambda')$ provided that the density is normalised and that it is zero beyond the position of the rightmost charge $\lambda$. In the large $N$ limit, this functional is dominated by the distribution of charge $\rho^*(\lambda')$ that minimises the energy functional. The large deviation function to the left of the typical regime $\lambda<2$ can be obtained by evaluating the difference of energy between the free Coulomb (or log) gas, i.e. for which the positions of charges can take value in $(-\infty,\infty)$, and the compressed Coulomb gas, i.e. for which the positions of charges can take value in $(-\infty,\lambda]$. In the latter, there is a macroscopic rearrangement of charges whose energy cost is $O(N^2)$ (c.f. Fig. \ref{Fig_pushed_pulled}). On the contrary, the large deviation function to the right of the typical regime $\lambda>2$ can be obtained by evaluating the energy for a single charge to be pulled out of the density and placed at position $\lambda$. In this case, there is no macroscopic rearrangement of charges and the energy cost is $O(N)$ (c.f. Fig. \ref{Fig_pushed_pulled}). The left large deviation function behaves for $\lambda\to \sqrt{2}_-$ as $\Phi_-^{{\rm G}\beta{\rm E}}(\lambda)\approx (\sqrt{2}-\lambda)^3 /(6\sqrt{2})$ \cite{dean2006large, dean2008extreme}, which yields
\be
\beta N^2 \Phi_-^{{\rm G}\beta{\rm E}}(\lambda)=\beta N^2 \frac{(\sqrt{2}-\lambda)^3}{6\sqrt{2}}=\frac{\beta }{24}\left(\sqrt{2} N^{2/3}(\sqrt{2}-\lambda)\right)^3\;,\;\;\lambda\to \sqrt{2}_-\;,
\ee
allowing a smooth matching with the left tail of the Tracy-Widom distribution in the first line of Eq. \eqref{TW_tails}. This matching between large deviations and the regime of typical fluctuations will be of major importance in the following and we therefore emphasise this point here. Similarly, the right tail behaves for $\lambda\to \sqrt{2}_+$ as $\Phi_+^{{\rm G}\beta{\rm E}}(\lambda)\approx 2^{7/4}(\lambda-\sqrt{2})^{3/2}/3$ \cite{majumdar2009large}, which yields
\be
\beta N \Phi_+^{{\rm G}\beta{\rm E}}(\lambda)=\beta N \frac{2^{7/4}}{3}(\lambda-\sqrt{2})^{\frac{3}{2}}=\frac{2\beta}{3}\left(\sqrt{2} N^{2/3}(\lambda-\sqrt{2})\right)^{\frac{3}{2}}\;,\;\;\lambda\to \sqrt{2}_+\;,
\ee
allowing a smooth matching with the right tail of the Tracy-Widom distribution in the second line of Eq. \eqref{TW_tails}. We conclude this section by mentioning that one can associate to the behaviour of the right large deviation function a third order phase transition \cite{majumdar2014top} (in the sense of the Ehrenfest classification of phase transitions). Indeed, defining the free-energy of the gas as
\be
F(\lambda)=-\lim_{N\to \infty}\frac{1}{\beta N^2}\ln \Prob\left[ \lambda_{\max}\leq \lambda\right]=\begin{cases}
\displaystyle \Phi_-^{{\rm G}\beta{\rm E}}(\lambda)&\;,\;\;\lambda\leq \sqrt{2}\;,\\
&\\
\displaystyle 0&\;,\;\;\lambda>\sqrt{2}\;,
\end{cases}
\ee
we see that $F(\sqrt{2})=F'(\sqrt{2})=F''(\sqrt{2})=0$ while $F'''(\sqrt{2}_-)=2^{-1/2}$. This mechanism of third order phase transition, together with the profile of the density $\rho(\lambda) \sim\sqrt{\lambda_{\rm e}-\lambda}$ was advanced to explain the ubiquity of the Tracy-Widom distribution \cite{majumdar2014top}.

%
%
%
%
%
%
%
%
%
%
%

We have seen that the exact mapping between GUE and the problem of fermions in a harmonic potential allows to extract the spatial statistics close to the soft edge.
We conclude this section by mentioning that for a different potential $V(x)$, the mapping to GUE no longer holds. However, we will now see that the sine scaling form \eqref{k_sin} for the kernel in the bulk and the Airy scaling form \eqref{k_airy} at the edge are universal and hold for a large class of smooth potential $V(x)$ (for instance $V(x)\sim |x|^p$ with $p>0$).

\subsection{Universality at the soft edge} \label{univ_prop_1d}

\subsubsection{Universality of the correlation kernel}

The universality of the bulk and edge scaling form were first shown for smooth potentials, i.e. $V(x)\in {\cal C}_1(\mathbb{R})$, in \cite{eisler2013universality} using a refined WKB approach to describe the wave-functions. Dean et al. introduced a different method \cite{dean2015universal}, allowing to obtain the correlation kernel in more general settings. This method will be useful in the following and we therefore introduce it briefly. It relies on the connection between the correlation kernel and the single particle Euclidean (imaginary time) propagator $G(y,t|x,0)$. To obtain this relation, we first rewrite the correlation kernel as
\be
K_N({ x},{ y})=K_{\mu}({ x},{ y})=\sum_{k=0}^{\infty}\overline{\phi}_{ k}({ x})\phi_{ k}({ y})\Theta(\mu-\epsilon_{k})\;,
\ee
where $\mu=\epsilon_F=\epsilon_{N-1}$ is the Fermi energy and $\Theta(x)$ is the Heaviside step function. The Euclidean propagator can then be expressed as \cite{dean2015universal, dean2016noninteracting}
\be
G(y,t|x,0)=\langle y|e^{-\frac{\hat H t}{\hbar}}|x\rangle=\sum_{k=0}^{\infty}\overline{\phi}_{ k}({ x})\phi_{ k}({ y})e^{-\frac{\epsilon_k t}{\hbar}}=\frac{t}{\hbar}\int_0^{\infty}e^{-\frac{\mu t}{\hbar}}K_{\mu}(x,y)d\mu\;.\label{prop_K_1d}
\ee
Note that from this formula the large $\mu$ (or $N$) limit of the correlation kernel corresponds to the short time $t$ behaviour of the propagator. The Euclidean propagator is solution of the partial differential equation
\be
\hbar\partial_t G(y,t|x,0)=\frac{\hbar^2}{2m}\partial_y^2 G(y,t|x,0)-V(y)G(y,t|x,0)\;,\;\;{\rm with}\;\;G(y,0|x,0)=\delta(x-y)\;.
\ee
The solution of this equation can be obtained as a path integral. Using a short time expansion of this path integral (corresponding to a diagrammatic expansion) \cite{makri1988correct, dean2016noninteracting} Dean et al. were able to recover the sine scaling form in the bulk \cite{dean2015universal}
\be\label{k_sine}
K_N(x,y)=\frac{1}{\ell_N(u)}K_{\sin}\left(\frac{x-y}{\ell_N(u)}\right)\;,\;\;u=\frac{x+y}{2}\;,\;\;\ell_N(u)=\left[N\pi \rho_N(u)\right]^{-1}\;.
\ee
In terms of the propagator, this result amounts roughly to rescale the propagator close to the point $u$ in the bulk and neglect the spatial variations of the potential on the scale $\ell_N(u)$. This result allowed to put the LDA description on firmer ground. Furthermore, close to the edge in $r_{\rm e}$ such that $V(r_{\rm e})=\mu$, Dean et al. found that the Airy kernel was also universal \cite{dean2015universal} 
\be
K_N(x,y)=\frac{1}{w_N}K_{\Ai}\left(\frac{x-r_{\rm e}}{w_N},\frac{y-r_{\rm e}}{w_N}\right)\;,\;\;w_N=\left(\frac{\hbar^2}{2m V'(r_{\rm e})}\right)^{\frac{1}{3}}\;.\label{w_N}
\ee
The detailed conditions of validity can be found in \cite{dean2015universal, dean2016noninteracting} and a rigorous proof is given in \cite{bornemann2011scaling}.
In terms of the propagator, this result amounts roughly to rescale the diffusion close to the edge $r_{\rm e}$ and linearise the potential on the scale $w_N$. 
We emphasise that the propagator method exposed here will allow to obtain the correlation kernel in situations were there is no explicit connection to random matrices -- for instance in $d>1$ -- and will be useful to derive many results in the following.

We now detail the implication of these results for the extreme value statistics of the Fermi gas. 


\subsubsection{Universality of the statistics of $x_{\max}$}

This result on the universality of the Airy scaling form of the kernel at the edge implies from Eq. \eqref{x_max_Det} the universality of the Tracy-Widom distribution for the typical distribution of $x_{\max}$  for any smooth confining potential $V(x)$
\be
\lim_{N\to \infty}\Prob\left[x_{\max}\leq r_{\rm e}+w_N s\right]={\cal F}_2(s)=\Det\left[\mathbb{I}-P_{[s,\infty)}K_{\Ai}P_{[s,\infty)}\right]\;.
\ee
As noted in \cite{majumdar2014top} and explained in section \ref{TW_3rd}, the Tracy-Widom distribution describes the universal properties emerging in a third order phase transition and with an order parameter vanishing as a square-root at the transition. We already obtained from LDA that the density of fermions presents for any smooth confining potential a profile that vanishes at the edge as $\rho_N(x)\sim \sqrt{r_{\rm e}-x}$.
It is then natural to ask if this universality is associated with a third order phase transition.  We expect, as it is always the case in RMT, that the atypical fluctuations to the left of the regime of typical fluctuations follow a large deviation principle with rapidity $N^2$,
\be
\Prob\left[x_{\max}\leq x\right]=e^{-N^2 \Phi_-\left(x\right)}\;,\;\;|x-r_{\rm e}|=O(r_{\rm e})\;.
\ee
This large deviation rate function $\Phi_-(x)$ should match the left tail asymptotic behaviour of the Tracy-Widom distribution in Eq. \eqref{TW_tails}, and we therefore expect that $\Phi_-\left(x\right)\sim (r_{\rm e}-x)^3$ as $x\to {r_{\rm e}}_-$. From this behaviour, it is clear that there is indeed a third order phase transition associated to this distribution of $x_{\max}$ for any smooth potential $V(x)$.

We have seen in this section that the fluctuations close to ``soft edges'' created by smooth confining potentials are universal in dimension one. We will now see how this framework and the results obtained in the one-dimensional case extend to higher dimension $d>1$.

%
  %
%
\section{Higher dimension $d>1$ system at zero temperature}\label{sec_d_soft}

In higher dimension and for a system with non-degenerate ground state, one can define a unique set of quantum numbers ${\bf k}_0,{\bf k}_1,\cdots,{\bf k}_{N-1}$ (where ${\bf k}$ is a $d$-dimensional vector) with energy $\epsilon_{{\bf k}_0}\leq \epsilon_{{\bf k}_1}\leq \cdots\leq \epsilon_{{\bf k}_{N-1}}=\mu$. The joint quantum probability of the positions ${\bf x}_i$'s of the fermions reads
\begin{align}
|\Psi_0({\bf x}_1,\cdots,{\bf x}_N)|^2&=\frac{1}{N!}\det_{1\leq i,j\leq N} \phi_{{\bf k}_{j-1}}({\bf x}_i)\det_{1\leq n,m\leq N} \overline{\phi}_{{\bf k}_{n-1}}({\bf x}_m)\\
&=\frac{1}{N!}\det_{1\leq i,j\leq N}K_N({\bf x}_i,{\bf x}_j)\;.
\end{align}
Using the orthonormality of wave functions in Eq. \eqref{ortho}, one can then prove that the positions of fermions also form at zero temperature a $d$-dimensional determinantal point process of correlation kernel  \cite{dean2015universal}
\be
K_N({\bf x},{\bf y})=\sum_{l=0}^{N-1} \overline{\phi}_{{\bf k}_l}({\bf x})\phi_{{\bf k}_l}({\bf y})=K_{\mu}({\bf x},{\bf y})=\sum_{\bf k}\overline{\phi}_{\bf k}({\bf x})\phi_{\bf k}({\bf y})\Theta(\mu-\epsilon_{\bf k})\;.
\ee
The effects of ground state degeneracies are sub-leading in the large $N$ limit \cite{dean2016noninteracting}, and the results presented in the following still hold in this limit.

As in the case of dimension $d=1$, one could have considered the problem in momentum space and obtained that the process is also determinantal with a kernel given by the Fourier transform of $K_N({\bf x},{\bf y})$ in ${\bf x}$ and ${\bf y}$. We will only restrict our study to the case of rotationally symmetric potentials $V({\bf x})=v(|{\bf x}|)$. In this case, using the decomposition of the Laplace operator as
\be
\Delta_{\bf x}=\left[r^{\frac{1-d}{2}}\partial_r^2\left(r^{\frac{d-1}{2}}\right)+\frac{(d-1)(d-3)}{4r^2}-\frac{\hat{\bf L}^2}{\hbar^2 r^2}\right]_{r=|{\bf x}|}\;,
\ee
we obtain that the squared angular momentum operator $\hat{\bf L}^2$ will be the only operator appearing in the Hamiltonian that depends on the angular coordinate ${\bf u}={\bf x}/|{\bf x}|$. We can introduce an analogous decomposition of the wave functions and their associated quantum number ${\bf k}=(n,l,{\bf m})$, which are decoupled into a radial and an angular part
\be\label{decomp_d_wf}
\phi_{\bf k}({\bf x})=|{\bf x}|^{\frac{1-d}{2}}\chi_{n,l}(|{\bf x}|)Y_{l,{\bf m}}\left(\frac{{\bf x}}{|{\bf x}|}\right)\;.
\ee

Next, we analyse separately the radial and angular part of the Hamiltonian. 

\subsubsection{Radial dependence}

The wave functions $\chi_{n,l}(r)$ are eigenfunctions of the effective one-dimensional Hamiltonian
\begin{align}
&\hat H_l\chi_{n,l}(r)=-\frac{\hbar^2}{2m}\partial_r^2 \chi_{n,l}(r)-v_{l,d}(r) \chi_{n,l}(r)=\epsilon_{n,l} \chi_{n,l}(r)\;,\\
&{\rm with}\;\;v_{l,d}(r)=v(r)+\frac{\hbar^2}{8m r^2}(2l+d-1)(2l+d-3)\;,\label{v_eff_l}\\
&{\rm and}\;\;\int_0^{\infty}dr\, \overline{\chi}_{n,l}(r)\chi_{n',l'}(r)=\delta_{n,n'}\delta_{l,l'}\;,
\end{align}
and where the energy $\epsilon_{n,l}$ depends only on the principal $n$ and angular $l$ quantum numbers. It has an associated degeneracy that depends explicitly on the orbital quantum number $l$ \cite{moshinsky1996contemporary}
\be
g_d(l)=\frac{(2l+d-2)(l+d-3)!}{l!(d-2)!}\;.\label{degen_l}
\ee
Note that in this zero temperature setting, there is for each value of $l$ a maximal value $m_{l}$ of $n$ such that $\epsilon_{l,m_l}\leq \mu$ and $\epsilon_{l,m_l+1}>\mu$. Furthermore, there exists a maximal value $l^*$ of $l$ such that $\epsilon_{l^*,0}\leq \mu$ and $\epsilon_{l^*,1}>\mu$.  Using these two results, we may decompose the number of particles into subset with same orbital quantum number $l$ as
\be
N=\sum_{l=0}^{l^*} g_d(l)m_{l}\;.
\ee
In each of these subsets of size $m_l$, one can prove using the orthonormality of the effective wave functions $\chi_{n,l}(r)$ that the radii $r_i$ of the fermions form a determinantal point process with kernel \cite{dean2017statistics}
\be\label{k_r_eff}
K_l(r,r')=\sum_{n=0}^{m_l}\chi_{n,l}(r)\chi_{n,l}(r')\;.
\ee

We now turn to the angular part of the Hamiltonian.

\subsubsection{Angular dependence}

The functions $Y_{l,{\bf m}}({\bf u})$ are the $d$-dimensional spherical harmonics and form a basis of $\hat{\bf L}^2$
\be
\hat{\bf L}^2Y_{l,{\bf m}}({\bf u})=\hbar^2l(l-d+2)Y_{l,{\bf m}}({\bf u})\;,\;\;{\rm with}\;\;\int_{|{\bf u}|=1} d^{d}{\bf u}\, \overline{Y}_{l,{\bf m}}({\bf u})Y_{l',{\bf m}'}({\bf u})=\delta_{l,l'}\delta_{{\bf m},{\bf m}'}\;,\label{ortho_spherharm}
\ee
where the orbital quantum number $l=0,1$ in $d=1$ (corresponding to even and odd states) while $l\in \mathbb{N}$ for $d\geq 2$. The $d-2$ quantum numbers that do not intervene in the energy are regrouped in the vector ${\bf m}$. Note that there exists a summation formula for these spherical harmonics, valid for $d\geq 3$,
\be
\sum_{{\bf m}} \overline{Y_{l,{\bf m}}}({\bf u})Y_{l,{\bf m}}({\bf v})=\frac{2l+d-2}{(d-2)S_d}C_l^{\frac{d-2}{2}}\left({\bf u}\cdot {\bf v}\right)\;,
\ee
where $S_d=2\pi^{d/2}/\Gamma(d/2)$ is the surface of the sphere in $d$-dimension and $C_l^m(\eta)$ is a Gengenbauer polynomial, solution of the differential equation
\be
(1-\eta^2)\partial_{\eta}^2 {C_l^m}(\eta)-(2m+1)\eta\,\partial_{\eta}{C_l^m}(\eta)+l(l+2m)C_l^m(\eta)=0 \;, \label{eq:Legendre} 
\ee
with the conditions $C_l^m(\eta)=(-1)^l C_l^m(\eta)$ and $C_{l}^m(1)={{l+d-3}\choose{l}}$\;. In dimension $d=2$, the equivalent summation formula reads
\be
\sum_{m=\pm 1}\overline{Y_{l,m}}({\bf u})Y_{l,m}({\bf v})=\frac{T_l({\bf u}\cdot{\bf v})}{\pi}=\frac{1}{\pi}\cos(l\theta)\;,\;\;{\rm with}\;\;{\bf u}\cdot{\bf v}=\cos(\theta)\;,
\ee
and where $T_l(x)$ is the Tchebychev polynomial of first kind of order $l$. 

We will now see how to use the radial and angular decomposition of the Hamiltonian to simplify the extreme value statistics of this problem.

\subsubsection{Extreme value statistics}

In this $d$-dimensional setting, we define the position $\displaystyle r_{\max}=\max_{1\leq i\leq N}|{\bf x}_i|$ of the particle the farthest away from the centre of the trap. As in the one-dimensional case, we may obtain alternative formulations for this probability. Using the $d$-dimensional determinantal framework, it can be expressed as a Fredholm determinant of the $d$-dimensional kernel $K_N({\bf x},{\bf y})$,
\be\label{d_dim_fred_tr}
\Prob\left[r_{\max}\leq r\right]=\Det\left[\mathbb{I}-P_{|{\bf x}|\geq r}K_N P_{|{\bf x}|\geq r}\right]=\exp\left(-\sum_{p=1}^{\infty}\frac{1}{p}\Tr\left[(P_{|{\bf x}|\geq r}K_N P_{|{\bf x}|\geq r})^p\right]\right)\;.
\ee
Furthermore, using the Cauchy-Binet-Andr{\'e}ief formula in Eq. \eqref{Cauchy_Binet}, this probability can be expressed as
\begin{align}
\Prob\left[r_{\max}\leq r\right]&=\frac{1}{N!}\int_{|{\bf x}_1|\leq r} d^d {\bf x}_1\cdots \int_{|{\bf x}_N|\leq r}d^d {\bf x}_N\det_{1\leq i,j\leq N} \phi_{{\bf k}_{j-1}}({\bf x}_i)\det_{1\leq n,m\leq N} \overline{\phi}_{{\bf k}_{n-1}}({\bf x}_m)\nn\\
&=\det_{0\leq i,j\leq N-1}\left[\delta_{{\bf k}_i,{\bf k}_j}-\int_{|{\bf x}|\geq r}d^d {\bf x}\, \overline{\phi}_{{\bf k}_{i}}({\bf x})\phi_{{\bf k}_{j}}({\bf x})\right]\;.
\end{align}
By integrating over the angular degrees of freedom and using Eq. \eqref{ortho_spherharm}, the term of overlap between wave-functions for different values of $l$ or ${\bf m}$ vanishes, and the determinant becomes block diagonal. The block for the value $l$ will appear $g_d(l)$ times and has a size $m_l\times m_l$, such that the CDF reads \cite{dean2017statistics}
\be\label{P_max_decomp}
\Prob\left[r_{\max}\leq r\right]=\prod_{l=0}^{l^*}\left[Q_l(r)\right]^{g_d(l)}\;,\;\;{\rm with}\;\;Q_l(r)=\det_{0\leq i,j\leq m_l}\left(\delta_{i,j}-\int_r^{\infty} ds\, \overline{\chi}_{i,l}(s)\chi_{j,l}(s)\right)\;.
\ee
The function $Q_l(r)$ can itself be expressed as a Fredholm determinant of the one-dimensional kernel $K_l(r,r')$ in Eq. \eqref{k_r_eff}
\be\label{P_max_prod_Fred}
Q_l(r)=\Det\left[\mathbb{I}-P_{[r,\infty)}K_l P_{[r,\infty)}\right]\;.
\ee
From Eq. \eqref{P_max_decomp} we obtain that in dimension $d>1$, the maximal radius is the maximum of a set of independent but not identically distributed random variables. This decomposition reflects the fact that the radii of fermions with different orbital quantum numbers $l$ are independent. Note that a similar expression can be obtained for the CDF of the fermion with minimal distance to the centre of the trap $\displaystyle r_{\min}=\min_{1\leq i\leq N}|{\bf x}_i|$
\be\label{P_min_prod_Fred}
\Prob\left[r_{\min}\geq r\right]=\prod_{l=0}^{l^*}\left[\bar{Q_l}(r)\right]^{g_d(l)}\;,\;\;{\rm with}\;\;\bar{Q_l}(r)=\Det\left[\mathbb{I}-P_{[0,r]}K_l P_{[0,r]}\right]\;.
\ee  

The results considered up to now are exact for any finite value of $N$. We will now consider the large $N$ limit.

\subsection{Local density approximation and bulk results}

The local density approximation (LDA) explained above in section \ref{sec_LDA} holds in the higher-dimensional setting. The $d$-dimensional Wigner function associated to this system takes in this approximation and in the large $N$ limit the scaling form
\be
W_N({\bf x},{\bf p})\approx \frac{\Theta(\mu-H({\bf x},{\bf p}))}{(2\pi \hbar)^d}=\frac{1}{(2\pi \hbar)^d}\Theta\left(\mu-\frac{{\bf p}^2}{2m}-v(|{\bf x}|)\right)\;.
\ee
The average density in position space is then obtained by integration  over ${\bf p}$
\be\label{LDA_d}
\rho_N({\bf x})=\int \frac{d^d {\bf p}}{N(2\pi \hbar)^d}\Theta\left[\mu-\frac{{\bf p}^2}{2m}-v\left(|{\bf x}|\right)\right]=\frac{\Omega_d}{N(2\pi)^d}\left(\frac{2m\left[\mu-v\left(|{\bf x}|\right)\right]}{\hbar^2}\right)^{\frac{d}{2}}\;,
\ee
where $\Omega_d=\pi^{d/2}/\Gamma(d/2+1)$ is the volume of the unit ball in dimension $d$.

\begin{figure}
\centering
\includegraphics[width=0.6\textwidth]{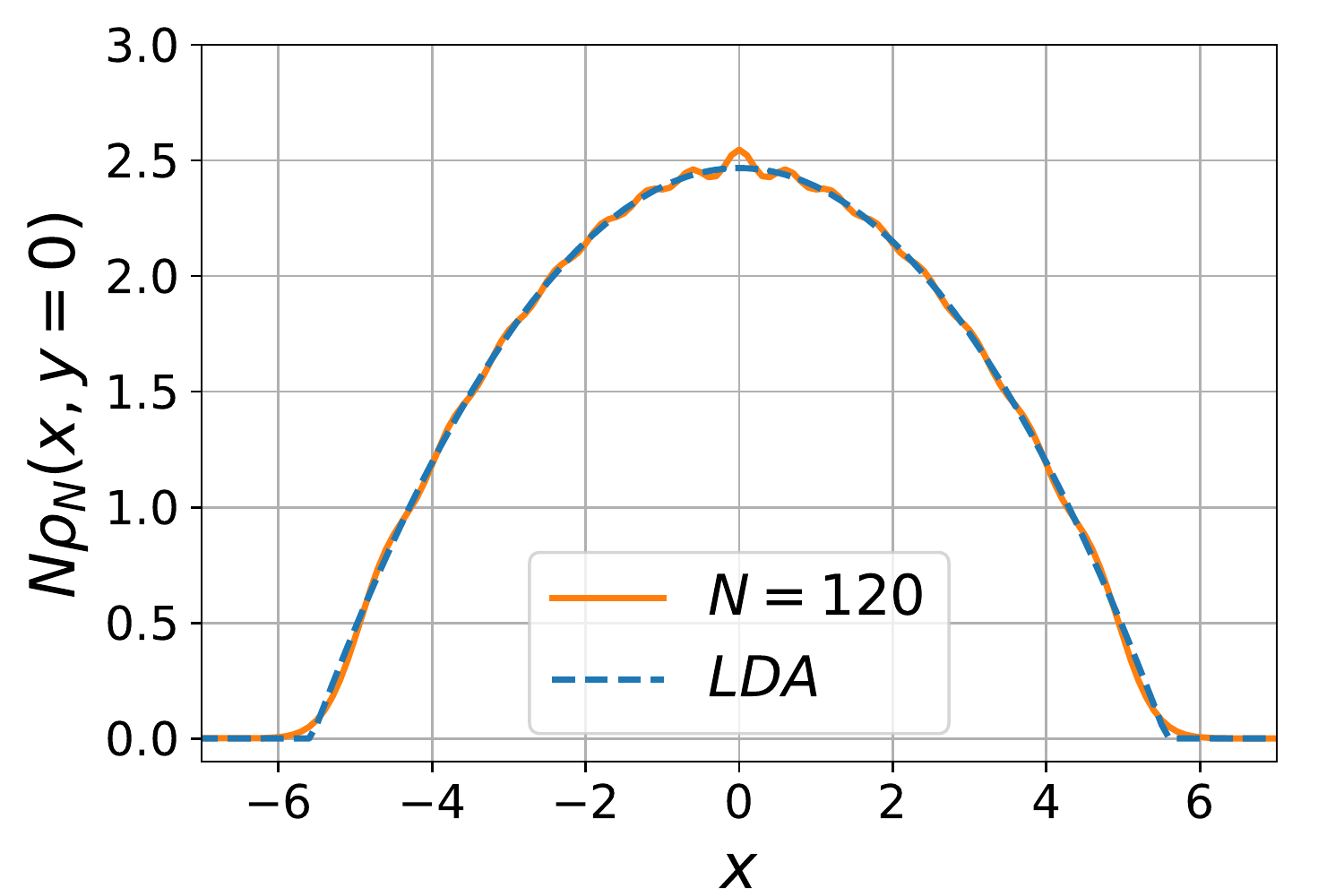}
\caption{Comparison between the bulk density obtained from LDA (blue) and the exact density profile for $N=120$ fermions (orange) in a two-dimensional harmonic potential $V(x)=\frac{1}{2}m\omega^2 (x^2+y^2)$ at zero temperature plotted as a function of $\alpha x$ for $y=0$ where $\alpha=\sqrt{m\omega/\hbar}$.}\label{Fig_dens_oh_2d}
\end{figure}

The correlation kernel $K_N({\bf x},{\bf y})$ in the bulk can also be obtained from this expression,
\be
K_N({\bf  x},{\bf  y})\approx \int \frac{d^d {\bf p}}{2\pi \hbar}e^{-\frac{\I {\bf p}\cdot({\bf x}-{\bf y})}{\hbar}}\Theta\left[\mu-\frac{{\bf p}^2}{2m}-v\left(u\right)\right]
=k_F(u)^d K_{\rm b}^{d}\left(k_F(u)|{\bf x}-{\bf y}|\right)\;,
\ee
where $u=|{\bf x}+{\bf y}|/2$ is the radius of the centre of mass of ${\bf x}$ and ${\bf y}$. The scaling function $K_{\rm b}^{d}(r)$ reads \cite{castin2006basic, scardicchio2009statistical}
\be\label{LDA_d_ker}
K_{\rm b}^{d}(r)=\frac{\J_{d/2}(r)}{(2\pi r)^{d/2}}\;,\,\;{\rm and}\;\;k_F(u)=\sqrt{\frac{2m\left[\mu-v\left(u\right)\right]}{\hbar^2}}\;.
\ee
As for the one-dimensional case, this scaling function for the kernel in the bulk can be obtained more rigorously from the short time expansion of the $d$-dimensional Euclidean propagator \cite{dean2015universal}. The density $\rho_N({\bf x})$ still has finite edges for $|{\bf x}|=r_{\rm e}$ such that $v(r_{\rm e})=\mu$ (c.f. Fig. \ref{Fig_dens_oh_2d}). At these edges, the LDA description breaks down and we will now consider the behaviour of the kernel in the large $N$ limit close to this edge.

\subsection{Large $N$ limit of the correlation kernel at the edge}

The scaling function for the large $N$ limit deep in the bulk was obtained in \cite{dean2015universal} using the Euclidean propagator method and confirmed the results obtained by the LDA.
%
%
%
%
%
%
The single particle Euclidean quantum propagator $G_d({\bf y},t|{\bf x},0)$ associated to the $d$-dimensional single particle Hamiltonian $\hat H$ is related to the $d$-dimensional $N$ particles correlation kernel in the same manner as in the case dimension one in Eq. \eqref{prop_K_1d} \cite{dean2015universal, dean2016noninteracting}
\be
G_d({\bf y},t|{\bf x},0)=\langle{\bf y}|e^{-\frac{\hat{H} t}{\hbar}}|{\bf x}\rangle=\sum_{{\bf k}}\overline{\phi}_{\bf k}({\bf x})\phi_{\bf k}({\bf y})e^{-\frac{\epsilon_{\bf k}t}{\hbar}}=\frac{t}{\hbar}\int_{0}^{\infty}e^{-\frac{\mu t}{\hbar}}K_{\mu}({\bf x},{\bf y})d\mu\;,\label{G_to_K}
\ee
where this propagator is now solution of the $d$-dimensional partial differential equation
\be
\hbar\partial_t G_d({\bf y},t|{\bf x},0)=\frac{\hbar^2}{2m}\Delta_{\bf y} G_d({\bf y},t|{\bf x},0)-v(|{\bf y}|)G_d({\bf y},t|{\bf x},0)\;,\;\;{\rm with}\;\;G({\bf y},0|{\bf x},0)=\delta^d({\bf x}-{\bf y})\;.
\ee
Note that it reduces to a scattering problem in the rotationally symmetric potential $v(|{\bf y}|)$.

At the edge, using again a general short time expansion \cite{makri1988correct, dean2016noninteracting} of the path integral representation of the propagator Dean et al. were able to find the $d$ dimensional scaling form of the correlation kernel close to a point ${\bf u}_{\rm e}$ such that $|{\bf u}_{\rm e}|=r_{\rm e}$ \cite{dean2015universal}
\be\label{K_airy_scal}
K_{\mu}({\bf x},{\bf y})\approx \frac{1}{w_N^d}K_{d}^{\rm s}\left(\frac{{\bf x}-{\bf u}_{\rm e}}{w_N},\frac{{\bf y}-{\bf u}_{\rm e}}{w_N}\right)\;,
\ee
where the superscript $^{\rm s}$ refers to the soft edge.
Defining the set of coordinates 
\be
u_n={\bf u}\cdot \frac{{\bf u}_{\rm e}}{r_{\rm e}}-r_{\rm e}\;,\;\;{\bf u}_t={\bf u}-(r_{\rm e}+u_n)\frac{{\bf u}_{\rm e}}{r_{\rm e}}
\ee
and with similar notations for ${\bf v}$, we can then express the scaling function $K_{d}^{\rm s}({\bf u},{\bf v})$ as
\be\label{K_d_soft}
K_{d}^{\rm s}({\bf u},{\bf v})=\int \frac{d^{d-1}{\bf l}}{(2\pi)^{d-1}}e^{\I ({\bf u}_t-{\bf v}_t)\cdot {\bf l}}K_{\Ai}(u_n+{\bf l}^2,v_n+{\bf l}^2)\;.
\ee
This scaling form generalises the Airy kernel in Eq. \eqref{k_airy} to higher dimension. It controls all the fluctuations of the Fermi gas close to the edge. In particular, we will now consider the fluctuations of the largest radius $r_{\max}$.

\subsubsection{Statistics of the maximal radius}

The CDF $\Prob\left[r_{\max}\leq r\right]$ of the largest radius $r_{\max}$ can be expressed in the large $N$ limit as the Fredholm determinant of the $d$-dimensional scaling form in Eq. \eqref{K_d_soft}. This expression is however not the most convenient to extract the large $N$ asymptotic result. Instead, the decomposition in orbital quantum numbers in Eq. \eqref{P_max_decomp} allows a deeper understanding. For each quantum number $l$, the radii of fermions form a determinantal point process whose average density can be obtained via the LDA in Eq. \eqref{LDA_1d} as
\be
\rho_{m_l}(r)\approx\frac{1}{\pi\hbar m_l}\sqrt{2m\left[\mu-v_{l,d}(r)\right]}\;.
\ee
In particular, it vanishes as $\rho_{m_l}(r)\sim \sqrt{r_{{\rm e},l}-r}$ at an $l$-dependent edge $r_{{\rm e},l}$ such that $v_{l,d}(r_{{\rm e},l})=\mu$. Close to this edge, the fluctuations are controlled by the one-dimensional Airy kernel in Eq. \eqref{k_airy}. In particular, the statistics of the largest radius $r_{\max,l}$ for a given value of $l$ is given by the 
%
%
%
%
%
%
%
Tracy-Widom distribution \cite{dean2017statistics}
\be
\Prob\left[r_{\max,l}\leq r\right]\approx {\cal F}_2\left(\frac{r-r_{{\rm e},l}}{w_{N,l}}\right)\;,
\ee
where both the positions of the edge $r_{{\rm e},l}$ and the typical scale $w_{N,l}$ depend explicitly on $l$. 
In the case of the $d$-dimensional harmonic potential, Dean et al. analysed in \cite{dean2017statistics} the explicit scaling of these two parameters with $\mu$ and $l$. Using the product form of the PDF
\be
\Prob\left[r_{\max}\leq r\right]\approx \prod_{l=0}^{l^*}\left[{\cal F}_2\left(\frac{r-r_{{\rm e},l}}{w_{N,l}}\right)\right]^{g_d(l)}\;,
\ee
they were then able to show that in the limit of large $\mu$ with $l\sim \mu$, the typical regime of fluctuation of $r_{\max}$ is given by a Gumbel distribution
\be
\lim_{\mu\to \infty}\Prob\left[r_{\max}\leq a_{\mu}+b_{\mu} r\right]=G_{\rm I}(r)=\exp\left(-e^{-r}\right)\;,\label{max_rad_soft}
\ee
where the coefficients $a_{\mu},b_{\mu}$ can be obtained exactly \cite{dean2017statistics}. In particular the typical fluctuations are centred rather far from the edge of the density $|a_{\mu}-r_{\rm e}|\gg w_N$. In dimension $d>1$, the typical fluctuations of $r_{\max}$ become trivial and fall into one of the three universality classes for the extreme value statistics of i.i.d. random variables.

One can safely assume that this Gumbel scaling form will hold universally for other smooth rotationally invariant potentials $v(|{\bf x}|)$ as this Gumbel CDF reflects the independence of the radii $r_{\max,l}$ for different values of $l$. In particular, only a fraction of these radii $r_{\max,l}$ for $l$ close to the largest value $l^*$ contribute to this regime of typical fluctuations in the large $\mu$ (or equivalently $N$) limit \cite{dean2017statistics}. 

In this higher-dimensional setting, it becomes quite difficult to analyse the regimes of atypical fluctuations as there exists no analogous method to the Coulomb gas, explained in the context of RMT in Eq. \eqref{CG_herm}. This result closes the discussion of smooth potentials in dimension $d>1$ at zero temperature and we will now consider the effects of thermal fluctuations on the Fermi gas.

%

\section{Finite temperature systems in dimension $d\geq 1$}\label{finite_T_smooth}

\begin{figure}
\centering
\includegraphics[width=0.6\textwidth]{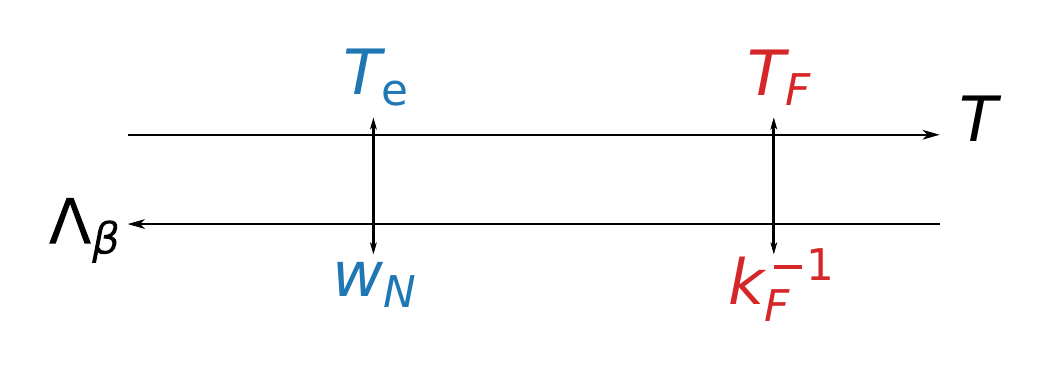}
\caption{Schemes of the typical temperature and  corresponding length scales in the bulk (red) and at the edge (blue).}\label{Fig_sacles_temp_length}
\end{figure}

Reaching ultra-low temperatures has been a goal for many decades to probe the effects of strong quantum correlations. It is currently possible in Fermi gases to reach temperatures of the order of a few $ \sim nK$ \cite{ketterle2008making}. However, these temperatures are not sufficiently small with respect to the Fermi energy to neglect completely the effects of thermal fluctuations and we will now see how they can be introduced. One naturally expects the effects of thermal fluctuations to become relevant in the spatial statistics when the de Broglie thermal length scale defined as $\Lambda_{\beta}=\sqrt{\frac{2\pi\hbar^2 \beta}{m}}$ becomes of the order of the typical inter-particle distance. In the bulk of the system, this distance is given by $k_F^{-1}\sim \ell_N$ where $k_F=\sqrt{2m\mu}/\hbar$ (with e.g. $k_F^{-1}\sim N^{-1/2}$ for the harmonic oscillator) is the Fermi wave-vector, while at the edge it is set by $w_N=\hbar^{2/3}/(2m V'(r_{\rm e}))^{1/3}$ \eqref{w_N} (with e.g. $w_N\sim N^{-1/6}$ for the harmonic oscillator) \cite{dean2015finite} (c.f. Fig. \ref{Fig_sacles_temp_length}). The effects of the thermal fluctuations will appear at the edge at lower temperature $T\sim T_{\rm e}$ than in the bulk and at temperatures of order $T\sim T_F$, the edge is effectively washed out by thermal fluctuations. 

At finite temperature, one needs to consider all the single particle excited states ${\bf k}$. A many body state with $N$ fermions can be uniquely defined by the set $\{{\bf k}_{i_1},\cdots,{\bf k}_{i_N}\}$ of its occupied states or by the set of occupation numbers $\{n_{\bf k}\}=\{n_{{\bf k}_1},n_{{\bf k}_2},\cdots\}$ for all states ${\bf k}$. For each given of these many-body state, the associated many body wave function is built as the Slater determinant
\be
\Psi_{\{n_{\bf k}\}}({\bf x}_1,\cdots,{\bf x}_N)=\frac{1}{\sqrt{N!}}\det_{1\leq l,m\leq N}\phi_{{\bf k}_{i_l}}({\bf x}_m)\;.
\ee
The quantum probability associated to this state then reads
\be
\left|\Psi_{\{n_{\bf k}\}}({\bf x}_1,\cdots,{\bf x}_N)\right|^2=\frac{1}{N!}\det_{1\leq i,j\leq N}K_{N}({\bf x}_i,{\bf x}_j,\{n_{\bf k}\})\;,
\ee
with the correlation kernel
\be
K_{N}({\bf x},{\bf y},\{n_{\bf k}\})=\sum_{l=1}^N \overline{\phi}_{{\bf k}_{i_l}}({\bf x})\phi_{{\bf k}_{i_l}}({\bf y})=\sum_{\bf k} n_{\bf k}\overline{\phi}_{\bf k}({\bf x})\phi_{\bf k}({\bf y})\;.
\ee
Note that as $n_{\bf k}^2=n_{\bf k}$, one can show that this kernel is reproducible, i.e. fulfils the property 
\begin{align}
\int d^d{\bf y}\, K_{N}({\bf x},{\bf y},\{n_{\bf k}\})K_{N}({\bf y},{\bf z},\{n_{\bf k}\})&=\sum_{{\bf k}_1,{\bf k}_2}n_{{\bf k}_1}n_{{\bf k}_2}\overline{\phi}_{{\bf k}_1}({\bf x})\int d^d{\bf y}\,\phi_{{\bf k}_1}({\bf y})\overline{\phi}_{{\bf k}_2}({\bf y})\phi_{{\bf k}_2}({\bf z})\nn\\
&=\sum_{{\bf k}}n_{{\bf k}}^2 \overline{\phi}_{{\bf k}}({\bf x})\phi_{{\bf k}}({\bf z})=K_{N}({\bf x},{\bf z},\{n_{\bf k}\})\;,
\end{align}
and fermions form a determinantal point process for any {\it fixed} many body state $\{n_{\bf k}\}$. To obtain the joint PDF of the positions at equilibrium, one then needs to compute the thermal average over the occupation numbers $n_{\bf k}$'s. 

If the number of fermions $N=\sum_{\bf k}n_{\bf k}$ is fixed, one needs to take the average in the canonical ensemble in which these occupation numbers are correlated \cite{giraud2018correlations}. The quantum and thermal joint probability of the positions reads in the canonical ensemble
\be
P_{\rm joint}^{\rm C}({\bf x}_1,\cdots,{\bf x}_N)=\frac{1}{Z_N^{\rm C}(\beta)}\sum_{\{n_{\bf k}\}=0,1}e^{-\beta\sum_{{\bf k}}n_{\bf k}\epsilon_{\bf k}}\left|\Psi_{\{n_{\bf k}\}}({\bf x}_1,\cdots,{\bf x}_N)\right|^2 \delta_{N,\sum_{\bf k}n_{\bf k}}\;,
\ee
where the superscript $^{\rm C}$ refers to the canonical ensemble, $Z_N^{\rm C}(\beta)$ is the canonical partition function and we recall that $\{n_{\bf k}\}=\{n_{{\bf k}_1},n_{{\bf k}_2},\cdots\}$. One can then show that because of the correlations between occupation numbers, the positions of fermions do not form a determinantal point process at finite value of $N$ \cite{dean2016noninteracting}.

To circumvent this difficulty, one can describe the system in the grand canonical ensemble where the temperature and the chemical potential $\mu$ (which is different from $\epsilon_F$ at finite temperature) is fixed while the number of particles fluctuates. In this ensemble, the occupation numbers become completely uncorrelated, and are Poisson distributed with a mean value given by the Fermi-Dirac statistics
\be
\overline{n_{\bf k}}=n_{\rm FD}(\epsilon_{\bf k})=\frac{1}{1+e^{\beta(\epsilon_{\bf k}-\mu)}}=\frac{\zeta}{\zeta+e^{\beta\epsilon_{\bf k}}}\;,
\ee 
where $\zeta=e^{\beta\mu}$ is the fugacity.
One can then show that the positions of fermions in the grand-canonical ensemble form a determinantal point process \cite{dean2016noninteracting}
\be
P_{\rm joint}^{\rm GC}({\bf x}_1,\cdots,{\bf x}_N)=\frac{1}{N}\det_{1\leq i,j\leq N}K_{\mu}^{\beta}({\bf x}_i,{\bf x}_j)\;,
\ee
where the superscript $^{\rm GC}$ refers to the grand-canonical ensemble. The correlation kernel $K_{\mu}^{\beta}({\bf x},{\bf y})$ is reproducible and reads
\be
K_{\mu}^{\beta}({\bf x},{\bf y})=\sum_{\bf k}n_{\rm FD}(\epsilon_{\bf k})\overline{\phi}_{\bf k}({\bf x})\phi_{\bf k}({\bf y})=\sum_{\bf k}\frac{\overline{\phi}_{\bf k}({\bf x})\phi_{\bf k}({\bf y})}{1+e^{\beta(\epsilon_{\bf k}-\mu)}}\;.\label{K_GC}
\ee

A basic principle of statistical mechanics is that the mean value of observables coincide in all the statistical ensemble in the thermodynamic limit, which corresponds for the canonical ensemble to $N,V \to \infty$ with $N/V$ fixed. For instance,
\be\label{N_mu}
N=\sum_{\bf k}n_{\rm FD}(\epsilon_{\bf k})=\sum_{\bf k}\frac{1}{1+e^{\beta(\epsilon_{\bf k}-\mu)}}\;.
\ee 
However, in general the fluctuations in different ensembles are different \cite{texier2017physique, pathria, giraud2018correlations}. It was shown by Dean et al. that by taking the large $N$ limit in the canonical ensemble, one recovers the local correlations of the grand-canonical ensemble \cite{dean2016noninteracting}. It is however not the case for global observables over the whole system \cite{grela2017kinetic,grabsch2018fluctuations}.
One can then use the determinantal structure with the correlation kernel of the grand-canonical ensemble given in Eq. \eqref{K_GC}.

We close this section by mentioning a useful relation between the finite temperature correlation kernel $K_{\mu}^{\beta}$ and its zero-temperature counterpart $K_{\mu}({\bf x},{\bf y})$. This relation is obtained by observing that \cite{dean2016noninteracting}
\be
\partial_{\mu}K_{\mu}({\bf x},{\bf y})=\sum_{\bf k}\overline{\phi}_{\bf k}({\bf x})\phi_{\bf k}({\bf y})\delta(\mu-\epsilon_{\bf k}),
\ee
such that the finite temperature kernel is expressed as
\be
K_{\mu}^{\beta}({\bf x},{\bf y})=\int_{0}^{\infty}\frac{d\epsilon }{1+e^{\beta(\epsilon-\mu)}}\partial_{\epsilon}K_{\epsilon}({\bf x},{\bf y})=\sum_{\bf k}\frac{\overline{\phi}_{\bf k}({\bf x})\phi_{\bf k}({\bf y})}{1+e^{\beta(\epsilon_{\bf k}-\mu)}}=\int_{0}^{\infty}\frac{K_{\epsilon}({\bf x},{\bf y})d\epsilon}{4\cosh^2\left(\frac{\beta}{2}(\epsilon-\mu)\right)}\;.\label{K_mu_to_K_b}
\ee
Note that the chemical potential appears explicitly in this equation and it will be essential in our analysis to consider its variation with the temperature.

\subsubsection{Thermodynamics of the gas: Chemical potential in the large $N$ limit}

To study the large $N$ limit, we first consider the variations of the finite temperature chemical potential $\mu$ with $N$ and $\beta$. The sum in Eq. \eqref{N_mu} can be replaced by an integral over $\epsilon$ by introducing the density of states $\hat \rho_{\rm st}(\epsilon)$
\be
N=\int_0^{\infty} \frac{ \hat \rho_{\rm st}(\epsilon)}{1+e^{\beta(\epsilon-\mu)}}d\epsilon\;,\;\;{\rm with}\;\;\hat \rho_{\rm st}(\epsilon)=\sum_{\bf k}\delta(\epsilon-\epsilon_{\bf k})\;,
\ee
where the density of states $\hat \rho_{\rm st}(\epsilon)$ is a distribution that depends explicitly on the spectrum, and therefore the confining potential. In the large $N$ limit, the density of states can be replaced by a continuous function $\hat \rho_{\rm st}(\epsilon)\to \rho_{\rm st}(\epsilon)$. One can then obtain an explicit expression for the variation of $N$ as a function of $\mu$ and $\beta$. For instance, in the case of the harmonic potential, one has
\be
\rho_{\rm st}(\epsilon)=\frac{1}{(d-1)! \hbar \omega}\left(\frac{\epsilon}{\hbar \omega}\right)^{d-1}\;,\;\;N=-\frac{1}{(\beta\hbar \omega)^d}\Li_{d}\left(-\zeta\right)\;,
\ee
where $\Li_d(s)=\sum_{k=1}^{\infty}k^{-d} s^{k}$ is the polylogarithm function and we recall that $\zeta=e^{\beta\mu}$ is the fugacity.
This result will be useful to obtain the behaviour of the kernel in the large $N$ limit as we now analyse.

\subsubsection{Kernel in the bulk: finite temperature LDA}

In the bulk, the exact results in the large $N$ limit still coincide with the predictions from the finite temperature local density approximation \cite{dean2016noninteracting}. In this approximation, we replace the Wigner function in the large $N$ limit by the Fermi-Dirac distribution evaluated at the value of the classical Hamiltonian at the local point $({\bf x},{\bf p})$ in the phase space \cite{bartel1985extended, castin2006basic}
\be
W_N({\bf x},{\bf p})=\frac{1}{(2\pi\hbar)^d}\frac{1}{1+e^{\beta\left(H({\bf p},{\bf x})-\mu\right)}}=\frac{1}{(2\pi\hbar)^d}\frac{\zeta}{\zeta+e^{\beta\left(\frac{{\bf p}^2}{2m}+V({\bf x})\right)}}\;.
\ee
The average density is then obtained by integrating over the momentum ${\bf p}$ as
\be
\rho_N({\bf x})\approx \int \frac{d^{d}{\bf p}}{N(2\pi\hbar)^d}\left(1+e^{\beta\left(\frac{{\bf p}^2}{2m}+V({\bf x})-\mu\right)}\right)^{-1}=-\frac{1}{N\Lambda_{\beta}^{d}}\Li_{\frac{d}{2}}\left(-\zeta e^{-\beta V({\bf x})}\right)\;,\label{LDA_t}
\ee 
where we recall that $\Lambda_{\beta}=\sqrt{\frac{2\pi\hbar^2 \beta}{m}}$ is the de Broglie thermal wave-length. 

The correlation kernel is obtained in the bulk by inverting a Weyl transform as
\begin{align}
K_{\mu}^{\beta}({\bf x},{\bf y})&\approx\int \frac{d^d {\bf p}}{N(2\pi \hbar)^d}e^{-\frac{\I {\bf p}\cdot({\bf x}-{\bf y})}{\hbar}}\left(1+e^{\beta\left(\frac{{\bf p}^2}{2m}+V({\bf u})-\mu\right)}\right)^{-1}\nn\\
&=\Lambda_{\beta}^{-d} K_{d,\beta}^{\rm b}\left(\frac{|{\bf x}-{\bf y}|}{\Lambda_{\beta}}\right)\;,
\end{align}
where ${\bf u}=({\bf x}+{\bf y})/2$ and the scaling function $K_{d,b}^{\rm b}(r)$ reads \cite{dean2016noninteracting}
\be
K_{d,\beta}^{\rm b}(r)=\int_0^{\infty}\frac{\zeta dk}{\zeta+e^{\frac{k^2}{4\pi}+\beta V({\bf u})}}\left(\frac{k}{2\pi}\right)^{\frac{d}{2}}\frac{\J_{d/2-1}(k r)}{r^{\frac{d}{2}-1}}\;.\label{K_bulk_T}
\ee
Next, we consider the behaviour of the kernel at the edge, considering only the case $d=1$ for simplicity.

\subsubsection{Kernel at the edge in one dimension and connection with KPZ equation}

The kernel can be obtained from the propagator method at the soft edge in general dimension $d\geq 1$ \cite{dean2016noninteracting}. We only discuss here the result at the edge in  dimension $d=1$. At high temperature $T\sim T_F$ (with e.g. $T_F\sim N$ for the harmonic potential), the density is non-zero up to infinity and strictly speaking there is no edge. However, in the regime of low  temperature $T\sim T_{\rm e}\ll T_F$ (with e.g. $T_{\rm e}\sim N^{1/3}$ for the harmonic potential), the density close to the zero-temperature edge $r_{\rm e}$ is small and the correlations are still quite different from the bulk results. The scale of temperature to have non-trivial statistics at the edge is set such that  the typical scale at zero temperature $w_N$ and the de Broglie wave-length $\Lambda_{\beta}$ are of same order $w_N\sim \Lambda_{\beta}$ (c.f. Fig. \ref{Fig_sacles_temp_length}). Rescaling the kernel close to the zero-temperature edge, one obtains \cite{dean2015finite}
\be
K_{\mu}^{\beta}(x,y)\approx \frac{1}{w_N}K_{1,b}^{\rm s}\left(\frac{x-r_{\rm e}}{w_N},\frac{x-r_{\rm e}}{w_N}\right)\;,
\ee
where $b=T_{\rm e}/T=(\hbar V'(r_{\rm e}))^{2/3}/(2m)^{1/3}/T$ is the rescaled inverse temperature \cite{dean2016noninteracting}. The scaling function $K_{1,b}^{\rm s}(u,v)$ is a finite temperature extension of the Airy kernel (see the similar structure in Eq. \eqref{k_airy})
\be\label{k_1d_finite_t}
K_{1,b}^{\rm s}(u,v)=\int_{-\infty}^{\infty}\frac{ds}{1+e^{-b s}}\Ai(s+u)\Ai(s+v)\;.
\ee
Note that in the limit $b\to +\infty$, $(1+e^{-bs})^{-1}\to \Theta(s)$ and one recovers the Airy kernel \eqref{k_airy}. We consider finally the fluctuations of the position $x_{\max}$ of the rightmost fermion at finite temperature which has interesting connections with the Kardar-Parisi-Zhang (KPZ) equation \cite{kardar1986dynamic} .
The statistics of $x_{\max}$ at finite temperature has a smooth transition from the Tracy-Widom distribution ${\cal F}_2(s)$ at zero temperature to a Gumbel distribution at large temperature $T\gg T_{\rm e}$ \cite{johansson2007gumbel, dean2015finite}, with a scaling function universal with respect to the confining potential $V(x)$.
The crossover 
function is given by the Fredholm determinant \cite{dean2015finite}
\begin{align}
\Prob\left[x_{\max}\leq r_{\rm e}+w_N s\right]&=\Det\left[\mathbb{I}-P_{[s,\infty)}K_{1,b}^{\rm s}P_{[s,\infty)}\right]\nn\\
&=\exp\left(-\sum_{p=1}^{\infty}\frac{1}{p}\Tr\left[\left(P_{[s,\infty)}K_{1,b}^{\rm s}P_{[s,\infty)}\right)^p\right]\right)\;.\label{trace_exp_t_airy}
\end{align}
In the limit of large $T$, i.e. $b\ll 1$, a thorough analysis allows to obtain that this Fredholm determinant is dominated by the first trace, i.e. $p=1$, in the expansion of Eq \eqref{trace_exp_t_airy} and to recover the Gumbel distribution \cite{le2016exact}. The Fredholm determinant of this kernel appears in a seemingly unrelated problem: the height fluctuations of an interface described by the Kardar-Parisi-Zhang equation \cite{dean2015finite, le2016exact}. In this model, one considers a height field $h(x,t)$ in $1+1$ space and time dimension which satisfies the non linear stochastic differential equation \cite{kardar1986dynamic} 
\begin{align}
&\partial_t h=\partial_{x}^2 h+(\partial_x h)^2+\sqrt{2}\eta(x,t)\;,\\
&{\rm with}\;\;\moy{\eta(x,t)}=0\;\;\moy{\eta(x,t)\eta(x',t')}=\delta(x-x')\delta(t-t')\;,\nn
\end{align}
and $\eta(x,t)$ is a Gaussian white noise. For the {\it droplet} initial condition $h(x,0)=x/\delta$ with $\delta\ll 1$, the exponential generating function of the moments of the rescaled height $\tilde h(0,t)= (h(0,t)+t/12)t^{-1/3}$ is expressed as the Fredholm determinant of the kernel in Eq. \eqref{k_1d_finite_t} \cite{calabrese2010free,sasamoto2010one}
\be
\Big\langle\exp\left(-e^{t^{1/3}(\tilde h(0,t)-\xi)}\right)\Big\rangle=\Det\left[\mathbb{I}-P_{[\xi,\infty)}K_{1,t^{1/3}}^{\rm s}P_{[\xi,\infty)}\right]\;.
\ee
The role of the inverse temperature $b$ in the case of the free fermions is played by the time $t^{1/3}$ in the case of the fluctuating interface.

We close this section by mentioning that the regime of typical fluctuation of $r_{\max}$ in dimension $d>1$ is somewhat of less interest as it is given by a Gumbel distribution both at zero and high temperature. 

In this section, we have reviewed the description of the spatial fluctuations of fermions confined by a smoothly varying potential. In particular, we have seen that the fluctuations become universal in the large $N$ limit both in the bulk of the density and close to the edge, where the density vanishes smoothly. In the next section, we consider the fluctuations for a ``hard edge'', where the potential imposes an abrupt drop of the density and where the analysis of the edge behaviour presented here does not hold.

\chapter{Non-interacting fermions in hard-edge potentials}
\label{ch: ferm_hard_edge}

In this chapter, we consider the $d$-dimensional system formed by $N$ non-interacting, spin-less, identical fermions of mass $m$ in a hard edge potential. In particular, we will analyse in detail the edge properties of the Fermi gas trapped by the rotationally symmetric potentials of the form
\be\label{trunc_pot_intro}
V({\bf x})=\begin{cases}
v(|{\bf x}|)&\;,\;\;|{\bf x}|<R\;,\\
&\\
+\infty&\;,\;\;|{\bf x}|\geq R\;,
\end{cases}
\ee
where $v(r)$ is a smooth potential ($v(r)\in{\cal C}_1(\mathbb{R})$). For $v(r)=0$ this problem reduces to the $d$-dimensional spherical hard-box potential.
The general potential in Eq. \eqref{trunc_pot_intro} imposes that all the particles stay within the boundary of the wall $|{\bf x}|\leq R$. All the single particle wave-functions $\phi_{\bf k}({\bf x})$ associated to this problem are identically zero for $|{\bf x}|>R$. By continuity of these wave-functions, it imposes Dirichlet boundary conditions for $|{\bf x}|=R$. Applying the finite temperature local density approximation to this system, one obtains from Eq. \eqref{LDA_t}
\be
\rho_N({\bf x})=\begin{cases}
\displaystyle -\frac{1}{N\Lambda_{\beta}^{d}}\Li_{\frac{d}{2}}\left(-\zeta e^{-\beta v(|{\bf x}|)}\right)&\;,\;\;|{\bf x}|<R\\
&\\
0 &\;,\;\;|{\bf x}|\geq R\;.
\end{cases}
\ee
The density vanishes abruptly for $|{\bf x}|=R$ and for any value of the temperature, creating a {\it hard edge}. This situation is quite different to the case of smoothly varying potentials where a {\it soft edge} emerged only at zero temperature. 
%
The system forms a determinantal point process and the correlations \eqref{fermions_p_point} are therefore entirely determined by the knowledge of the associated correlation kernel defined in \eqref{K_N_gen}. The results obtained in the previous section for the large $N$ behaviour of the correlation kernel fail to describe the fluctuations at the hard edge. One needs to develop alternative methods to analyse this situation. The introduction of a hard edge in the system will also drastically change the behaviour of the extreme value statistics. It may already be seen in the limit of high temperature, where one recovers the classical limit (as seen in chapter \ref{chap:fermions_intro}), and the extreme value statistics is described by i.i.d. random variables. For a system without hard wall ($R=\infty$) and confined in the spherically symmetric potential $v(r)\gg \ln r$, as $r\to \infty$ this problem will fall in the Gumbel universality class. However, if a finite hard edge is imposed, the problem now falls into the Weibull universality class (c.f. chapter \ref{intro_iid}).
At lower temperature, the positions of fermions become strongly correlated and one might then wonder what will be the equivalent of the Tracy-Widom $\beta=2$ distribution in this new setting.
We first consider a specific choice of potential for which there is an exact mapping to the Laguerre Unitary Ensemble (LUE) of RMT and where a hard edge naturally occurs.

\section{Hard edge in models of fermions}\label{he_fermions}

\begin{figure}
\centering
\includegraphics[width=0.6\textwidth]{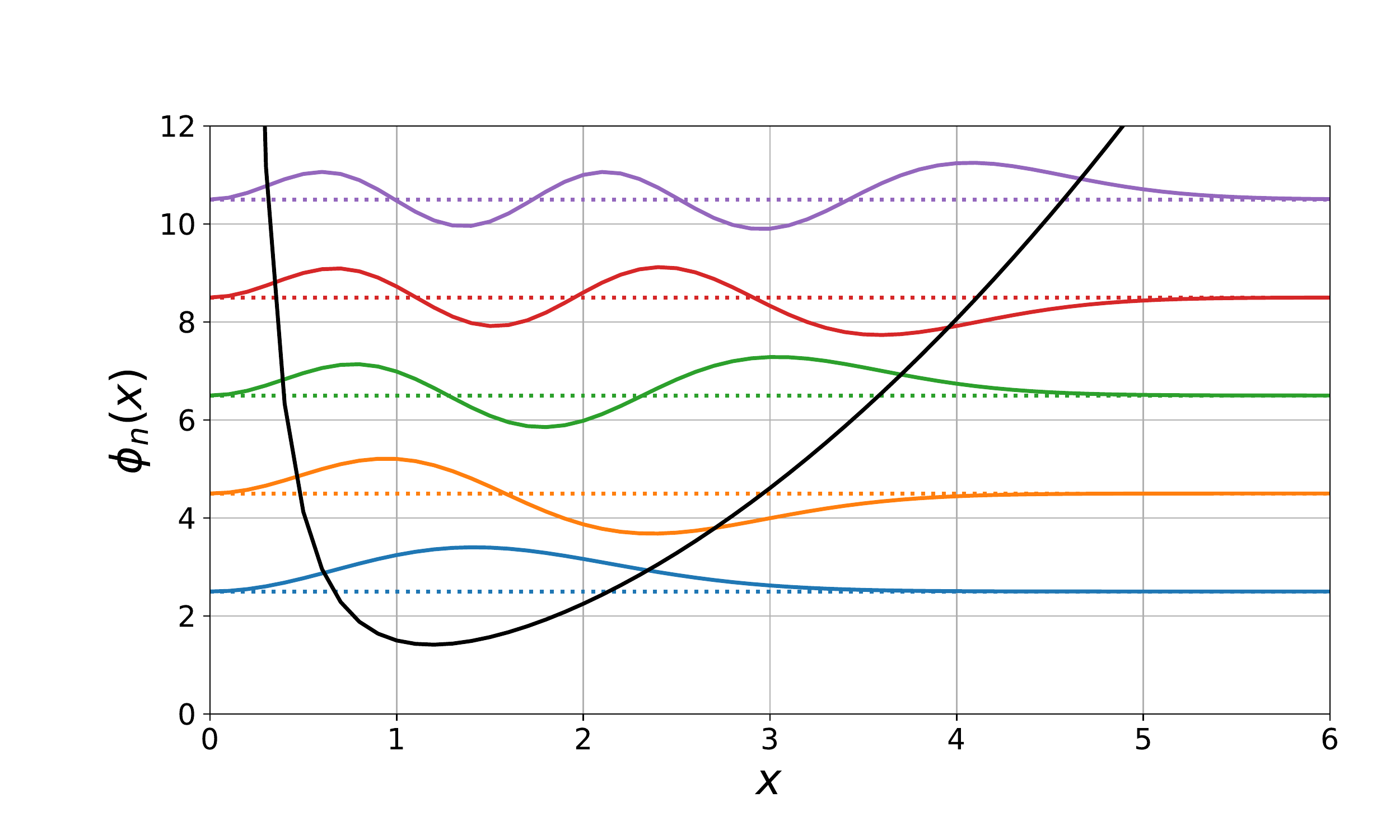}
\caption{Single-particle energies and wave-functions associated to a system of $N=5$ fermions at zero temperature in the potential described in Eq. \eqref{x2_1_x2_pot}.}\label{Fig_phi_lag}
\end{figure}

In this section, we consider $N$ non-interacting fermions confined by the trapping potential
\be
v(r)=\begin{cases}
\displaystyle +\infty &\;,\;\;r\leq 0\;.\\
&\\
\displaystyle \frac{1}{2}m\omega^2 r^2+ \frac{\hbar^2 a(a-1)}{2m r^2}&\;,\;\;r>0\;.
\end{cases}\label{x2_1_x2_pot}
\ee
Note that for $a=l+(d-1)/2$, this potential corresponds exactly to the effective potential applied to fermions in a $d$-dimensional spherically symmetric harmonic potential and with orbital quantum number $l$ (c.f. Eq. \eqref{v_eff_l}). For this potential, the Hamiltonian is exactly solvable and the wave-functions and energies read 
\be\label{wf_lag}
\phi_n(r)=\Theta(r) c_{a,k}L_{n}^{a-\frac{1}{2}}\left(\alpha^2 r^2\right) (\alpha r)^{a}e^{-\frac{\alpha^2 r^2}{2}}\;,\;\;\epsilon_n=\hbar\omega\left(2n+a+\frac{1}{2}\right)\;,\;\;\alpha=\sqrt{\frac{m\omega}{\hbar}}\;.
\ee 
where $L_{k}^{\nu}(x)=x^{-\nu}e^{x}\partial_x^k (e^{-x}x^{k+\nu})/k!$ is the Laguerre polynomial of degree $k$ and $c_{a,n}^2=2\, n!/\Gamma(n+a+1/2)$ is a constant ensuring the normalisation. The single particle wave functions associated to the ground state of a system with $N=5$ fermions are represented in Fig. \ref{Fig_phi_lag}. As in the case of the harmonic potential, the many body joint ground state probability of the positions $r_i$'s of the fermions can be computed exactly by using the Vandermonde identity in Eq. \eqref{VdM} \cite{nadal2009nonintersecting}
\be
|\Psi_0(r_1,\cdots,r_N)|^2=\frac{1}{Z_N(a)}\prod_{i<j}|r_i^2-r_j^2|^2 \prod_{k=1}^N r_k^{2a}e^{-\alpha^2 r_k^2}\;,\label{P_joint_x2_1_x2}
\ee 
where $Z_N(a)$ is a normalisation constant. The fermions form a determinantal point process with correlation kernel
\be
K_N(r,r')=\sum_{n=0}^{N-1}\overline{\phi}_n(r)\phi_n(r')\;,\label{kernel_1_x_2}
\ee
where $\phi_n(r)$ is given in Eq. \eqref{wf_lag}.
The joint distribution in Eq. \eqref{P_joint_x2_1_x2} also appears in RMT for the eigenvalues of the Wishart Unitary Ensemble, also called Laguerre Unitary Ensemble (LUE), which we now briefly defined.
Matrices belonging to this ensemble are obtained by first filling a matrix $X$ of size $M\times N$ with $M\geq N$ with i.i.d. complex Gaussian variables
\be
x_{ij}\sim {\cal N}\left(0,\frac{1}{2\sqrt{N}}\right)+\I\,{\cal N}\left(0,\frac{1}{2\sqrt{N}}\right)\;,\;\;i=1,\cdots,M\;,\;\;j=1,\cdots,N\;.
\ee
The covariance matrix of its random entries $W=X^{\dag}X$ will be a $N\times N$ Hermitian matrix with real and positive eigenvalues. The joint PDF of its eigenvalues reads \cite{mehta2004random,forrester2010log,akemann2011oxford} 
\be
P_{\rm joint}^{\rm LUE}(\lambda_1,\cdots,\lambda_N)=\frac{1}{Z_N^{\rm LUE}(\alpha)} \prod_{i<j}|\lambda_i-\lambda_j|^2 \prod_{i=1}^N \lambda_i^{\nu} e^{-N \lambda_i}\;,\label{P_joint_LUE}
\ee
where $\nu=M-N$.  
Introducing the change of variables $\lambda_i=\alpha^2 r_i^2/N$, the two distributions in Eq. \eqref{P_joint_x2_1_x2} and \eqref{P_joint_LUE} coincide exactly with $\nu=a-1/2$. 

\begin{figure}
\centering
\includegraphics[width=0.6\textwidth]{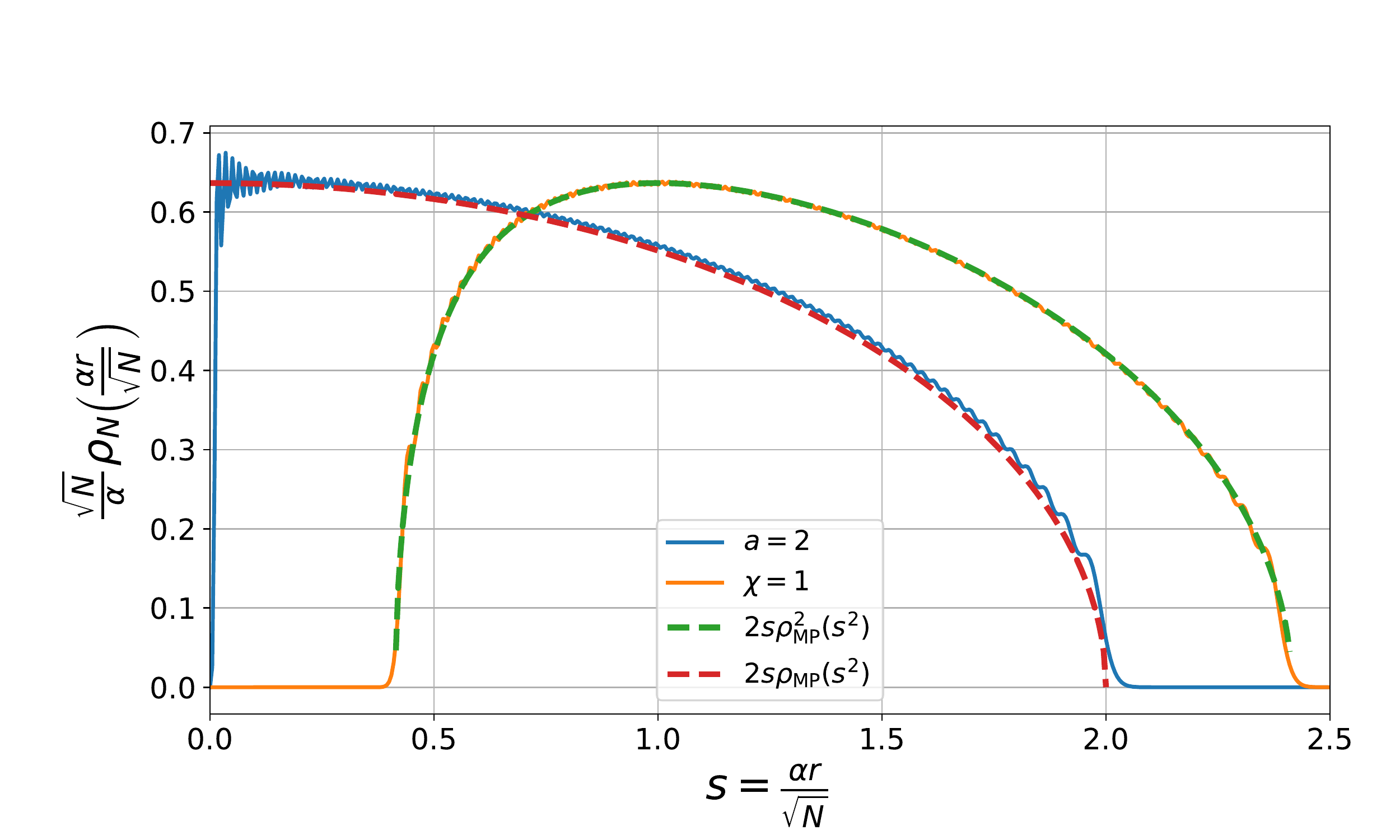}
\caption{Exact rescaled average density $\frac{\sqrt{N}}{\alpha}\rho_N\left(\frac{\alpha r}{\sqrt{N}}\right)$ for $N=100$ fermions in the potential $v(r)$ given in Eq. \eqref{x2_1_x2_pot} as a function of the rescaled position $s=\frac{\alpha r}{\sqrt{N}}$ for $a=2$ (in blue) and $\chi=a/N=1$ (in orange). These density are compared with  the large $N$ bulk density given for $a=2$ in Eq. \eqref{MP_a_1} (plotted in dashed red) and for $\chi=1$ in Eq. \eqref{MP_chi_1}
 (plotted in dashed green), showing a good agreement.}\label{Fig_wish_dens}
\end{figure}

In the large $N$ limit, the density in the system is obtained from the local density approximation in Eq. \eqref{LDA_1d} and reads
\be
\rho_N(x)\approx \Theta(r)\frac{\sqrt{k_F^2\, r^2- \alpha^2 r^4-a(a-1)}}{\pi N r}\;,\;\;{\rm with}\;\;k_F=\frac{\sqrt{2m\mu}}{\hbar}\approx\alpha\sqrt{4N+2a}\;.
\ee
In the limit $N\to \infty$ with $\chi=a/N$ fixed, the Fermi wave-vector $k_F=\alpha\sqrt{N(4+2\chi)}$ and the potential $v(r)\sim N^2 \chi^2/r^2$ becomes very strong close to the origin, preventing the fermions to reach $r=0$.
A finite gap opens between the origin and the left edge of the density, which reads in this limit
\be
\rho_N(r)\approx \frac{2\alpha^2 r}{N} \rho_{\rm MP}^{\chi}\left(\frac{\alpha^2 r^2}{N}\right)=\Theta(r)\frac{\sqrt{(N c-\alpha^2 r^2)(\alpha^2 r^2-N b)}}{N\pi r}\;,\label{MP_chi_1}
\ee
where $b=2+\chi-2\sqrt{1+\chi}$ and $c=2+\chi+2\sqrt{1+\chi}$. The density vanishes close to its edges in $r_{{\rm e}_{-}}=\alpha^{-1} N b$ and $r_{{\rm e}_{+}}=\alpha^{-1} N c$ as a square-root as seen in Fig. \ref{Fig_wish_dens}. This behaviour is characteristic of soft edges and one can indeed show that in the large $N$ limit, the fluctuations at this edge are described by the Airy kernel \cite{mehta2004random,forrester2010log,akemann2011oxford}.
%
%
%
Furthermore, taking the large $N$ limit with $a=O(1)$, $k_F\approx \alpha\sqrt{4N}$ and one can neglect the contribution from the potential close to the origin. One obtains the simple expression for the density
\be
\rho_N(r)\approx \frac{2\alpha^2 r}{N} \rho_{\rm MP}\left(\frac{\alpha^2 r^2}{N}\right)=\Theta(r)\frac{\alpha}{N\pi}\sqrt{4N-\alpha^2 r^2}\;,\label{MP_a_1}
\ee
where the function $\rho_{\rm MP}(\lambda)=\rho_{\rm MP}^{\chi=0}(\lambda)$ is the Mar\v cenko-Pastur distribution, associated to the eigenvalues of the LUE \cite{marvcenko1967distribution}.
In this case, while the average density vanishes as a square-root for $r=2\sqrt{N}/\alpha$, the boundary condition imposes an abrupt drop of this density for $r=0$ as seen in Fig. \ref{Fig_wish_dens}. This behaviour is neither captured by the LDA description nor by the {\it soft edge} description of the edge behaviour developed in the last chapter and is characteristic of a {\it hard edge}. We will now see how to obtain the associated correlation kernel. 

\subsubsection{Hard edge limit: Bessel kernel}

In the large $N$ limit, the typical inter-particle distance $\ell_N$ close to the hard edge can be evaluated by using that there are $O(1)$ particles in the interval $[0,\ell_N]$, i.e.
\be
\int_0^{\ell_N}\rho_{N}(x)dx \sim \frac{1}{N}\;\Rightarrow\;\ell_N\sim \frac{1}{\sqrt{N}}\sim k_F^{-1}\;.
\ee
Using the results from RMT, one can indeed show that on this typical scale the correlation kernel takes the scaling form \cite{mehta2004random,forrester2010log,akemann2011oxford}
\begin{align}
&K_{N}(r,r')=2k_F\sqrt{r\, r'} K_{\rm Be}^{a-1/2}(k_F^2 r^2,k_F^2 {r'}^2)\;,\;\;{\rm with}\nn\\
&2\sqrt{u v} K_{\rm Be}^{a-1/2}(u^2,v^2)=\int_0^{1}k \sqrt{u\,v}\J_{a-1/2}(k u)\J_{a-1/2}(k v)dk\;,\label{K_1_x_2}
\end{align}
where the scaling function $K_{\rm Be}^{\nu}\left(u,v\right)$ is called the Bessel kernel and reads
\be\label{k_bessel}
K_{\rm Be}^{\nu}(u,v)=\frac{1}{2}\int_{0}^1 k \J_{\nu}(k\sqrt{x})\J_{\nu}(k\sqrt{y})=\frac{\sqrt{v}\J_{\nu}(\sqrt{u})\J_{\nu-1}(\sqrt{v})-\sqrt{u}\J_{\nu-1}(\sqrt{v})\J_{\nu}(\sqrt{u})}{2(u-v)}\;,
\ee
and where $\J_{\nu}(u)$ is the Bessel function of first kind of index $\nu$.

\begin{figure}
\centering
\includegraphics[width=0.6\textwidth]{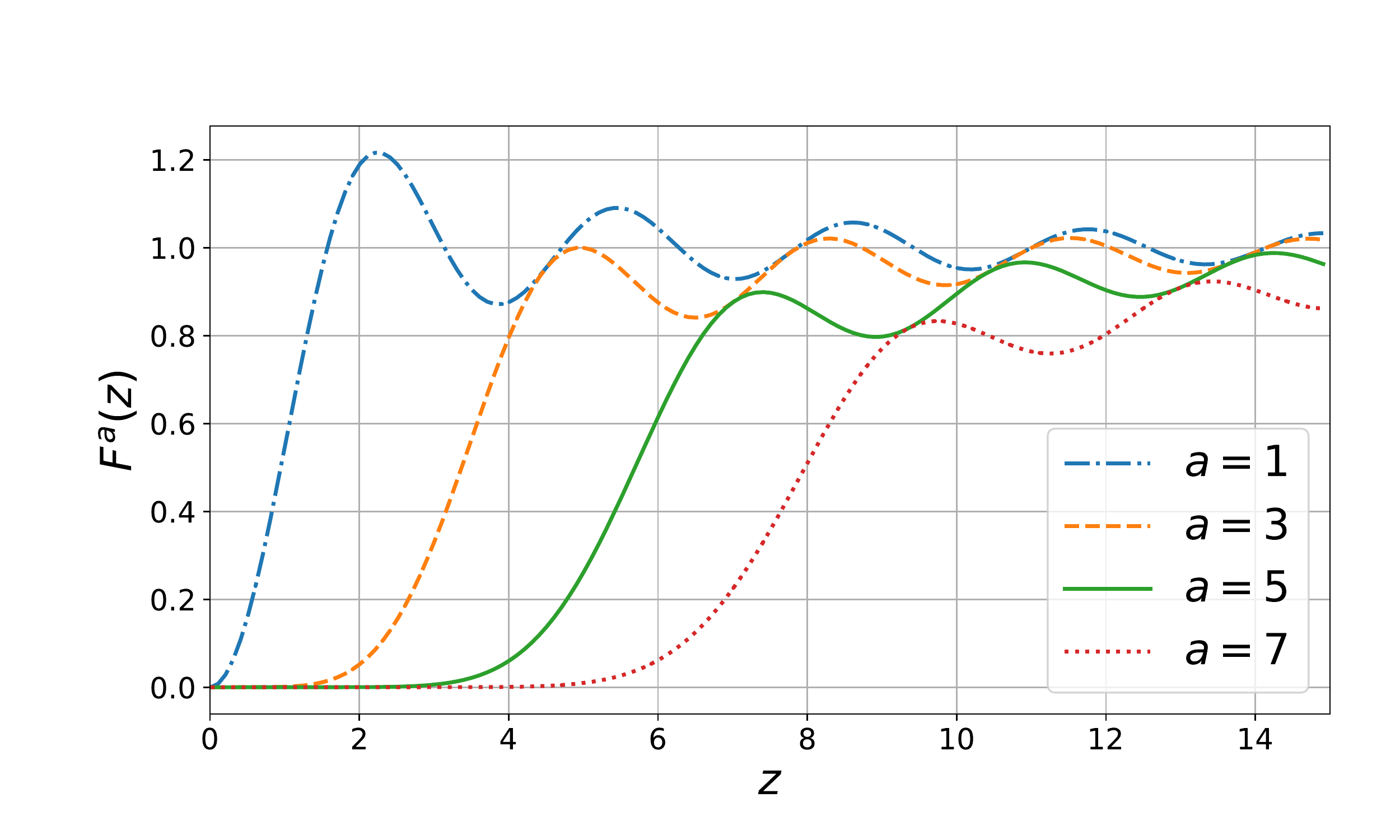}
\caption{Plot of the scaling function $F^a(z)$ in Eq. \eqref{F_a_z} representing the density profile close to the hard edge for $a=1,3,5,7$ respectively in blue, orange, green and red. As $a$ increases, a ``pseudo-gap'' opens between the edge of the density and the origin.}\label{Fig_F_alpha}
\end{figure}

From this result, we can extract the behaviour of the exact density profile close to the hard edge
\begin{align}
&\rho_N(x)=\frac{1}{N}K_N(x,x)\approx \frac{1}{\pi}F^a(k_F x)\;,\;\;{\rm with}\;,\nn\\
&F^a(z)=\frac{1}{2}\left[z \left(\J_{a-\frac{1}{2}}(z)^2+\J_{a+\frac{1}{2}}(z)^2\right)+(2a-1)\J_{a-\frac{1}{2}}(z)\J_{a+\frac{1}{2}}(z)\right]\;.\label{F_a_z}
\end{align}
The scaling function $F^a(z)$ vanishes at the origin with a power law depending explicitly on $a$
\be
F^a(z)\approx \frac{\pi}{\Gamma(a+1/2)\Gamma(a+3/2)}\left(\frac{z}{2}\right)^{2a}\;,\;\;z\to 0\;,
\ee
while it goes to $F^a(z)=1$ for $z\to \infty$, matching smoothly with the bulk density in Eq. \eqref{MP_a_1}. This density profile is plotted in Fig. \ref{Fig_F_alpha} for several values of $a$.

The Bessel kernel controls the fluctuations close to the hard edge and in particular the fluctuations of the leftmost fermion $r_{\min}=\displaystyle \min_{1\leq i\leq N}r_i$ that we now analyse.

\subsubsection{Extreme value statistics}
Before considering the case of $r_{\min}$, we mention that using the exact mapping onto the largest eigenvalue $\lambda_{\max}$ in the LUE, the regime of typical fluctuations of $r_{\max}$ is given by the Tracy-Widom distribution
as it was the case in the harmonic potential in connection to the GUE. But the large deviations associated to the LUE $\Phi_{\pm}^{{\rm L}\beta{\rm E}}(\lambda,\nu)$ (these functions are the same for all $\beta$ Laguerre ensembles), controlling the atypical fluctuations, depend explicitly on $\nu$
and are different from the case of the GUE in Eq. \eqref{suma_GUE}. The left rate function $\Phi_{-}^{{\rm L}\beta{\rm E}}(\lambda,\nu)$ was computed in Ref. \cite{vivo2007large} while the right large deviation function $\Phi_{+}^{{\rm L}\beta{\rm E}}(\lambda,\nu)$ was obtained in Ref. \cite{majumdar2009large}. 

\begin{figure}
\centering
\includegraphics[width=0.6\textwidth]{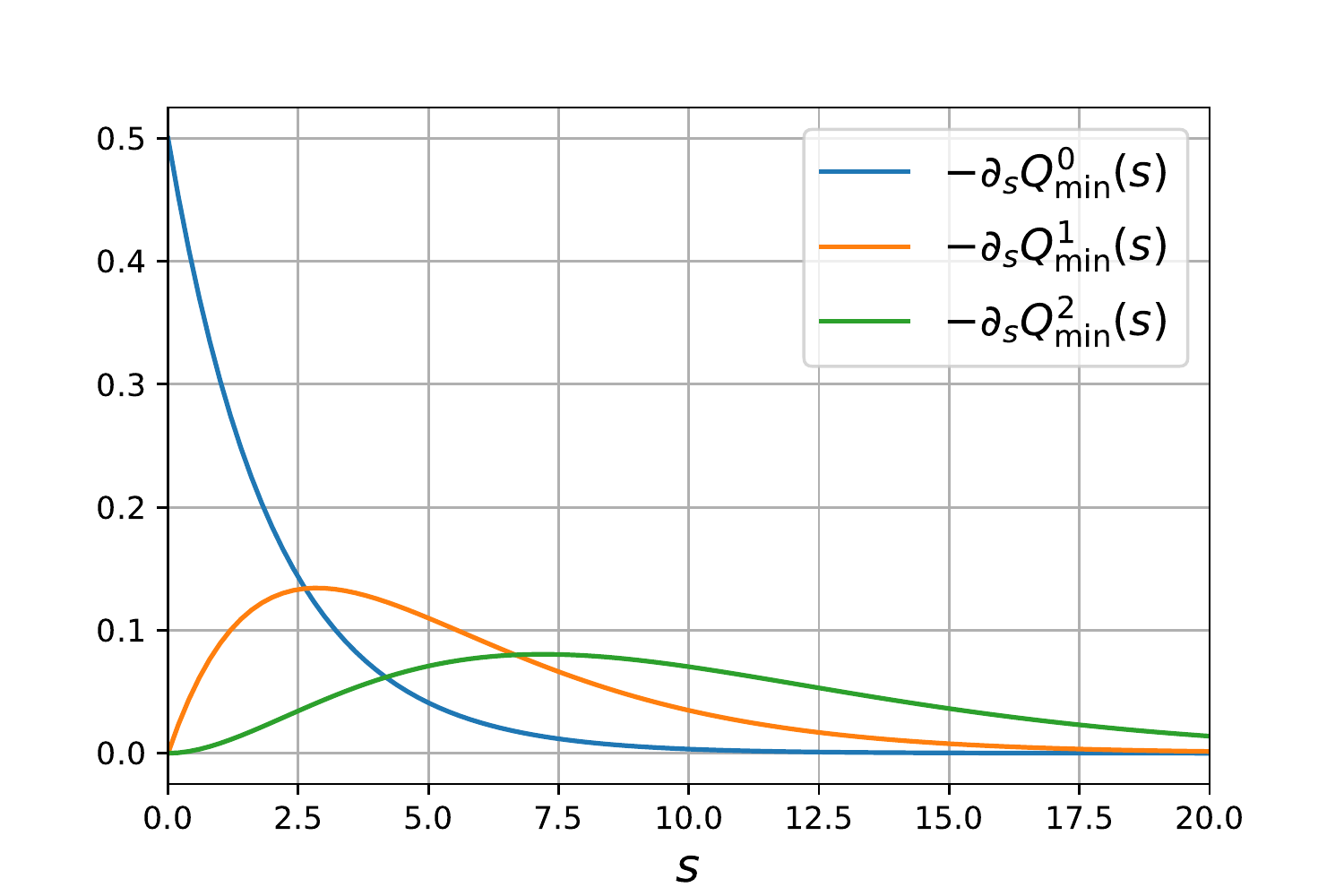}
\caption{Plot of the PDF $-\partial_s Q_{\min}^{\nu}(s)$ in Eq. \eqref{PDF_P_min_wish} for $\nu=0,1,2$ respectively in blue, orange and green.}\label{Fig_P_min_wish}
\end{figure}

Considering now the case of $r_{\min}$, we will again use the exact mapping $\lambda_{\min}=N\alpha^2 r_{\min}^2$. In the LUE, the symmetry between $\lambda_{\min}$ and $\lambda_{\max}$ is broken. There exists in the literature a number of expressions for the distribution of the typical fluctuations of $\lambda_{\min}$ \cite{forrester1994complex,forrester2010log,dumitriu2003eigenvalue,edelman1989eigenvalues,tracy1994fredholm}, (see also \cite{edelman2016beyond}). Here, we only reproduce the Fredholm determinant
\be\label{CDF_min_wish}
\lim_{N\to \infty}\Prob\left[\lambda_{\min}\leq \frac{s}{N^2}\right]=Q_{\min}^{\nu}(s)=\Det\left[\mathbb{I}-P_{[0,s]}K_{\rm Be}^{\nu}P_{[0,s]}\right]\;,
\ee
where $K_{\rm Be}^{\nu}(x,y)$ is given in Eq. \eqref{k_bessel} and the determinantal expression valid for $\nu=a-1/2 \in \mathbb{N}$
\be\label{PDF_P_min_wish}
-\partial_s Q_{\min}^{\nu}(s)=\frac{e^{-s/2}}{2}\det_{1\leq i,j\leq \nu }{\rm I}_{i-j+2}(\sqrt{2s})\;,
\ee
where ${\rm I}_\nu(x)$ is the modified Bessel function of first kind. This PDF is plotted in Fig. \ref{Fig_P_min_wish}. 
Using the exact mapping with LUE the CDF of the typical fluctuations of the position of the fermion the closest to the origin $r_{\min}$ are given by
\be
\lim_{N\to \infty}\Prob\left[r_{\min}\geq k_F^{-1} u\right]=Q_{\min}^{a-1/2}(u^2)=\Det\left[\mathbb{I}-P_{[0,u^2]}K_{\rm Be}^{a-1/2}P_{[0,u^2]}\right]\;.
\ee 
Note that the fluctuations of $r_{\min}$ are quite different from the Tracy-Widom distribution and enter in a different universality class. The CDF $ Q_{\min}^{\nu}(s)$ also has notable applications in the context of QCD \cite{akemann2011oxford} (see chapter 32 there).

This result concludes this introduction to hard edge potentials, for which we have seen that the fluctuations at the edge do not enter the same universality class as for soft edges. We refer to Article \ref{Art:ferm_long} for more information on the finite temperature extension of this model of fermions. In the next section, we study in detail another type of {\it hard edge} potential where exact results can be obtained: the hard box potential.

\section{Hard box potential in dimension $d=1$}

\begin{figure}
\centering
\includegraphics[width=0.6\textwidth]{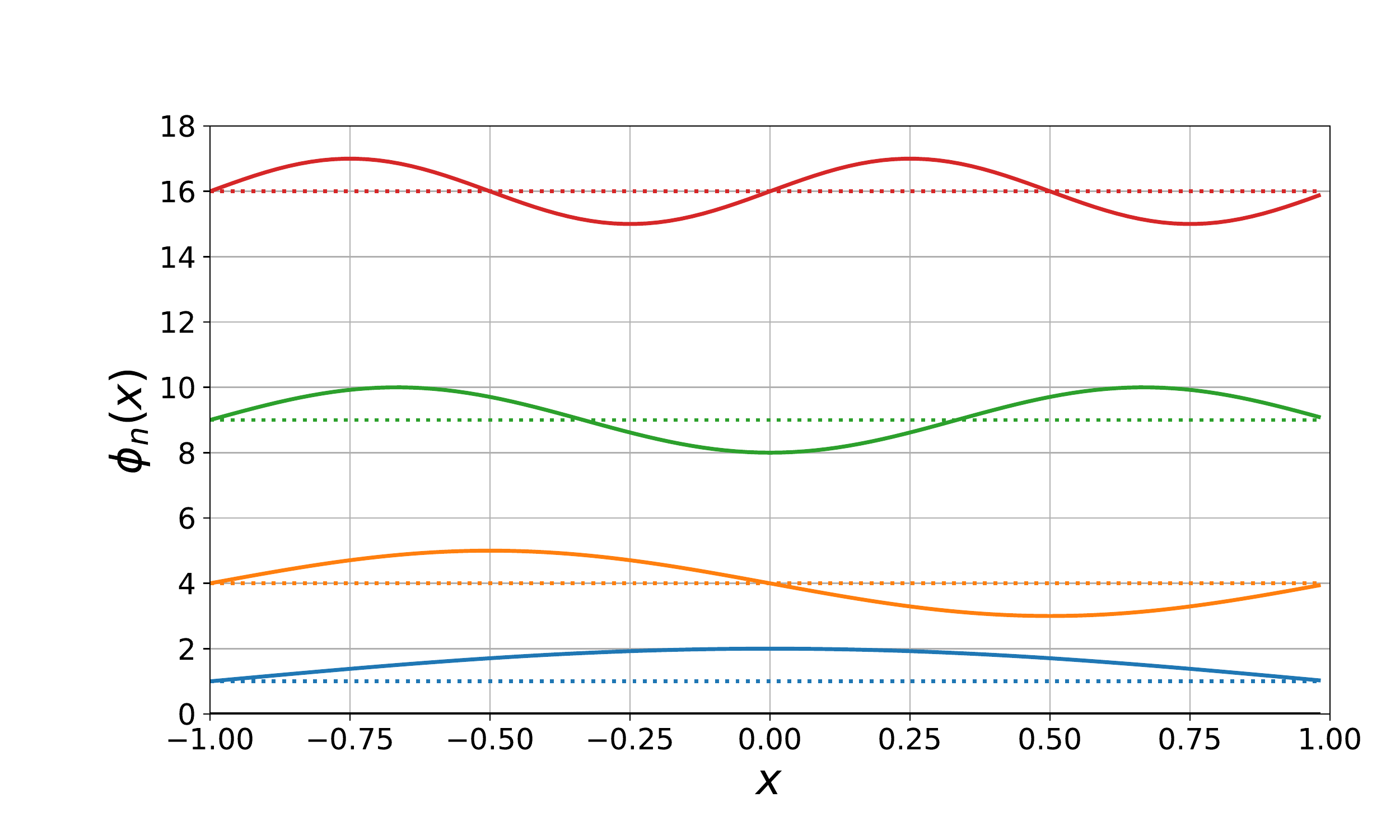}
\caption{Single-particle energies and wave-functions associated to a system of $N=4$ fermions at zero temperature in the hard box potential given in Eq. \eqref{hb_pot_1d}.}\label{Fig_phi_hb}
\end{figure}

We first consider the one-dimensional hard box. As we have seen in section \ref{sec_d_soft}, the effects of the quantum correlations are weaker in dimension $d>1$, where fermions with different orbital quantum numbers are independent. In dimension one, on the contrary, all fermions are strongly correlated and the effects of the Pauli exclusion principle are the strongest.
We will analyse in detail the spatial statistics for the Fermi gas in the one-dimensional hard box potential
\be\label{hb_pot_1d}
V(x)=\begin{cases}
0&\;,\;\;|x|<R\;,\\
&\\
+\infty&\;,\;\;|x|\geq R\;.
\end{cases}
\ee
From now on, we set $R=1$ (which amounts to rescale all the lengths by $R$).
In one dimension, the Hamiltonian of the system is simply given by
\be
\hat {\cal H}_N=\sum_{i=1}^N \hat{H}_i\;,\;\;{\rm with}\;\;\hat H=-\frac{\hbar^2 }{2m}\partial_x^2\;,
\ee
and we impose Dirichlet boundary conditions in $x=\pm 1$. The single particle wave-functions and energies are thus labelled by a non-zero integer $n\in \mathbb{N}_+$and given by 
\be
\phi_n(x)=\sin\left(\frac{\pi n}{2}(x+1)\right)\;,\;\;\epsilon_n=\frac{\hbar^2 \pi^2 n^2}{8m}\;.
\ee
These wave-functions are represented in Fig. \ref{Fig_phi_hb}

\subsection{Zero-temperature quantum PDF and Jacobi Unitary Ensemble}

At temperature $T=0$, only the $N$ lowest energy states $n=1,\cdots,N$ are occupied.
The many body wave function is given by the Slater determinant built from these $N$ levels
\be
\Psi_0(x_1,\cdots,x_N)=\frac{1}{\sqrt{N!}}\det_{1\leq k,l\leq N}\sin\left(\frac{\pi k}{2}(x_l+1)\right)\;.
\ee
The ground state joint probability of the positions of these fermions can be computed exactly in this case using $\sin(k u)=\sin(u)U_{k-1}(\cos(u))$ -- where $U_n(x)$ is the Tchebychev polynomial of second kind of order $n$ -- and the Vandermonde identity in Eq. \eqref{VdM}
\be
\boxed{|\Psi_0(x_1,\cdots,x_N)|^2=\frac{1}{Z_N}\prod_{i<j}\left|\sin\left(\frac{\pi x_i}{2}\right)-\sin\left(\frac{\pi x_j}{2}\right)\right|^2 \prod_{k=1}^N \cos^2\left(\frac{\pi x_k}{2}\right)\;.}\label{joint_PDF_fermions}
\ee
This joint probability appears also in in the so-called Jacobi Unitary ensemble (JUE). 

This ensemble can be built by first considering two independent Wishart (Laguerre) matrices \cite{dumitriu2008distributions} $W_1=M_1^{-1} X_1^{\dag}X_1$ and $W_2=M_2^{-1} X_2^{\dag}X_2$ where $X_1$ is a matrix of size $M_1\times N$ and $X_2$ is of size $M_2 \times N$, with $M_1,M_2\geq N$. The matrix defined as
\be
J=W_1(W_1+W_2)^{-1}\;,
\ee
is then a Jacobi matrix, where $^{-1}$ refers here to the matrix inversion. Its eigenvalues all belong to the interval $\lambda\in [0,1]$ and their joint PDF reads
\be
P_{\rm joint}^{\rm JUE}(\lambda_1,\cdots,\lambda_N)=\frac{1}{Z_N^{\rm JUE}(a,b)} \prod_{i<j}|\lambda_i-\lambda_j|^2 \prod_{i=1}^N (1-\lambda_i)^a \lambda_i^{b} \;,\label{P_joint_JUE}
\ee 
where $a=M_1-N$ and $b=M_2-N$. Taking now $\lambda_i=(1+\sin(\pi x_i/2))/2$ and for the specific values $a=b=1/2$, the joint PDF in Eq. \eqref{P_joint_JUE} is mapped exactly onto Eq. \eqref{joint_PDF_fermions}. Note that one can associate statistics of confined fermions for generic values of $a,b$ (see Article \ref{Art:ferm_long} for details).
%
%
%
%
%
%
%
%
%
%
%
From this exact mapping, it is clear that the positions of the fermions form a determinantal point process and the correlation kernel can be obtained exactly for finite $N$ as
\begin{align}
K_{N}(x,y)&=\sum_{k=1}^N \sin\left(\frac{\pi k}{2}(x+1)\right)\sin\left(\frac{\pi k}{2}(y+1)\right)\nn\\
&=\dfrac{\sin\left(\frac{(2N+1)\pi}{4}(x-y)\right)}{4\sin\left(\frac{\pi}{4}(x-y)\right)}-\dfrac{\sin\left(\frac{(2N+1)\pi}{4}(2+x+y)\right)}{4\sin\left(\frac{\pi}{4}(2+x+y)\right)}\;.\label{K_N_hb_1d}
\end{align}

For this system of $N$ particles, the density is uniform over the box $[-1,1]$ in the large $N$ limit and the typical inter-particle distance is given both in the bulk and at the edge by $k_F^{-1}$ where $k_F$ is the Fermi wave-vector 
\be
k_F=\sqrt{\frac{2m \mu}{\hbar^2}}=\frac{N\pi}{2}\;.
\ee
Using the exact formula in Eq. \eqref{K_N_hb_1d} and far from the edges of the density $k_F|x\pm 1|\gg 1$ the sine scaling form, already obtained in Eq. \eqref{k_sine}, is recovered in the large $N$ limit
\be
K_N(x,y)\approx k_F K_{1}^{\rm b}\left(k_F(x-y)\right)\;,\;\;{\rm with}\;\;K_{1}^{\rm b}(r)=K_{\sin}(r)=\frac{\sin(r)}{\pi r}\;.
\ee 
Note that this result coincides with the local density approximation prediction in Eq. \eqref{LDA_1d}. This scaling function is universal for the bulk statistics of one-dimensional Fermi gas in their ground state, irrespectively of the edge behaviour.

\begin{figure}
\centering
\includegraphics[width=0.6\textwidth]{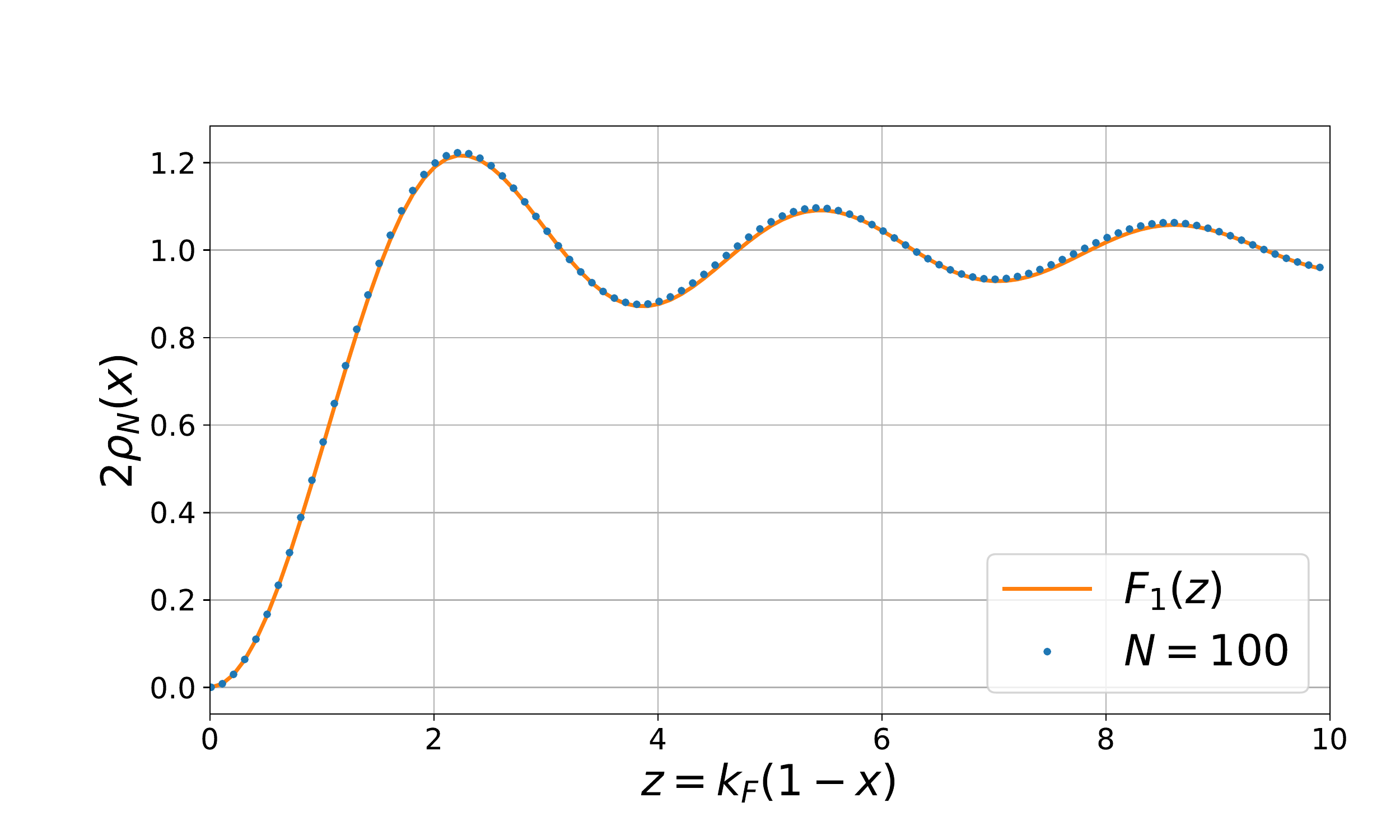}
\caption{Comparison between the exact rescaled density $2\rho_N(x)$ close to the hard edge as a function of $z=k_F(1-x)$ for $N=100$ fermions (blue dots) and the scaling function $F_1(z)$ given in Eq. \eqref{dens_1d_edge_T0} (in orange).}\label{Fig_F_1_hard}
\end{figure}

Taking now the edge scaling limit $k_F|1-x|\sim k_F|1-y|\sim 1$ of the correlation kernel in Eq. \eqref{K_N_hb_1d}, one obtains instead
\be
\boxed{\begin{array}{rl}
&\displaystyle K_N(x,y)\approx k_F K_{1}^{\rm e}\left(k_F(1-x),k_F(1-y)\right)\;,\vspace{0.1cm}\\
&\displaystyle K_{1}^{\rm e}(u,v)=\frac{\sin(u-v)}{\pi(u-v)}-\frac{\sin(u+v)}{\pi(u+v)}\;.\label{k_1d_hb_e}
\end{array}}
\ee
Note that this correlation kernel corresponds exactly to the case $a=1$ of Eq. \eqref{K_1_x_2}, where the potential term $\sim r^{-2}$ in Eq. \eqref{x2_1_x2_pot} disappears.
From this result, the average density $\rho_N(x)$ takes the scaling form at the edge
\be\label{dens_1d_edge_T0}
\boxed{\rho_N(x)=\frac{1}{N}K_N(x,x)\approx \frac{1}{2} F_1(k_F(1-x))\;,\;\;{\rm with}\;\;F_1(z)=1-\frac{\sin(2z)}{2z}\;,}
\ee
where we used here that $k_F/(\pi N)=1/2$. Note that this density vanishes quadratically close to the edge $F_1(z)\approx \frac{2}{3}z^2$ as $z\to 0$, while it matches the uniform density in the bulk $F_1(z)\approx 1$ for $z\to \infty$. The scaling function $F_1(z)$ is plotted in Fig. \ref{Fig_F_1_hard}.

\subsubsection{Method of images}

The structure of the edge kernel in Eq. \eqref{k_1d_hb_e} is reminiscent of a method of images. Indeed, the walls at $x=\pm 1$ impose that all wave functions $\phi_k(x)$ be identically zero outside the box. By continuity of these wave functions, it imposes Dirichlet boundary conditions $\phi_k(x=\pm 1)=0$. The same Dirichlet boundary conditions apply to the Euclidean propagator $G(y,t|x,0)$ defined as
\be
G(y,t|x,0)=\langle y|e^{-\frac{\hat H t}{\hbar}}|x\rangle=\sum_{k=1}^{\infty}e^{-\frac{\epsilon_{k} t}{\hbar}}\overline{\phi}_k(x)\phi_k(y)\;,
\ee
solution of the free diffusion equation in the box
\be
\partial_t G(y,t|x,0)=\frac{\hbar}{2m}\partial_y^2 G(y,t|x,0)\;,\;\;{\rm with}\;\;G(y,0|x,0)=\delta(x-y)\;.
\ee
This propagator can be obtained exactly from the free propagator using the method of images
\be
 G(y,t|x,0)=\sqrt{\frac{m}{2\pi \hbar t}}\sum_{n=-\infty}^{\infty}e^{-\frac{m}{2\hbar t}\left(4n+x-y\right)^2}-\sqrt{\frac{m}{2\pi \hbar t}}\sum_{n=-\infty}^{\infty}e^{-\frac{m}{2\hbar t}\left(4n+2+x+y\right)^2}\;.
\ee
Taking the inverse Laplace transform by using Eq. \eqref{LT_bessel} in the table of Appendix \ref{LT}, one obtains another exact representation for the correlation kernel~\footnote{To obtain this construction using the method of images, one can first take the Laplace transform with respect to the time by introducing $\tilde G(y,x;s)=\int_0^{\infty}e^{-\frac{s t}{\hbar}}G(y,t|x,0)dt$ and solve the ordinary differential equation in space with the correct boundary conditions. Taking the inverse Laplace transform, one obtains an infinite sum over the simple poles of the function $\tilde G(y,x;s)$ as in Eq. \eqref{K_im}.}
\begin{align}
K_{\mu}(x,y)&=\int_{\cal C}\frac{dt}{2i \pi t}e^{\frac{\mu t}{\hbar}}G(y,t|x,0)\nn\\
&=\sum_{n=-\infty}^{\infty}\frac{\sin\left(k_F(4n+x-y)\right)}{\pi(4n+x-y)}-\sum_{l=-\infty}^{\infty}\frac{\sin\left(k_F(4l+2+x+y)\right)}{\pi(4l+2+x+y)}\;.\label{K_im}
\end{align}
In the bulk limit, for $k_F|x\pm 1|\gg 1$ (or $k_F|y\pm 1|\gg 1$), all the terms of the sums in Eq. \eqref{K_im} (for all $l,n$) except $n=0$ give a vanishing contribution and one recovers the sine scaling form. At the edge, for $k_F|x\pm 1|\sim k_F|y\pm 1|=O(1)$, all the terms of the sum in Eq. \eqref{K_im} except $n=0$ and $l=-1$ give a vanishing contribution. This method gives an alternative derivation, via the so-called method of images, of the result for the hard edge scaling form in Eq. \eqref{k_1d_hb_e}. 

\subsubsection{Statistics of the rightmost fermion $x_{\max}$ at $T=0$}

\begin{figure}
\centering
\includegraphics[width=0.6\textwidth]{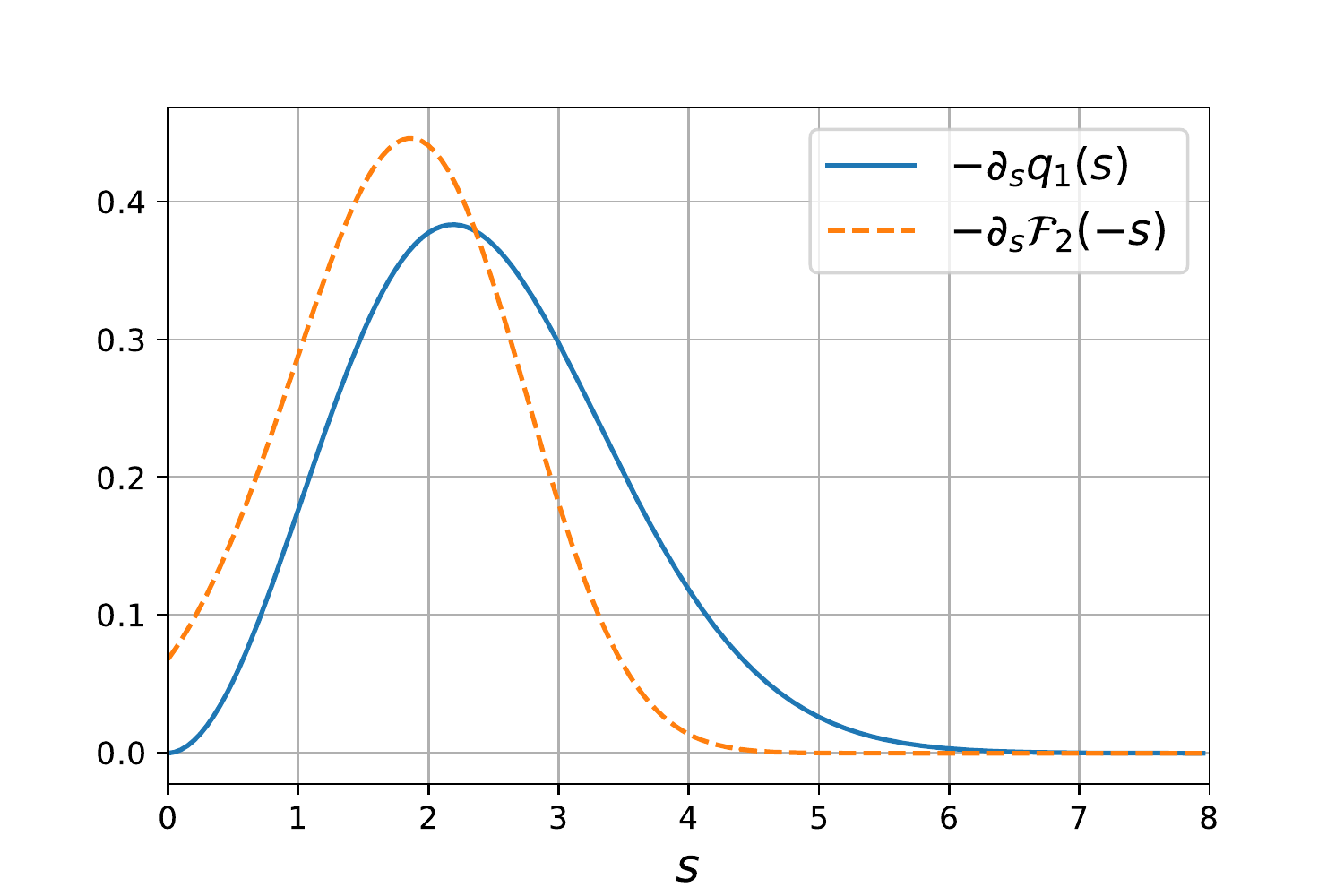}
\caption{Plot of the PDF $-\partial_s q_1(s)$ in Eq. \eqref{CDF_x_max_box_T0} representing the typical fluctuations of $x_{\max}$ (or $x_{\min}$) in a hard box potential at $T=0$. The Tracy-Widom distribution $-\partial_s {\cal F}_2(-s)$ is plotted (in dashed orange) for comparison. The function $-\partial_s q_1(s)$ was plotted using the algorithm developed in Ref. \cite{bornemann2010numerical}. }\label{Fig_x_max_box_T0}
\end{figure}

From our result on the scaling form of the correlation kernel at the edge in Eq. \eqref{k_1d_hb_e}, it is natural to expect that $k_F|1-x_{\max}|=O(1)$ is the typical scale of fluctuations for the position of the rightmost fermion $\displaystyle x_{\max}=\max_{1\leq i\leq N} x_i$. 
One can indeed express the CDF in this regime of typical fluctuation as
\be
\lim_{N\to \infty}\Prob\left[x_{\max}\leq 1-\frac{s}{k_F}\right]=q_1(s)\;,
\ee
where the scaling function $q_1(s)$ is a Fredholm determinant (see Appendix \ref{Fred_det_app} for a definition)
\be
\boxed{q_1(s)=\Det\left[\mathbb{I}-P_{[0,s]}K_{1}^{\rm e}P_{[0,s]}\right]=\exp\left(-\sum_{p=1}^{\infty}\frac{1}{p}\Tr\left[(P_{[0,s]}K_{1}^{\rm e}P_{[0,s]})^p\right]\right)\;.}\label{CDF_x_max_box_T0}
\ee
%
This scaling function $q_1(s)$ coincides exactly with the scaling function $Q_{\min}^{1/2}(s^2)$ for the typical fluctuations of the smallest eigenvalue $\lambda_{\min}$ in the Laguerre Unitary Ensemble for $\nu=1/2$ in Eq. \eqref{CDF_min_wish}. 
%
%
%
The scaling function $q_1(s)$ can be expressed in terms of the solution $\sigma(x)$ of a Painlev\'e equation, as for the Tracy-Widom distribution ${\cal F}_2(s)$ (c.f. Eq. \eqref{TW_2}),
\be\label{q_Painleve}
q_1(s)=\exp\left(\int_0^{2s}\left(\sigma(x)-\sqrt{\sigma(x)-x\sigma'(x)}\right)\frac{dx}{2x}\right)\;.
\ee
In this case the function $\sigma(x)$ is the solution of a Painlev\'e VI equation \cite{jimbo1980density,tracy1994fredholm,duenez2010lowest}
\be\label{Painleve}
(x\sigma'')^2+4(x\sigma'-\sigma)(x\sigma'-\sigma+\sigma'^2)=0\;,\;\;{\rm with}\;\;\sigma(x)\sim -\frac{x}{\pi}\;,\;\;{\rm for}\;\;x\to 0\;.
\ee 
Therefore, the distribution $q_1(s)$ is very different from the Tracy-Widom distribution (c.f. Fig. \ref{Fig_x_max_box_T0}).
The scaling function $q_1(s)$ can alternatively be expressed as
\be\label{q_Painleve_3}
q_1(s)=\exp\left(-\frac{1}{4}\int_0^{s^2}\ln\left(\frac{s^2}{z}\right)h(z)^2 dz\right)
\ee
in terms of the solution $h(z)$ to the Painlev\'e III equation \cite{tracy1994level2} 
\be\label{Painleve_3}
z(h^2-1)(z h''+h')=h(z h')^2+\frac{1}{4}(z-\frac{1}{2})h+\frac{z}{4} h^3(h^2-2)\;,
\ee
for which $h(z)\sim \sqrt{\frac{2}{\pi}}z^{1/4}$ as $z\to 0$.

In particular, these two representations allow to obtain the asymptotic behaviours of the CDF $q_1(s)$
\be
q_1(s) =\begin{cases}
\displaystyle 1-\frac{2}{9\pi}s^3-\frac{2}{75\pi}s^5 + O(s^7)&\;,\;\;s\to 0\\
&\\
\displaystyle s^{-\frac{1}{8}} e^{-\frac{s^2}{4}+\frac{s}{2} + O(1)}&\;,\;\; s\to \infty\;.
\end{cases} \label{eq: typ_fluc_1d}
\ee
These asymptotic results were first derived by Dyson in \cite{dyson1976fredholm}.
The PDF $-\partial_s q_1(s)$ in the regime of typical fluctuations is plotted in blue in Fig. \ref{Fig_x_max_box_T0}. Note that this PDF decays much slower than the Tracy-Widom distribution in the first line of Eq. \eqref{TW_tails} as the effects of the confining potential are weaker in the case of JUE.

\begin{figure}
\centering
\includegraphics[width=0.6\textwidth]{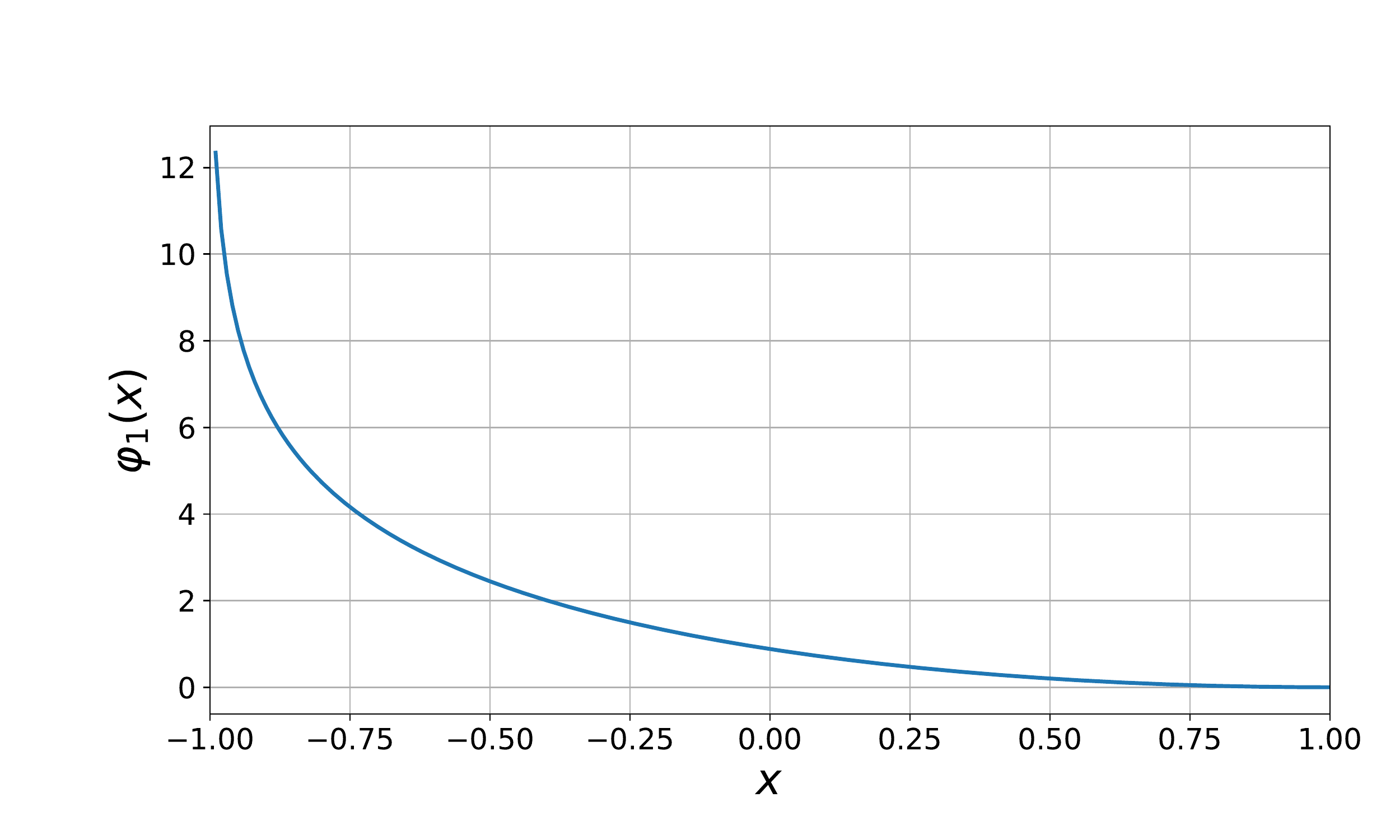}
\caption{Plot of the large deviation function $\varphi_1(x)$ in Eq. \eqref{LD_T_0_hb_1d} as a function of $x$.}\label{Fig_varphi_1}
\end{figure}

The atypical fluctuations can also be obtained exactly using the mapping with the Jacobi Unitary Ensemble. In this ensemble and for $a\sim b=O(1)$, the atypical fluctuations of the largest eigenvalue $\lambda_{\max}$ are characterised by the simple large deviation rate function
\be
\Prob\left[\lambda_{\max}\leq \lambda\right]\approx e^{-N^2\Phi^{\rm JUE}(\lambda)}\;,\;\;{\rm where}\;\;\Phi^{\rm JUE}(\lambda)=-\ln \lambda\;,\;\;\lambda=O(1)\;.
\ee
Using the mapping between our problem and the JUE, we obtain that $\lambda_{\max}=(1+\sin(\pi x_{\max}/2))/2$. Using additionally $k_F=N\pi/2$, we obtain the large deviation rate function for $x_{\max}$,
\be
\boxed{\Prob\left[x_{\max}\leq x\right]\approx e^{-k_F^2\varphi_1(x)}\;,\;\;{\rm where}\;\;\varphi_1(x)=-\frac{4}{\pi^2}\ln\left[\frac{1}{2}+\frac{1}{2}\sin\left(\frac{\pi x}{2}\right)\right]\;.}\label{LD_T_0_hb_1d}
\ee
Note that taking the limit $x\to 1$, one obtains $\varphi_1(x)\approx \frac{1}{4}(x-1)^2$, which matches smoothly the large $s=k_F(1-x)$ asymptotic behaviour of the CDF in the typical regime $q_1(s)\approx e^{-s^2/4}$, displayed in the second line of Eq. \eqref{eq: typ_fluc_1d}. The large deviation function $\varphi_1(x)$ is plotted in Fig. \ref{Fig_varphi_1}. It has a logarithmic singularity for $x\to -1$.

\begin{figure}
\centering
\includegraphics[width=0.6\textwidth]{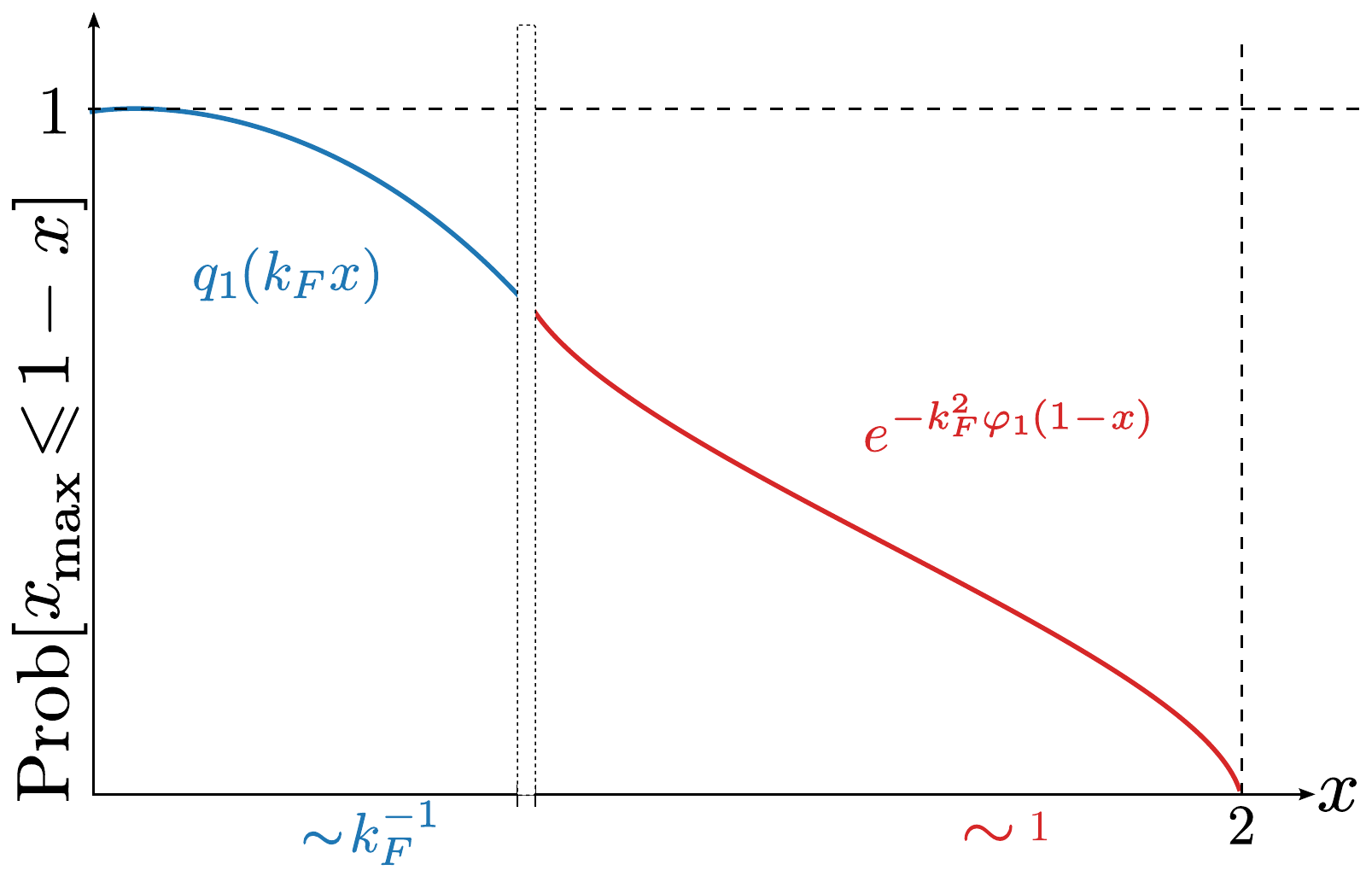}
\caption{Sketch of the typical (blue) and large (red) fluctuation regimes of the probability $\Prob\left[x_{\max}\leq 1-x\right]$ of the position $x_{\max}$ of the rightmost fermion (c.f. Eq. \eqref{fluc_x_max_hb_1d_t0_sum}).}\label{fluc_x_max_hb_1d_t0_fig}
\end{figure}

The different regimes of fluctuation of $x_{\max}$ at zero temperature can be summarised as (see also Fig. \ref{fluc_x_max_hb_1d_t0_fig})
\be
\boxed{
\Prob\left[x_{\max}\leq x\right]=\begin{cases}
q_1(k_F(1-x))&\;,\;\;k_F(1-x)=O(1)\;,\\
&\\
\displaystyle e^{-k_F^2 \varphi_1(x)}&\;,\;\;(1-x)=O(1)\;.
\end{cases}}\label{fluc_x_max_hb_1d_t0_sum}
\ee

We will now consider how the thermal fluctuations affect the system and in particular the behaviour close to the hard edge.


\subsection{One-dimensional hard box at finite temperature}\label{hb_1d_t}

At finite temperature $T=1/(k_B\beta)$, we expect the correlations between the positions of fermions to be weaker. The typical scale of temperature for this system is given by the Fermi temperature $T_F=\epsilon_F/k_B$. Indeed, in the limit $\beta\epsilon_F\gg 1$, the temperature will only allow fermions to occupy a few energy levels above the Fermi energy and one expects to recover the zero temperature results. Inversely, in the limit $\beta\epsilon_F\ll 1$, the temperature will be large, fermions will be able to occupy many energy levels, such that the mean occupation number of each level will be small $\overline{n}_{k}=n_{\rm FD}(\epsilon_k)\sim e^{-\beta\epsilon_k}$. As a result, the effect of the quantum statistics will be weak in this high temperature regime and one expects to recover the description of classical statistical mechanics. One can associate to the typical quantum energy scale $\epsilon_F$ a length scale of quantum origin $k_F^{-1}$ while associated to the temperature scale $k_B T$ is the de Broglie thermal length scale $\Lambda_{\beta}=\sqrt{\frac{2\pi\hbar^2 \beta}{m}}$. We expect that for $\Lambda_{\beta}\gg k_F^{-1}$, the quantum fluctuations will be dominating while for $\Lambda_{\beta}\ll k_F^{-1}$ the thermal fluctuations will be dominating in the spatial structure. Note that for this system, the typical temperature scale is the same in the bulk and at the edge.

We will consider in this section the regime where both of these scales are of the same order, defining the rescaled inverse temperature
\be
b=\beta\epsilon_F=\frac{(k_F\Lambda_{\beta})^2}{4\pi}=O(1)\;.
\ee
The local density approximation applies also at finite temperature \cite{bartel1985extended, castin2006basic}. In particular, it allows to obtain the finite temperature bulk correlation kernel
\be
K_{\mu}^{\beta}(x,y)\approx \Lambda_{\beta}^{-1} K_{1,b}^{\rm b}\left(\frac{{ x}-{ y}}{\Lambda_{\beta}}\right)\;,
\ee
with the scaling function \cite{garcia2000chiral, johansson2005random, dean2015finite}
\be
K_{1,b}^{\rm b}(r)=\frac{1}{\pi}\int_0^{\infty}\frac{\zeta dk}{\zeta+e^{\frac{k^2}{4\pi}}}\cos(k r)\;.\label{k_bulk_1d_t}
\ee
In this expression $\zeta=e^{\beta\mu}$ is the fugacity. Using the equivalence of ensembles one can obtain a closed form formula for $\zeta$ in the regime $b=\beta\epsilon_F=O(1)$,
\be
N=\sum_{k=1}^{\infty}\frac{\zeta}{\zeta+e^{\beta\frac{\hbar^2 \pi^2 k^2}{8m}}}\Rightarrow \int_0^{\infty}\frac{\zeta du}{\zeta+e^{b u^2}}=-\sqrt{\frac{\pi}{4b}}\Li_{\frac{1}{2}}\left(-\zeta\right)=1\;,\label{fugacity_1d}
\ee
where we recall that $\Li_{p}(x)=\sum_{k=1}^{\infty} k^{-p} x^{k}$ is the polylogarithm function. From its asymptotic behaviours,
\be\label{polylog_as}
-\Li_p(-z)\approx\begin{cases}
\displaystyle z&\;,\;\;z\to 0\;,\\
&\\
\displaystyle \frac{(\ln z)^p}{\Gamma(p+1)}&\;,\;\;z\to \infty\;,
\end{cases}
\ee 
one recovers $\mu\approx \epsilon_F=\frac{\hbar^2 k_F^2}{2m}>0$ in the low temperature limit $b=\beta\epsilon_F \gg 1$, and $\mu=\frac{1}{\beta}\ln\left(\frac{k_F\Lambda_{\beta}}{\pi}\right)<0$ in the high temperature limit $b\ll 1$. 

\begin{figure}
\centering
\includegraphics[width=0.6\textwidth]{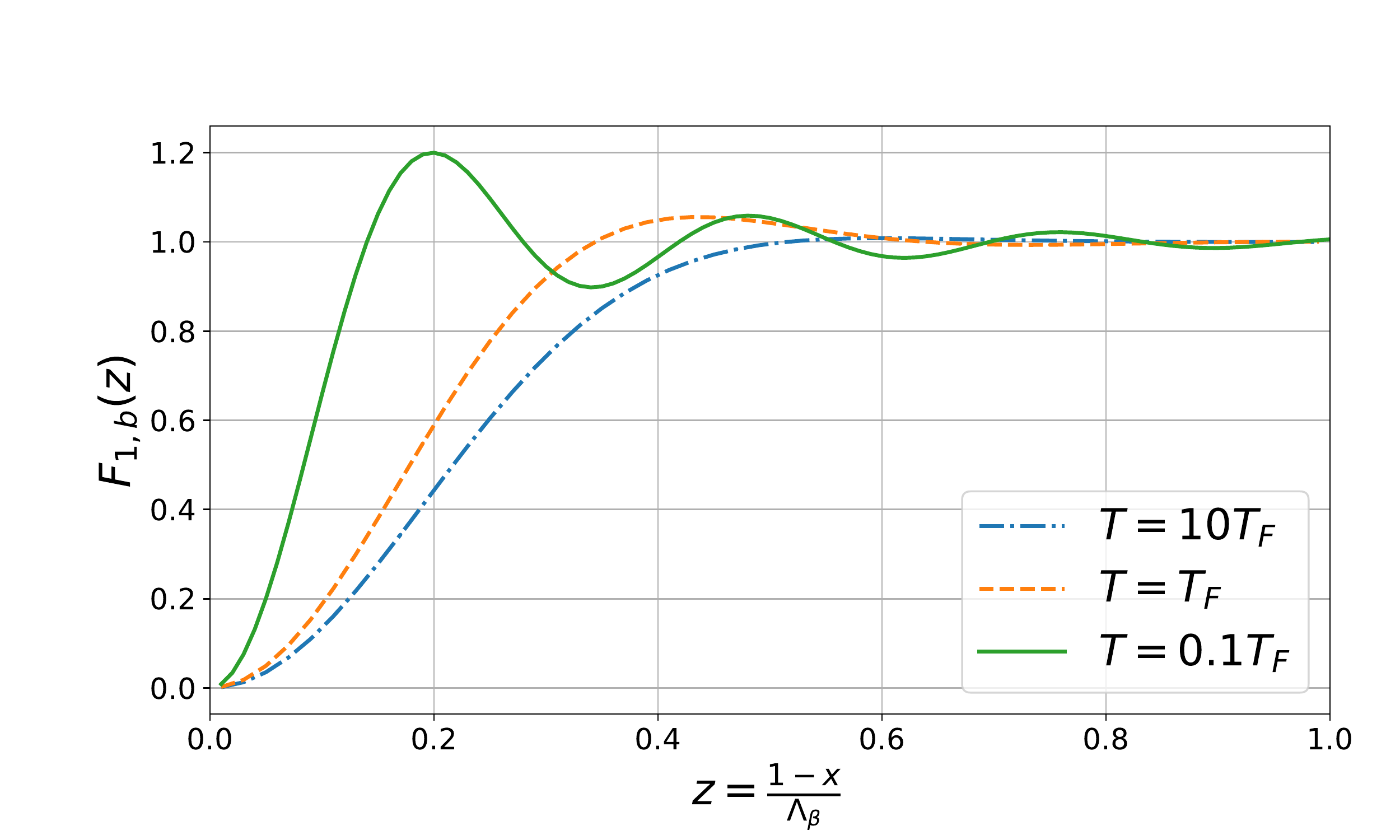}
\caption{Plot of the scaling function $F_{1,b}(z)$ given in Eq. \eqref{F_1b_eq} describing the average density close to the hard edge for $T/T_F=0.1,1,10$ respectively in green, orange and blue. The oscillations that are of quantum origin are smoothened at finite temperature.}\label{Fig_F_1_b}
\end{figure}

At the edge, the finite temperature correlation kernel $K_{\mu}^{\beta}(x,y)$ is again obtained from the bulk correlation kernel using the method of images
\be
\boxed{K_{\mu}^{\beta}(x,y)\approx \Lambda_{\beta}^{-1} K_{1,b}^{\rm e}\left(\frac{1-{ x}}{\Lambda_{\beta}},\frac{1-{ y}}{\Lambda_{\beta}}\right)\;,\;\;{\rm with}\;\;K_{1,b}^{\rm e}(u,v)=K_{1,b}^{\rm b}(u-v)-K_{1,b}^{\rm b}(u+v)\;.}\label{k_hb_1d_t}
\ee
Using this behaviour, one can obtain the average density close to the hard wall at finite temperature
\be
\boxed{\rho_N(x)=\frac{1}{2} F_{1,b}\left(\frac{1-x}{\Lambda_{\beta}}\right)\;,\;\;{\rm with}\;\;F_{1,b}(z)=1+\frac{1}{\pi\Li_{1/2}\left(-\zeta\right)}\int_0^{\infty}\frac{\zeta dk}{\zeta+e^{\frac{k^2}{4\pi}}}\cos(2k z)\;.}\label{F_1b_eq}
\ee
Taking the limit $z\to \infty$, the scaling function $F_{1,b}(z)\to 1$ matching smoothly the uniform density in the bulk, while for $z\to 0$, one obtains
\be
F_{1,b}(z)\approx \frac{4\pi \Li_{3/2}(-\zeta)}{\Li_{1/2}(-\zeta)}z^2\;,\;\;z\to 0\;.
\ee
The density vanishes quadratically close to the wall for any temperature. However,  the typical scale $\Lambda_{\beta}$ associated to this behaviour decreases with the temperature. In the low temperature limit, using $\zeta=e^{\beta \epsilon_F}\gg 1$ and the asymptotic behaviour in the second line of Eq. \eqref{polylog_as}, the average density close to the wall matches smoothly the zero temperature result given in Eq. \eqref{dens_1d_edge_T0},
\be
F_{1,b}(z)\approx \frac{8\pi b}{3}z^2=\frac{2}{3}\left(k_F\Lambda_{\beta} z\right)^2\approx F_1\left(k_F\Lambda_{\beta} z\right)\;.
\ee
This scaling function $F_{1,b}(z)$ is plotted in Fig. \ref{Fig_F_1_b} for several values of the temperature.

We will now use the result in Eq. \eqref{k_hb_1d_t} for the hard edge correlation kernel to analyse the statistics of the rightmost fermion $x_{\max}$ at finite temperature.

\subsubsection{Statistics of the rightmost fermion $x_{\max}$ at finite temperature $T$}

\begin{figure}
\centering
\includegraphics[width=0.6\textwidth]{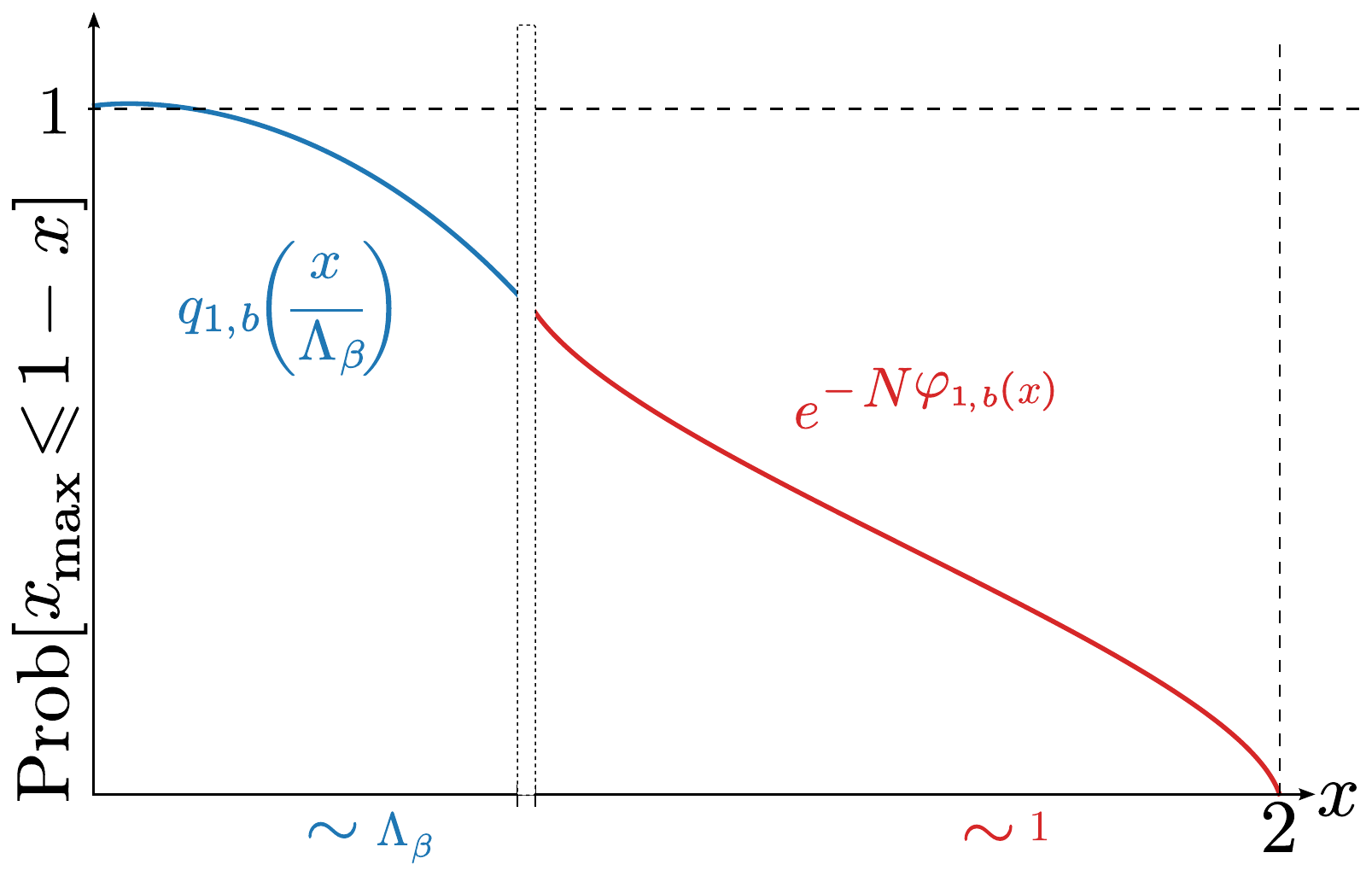}
\caption{Sketch of the typical (blue) and large (red) fluctuation regimes of the probability $\Prob\left[x_{\max}\leq 1-x\right]$ of the position $x_{\max}$ of the rightmost fermion at finite temperature for $b=\beta\epsilon_F=O(1)$ (c.f. Eq. \eqref{fluc_x_max_hb_1d_t_sum}).}\label{fluc_x_max_hb_1d_t_fig}
\end{figure}

The CDF of the position of the rightmost fermion $x_{\max}$ takes at finite temperature the scaling form
\be
\Prob\left[x_{\max}\leq 1-\Lambda_\beta s\right]\approx q_{1,b}(s)\;,\label{scal_temp}
\ee
where $q_{1,b}(s)$ is given in terms of the Fredholm determinant
\be\label{q_1_b}
\boxed{\frame{\boxed{q_{1,b}(s)=\Det\left[\mathbb{I}-P_{[0,s]}K_{1,b}^{\rm e}P_{[0,s]}\right]=\exp\left(-\sum_{p=1}^{\infty}\frac{1}{p} \Tr\left[(P_{[0,s]}K_{1,b}^{\rm e}P_{[0,s]})^p\right]\right)\;,}}}
\ee
where $K_{1,b}^{\rm e}(u,v)$ is the finite temperature one-dimensional kernel given in Eq. \eqref{k_hb_1d_t} (with $K_{1,b}^{\rm b}(r)$ in Eq. \eqref{k_bulk_1d_t}).
The asymptotic behaviours of this exact result for $s\to 0$ and $s\to \infty$ can be obtained more explicitly: 

\begin{itemize}
\item The small $s$ asymptotic behaviour of $q_{1,b}(s)$ can be evaluated using the development in terms of traces in Eq. \eqref{q_1_b}, keeping only the lowest order ($p=1$) term
\be\label{q_1_b_small}
q_{1,b}(s)\approx \exp\left(- \Tr\left[P_{[0,s]}K_{1,b}^{\rm e}P_{[0,s]}\right]\right)\approx 1-\Tr\left[P_{[0,s]}K_{1,b}^{\rm e}P_{[0,s]}\right]\;,\;\;s\to 0\;.
\ee
This first trace is expressed as the integral over the density scaling function in Eq. \eqref{F_1b_eq}
\be
\Tr\left[P_{[0,s]}K_{1,b}^{\rm e}P_{[0,s]}\right]= -\Li_{1/2}(-\zeta)\int_0^s F_{1,b}(r)dr\approx -\frac{4\pi \Li_{3/2}(-\zeta)}{3}s^3\;.
\ee
Inserting this expression in Eq. \eqref{q_1_b_small}, this yields
\be\label{q_1_b_small}
\boxed{\frame{\boxed{q_{1,b}(s)\approx 1+\frac{4\pi \Li_{3/2}(-\zeta)}{3}s^3\;,\;\;s\to 0\;.}}}
\ee
Using again $\zeta=e^{\beta \epsilon_F}\gg 1$ together with the second line of Eq. \eqref{polylog_as}, the zero temperature result given in the first line of Eq. \eqref{eq: typ_fluc_1d} is recovered
\be
q_{1,b}(s)\approx 1-\frac{2}{9\pi}\left(k_F \Lambda_{\beta} s\right)^3\approx q_1\left(k_F \Lambda_{\beta} s\right)\;,\;\;s\to 0\;.
\ee
\item In order to obtain the large $s\gg 1$ asymptotic behaviour, we first consider a general term of order $p$ in the development in traces in Eq. \eqref{q_1_b} \cite{le2016exact}
\be\label{trace_p}
\Tr\left[(P_{[0,s]}K_{1,b}^{\rm e}P_{[0,s]})^p\right]=\int_0^{\infty}\frac{\zeta dk_1}{\zeta + e^{\frac{k_1^2}{4\pi}}}\cdots\int_0^{\infty}\frac{\zeta dk_p}{\zeta + e^{\frac{k_p^2}{4\pi}}}\prod_{j=1}^p \frac{1}{\pi}\int_0^{s}\cos(k_j u)\cos(k_{j+1} u)du\;,
\ee
with periodic boundary conditions $k_{p+1}=k_1$. The integral over $u$ yields
\be\label{k_trace_1d_hb}
\frac{1}{\pi}\int_0^{s}\cos(k_j u)\cos(k_{j+1} u)du=\frac{\sin(s(k_j-k_{j+1}))}{\pi(k_j-k_{j+1})}-\frac{\sin(s(k_j+k_{j+1}))}{\pi(k_j+k_{j+1})}\;.
\ee
In the limit of large $s$, the right hand side is dominated by the first term, which is of order $s$ for $k_j\sim k_{j+1}$ while the second term only gives an oscillatory contribution of order $O(1)$. Inserting the first term in the right hand side of Eq. \eqref{k_trace_1d_hb} in Eq. \eqref{trace_p}, it yields
\be
\Tr\left[(P_{[0,s]}K_{1,b}^{\rm e}P_{[0,s]})^p\right]\approx \int_0^{\infty}\frac{\zeta dk_1}{\zeta + e^{\frac{k_1^2}{4\pi}}}\cdots\int_0^{\infty}\frac{\zeta dk_p}{\zeta + e^{\frac{k_p^2}{4\pi}}}\prod_{j=1}^p \frac{\sin(s(k_j-k_{j+1}))}{\pi(k_j-k_{j+1})}\;.
\ee
We introduce in this integral the change of variables $K=\frac{1}{p}\sum_{i=1}^p k_i$ and $\omega_j=s(k_j-k_{j+1})$ and take the large $s$ limit. In this limit, the expression reads
\be
\Tr\left[(P_{[0,s]}K_{1,b}^{\rm e}P_{[0,s]})^p\right]\approx s\int_0^{\infty}dK \left(\frac{\zeta}{\zeta + e^{\frac{K^2}{4\pi}}}\right)^p I_p\;,\label{trace_order_p_final}
\ee 
where $I_p$ is a $p$-fold integral that can be computed explicitly
\begin{align}
I_p=&\prod_{j=1}^{p}\int_{-\infty}^{\infty}d\omega_i \frac{\sin(\omega_i)}{\pi \omega_i} \delta\left(\sum_{l=1}^p \omega_l\right)=\int_{-\infty}^{\infty}\frac{d\lambda}{2\pi} \prod_{j=1}^{p}\int_{-\infty}^{\infty}d\omega_i \frac{\sin(\omega_i)}{\pi \omega_i}e^{i \lambda \omega_i}\\
=&\int_{-\infty}^{\infty}\frac{d\lambda}{2\pi}\Theta(1-|\lambda|)=\frac{1}{\pi}\;.
\end{align}
In this regime, all terms of arbitrary order $p$ in Eq. \eqref{trace_order_p_final} are proportional to $s$. Inserting Eq. \eqref{trace_order_p_final} in Eq. \eqref{q_1_b} and resumming all the terms of arbitrary order $p$ in the expansion, one obtains
\be
q_{1,b}(s)\approx\exp\left(-\frac{s}{\pi}\int_0^{\infty}dK \ln\left(1+\zeta e^{-\frac{K^2}{4\pi}}\right)\right)\;,\;\;s\to\infty\;.
\ee
The integral over $K$ can be computed explicitly, which yields
\be\label{q_1_b_large}
\boxed{\frame{\boxed{q_{1,b}(s)\approx \exp\left(\Li_{3/2}(-\zeta)s\right)\;,\;\;s\to\infty\;.}}}
\ee
This exponential decay of the finite temperature CDF as $(1-x_{\max})\gg \Lambda_{\beta}$ is quite different from the Gaussian decay of the zero-temperature CDF $q_1(s)$ in \eqref{eq: typ_fluc_1d} as $(1-x_{\max})\gg k_F^{-1}$. One naturally expects that there must be a crossover function for the intermediate scale $k_F^{-1}\ll (1-x_{\max})\ll \Lambda_{\beta}$ matching both of these behaviours but this problem turns out to be quite hard to solve and is therefore left for future studies.
\end{itemize}
Another interesting problem would be to compute the large deviation function in the regime $T\sim T_F$. We obtained in Eq. \eqref{LD_T_0_hb_1d} the rate function in the zero temperature regime. Furthermore, in the classical regime using i.i.d. variables for the positions of the fermions, it is trivial to obtain the large deviation rate function controlling the atypical fluctuations
\be
\Prob\left[x_{\max}\leq x\right]=\left(\frac{1+x}{2}\right)^N=\exp\left(N\ln\left(\frac{1+x}{2}\right)\right)\;,\;\;x=O(1)\;.\label{LD_class_1d_t}
\ee
It is therefore natural to expect that in the regime of the temperature $T\sim T_F$, i.e. $b=O(1)$, there is a large deviation form $\Prob\left[x_{\max}\leq w\right]\approx \exp\left(-N \varphi_{1,b}(x)\right)$, which reduces at high temperature where $b\to 0$ to this classical result. Inserting $x=1-\Lambda_{\beta}s$ in Eq. \eqref{LD_class_1d_t} and taking the limit $s\to 0$, with $N\Lambda_{\beta}s=O(1)$, one obtains 
\be
\Prob\left[x_{\max}\leq 1-\Lambda_{\beta}s\right]\approx \exp\left(-\frac{N\Lambda_{\beta}s}{2}\right)\;.
\ee
It is then easy to obtain by inserting in Eq. \eqref{q_1_b_large} the behaviour in the first line of Eq. \eqref{polylog_as}, valid in the high temperature limit $\zeta=e^{\beta \mu}\approx N\Lambda_{\beta}/2 \ll 1$, that it matches smoothly with the tail of the regime of typical fluctuations. 
%
%
%
%
%
%
%
One therefore expects at finite temperature, for $b=\beta\epsilon_F=O(1)$, two different regimes of fluctuation of $x_{\max}$ summarised as (see also Fig. \ref{fluc_x_max_hb_1d_t_fig})
\be
\boxed{
\Prob\left[x_{\max}\leq x\right]=\begin{cases}
q_{1,b}\left(\frac{1-x}{\Lambda_{\beta}}\right)&\;,\;\;(1-x)=O(\Lambda_{\beta})\;,\\
&\\
\displaystyle e^{-N\varphi_{1,b}(1-x)}&\;,\;\;(1-x)=O(1)\;.
\end{cases}}\label{fluc_x_max_hb_1d_t_sum}
\ee

We will now analyse to which extent the results obtained for the hard box potential close to the hard edge hold for a more general class of potentials.

\section{Universality close to the hard edge}

In this section, we study the universality class associated to the hard edge. We first consider ``truncated potentials'' whose behaviour at the edge crosses over smoothly between soft and hard edge.

\subsection{Truncated potentials}

We first consider the case where the potential $V(x)$ has two components, a hard box potential of size $R$ and a smooth non-uniform potential $v(x)$,
\be\label{trunc_pot}
V({ x})=\begin{cases}
v({ x})&\;,\;\;|{ x}|<R\;,\\
&\\
+\infty&\;,\;\;|{ x}|\geq R\;.
\end{cases}
\ee

We call this potential a ``truncated potential''. For simplicity, we will focus on the case of a truncated linear potential with $v(x)=\mu|x|/r_{\rm e}$. Applying the local density approximation at zero temperature, we obtain the average density
\be\label{rho_trunc_pot_bulk}
\rho_N(x)=\frac{1}{N \pi}\sqrt{\frac{2m\mu(r_{\rm e}-|x|)}{\hbar^2 r_{\rm e}}}\Theta(R-|x|)=\frac{k_F}{N \pi}\sqrt{\frac{r_{\rm e}-|x|}{r_{\rm e}}}\Theta(R-|x|)\;.
\ee
The behaviour of this density will depend on the respective scales $R$ and $r_{\rm e}$. 

\begin{figure}
\centering
\includegraphics[width=0.6\textwidth]{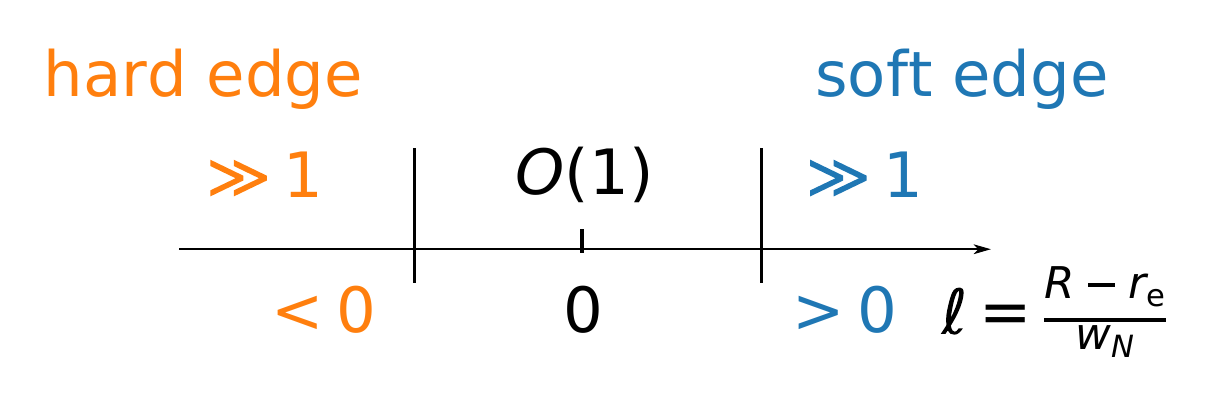}
\caption{Scheme of the length scales of the problem. In the regime $\ell=(R-r_{\rm e})/w_N>0$ and $\ell\gg 1$ (in blue), a soft edge behaviour is recovered for $|x-r_{\rm e}|\sim w_N$. On the contrary, in the regime $\ell<0$ and $|\ell|\gg 1$ (in orange), a hard edge behaviour is recovered for $|R-r_{\rm e}|\sim k_F^{-1}$. The transition between these two regimes appear for $\ell =O(1)$.}\label{Fig_scales_trunc}
\end{figure}

\begin{figure}[h]
 \centering
 \includegraphics[width=1\textwidth]{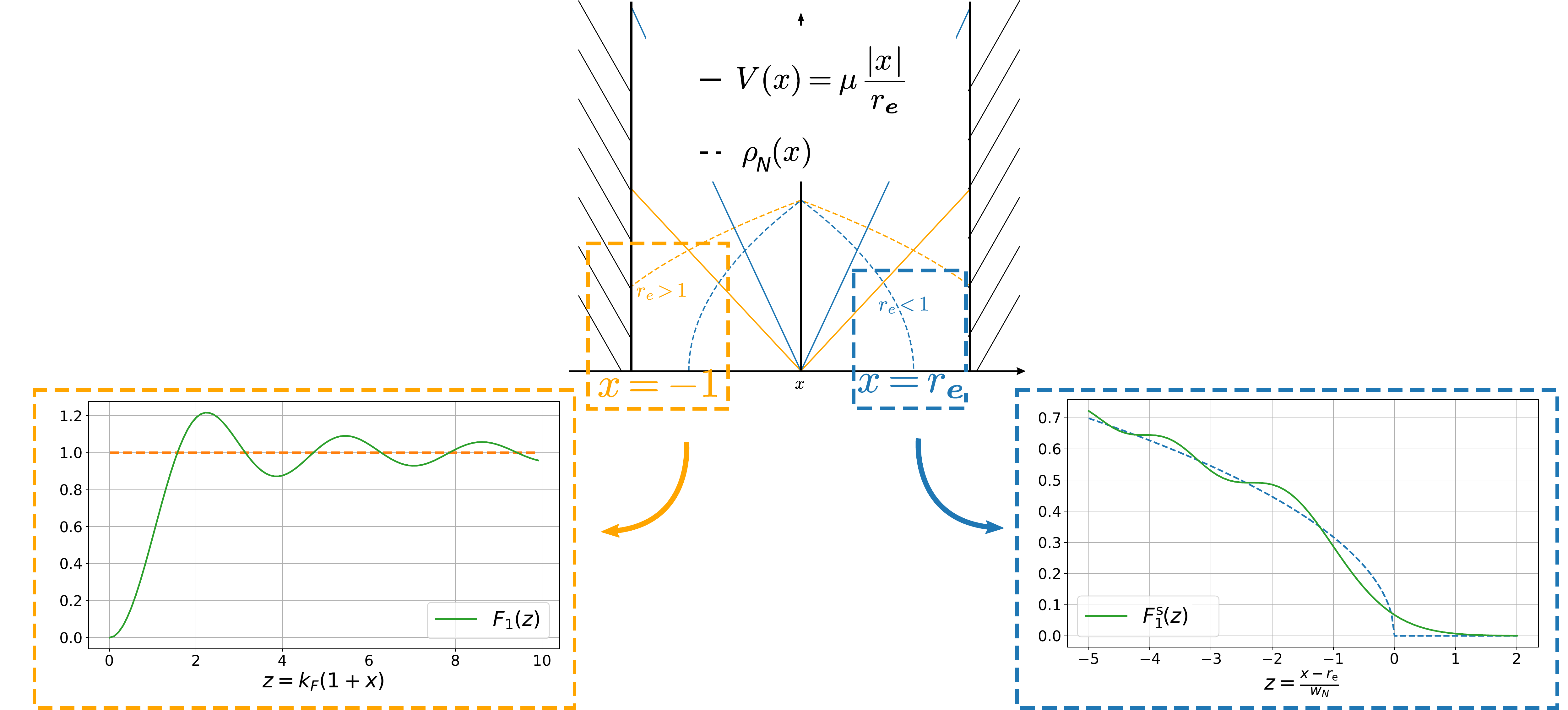}
 \caption{Sketch of two extreme situations for a linear potential of increasing slope in the bulk. In the first situation -- in orange -- the potential $V(x)$ inside the box is not strong enough to create an edge $r_e$ of the density $\rho_N(x)\propto \sqrt{r_{\rm e}-|x|}$ given in Eq. \eqref{rho_trunc_pot_bulk} that lies in the box, i.e. $r_{\rm e}>1$. On a scale $s=k_F(1+x)=O(1)$ close to the hard edge (with $x<0$), the density takes the hard edge scaling form, with the scaling function $F_1(z)$ given in Eq. \eqref{dens_1d_edge_T0}. In the second situation -- in blue -- the potential creates an edge $r_{\rm e}<1$ that lies in the box, but far enough from the walls for the fermions to be impacted by their presence. On a scale $z=(r_{\rm e}-x)/w_N=O(1)$ -- where $w_N=k_F^{-2/3}r_{\rm e}^{1/3}$ -- the density takes the soft edge scaling form with the scaling function $F_{1}^{\rm s}(z)$ given in Eq. \eqref{F_1_soft}. Note that the solid orange and blue lines in the top
 figure are the plots of $V(x)$ as a function $x$ and that their dotted counterparts are the corresponding bulk densities $\rho_N(x)$ as a function of $x$.}\label{Fig_distance}
\end{figure}

If $r_{\rm e}<R$, one recovers the case of a {\it soft edge} where the density vanishes smoothly as $\rho_N(x)\sim \sqrt{r_{\rm e}-x}$ for $x\to r_{\rm e}$ (c.f. Figs. \ref{Fig_distance} and \ref{Fig_scales_trunc}). In this case, the hard box potential does not play any role and one expects that the fluctuations at the soft edge are controlled by the Airy kernel in Eq. \eqref{k_airy}. 

Furthermore, if $r_{\rm e}>R$, one recovers the case of a {\it hard edge} where the density vanishes abruptly $\rho_N(x)\sim \Theta(R-x)$ as $x\to R$ (c.f. Figs. \ref{Fig_distance} and \ref{Fig_scales_trunc}). In this case the potential appears locally uniform at the edge and the fluctuations should be controlled by the hard edge correlation kernel in Eq. \eqref{k_1d_hb_e}.

The transition from soft to hard edge will occur when the algebraic distance between the wall and the soft edge is of order of the typical scale at the soft edge (c.f. Fig. \ref{Fig_scales_trunc}), i.e. for
\be
\ell=\frac{R-r_{\rm e}}{w_N}=O(1)\;,\;\;{\rm with}\;\;w_N=\left(\frac{\hbar^2}{2m v'(r_{\rm e})}\right)^{\frac{1}{3}}\;,
\ee
with in particular $w_N=k_F^{-2/3} r_{\rm e}^{1/3}$ in the case of the linear potential.

In order to compute the correlation kernel for the intermediate case $|\ell|=O(1)$, we will use the method of the propagator introduced in section \ref{univ_prop_1d}.
We now consider in detail the behaviour of the Euclidean propagator associated to the truncated linear potential in Eq. \eqref{trunc_pot}.
We introduce the single-body Euclidean propagator associated to this problem $G(y,t|x,0)$
\be
G(y,t|x,0)=\langle y|e^{-\frac{\hat H t}{\hbar}}|x\rangle = \frac{t}{\hbar}\int_0^{\infty}d\mu\,e^{-\frac{\mu t}{\hbar}}K_{\mu}(x,y)\;.
\ee
This propagator is solution of the partial differential equation
\be
\hbar \partial_t G(y,t|x,0)=\frac{\hbar^2}{2m}\partial_y^2 G(y,t|x,0)-\mu\frac{|y|}{r_{\rm e}} G(y,t|x,0)\;,\;\;{\rm with}\;\;G(y,0|x,0)=\delta(x-y)\;,
\ee
and with Dirichlet boundary conditions for $y=\pm R$. To obtain the solution of this equation, we may introduce the inverse Laplace transform of the Euclidean propagator
\be
G(y,t|x,0)=\int_0^{\infty}d\epsilon\, e^{-\frac{\epsilon t}{\hbar}}\tilde G(y|x;\epsilon)\;,
\ee
where the function $\tilde G(y|x;\epsilon)$ is solution of the ordinary differential equation
\begin{align}
&\frac{\hbar^2}{2m}\partial_y^2 \tilde G(y|x;\epsilon)=\left(\mu\frac{|y|}{r_{\rm e}}-\epsilon\right)\tilde G(y|x;\epsilon)\;,\;\;{\rm with}\;\;G(y= \pm R|x;\epsilon)=0\;,\nn\\
&{\rm and}\;\;\int_0^{\infty} \tilde G(y|x;\epsilon)d\epsilon=\delta(x-y)\;.\label{prop_lin}
\end{align}
Using Eq. \eqref{LT_step} in the table of Laplace transform of Appendix \ref{LT}, one may realise that $\tilde G(y|x;\epsilon)$ is closely related to the correlation kernel $K_{\mu}(x,y)$, 
\be\label{prop_G}
K_{\mu}(x,y)=\int_0^{\infty}d\epsilon\int_{\cal C}\frac{dt}{2i\pi t}\exp\left(\frac{(\mu-\epsilon) t}{\hbar}\right)\tilde G(y|x;\epsilon)=\int_0^{\mu}d\epsilon\, \tilde G(y|x;\epsilon)\;.
\ee

Let us first consider the case where $R\to \infty$, which will be useful to understand the general case. Introducing the rescaled function close to the edge
\be
\tilde G(y|x;\epsilon)=\frac{1}{w_N}\tilde G_{\infty}^{\rm e}\left(\frac{r_{\rm e}-x}{w_N}\Big|\frac{r_{\rm e}-y}{w_N};\frac{r_{\rm e}}{w_N}(\mu-\epsilon)\right)\;,\;\;{\rm with}\;\;w_N=\left(\frac{2m\mu}{\hbar^2 r_{\rm e}}\right)^{-1/3}\;,
\ee
in Eq. \eqref{prop_lin} and taking the large $\mu$ limit, one obtains
\begin{align}
&\partial_v^2 \tilde  G_{\infty}^{\rm e}(v|u;\lambda)=(v+\lambda)\tilde  G_{\infty}^{\rm e}(v|u;\lambda)\;,\;\;\tilde  G_{\infty}^{\rm e}(v\to +\infty|u;\lambda)=0\nn\\
&{\rm and}\;\;\int_{-\infty}^{\infty} \tilde  G_{\infty}^{\rm e}(v|u;\lambda)d\lambda=\delta(v-u)\;.
\end{align}
The solution $\tilde  G_{\infty}^{\rm e}(v|u,\lambda)$ of this equation is then simply expressed in terms of the Airy function $\Ai(x)$, solution of the equation $f''(x)=x f(x)$ which vanishes at $x\to \infty$,
\be
\tilde  G_{\infty}^{\rm e}(v|u,\lambda)=\Ai(v+\lambda)\Ai(u+\lambda)\;.
\ee
Inserting this scaling form in Eq. \eqref{prop_G}, one recovers in the large $\mu \to\infty$ limit the standard Airy kernel at the soft edge
\be
K_{\mu}(x,y)=\frac{1}{w_N}K_{\Ai}\left(\frac{x-r_{\rm e}}{w_N},\frac{y-r_{\rm e}}{w_N}\right)\;,\;\;{\rm with}\;\;K_{\Ai}(u,v)=\int_0^{\infty}\Ai(u+s)\Ai(v+s)ds\;.
\ee

Considering now the general case of a finite value of $R$, we rescale the function $G(y|x;\epsilon)$ close to the wall in $R$ as
\be
\tilde G(y|x;\epsilon)=\frac{1}{w_N}\tilde  G_{\ell}^{\rm e}\left(\frac{r_{\rm e}-x}{w_N}\Big|\frac{r_{\rm e}-y}{w_N};\frac{r_{\rm e}}{w_N}(\mu-\epsilon)\right)\;,\;\;{\rm with}\;\;\ell=\frac{R-r_{\rm e}}{w_N}\;.
\ee
Inserting this scaling form in Eq. \eqref{prop_lin}, and taking the limit $\mu\to \infty$, the function $\tilde  G_{\ell}^{\rm e}(v|u;\lambda)$ is solution of the ordinary differential equation
\begin{align}
&\partial_v^2 \tilde G_{\ell}^{\rm e}(v|u;\lambda)=(\ell+v+\lambda)\tilde G_{\ell}^{\rm e}(v|u;\lambda)\;,\;\;u,v\leq 0\;,\;\;\tilde G_{\ell}^{\rm e}(v=0|u;\lambda)=0\nn\\
&{\rm and}\;\;\int_{-\infty}^{\infty} \tilde G_{\ell}^{\rm e}(v|u;\lambda)d\lambda=\delta(v-u)\;.
\end{align}
This problem can be solved exactly \cite{ferrari2013spatial, martin1998final} and the solution $\tilde G_{\ell}^{\rm e}(v|u;\lambda)$ for the Dirichlet boundary condition in $v=0$ is built from the two linearly independent solutions $\Ai(x)$ and $\Bi(x)$ of the equation $f''(x)=x f(x)$,
\begin{align}
&\tilde G_{\ell}^{\rm e}(v|u;\lambda)=\Theta(-u)\Theta(-v)\sigma(\lambda+\ell,v+\lambda+\ell)\sigma(\lambda+\ell,u+\lambda+\ell)\;,\nn\\
&{\rm with}\;\;\sigma(r,s)=\frac{\Ai(r)\Bi(s)-\Ai(r)\Bi(s)}{\left(\Ai(r)^2+\Bi(r)^2\right)^{1/2}}\;.
\end{align}
Note that in the limit $\ell\to \infty$ one recovers $\tilde G_{\infty}^{\rm e}(v|u;\lambda)=\lim_{\ell\to +\infty}\tilde G_{\ell}^{\rm e}(v-\ell|u-\ell;\lambda)$. Inserting the scaling form for $\tilde G(y|x;\epsilon)$ in Eq. \eqref{prop_G}, one obtains in the large $\mu \to\infty$ limit the scaling form for the correlation kernel
\be
K_{\mu}(x,y)=\frac{1}{w_N}K_{1}^{\ell}\left(\frac{x-r_{\rm e}}{w_N},\frac{y-r_{\rm e}}{w_N}\right)\;.
\ee
The scaling function $K_{1}^{\ell}(u,v)$ is only non zero for $u,v\leq 0$, and reads
\be
\boxed{\begin{array}{rl}
&\displaystyle K_{1}^{\ell}(u,v)=\Theta(-u)\Theta(-v)\int_{\ell}^{\infty}\sigma(s,s+u)\sigma(s,s+v)ds\;,\vspace{0.2cm}\\
&\displaystyle \sigma(r,s)=\frac{\Ai(r)\Bi(s)-\Ai(r)\Bi(s)}{\left(\Ai(r)^2+\Bi(r)^2\right)^{1/2}}\;.\label{trunc_ker}
\end{array}}
\ee
This new correlation kernel $K_1^{\ell}(u,v)$ depends continuously on the parameter $\ell$, which is the rescaled algebraic distance between the wall and the soft edge. In particular, one can show that in the limit $R\gg r_{\rm e}$ where the wall is much farther than the soft edge, i.e. $\ell \to +\infty$ (c.f. Fig. \ref{Fig_scales_trunc}), one recovers after a simple rescaling the standard Airy kernel
\be
\lim_{\ell \to \infty} K_{1}^{\ell}(u-\ell,v-\ell)=K_{\Ai}(u,v)=\int_0^{\infty}ds \Ai(s+u)\Ai(s+v)\;.\label{K_l_to_K_ai}
\ee
Furthermore, in the limit $R\ll r_{\rm e}$ where the soft edge is much farther than the wall , i.e. $\ell \to -\infty$ (c.f. Fig. \ref{Fig_scales_trunc}), one obtains after a slight rescaling the hard box correlation kernel
\be
\lim_{\ell \to -\infty} \frac{1}{\sqrt{-\ell}}K_{1}^{\ell}\left(\frac{u}{\sqrt{-\ell}},\frac{v}{\sqrt{-\ell}}\right)=K_{1}^{\rm e}(u,v)=\frac{2}{\pi}\int_0^{1}dk \sin(k u)\sin(k v)\;.\label{K_l_to_K_hb}
\ee
In particular, the density scaling function associated to $K_{1}^{\ell}(u,v)$ reads
\be
\boxed{F_{1}^{\ell}(z)=K_{1}^{\ell}(z,z)=\Theta(-z)\int_{\ell}^{\infty}\frac{(\Ai(s+z)\Bi(s)-\Bi(s+z)\Ai(s))^2}{\Ai^2(s)+\Bi^2(s)}ds\;.}\label{rho_ell_eq}
\ee
This function vanishes quadratically close to the wall, as in the case of the hard box
\be
F_{1}^{\ell}(z)=\alpha(\ell)z^2\;,\;\;z\to 0\;,
\ee
where the coefficient $\alpha(\ell)$ depends smoothly on $\ell$ and has an exact expression \cite{vallee2010airy}
\be
\alpha(\ell)=\frac{1}{\pi^2}\int_\ell^{\infty}\frac{ds}{\Ai^2(s)+\Bi^2(s)}=\frac{1}{\pi}\arctan\left(\frac{\Ai(\ell)}{\Bi(\ell)}\right)+\sum_{k=1}^{\infty}\Theta(a_k-\ell)\;.
\ee
The $a_k$'s are the zeroes of the Airy function, which are all negative and behave as $a_k\sim -(3\pi k/2)^{2/3}$ for $k\gg 1$. For the particular case $\ell=0$, one obtains the simple expression $\alpha(0)=1/6$. The asymptotic behaviours of $\alpha(\ell)$ read
\be
\alpha(\ell)\approx\begin{cases}
\displaystyle \frac{2}{3\pi}|\ell|^{\frac{3}{2}}&\;,\;\;\ell\to-\infty\;,\\
&\\
\displaystyle \frac{e^{-\frac{4}{3}\ell^{3/2}}}{2}&\;,\;\;\ell\to +\infty\;.
\end{cases}
\ee

\begin{figure}
\centering
\includegraphics[width=0.7\textwidth]{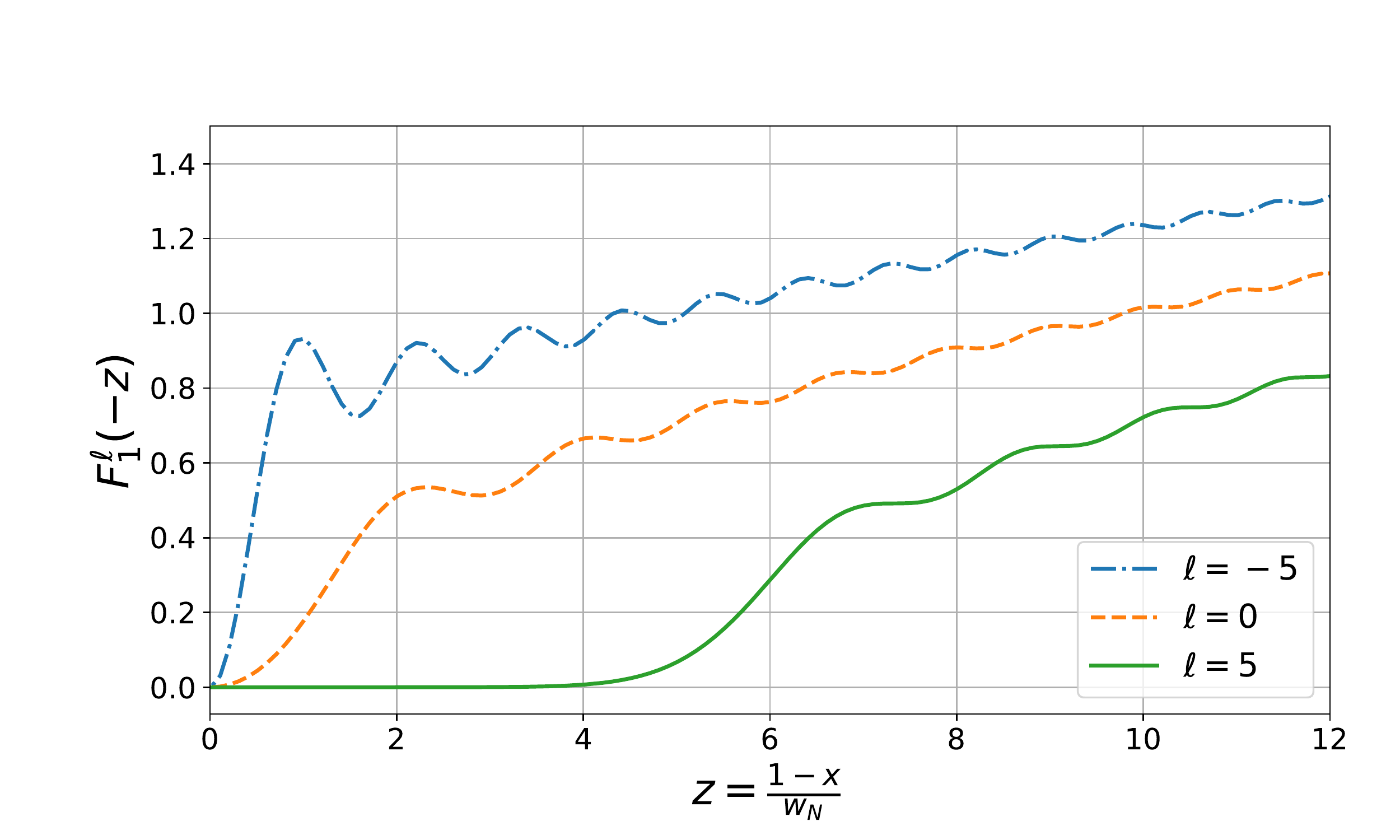}
\caption{Plot of the scaling function $F_{1}^{\ell}(z)$ given in Eq. \eqref{rho_ell_eq} for $\ell=-5,0,5$ respectively in blue, orange and green.}\label{Fig_dens_trunc}
\end{figure}

In the limit $|z|\sim|\ell|\gg 1$, one obtains instead the asymptotic behaviour
\be
F_{1}^{\ell}(z)\approx \Theta(-z)\Theta(|z|-\ell)\frac{\sqrt{|z|-\ell}}{\pi}\;,\;\;|z|\sim|\ell|\to\infty\;.
\ee
In particular, in the limit where $\ell$ is large and positive, one recovers close to $|z|=\ell$ the square root profile characteristic of the soft edge behaviour (c.f. Fig. \ref{Fig_distance}). In the limit where $\ell$ is large and negative, the density profile goes to a constant value, which is characteristic of hard edge (c.f. Fig. \ref{Fig_distance}). The scaling function $F_1^{\ell}(z)$ is plotted in Fig. \ref{Fig_dens_trunc} for $\ell =-5,0,5$.

\subsubsection{Fluctuations of the position $x_{\max}$ of the rightmost fermion}

\begin{figure}
\centering
\includegraphics[width=0.6\textwidth]{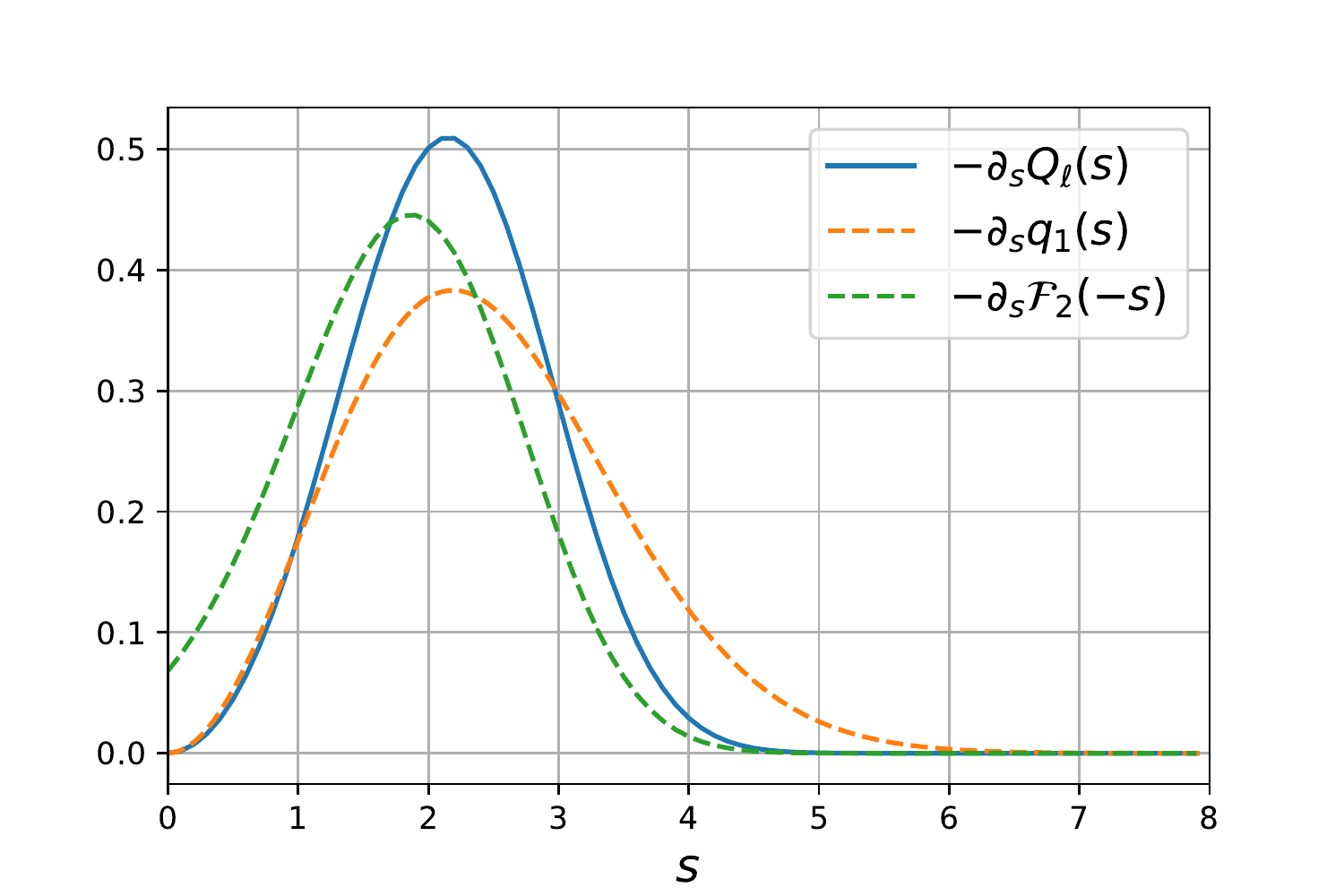}
\caption{Plot of the PDF $-\partial_s Q_{\ell}(s)$ (in blue) where $Q_{\ell}(s)$ is given in Eq. \eqref{CDF_K_l} representing the typical fluctuations of $x_{\max}$ (or $x_{\min}$) in a truncated linear potential at $T=0$. We have reproduced here the hard box scaling function $-\partial_s q_1(s)$ (in dashed orange) obtained from Eq. \eqref{CDF_x_max_box_T0} and the Tracy-Widom distribution $-\partial_s {\cal F}_2(-s)$ (in dashed green) for the sake of comparison between these distributions. The PDFs $-\partial_s Q_{\ell}(s)$ and $-\partial_s q_{1}(s)$ were obtained using the algorithm developed in \cite{bornemann2010numerical}. }\label{Fig_x_max_trunc}
\end{figure}

For the truncated linear potential, the typical scale of fluctuation of $x_{\max}$ close to the wall in $R$ is given by the soft edge typical scale $w_N$. In this 
typical regime of fluctuation, the CDF of the rightmost fermion $x_{\max}$ is given by the Fredholm determinant
\be\label{CDF_K_l}
\lim_{N\to\infty}\Prob\left[x_{\max}\leq R-w_N s\right]=Q_{\ell}(s)=\Det\left[\mathbb{I}-P_{[-s,0]}K_{1}^{\ell}P_{[-s,0]}\right]\;,
\ee
where the scaling function for the kernel $K_1^{\ell}(u,v)$ given in Eq. \eqref{trunc_ker} depends explicitly on the parameter $\ell$. Using the convergence properties of the kernel scaling function for $\ell \to +\infty$ in Eq. \eqref{K_l_to_K_ai} and $\ell \to -\infty$ in \eqref{K_l_to_K_hb}, the function $Q_{\ell}(s)$ allows a smooth matching between the Tracy-Widom $\beta=2$ distribution ${\cal F}_2(s+\ell)$ and the hard-box CDF $q_1(s/\sqrt{-\ell})$. The PDF $-\partial_s Q_{\ell}(s)$ is plotted in Fig. \ref{Fig_x_max_trunc} for the special case $\ell=0$ together with a comparison with the Tracy-Widom distribution and the hard edge PDF $-\partial_s q_1(s)$. The question of whether the CDF $Q_{\ell}(s)$ is associated to a Painlev\'e transcendent as it is the case for both the Tracy-Widom distribution ${\cal F}_2(s)$ (c.f. Eq. \eqref{TW_2}) and $q_1(s)$ (c.f. Eq. \eqref{q_Painleve_3}) remains open.

We close this section by mentioning that the framework used here is not restricted to a linear potential within the box and the results should be universal with respect to the potential $v(x)$. Heuristically, one may always linearise the smooth potential $v(x)$ close to the soft edge $r_{\rm e}$ and recover the results obtained here. We refer to Article \ref{Art:ferm_long} for more details on this matter (and for its generalisation to higher dimension $d$).

As seen in section \ref{he_fermions}, some inverse power-law potentials lead to a hard edge behaviour. We will now analyse the spatial statistics close to the origin for these potential and in particular whether the results for the hard-edge kernel in Eq. \eqref{k_1d_hb_e} are retrieved.

\subsection{Inverse power law potentials}

We now consider inverse power-law potentials of the type
\be\label{inv_power}
V({ x})=\begin{cases}
+\infty &\;,\;\;x\leq 0\;,\\
&\\
\frac{\hbar^2 a(a-1)}{2m x^{\gamma}}&\;,\;\;x>0\;,
\end{cases}
\ee
where for $\gamma,\,a\geq 1$ one naturally expects a hard edge to occur close to the origin. For these inverse power law potentials, the quantum states form a continuum indexed by $k\in[0,\infty)$, with an associated set of orthonormal wave functions
\be
\int_0^{\infty}\overline{\phi_k}(x)\phi_l(x)dx=\delta(k-l)\;.
\ee
Ensuring the finiteness of the potential energy for all states 
\be
\int_0^{\infty}|\phi_k(x)|^2 V(x)dx<\infty
\ee
imposes Dirichlet boundary conditions for the wave functions $\phi_k(0)=0$ \cite{andrews1976singular}. If one considers instead weaker inverse power law potentials $V(x)\sim x^{-\gamma}$ with $0<\gamma<1$, the finiteness of potential energy does not impose a Dirichlet boundary condition. Note that one could add another confining potential to this problem (for instance $v(x)=\frac{1}{2}m\omega^2 x^2$), to ensure the discreteness of the quantum states without changing in the large $\mu$ limit the physics close to the hard edge in $x=0$. 
We introduce the single body Euclidean propagator $G(y,t|x,0)$, solution of the equation
\be
\hbar \partial_t G(y,t|x,0)=\frac{\hbar^2}{2m}\partial_y^2 G(y,t|x,0)-\frac{\hbar^2 a(a-1)}{2m y^{\gamma}}G(y,t|x,0)\;,\;\;{\rm with}\;\;G(y,0|x,0)=\delta(x-y)\;,
\ee
and with Dirichlet boundary condition for $y=0$. We may solve this equation by introducing its inverse Laplace transform form $t$ to $\epsilon_q$, evaluated for $\epsilon_q=\frac{\hbar^2 q^2}{2m}$,
\be
G(y,t|x,0)=\int_0^{\infty}\tilde G\left(y|x;\epsilon_q=\frac{\hbar^2 q^2}{2m}\right)\exp\left(-\frac{\hbar q^2 t}{2m}\right)\frac{\hbar^2 q dq}{m}\;,
\ee
which satisfies the ordinary differential equation
\be
\partial_y^2 \tilde G(y|x;\epsilon_q)+\left(q^2-\frac{a(a-1)}{y^{\gamma}}\right)\tilde G(y|x;\epsilon_q)=0,\;\;{\rm with}\;\;\int_0^{\infty} \tilde G(y|x;\epsilon_q) \frac{\hbar^2 qdq}{m}=\delta(x-y)\;,
\ee
and Dirichlet boundary condition for $y=0$. This function $\tilde G(y|x;\epsilon_q)$ allows us to obtain the correlation kernel as
\be
K_{\mu}(x,y)=\int_0^{\infty}dq\int_{\cal C}\frac{dt}{2i\pi t}\exp\left(\frac{\hbar (k_F^2-q^2)t}{2m}\right)\frac{\hbar q}{m}\tilde G(y|x;\epsilon_q)=\int_0^{k_F}dq\, \tilde G(y|x;\epsilon_q)\;,
\ee
where we used $\mu=\frac{\hbar^2 k_F^2}{2m}$. Rescaling the function $\tilde G(y|x;\epsilon_q)$ close to the origin,
\be
\tilde G(y|x;\epsilon_q)=k_F \tilde G^{\rm e}\left(k_F y|k_F x;\frac{\epsilon_q}{\mu}\right)\;,
\ee
it satisfies the equation
\be
\partial_v^2\tilde G^{\rm e}(v|u;\kappa^2)+\left(\kappa^2-k_F^{\gamma-2}\frac{a(a-1)}{v^{\gamma}}\right)\tilde G^{\rm e}(v|u;\kappa^2)=0\;,\;\;\int_0^{\infty}\tilde G^{\rm e}(v|u;\kappa^2)d\kappa=\delta(u-v)\;.\label{prop_inv}
\ee
Note that the correlation kernel at the hard edge is obtained from this rescaled propagator $\tilde G^{\rm e}(v|u;\kappa^2)$ in the regime $\kappa=\frac{\epsilon_q}{\mu}<1$.

It is then clear that if $1<\gamma<2$, the potential term in Eq. \eqref{prop_inv} becomes irrelevant in the limit $k_F\to\infty$. This yields
\be
\tilde G^{\rm e}(v|u;\kappa^2)=\frac{2}{\pi}\sin(\kappa x)\sin(\kappa y)\;.
\ee
In this case one recovers, on the typical scale $k_F^{-1}$ close to the origin, the hard box correlation kernel \eqref{k_1d_hb_e}
\be
K_{\mu}(x,y)=k_F K_1^{\rm e}(k_Fx,k_F y)\;,\;\;{\rm with}\;\;K_1^{\rm e}(u,v)=\frac{2}{\pi}\int_0^{1}\sin(\kappa x)\sin(\kappa y)d\kappa\;.
\ee
For $\gamma>2$, the potential term in Eq. \eqref{prop_inv} dominates and the correlation kernel becomes exponentially small close to the origin in the limit $k_F\to \infty$. In this case, there is a finite edge $r_{\rm e}\sim k_F^{-2/\gamma}$ away from the origin where the density vanishes. Close to this edge, one recovers the soft edge scaling form on a typical scale $w_N\sim k_F^{-2(1+\gamma)/(3\gamma)}$ 
\be
K_{\mu}(x,y)=\frac{1}{w_N}K_{\Ai}\left(\frac{x-r_{\rm e}}{w_N},\frac{y-r_{\rm e}}{w_N}\right)\;,\;\;{\rm with}\;\;K_{\Ai}(u,v)=\int_0^{\infty}\Ai(u+s)\Ai(v+s)ds\;.
\ee

Finally in the case $\gamma=2$, the terms in $\kappa^2$ and of the potential in Eq. \eqref{prop_inv} are both of the same order. Close to the hard edge, we recover the correlation kernel in Eq. \eqref{K_1_x_2} that depends explicitly on $a$.

To conclude this section, we have seen that the hard edge correlation kernel $K_1^{\rm e}(u,v)$ is relevant to describe both a hard box with a non-uniform potential within the box and inverse power law potentials $v(x)\sim x^{-\gamma}$ with $1<\gamma<2$. In addition, we have obtained a complete classification of the edge behaviour for power law potentials $v(x)\sim |x|^p$ for any positive or negative value of $p$. In the next section, we consider the hard box potential in dimension $d>1$.

\section{Higher-dimensional hard box}

\begin{figure}
\centering
\includegraphics[width=0.4\textwidth]{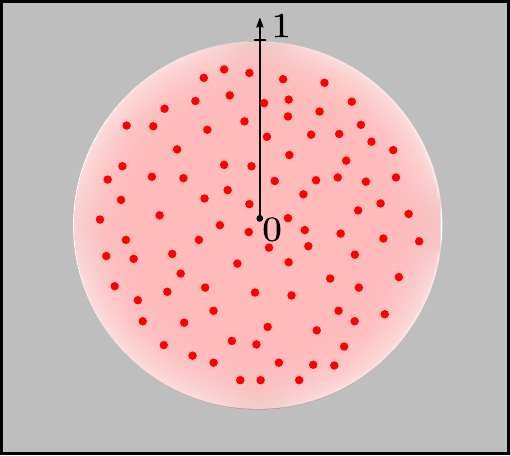}
\caption{Sketch of a configuration of positions for fermions inside a two-dimensional spherical hard box.}\label{Fig_box_2d}
\end{figure}

We now consider the spherically symmetric hard box potential in dimension $d>1$ defined as (see also Fig. \ref{Fig_box_2d})
\be\label{trunc_pot_d}
V({\bf x})=\begin{cases}
0&\;,\;\;|{\bf x}|<R\;,\\
&\\
+\infty&\;,\;\;|{\bf x}|\geq R\;,
\end{cases}
\ee
Note that in dimension $d>1$, we also considered in Article \ref{Art:fermions_lett} different shapes for the hard box potential. We set $R=1$ in the following. We first consider the zero temperature limit.

\subsection{Spherical hard box in dimension $d$ and at zero temperature}

\begin{figure}
\centering
\includegraphics[width=0.2\textwidth]{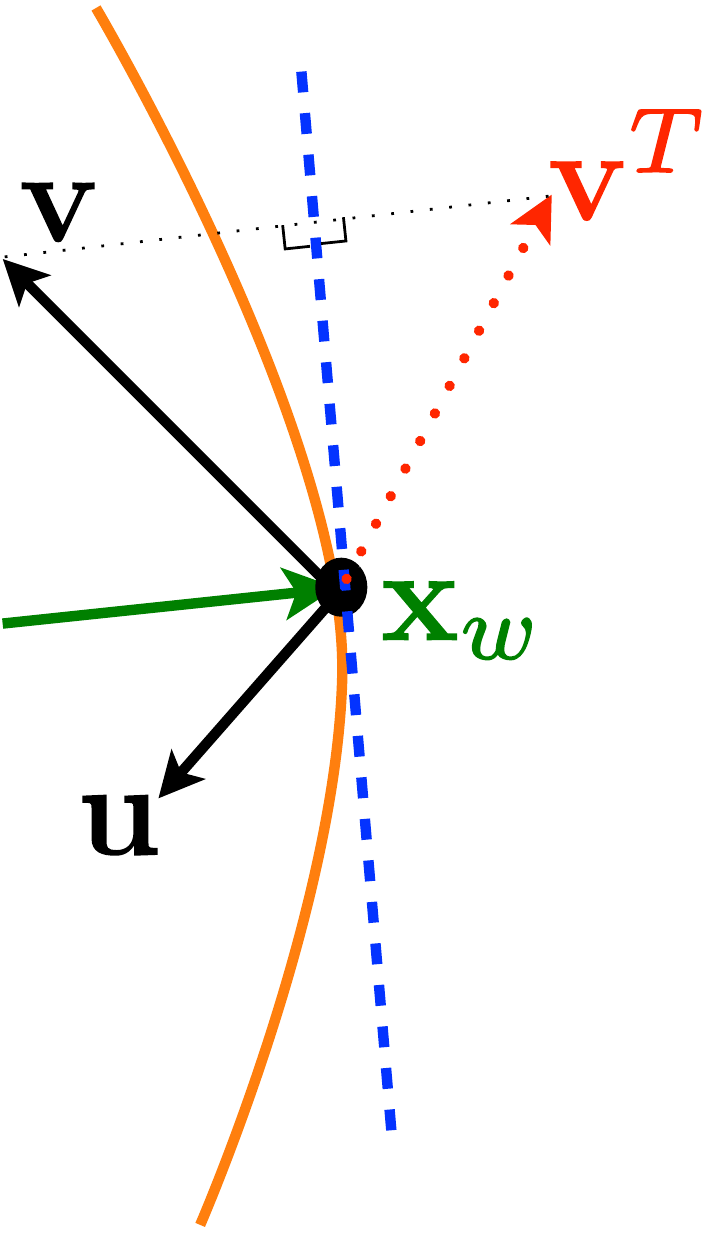}
\caption{Sketch of the construction of the image ${\bf v}^T$ of ${\bf v}$ that is orthogonal to ${\bf x}_w$.}\label{Fig_im_d}
\end{figure}

At zero temperature, we first compute the $d$-dimensional correlation kernel using the method of the propagator. We recall that the single particle Euclidean propagator can be expressed in terms of the correlation kernel as
\be
G_d({\bf y},t|{\bf x},0)=\frac{t}{\hbar}\int_0^{\infty}d\mu\,e^{-\frac{\mu t}{\hbar}}K_{\mu}({\bf x},{\bf y})\;.
\ee
It is solution of the free diffusion equation in dimension $d$
\be\label{diff_eq_prop_1}
\partial_t G_d({\bf y},t|{\bf x},0)=\frac{\hbar}{2m}\Delta_{\bf y} G_d({\bf y},t|{\bf x},0)\;,\;\;{\rm with}\;\;G_d({\bf y},0|{\bf x},0)=\delta^d({\bf x}-{\bf y})\;,
\ee 
and with Dirichlet boundary conditions for $|{\bf y}|=1$. To obtain the edge scaling form, we rescale the propagator close to a point ${\bf x}_w$ situated on the hard wall $|{\bf x}_w|=1$. The typical scale of fluctuations close to the edge is again given by the inverse of the Fermi wave vector $k_F$. This scale can be expressed in terms of the number of particles $N$ using the local density approximation \eqref{LDA_d} 
\be
\rho_N({\bf x})\approx\frac{\Omega_d}{N}\left(\frac{k_F}{2\pi}\right)^d=\frac{1}{\Omega_d}\Rightarrow N=\Omega_d^2\left(\frac{k_F}{2\pi}\right)^d\Rightarrow k_F=\frac{(N\Omega_d^2)^{\frac{1}{d}}}{2\pi}\;,\label{N_t0}
\ee
where we recall that $\Omega_d=\pi^{d/2}/\Gamma(d/2+1)$ is the volume of the $d$-dimensional ball.
As the typical scale of fluctuations is $k_F^{-1}$, one expects the scaling form
\be
G_d({\bf y},t|{\bf x},0)=k_F^d G_{d}^{\rm e}\left(k_F({\bf x}_w-{\bf y}),\frac{\mu t}{\hbar}\Big|k_F({\bf x}_w-{\bf y}),0\right)\;,
\ee
where the scaling function $G_{d}^{\rm e}({\bf v},\tau|{\bf u},0)$ also satisfies a free diffusion equation
\be\label{diff_eq_prop_2}
\partial_\tau G_d^{\rm e}({\bf v},\tau |{\bf u},0)=\Delta_{\bf v} G_d^{\rm e}({\bf v},\tau |{\bf u},0)\;,\;\;{\rm with}\;\;G_d^{\rm e}({\bf v},\tau |{\bf u},0)=\delta^d({\bf u}-{\bf v})\;,
\ee 
and with Dirichlet boundary conditions for 
\be
\left({\bf x}_w+\frac{1}{k_F}{\bf v}\right)^2=1+\frac{2}{k_F}{\bf x}_w\cdot {\bf v}+\frac{{\bf v}^2}{k_F^2}=1\;,\;\;{\rm i.e.}\;\;{\bf x}_w\cdot {\bf v}=-\frac{{\bf v}^2}{2k_F}\;.
\ee
In the limit $k_F\to \infty$, this boundary condition only applies to the hyperplane orthogonal to the vector ${\bf x}_w$, i.e ${\bf x}_w\cdot {\bf v}=0$ (c.f. Fig. \ref{Fig_im_d}).

For a boundary condition on an hyperplane, the solution can again be built from the free solution using the standard method of images and reads
\be
G_{d}^{\rm e}({\bf u},t|{\bf v},0)=\frac{\displaystyle e^{-\frac{\left({\bf u}-{\bf v}\right)^2}{4\tau}}}{(4\pi \tau)^{d/2}}-\frac{\displaystyle e^{-\frac{({\bf u}-{\bf v}^T)^2}{4\tau}}}{(4\pi \tau)^{d/2}}\;,
\ee
where ${\bf v}^T={\bf v}-2({\bf v}\cdot{\bf x}_w){\bf x}_w$ is the image of ${\bf v}$ by reflection with respect to the hyperplane ${\bf v}\cdot{\bf x}_{w}=0$ as seen in Fig. \ref{Fig_im_d}. 
To obtain the scaling form of the correlation kernel close to the edge, one needs to compute the inverse Laplace transform 
\be
K_{d}^{\rm e}({\bf u},{\bf v})=\int_{\cal C}\frac{d\tau}{2i\pi \tau} e^{\tau} G_{d}^{\rm e}({\bf u},\tau|{\bf v},0)\;,
\ee
where ${\cal C}$ is the Bromwich contour. Using Eq. \eqref{LT_bessel} in the table of Appendix \ref{LT}, this Laplace transform can be inverted explicitly as
\be
\boxed{K_{d}^{\rm e}({\bf u},{\bf v})=K_{d}^{\rm b}(|{\bf u}-{\bf v}|)-K_{d}^{\rm b}(|{\bf u}-{\bf v}^T|)=\frac{\J_{\frac{d}{2}}\left(|{\bf u}-{\bf v}|\right)}{\left(2\pi |{\bf u}-{\bf v}|\right)^{d/2}}-\frac{\J_{\frac{d}{2}}(|{\bf u}-{\bf v}^T|)}{\left(2\pi |{\bf u}-{\bf v}^T|\right)^{d/2}}\;.}\label{K_d_ell_t0}
\ee
Note that taking $d=1$ in this equation and using $\J_{1/2}(x)=\sqrt{2/(\pi x)}\,\sin(x)$, the result for the one-dimensional correlation kernel in Eq. \eqref{k_1d_hb_e} is recovered. In Articles \ref{Art:fermions_lett} and \ref{Art:ferm_long}, we developed alternative methods to derive this result. 

\begin{figure}
\centering
\includegraphics[width=0.6\textwidth]{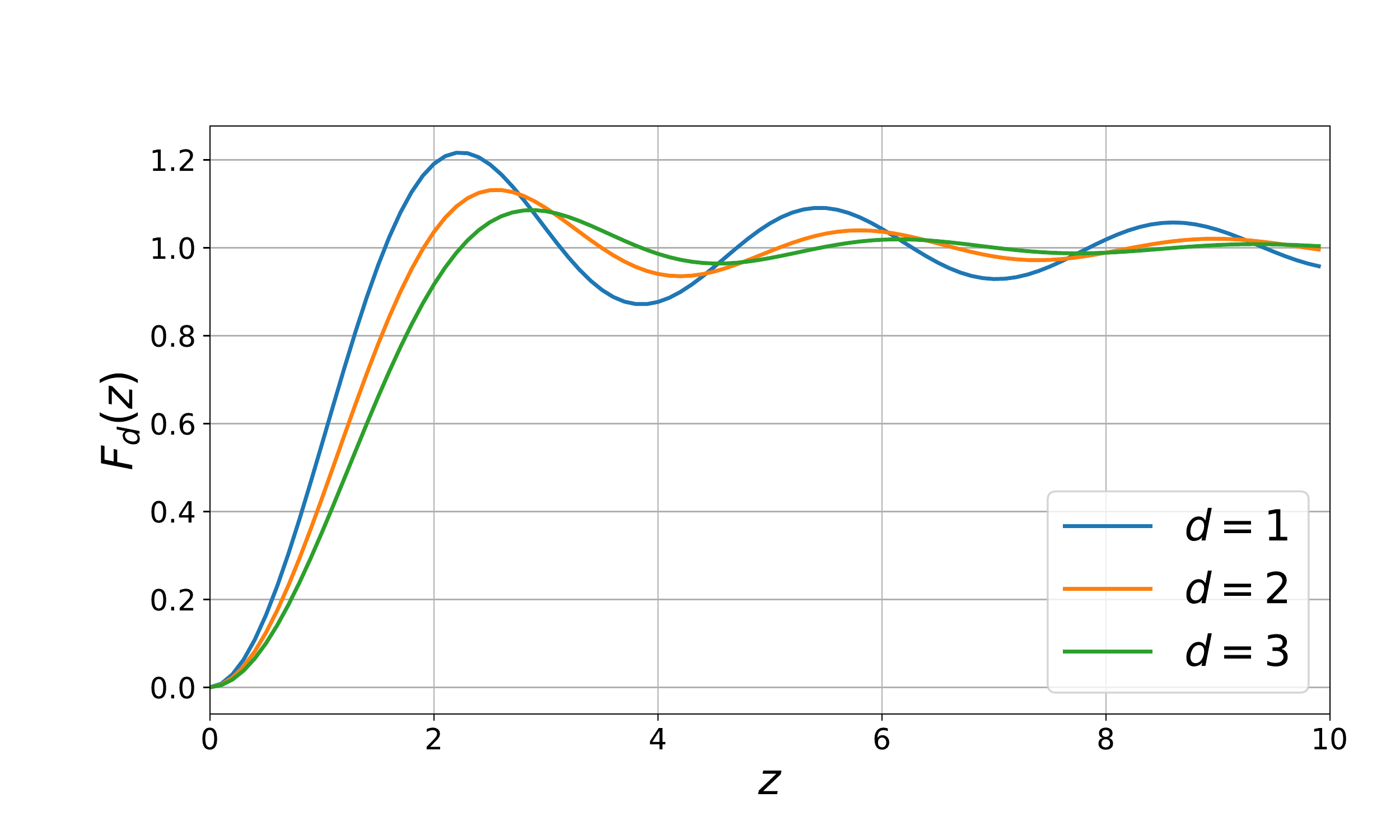}
\caption{Plot of the scaling function $F_{d}(z)$ given in Eq. \eqref{dens_d_t0} for $d=1,2,3$ respectively in blue, orange and green. The oscillations that are of quantum origin are smaller in higher dimension, where the effects of the Pauli principle are weaker.}\label{Fig_dens_d}
\end{figure}

From this result, we obtain the behaviour of the density close to the hard wall as
\be
\boxed{\begin{array}{rl}
&\displaystyle \rho_N({\bf x})\approx \frac{k_F^d}{N}K_{d}^{\rm e}(k_F({\bf x}_w-{\bf x}),k_F({\bf x}_w-{\bf x}))=\frac{1}{\Omega_d} F_{d}(k_F(1-|{\bf x}|))\;,\vspace{0.2cm}\\
&{\rm with}\;\;\displaystyle F_d(z)=1-\Gamma\left(\frac{d}{2}+1\right)z^{-\frac{d}{2}}\J_{\frac{d}{2}}(2z)\;,\label{dens_d_t0}
\end{array}}
\ee
where we used that $K_{d}^{\rm b}(0)=\Omega_d/(2\pi)^d=N/(\Omega_d k_F^d)$. This average density vanishes quadratically close to the wall in all dimensions $d$,
\be\label{F_d_small}
F_d(z)\approx \frac{2z^2}{d+2}\;,\;\;z\to 0\;,
\ee
and matches smoothly the uniform density in the box away from the boundaries $F_d(z)\to 1$ as $r\to \infty$. The scaling function $F_d(z)$ is plotted for $d=1,2,3$ in Fig. \ref{Fig_dens_d}.

\begin{figure}
\centering
\includegraphics[width=0.6\textwidth]{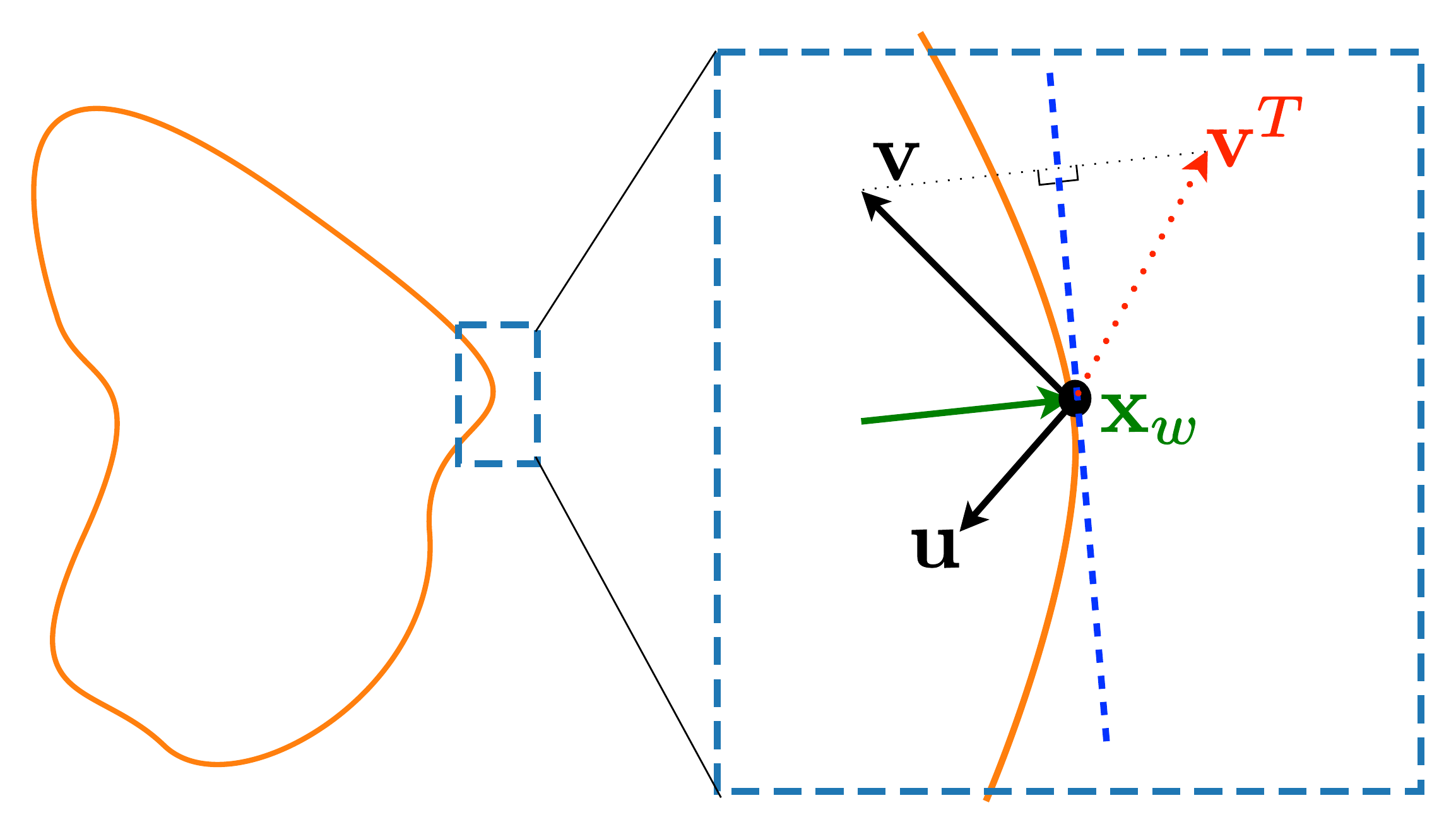}
\caption{For any smoothly varying boundary domain, the fluctuations on a small scale of order $O(k_F^{-1})$ are obtained, as for the spherical box using the method of images, with a similar construction as in Fig. \ref{Fig_im_d}.}\label{Fig_im_d_shape}
\end{figure}

Note that the method used in this section can be extended to any smoothly varying boundary (without wedge) and is not restricted to the spherical hard box as explained in Fig. \ref{Fig_im_d_shape}. In the case of a wedge forming an angle $\theta$, the density vanishes at the apex with an exponent depending smoothly on the value of $\theta$ (see Article \ref{Art:fermions_lett} for the case of a wedge in a $2d$ plane).

\subsubsection{Effective one-dimensional kernel at zero temperature}

As we have seen in section \ref{sec_d_soft}, for a rotationally symmetric potential, the positions of fermions with different orbital quantum numbers $l$ are independent. This decomposition will be quite convenient in order to obtain the statistics of the largest radius $r_{\max}$. We therefore first consider the process of the radii of fermions for a fixed value of $l$. To this purpose, we recall the effective one-dimensional correlation kernel given in Eq. \eqref{k_r_eff}
\be
K_l(r,r')=\sum_{n}\Theta(\mu-\epsilon_{n,l})\overline{\chi}_{n,l}(r)\chi_{n,l}(r)\;,
\ee
that we will now compute for the specific case of the hard box potential. We associate to this correlation kernel a single body quantum propagator
\be
G_l(r,t|r',0)=\frac{t}{\hbar}\int_0^{\infty}d\mu\,e^{-\frac{\mu t}{\hbar}}K_l(r',r)\;.
\ee
This propagator is solution of the equation 
\begin{align}
&\partial_t G_l(r,t|r',0)=\frac{\hbar}{2m}\partial_r^2 G_l(r,t|r',0)+\frac{\hbar}{8m r^2}(2l+d-1)(2l+d-3)G_l(r,t|r',0)\;,\nn\\
&{\rm with}\;\;G_l(r,0|r',0)=\delta(r-r')\;,
\end{align}
and with Dirichlet boundary conditions for $r=1$. Introducing the rescaled propagator
\be
G_l(r,t|r',0)=k_F G_{\frac{l}{k_F}}^{\rm e}\left(k_F(1-r),\frac{\mu t}{\hbar}\Big|k_F(1-r'),0\right)\;,
\ee
it will be solution in the large $k_F$ limit and for $0<\ell=l/k_F<1$ of the equation
\begin{align}
&\partial_\tau G_\ell^{\rm e}(s,\tau|s',0)=\partial_s^2 G_\ell^{\rm e}(s,\tau|s',0)+\ell^2 G_\ell^{\rm e}(s,\tau|s',0)\;,\nn\\
&{\rm with}\;\;G_\ell^{\rm e}(s,0|s',0)=\delta(s-s')\;,
\end{align}
and Dirichlet boundary condition in $s=0$. This equation is a diffusion equation in a uniform and constant potential, whose solution reads
\be
G_\ell^{\rm e}(s,\tau|s',0)=\left[\frac{\displaystyle e^{-\frac{(s-s')^2}{4\tau}}}{\sqrt{4\tau}}-\frac{\displaystyle e^{-\frac{(s+s')^2}{4\tau}}}{\sqrt{4\tau}}\right]e^{\ell^2 \tau}\;.
\ee
Taking the inverse Laplace transform using  Eq. \eqref{LT_bessel} in the table of Appendix \ref{LT}, one obtains the scaling form at the edge
\be
\boxed{K_l(r',r)\approx k_F K_1^{\rm e}\left(k_F\sqrt{1-\ell^2}(1-r),k_F\sqrt{1-\ell^2}(1-r')\right)\;,}\label{K_l_edge}
\ee
where $K_1^{\rm e}(u,v)$ is given in Eq. \eqref{k_1d_hb_e}. We will now use Eq. \eqref{K_l_edge} and \eqref{K_d_ell_t0} to obtain the statistics of the largest radius $r_{\max}$.

\subsubsection{Statistics of $r_{\max}$ at zero temperature}\label{stat_zero_t}

\begin{figure}
\centering
\includegraphics[width=0.6\textwidth]{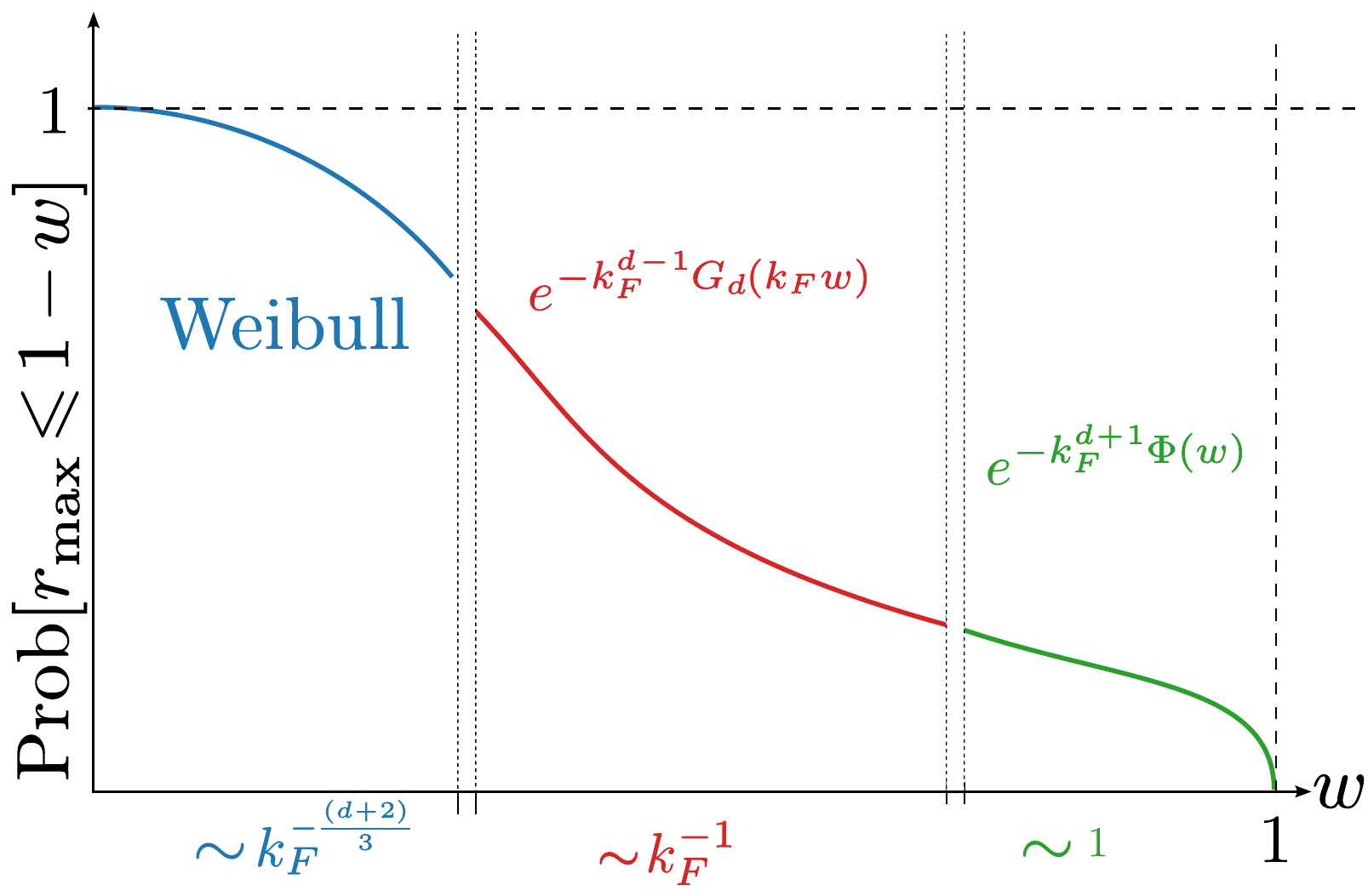}
\caption{Sketch of the typical (blue), intermediate (red) and large (green) fluctuation regimes of the probability $\Prob\left[r_{\max}\leq 1-w\right]$ of the radius $r_{\max}$ of the farthest fermion from the centre of the hard box at zero temperature (c.f. Eq. \eqref{sum_r_max_t_0}).}\label{fig_r_max_t_0}
\end{figure}

We consider now the statistics of the largest radius $\displaystyle r_{\max}=\max_{1\leq i\leq N}|{\bf x}_i|$ in dimension $d>1$ and at zero temperature.
One can show that in this setting there exists three scales of fluctuations of $r_{\max}$ that are summarised as (see also Fig. \ref{fig_r_max_t_0})
\be\label{sum_r_max_t_0}
\boxed{\Prob\left[r_{\max}\leq w\right]\approx\begin{cases}
\exp\left[- \frac{2}{3(d+2)}\frac{\Omega_d S_d}{(2\pi)^d}\left(k_F^{\frac{(d+2)}{3}}(1-w)\right)^3\right]&\;,\;\;|1-w|\sim k_F^{-\frac{(d+2)}{3}}\;,\\
&\\
\exp\left(-k_F^{d-1}G_d(k_F(1-w))\right)&\;,\;\;|1-w|\sim k_F^{-1}\;,\\
&\\
\exp\left(-k_F^{d+1}\Phi_d(1-w)\right)&\;,\;\;|1-w|=O(1)\;.
\end{cases}}
\ee

Interestingly, in the $d$-dimensional case we showed that the large deviation regime in the third line of Eq. \eqref{sum_r_max_t_0} does not match with the tail of the typical regime in the first line of Eq. \eqref{sum_r_max_t_0}. The effects of correlations are weak in the typical regime of fluctuation and one obtains a Weibull distribution as in the case of i.i.d. random variables (c.f. the discussion above). An {\it intermediate deviation regime} emerges for scales of the order of the inter-particle distance $k_F^{-1}$ as indicated in Eq. \eqref{sum_r_max_t_0} that is characteristic of the strong correlations between positions and one can obtain the associated rate function $G_d(s)$ exactly in Eq. \eqref{int_dev_box_d_t0} below. As in the one-dimensional case (c.f.  Eq. \eqref{q_1_b}), the fluctuations of $r_{\max}$ are quite different from the case of a soft edge treated in \eqref{max_rad_soft}.

We will now treat separately the different regimes of fluctuations.

\begin{itemize}

\item The regime of typical fluctuations for $r_{\max}$ in dimension $d>1$ given in the first line of \eqref{sum_r_max_t_0} $k_F^{(d+2)/3}(1-r)=O(1)$ corresponds to a length scale much smaller than the typical scale of fluctuation of the density given by $k_F(1-r)=O(1)$. Expressing the probability as a Fredholm determinant of the $d$-dimensional kernel, one can show that as $k_F^{\frac{2+d}{3}}\ll k_F^{-1}$, the usual development of the Fredholm determinant in terms of the traces of the projected kernel can be restricted to its first term. This yields
\be
\Prob\left[r_{\max}\leq 1-k_F^{-\frac{d+2}{3}} z\right]\approx\exp\left(-\frac{\Omega_d S_d}{(2\pi)^d} k_F^{d}\int_{1-k_F^{-(d+2)/3} z}^{1}r^{d-1}F_d(k_F(1-r))dr \right)\;,
\ee
where we used that $K_d^{\rm e}({\bf u},{\bf u})=\Omega_d/(2\pi)^d F_d(|{\bf u}|)$ together with the spherical symmetry of the problem. After a change of variable $r\to u=k_F^{(d+2)/3}(1-r)$ and using the small $z$ asymptotic behaviour of $F_d(z)$ in Eq. \eqref{F_d_small}, one obtains the Weibull distribution described by the first line of Eq. \eqref{sum_r_max_t_0}. This result is an occurrence of extension of the universality classes of i.i.d. random variables when the random variables $r_{\max,l}$, corresponding to all possible values of the orbital quantum numbers $l$, are independent but not identically distributed (c.f. chapter \ref{intro_iid}).

\item The intermediate deviation regime can be obtained using the product structure of the CDF,
\begin{align}
\Prob\left[r_{\max}\leq r\right]&=\prod_{l=0}^{l^*}\left(\Det\left[\mathbb{I}-P_{[r,1]}K_l P_{[r,1]}\right]^{g_d(l)}\right)\nn\\
&=\exp\left(\sum_{l=0}^{l^*}g_d(l)\ln\Det\left[\mathbb{I}-P_{[r,1]}K_l P_{[r,1]}\right]\right)\;.
\end{align}
In the large $N$ limit and in the regime $k_F(1-r)=w=O(1)$, we may replace the finite $N$ and $l$ kernel by its scaling form at the edge in the limit $k_F\gg 1$ with $\ell=l/k_F=O(1)$ given in Eq. \eqref{K_l_edge}. In this regime, $l^*\to k_F$ and the finite sum over $l$ may be replaced by an integral over $\ell$. Finally, using the large $l$ scaling form of the degeneracy in Eq. \eqref{degen_l}
\be
g_d(l)\approx \frac{2l^{d-2}}{\Gamma(d-1)}\;,\;\;l\to\infty\;,
\ee
one obtains the scaling form given in the second line of Eq. \eqref{sum_r_max_t_0} with the scaling function $G_d(s)$ given by
\be
\boxed{G_d(s)=-\int_0^1 \frac{2\ell^{d-2} d\ell}{\Gamma(d-1)} \ln q_1\left(s\sqrt{1-\ell^2}\right)\;.}\label{int_dev_box_d_t0}
\ee
One can obtain from Eq. \eqref{int_dev_box_d_t0} and using the asymptotic behaviours of $q_1(s)$ given in Eq. \eqref{eq: typ_fluc_1d}, the asymptotic behaviours 
\be
G_d(s)\approx\begin{cases}
\displaystyle \frac{2}{3(d+2)}\frac{\Omega_d S_d}{(2\pi)^d}s^3&\;,\;\;s\to 0\;.\\
&\\
\displaystyle \frac{d}{(d+1)!}s^2&\;,\;\;s\to \infty\;.
\end{cases}
\ee
In particular the small $s=z k_F^{(1-d)/3}$ behaviour matches smoothly with the regime of typical fluctuations. Furthermore, the large $s$ behaviour indicates the presence of a large deviation form as expressed in the third line of Eq. \eqref{sum_r_max_t_0}.
\item The large deviation function $\Phi_d(r)$ cannot be obtained for any value of $r=O(1)$ but one can obtain its asymptotic behaviour for $r\to 0$. It can be obtained by computing first the probability that there is no fermion in the interval $[r,1]$ for $r\to 0$ independently for each values of $l$. In this limit, the problem for each value of $l$ can be mapped onto the JUE given in Eq. \eqref{P_joint_JUE} with an index $b_l\sim l+(d-2)/2$ that depends explicitly on $l$, while $a=0$. After a few steps of computations, (see Article \ref{Art:fermions_lett} for details) one obtains
\begin{align}
&\Phi_d(r)\approx -\kappa_d \ln r\;,\;\;r\to 0\;,\nn\\
&\kappa_d=\int_0^{\pi/2}\frac{4\cos(t)^{d-2}dt }{\pi^2\Gamma(d-1)}\sin(t)(\sin(t)-t\cos(t))(\sin(t)+(\pi-t)\cos(t))dt\;.
\end{align}
This result confirms in particular the scaling of the rapidity $\sim k_F^{d+1}$ and therefore the presence of this large deviation regime.
\end{itemize}
These results on the three regimes of fluctuation of the largest radius $r_{\max}$ concludes our study of the $d$-dimensional spherical hard box at zero temperature, and we will now consider the effects of the thermal fluctuations on the system.
\subsection{Spherical hard box in dimension $d$ and at finite temperature}

\begin{figure}
\centering
\includegraphics[width=0.6\textwidth]{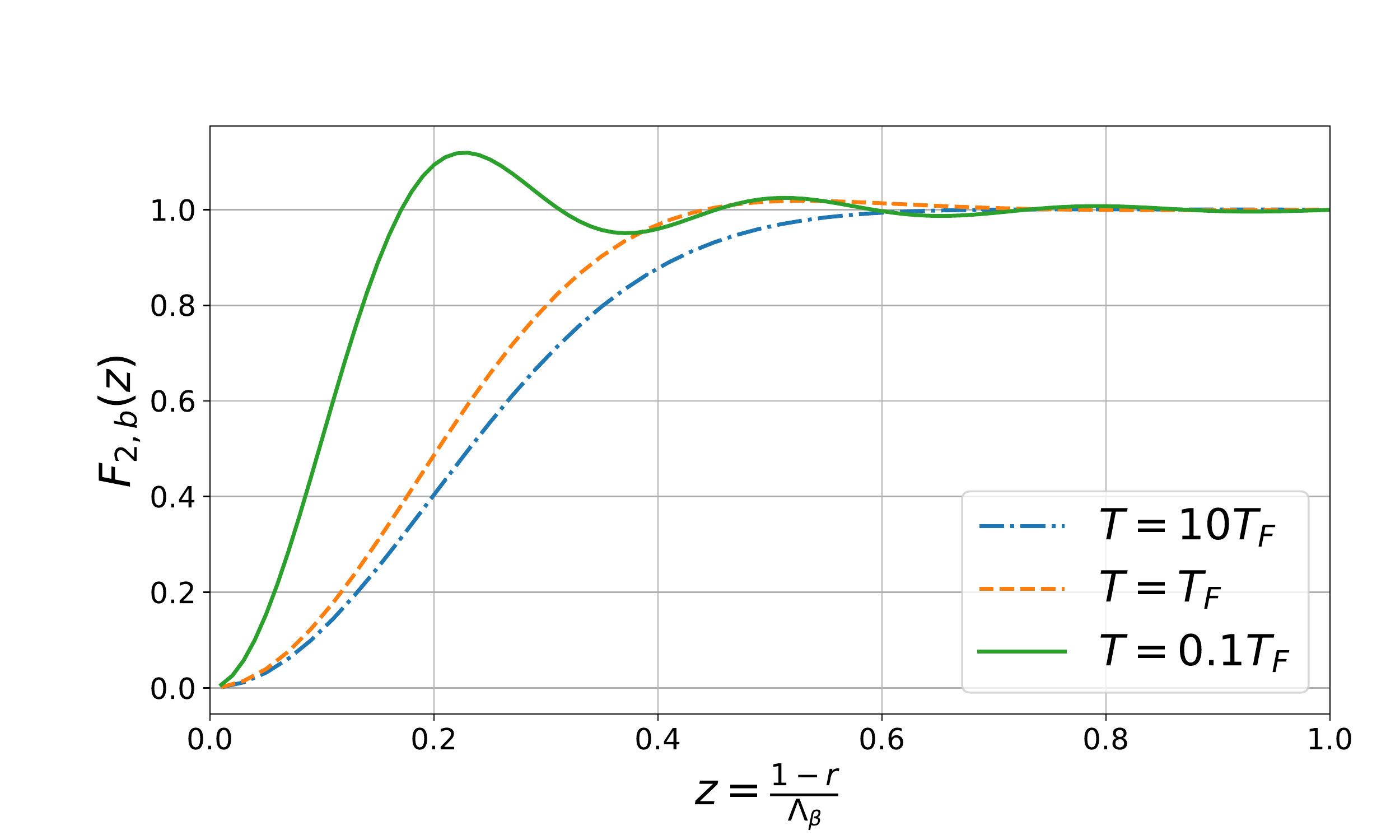}
\caption{Plot of the scaling function $F_{2,b}(z)$ given in Eq. \eqref{F_1b_eq} describing the average density close to the hard edge for $T/T_F=0.1,1,10$ in dimension $d=2$ respectively in green, orange and blue. The oscillations that are of quantum origin are smoothened at finite temperature.}\label{Fig_F_2_b}
\end{figure}

At finite temperature $T=1/(k_B \beta)$, we expect the effects of quantum and thermal fluctuations to be of the same order for 
\be
b=\beta\epsilon_F=\frac{(k_F\Lambda_{\beta})^2}{4\pi}=O(1)\;,
\ee
with $\Lambda_{\beta}=\sqrt{\frac{2\pi\hbar^2 \beta}{m}}$ the de Broglie thermal wave-length (see discussion in section \ref{hb_1d_t}).
In the large $N$ limit, the positions of the fermions in the canonical ensemble (for fixed $N$) will locally form a determinantal point process. The correlation kernel at the edge can be obtained from the zero temperature result using Eq. \eqref{K_mu_to_K_b}. Close to a point ${\bf x}_w$ on the boundary, one obtains the scaling form  
\be
K_{\mu}^{\beta}({\bf x},{\bf y})=\Lambda_{\beta}^{-d}K_{d,b}^{\rm e}\left(\frac{{\bf x}_w-{\bf x}}{\Lambda_{\beta}},\frac{{\bf x}_w-{\bf y}}{\Lambda_{\beta}}\right)\;,
\ee
with the scaling function
\be
\boxed{\begin{array}{rl}
&\displaystyle K_{d,b}^{\rm b}({\bf u},{\bf v})=K_{d,b}^{\rm b}(|{\bf u}-{\bf v}|)-K_{d,b}^{\rm b}(|{\bf u}-{\bf v}^T|)\;,\vspace{0.2cm}\\
&\displaystyle K_{d,b}^{\rm b}({r})=\int_0^{\infty}\frac{\zeta dk}{\zeta+e^{\frac{k^2}{4\pi}}}\left(\frac{k}{2\pi}\right)^{\frac{d}{2}}\frac{\J_{\frac{d}{2}-1}\left(k\,r\right)}{{r}^{\frac{d}{2}-1}}\;,\label{k_e_d_t}
\end{array}}
\ee
where $\zeta=e^{\beta\mu}$ is the fugacity. One can obtain a closed form equation for $\zeta$ by first using the finite temperature local density approximation in Eq. \eqref{LDA_t},
\be\label{dens_t_d}
\rho_N({\bf x})\approx\frac{1}{\Omega_d}=-\frac{1}{N\Lambda_{\beta}^d}\Li_{\frac{d}{2}}(-\zeta)\;.
\ee
Using then that the value of $N$ at finite temperature coincides with its value at zero temperature given in Eq. \eqref{N_t0}, one obtains the result
\be
-\Li_{\frac{d}{2}}(-\zeta)=\Omega_d \left(\frac{b}{\pi}\right)^\frac{d}{2}\;.
\ee
From Eq. \eqref{k_e_d_t}, we obtain the behaviour of the density close to the boundary
\be
\boxed{\begin{array}{rl}
&\displaystyle \rho_N(x)\approx  \frac{1}{N\Lambda_{\beta}^d} K_{d}^{\rm e}\left(\frac{{\bf x}_w-{\bf x}}{\Lambda_{\beta}},\frac{{\bf x}_w-{\bf x}}{\Lambda_{\beta}}\right)=\frac{1}{\Omega_d}F_{d,b}\left(\frac{1-|{\bf x}|}{\Lambda_{\beta}}\right)\;,\vspace{0.1cm}\\
&\displaystyle F_{d,b}(z)=1+\frac{1}{\Li_{\frac{d}{2}}(-\zeta)}\int_0^{\infty}\frac{\zeta dk}{\zeta+e^{\frac{k^2}{4\pi}}}\left(\frac{k}{2\pi}\right)^{\frac{d}{2}}\frac{\J_{\frac{d}{2}-1}(2k\, z)}{(2z)^{\frac{d}{2}-1}}\;,\label{F_db}
\end{array}}
\ee
where we used that $K_{d,b}^{\rm b}(0)=-\Li_{d/2}(-\zeta)=(N\Lambda_{\beta}^d)/\Omega_d$.
Note that the density matches smoothly for $z\to \infty$ with the uniform density in the bulk $F_{d,b}(z)\to 1$, while it vanishes quadratically at any temperature and for any dimension $d$ at the boundary
\be\label{F_d_b_small}
F_{d,b}(z)\approx 4\pi \frac{\Li_{\frac{d}{2}+1}(-\zeta)}{\Li_{\frac{d}{2}}(-\zeta)}z^2\;,\;\;z\to 0\;.
\ee
In the low temperature limit $b=\beta\epsilon_F\gg 1$, using the asymptotic behaviour for $\zeta=e^{b}$ of the polylogarithm in Eq. \eqref{polylog_as}, one obtains
\be
F_{d,b}(z)\approx \frac{8\pi b}{d+2}z^2\approx F_d(\Lambda_{\beta}k_F z)\;,\;\;z\to 0\;, 
\ee
matching smoothly with the zero-temperature result given in Eq. \eqref{dens_d_t0}. Furthermore, at high temperature $F_{d,b}(z)\approx 4\pi z^2$ for $z\to 0$. The scaling form $F_{2,b}(r)$ is plotted in Fig. \ref{Fig_F_2_b}.

The effective one-dimensional kernel $K_{l}^{\beta}(r,r')$ can be obtained in the regime $b=\beta\epsilon_F=O(1)$ and for $l=\ell/\Lambda_{\beta}$ close to the wall on the scale $(1-r)\sim \Lambda_{\beta}$ by combining the description from the finite temperature local density approximation \eqref{K_bulk_T} setting in this regime $\beta v_{l,d}(r=1)\approx \beta\hbar^2 l^2/(2m)\approx\ell^2/(4\pi)$ and the method of images as
\be\label{K_l_t}
K_{l}^{\beta}(r,r')=\Lambda_{\beta}^{-1} K_{1,b_{\ell }}^{\rm e}\left(\frac{1-r}{\Lambda_{\beta}},\frac{1-r'}{\Lambda_{\beta}}\right)\;,
\ee
where the scaling function $K_{1,b_{\ell }}^{\rm e}(u,v)$ is the one-dimensional edge scaling function at finite temperature obtained by setting $d=1$ in Eq. \eqref{k_e_d_t} and replacing $\zeta$ by $\zeta_\ell=\zeta e^{-\ell^2/(4\pi)}$.

\subsubsection{Statistics of $r_{\max}$ at finite temperature}

At finite temperature, and in the regime $b=\beta\epsilon_F=O(1)$ one still expects to observe three regimes of fluctuations as in the zero temperature case in Eq. \eqref{sum_r_max_t_0} (c.f. Fig. \ref{fig_r_max_t})
\be\label{sum_r_max_t}
\boxed{\frame{\boxed{\Prob\left[r_{\max}\leq w\right]\approx\begin{cases}
\exp\left[\frac{4\pi}{3}S_d\Li_{\frac{d}{2}+1}(-\zeta)\left(\frac{(1-w)}{\Lambda_{\beta}^{(d+2)/3}}\right)^3\right]&\;,\;\;|1-w|\sim \Lambda_{\beta}^{\frac{(d+2)}{3}}\;,\\
&\\
\exp\left(-\Lambda_{\beta}^{-(d-1)} G_{d,b}\left(\frac{(1-w)}{\Lambda_{\beta}}\right)\right)&\;,\;\;|1-w|\sim\Lambda_{\beta}\;,\\
&\\
\exp\left(-\Lambda_{\beta}^{-d}\Phi_{d,b}(1-w)\right)&\;,\;\;|1-w|=O(1)\;.
\end{cases}}}}
\ee
 We separate again the analysis for each of these regimes.

\begin{itemize}
\item In the typical regime of fluctuations, the Weibull scaling form in the first line of Eq. \eqref{sum_r_max_t} can be obtained by expressing the CDF as a Fredholm determinant. Restricting the development in terms of trace of this determinant to the first order, one obtains
\be
\Prob\left[r_{\max}\leq 1-\Lambda_{\beta}^{\frac{d+2}{3}} z\right]\approx\exp\left(\frac{S_d\Li_{\frac{d}{2}}(-\zeta)}{\Lambda_{\beta}^{d}}\int_{1-\Lambda_{\beta}^{(d+2)/3} z}^{1}r^{d-1}F_{d,b}\left(\frac{(1-r)}{\Lambda_{\beta}}\right)dr \right)\;.
\ee
Changing the integration variable $r\to u=\Lambda_{\beta}^{(d+2)/3}(1-r)$, and using the small $z$ behaviour of $F_{d,b}(z)$ in Eq. \eqref{F_d_b_small} one obtains the first line of Eq. \eqref{sum_r_max_t}. Note that in the regime of low temperature $b=\beta\epsilon_F\gg 1$ using $\zeta\approx e^{b}$ together with Eq. \eqref{polylog_as}, this result matches smoothly with the first line of \eqref{sum_r_max_t_0}
\begin{align}
-\frac{4\pi}{3}S_d\Li_{\frac{d}{2}+1}(-\zeta)\left(\frac{(1-w)}{\Lambda_{\beta}^{(d+2)/3}}\right)^3&\approx \frac{4\pi}{3}S_d \frac{2 b^{\frac{d}{2}+1}}{(d+2)\Gamma\left(\frac{d}{2}+1\right)}(4\pi b)^{-\frac{d}{2}-1}\left(k_F(1-w)\right)^3\nn\\
&=\frac{2}{3(d+2)}\frac{S_d \Omega_d}{(2\pi)^d}\left(k_F(1-w)\right)^3\;,
\end{align}
where we used $\Gamma(z+1)=z\Gamma(z)$ and $\Omega_d=\pi^{d/2}/\Gamma(d/2+1)$.

\begin{figure}[h]
\centering
\includegraphics[width=0.6\textwidth]{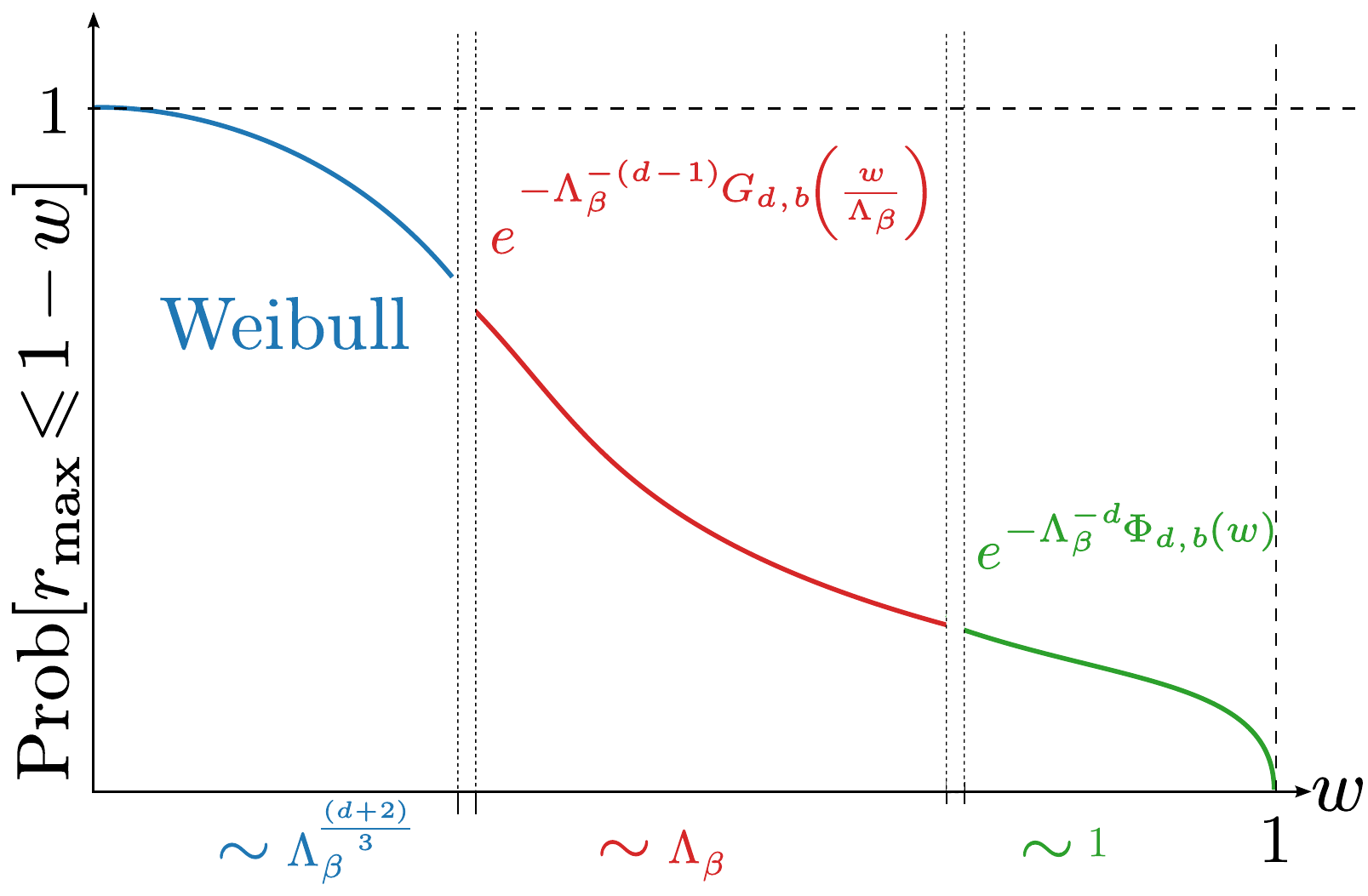}
\caption{Sketch of the typical (blue), intermediate (red) and large (green) fluctuation regimes of the probability $\Prob\left[r_{\max}\leq 1-w\right]$ of the radius $r_{\max}$ of the farthest fermion from the centre of the hard box at finite temperature, for $b=\beta\epsilon_F=O(1)$ (c.f. Eq. \eqref{sum_r_max_t}).}\label{fig_r_max_t}
\end{figure}

\item The intermediate deviation result in the second line of Eq. \eqref{sum_r_max_t} can be obtained using the product structure of the CDF. At finite temperature, it reads
\begin{align}
\Prob\left[r_{\max}\leq r\right]&=\prod_{l=0}^{\infty}\left(\Det\left[\mathbb{I}-P_{[r,1]}K_l^{\beta} P_{[r,1]}\right]^{g_d(l)}\right)\nn\\
&=\exp\left(\sum_{l=0}^{\infty}g_d(l)\ln\Det\left[\mathbb{I}-P_{[r,1]}K_l^{\beta} P_{[r,1]}\right]\right)\;.
\end{align}
In the regime $|1-r|\sim \Lambda_{\beta}$, one can replace the correlation kernel $K_l^{\beta}$ by its scaling function given in Eq. \eqref{K_l_t}. Replacing in the regime $b=O(1)$ the discrete sum over $l$ by an integral over $\ell =l\Lambda_{\beta}$, one obtains the scaling form given in the second line of Eq. \eqref{sum_r_max_t}, with the scaling function
\be
\boxed{\frame{\boxed{G_{d,b}(s)=-\int_0^{\infty}\frac{2\ell^{d-2}}{\Gamma(d-1)}\ln q_{1,b_{\ell}}(s)d\ell\;,}}}
\ee
where $q_{1,b_{\ell}}(s)$ is the one-dimensional scaling function given in Eq. \eqref{q_1_b} with the replacement $\zeta\to \zeta_{\ell}=\zeta e^{-\ell^2/(4\pi)}$. From the asymptotic behaviours of $q_{1,b}(s)$, given respectively in Eq. \eqref{q_1_b_small} for $s\to 0$ and \eqref{q_1_b_large} for $s\to \infty$, one obtains the asymptotic behaviours
\be\label{G_db_as}
G_{d,b}(s)\approx\begin{cases}
\displaystyle -\frac{4\pi}{3}S_d \Li_{\frac{d}{2}+1}(-\zeta)s^3&\;,\;\;s\to 0\;,\\
&\\
-S_d \Li_{\frac{d}{2}+1}(-\zeta)s \displaystyle &\;,\;\;s\to \infty\;,
\end{cases}
\ee
where we used that
\be
\int_0^{\infty}\frac{2\ell^{d-2} d\ell}{\Gamma(d-1)}\Li_{\frac{3}{2}}(-\zeta e^{-\frac{\ell^2}{4\pi}})=\sum_{p=1}^{\infty} p^{-3/2} (-\zeta)^{p}\int_0^{\infty}\frac{2\ell^{d-2} d\ell}{\Gamma(d-1)}e^{-p\frac{\ell^2}{4\pi}}=S_d \Li_{\frac{d}{2}+1}(-\zeta)\;.
\ee
In particular, the first line of Eq. \eqref{G_db_as} matches smoothly with the typical fluctuations. Using the second line of Eq. \eqref{G_db_as} leads naturally to expect a large deviation regime as given in the third line of Eq. \eqref{sum_r_max_t}.
\item The large deviation regime in Eq. \eqref{sum_r_max_t} cannot be obtained for arbitrary values of $b=\beta\epsilon_F$. In the limit $b\ll 1$, it should match smoothly with the classical result
\be
\Prob\left[r_{\max}\leq r\right]=\left(\frac{S_d}{\Omega_d}\int_0^{r} u^{d-1}du \right)^N=\exp\left(dN\ln r\right)\;,
\ee
On the one hand, using this classical regime in the limit $|1-r|=O(N^{-1})$, one obtains
\be
\Prob\left[r_{\max}\leq r\right]\approx \exp\left(-dN(1-r)\right)\;.\label{class_ld_d}
\ee
On the other hand, using now the second line of Eq. \eqref{G_db_as}, one obtains in the regime $b\ll 1$
\be
\Prob\left[r_{\max}\leq r\right]\approx \exp \left(\frac{S_d}{\Lambda_{\beta}^d} \Li_{\frac{d}{2}+1}(-\zeta)(1-r)\right)\approx \exp \left(-\frac{S_d}{\Omega_d}(1-r)\right)\;,
\ee
where we used that $\zeta\to 0$ in the limit $b\ll 1$, such that $-\Li_{d/2+1}(-\zeta)\sim -\Li_{d/2}(-\zeta)\sim \zeta$ together with the value of $N$ as a function of $\Lambda_{\beta}$ in Eq. \eqref{dens_t_d}. Using finally $S_d=d \Omega_d$ we obtain a smooth matching between the large deviation function in the classical regime for $|1-r|=O(N^{-1})$ in Eq. \eqref{class_ld_d} and the intermediate deviation function in the second line of Eq. \eqref{G_db_as} and for $b=\beta\epsilon_F\ll 1$.
\end{itemize}
This last result concludes this chapter on the statistics of fermions in hard edge potentials.

\section{Summary of the results for fermions in hard edges}

In this section, we extended the spatial description of fermions obtained for smooth confining potentials in \cite{dean2016noninteracting} to the case of {\it hard edges} where the density drops abruptly to zero at the edge. We obtained that there is a new universality class associated to the local correlations close to a {\it hard edge}, with fluctuations that do not depend on the overall shape of the trapping potential but rather on its specific behaviour close to the hard edge. The edge statistics are quite different from the case of {\it soft edges} and are controlled by the correlation kernel in Eq. \eqref{K_d_ell_t0} for $T=0$ and Eq. \eqref{k_e_d_t} for $T>0$. In particular, the extreme value statistics fall into a different universality class. The fluctuations of $x_{\max}$ for $d=1$ are given at zero temperature in Eq. \eqref{CDF_x_max_box_T0} and exhibit a regime of typical fluctuations and a large deviation regime. In  higher dimension $d\geq 2$, the fluctuations of $r_{\max}$ given in Eq. \eqref{sum_r_max_t_0} exhibit three regimes of fluctuations: typical, intermediate and large. In dimension $d>1$, while the typical behaviour only depends on the behaviour of the density close to the wall, the new {\it intermediate regime} depends on all the correlations in the gas close to the edge. These results are extended to finite temperature in Eq. \eqref{sum_r_max_t}.  We will show in the following that this type of {\it intermediate regime} emerges in various contexts.

%
%
%
%
%
%
%
%
%
%
%



\AddArticle{Art:fermions_lett}{Statistics of fermions in a d-dimensional box near a hard wall}

 \begin{center}
   {\large \textbf{Statistics of fermions in a d-dimensional box near a hard wall}}
 \end{center}


 \vspace{2cm}

 \noindent B. Lacroix-A-Chez-Toine, P. Le Doussal, S.~N. Majumdar, G. Schehr,\\
 Europhys. Lett. {\bf 120} (1), 10006 (2018).\\

 \ding{43}
 \href{https://iopscience.iop.org/article/10.1209/0295-5075/120/10006/meta}{https://iopscience.iop.org/article/10.1209/0295-5075/120/10006/meta}

 \ding{43}
 \href{https://arxiv.org/abs/1706.03598}{https://arxiv.org/abs/1706.03598}

\begin{abstract}
We study $N$ noninteracting fermions in a domain bounded by a hard wall potential
in $d \geq 1$ dimensions. We show that for large $N$, the correlations at the edge of the Fermi gas
(near the wall) at zero temperature are described by a universal
kernel, different from the universal edge kernel valid for smooth confining potentials. 
We compute this $d$ dimensional hard edge kernel exactly for a spherical domain and argue, using 
a generalized method of images, that it holds close to any sufficiently smooth boundary.
As an application we compute the quantum statistics of the position of the fermion closest
to the hard wall. Our results are then extended in several directions, including
non-smooth boundaries such as a wedge, and also to finite temperature. 
\end{abstract}


 \AddArticle{Art:ferm_long}{Non-interacting fermions in hard-edge potentials}

 \begin{center}
   {\large \textbf{Non-interacting fermions in hard-edge potentials}}
 \end{center}


 \vspace{2cm}

 \noindent B. Lacroix-A-Chez-Toine, P. Le Doussal, S.~N. Majumdar, G. Schehr,\\
 J. Stat. Mech {\bf 12}, 123103 (2018).\\

 \ding{43}
 \href{https://doi.org/10.1088/1742-5468/aa9bb2}{https://doi.org/10.1088/1742-5468/aa9bb2}

 \ding{43}
 \href{https://arxiv.org/abs/1806.07481}{https://arxiv.org/abs/1806.07481}

\begin{abstract}
We consider the spatial quantum and thermal fluctuations of non-interacting Fermi gases of $N$ particles confined in $d$-dimensional non-smooth potentials. We first present a thorough study of the spherically symmetric pure hard-box potential, with vanishing potential inside the box, both at $T=0$ and $T>0$. We find that the correlations near the wall are described by a ``hard edge'' kernel, which depend both on $d$ and $T$, and which is different from the ``soft edge'' Airy kernel, and its higher $d$ generalizations, found for smooth potentials. We extend these results to the case where the potential is non-uniform inside the box, and find that there exists a family of kernels which interpolate between the above ``hard edge'' kernel and the ``soft edge'' kernels. Finally, we consider one-dimensional singular potentials of the form $V(x)\sim x^{-\gamma}$ with $\gamma>0$. We show that the correlations close to the singularity at $x=0$ are described by this ``hard edge'' kernel for $1\leq\gamma<2$ while they are described by a broader family of ``hard edge'' kernels known
as the Bessel kernel for $\gamma=2$ and, finally by the Airy kernel for $\gamma>2$. These one-dimensional kernels also appear in random matrix theory, and we provide here the mapping between the $1d$ fermion
models and the corresponding random matrix ensembles.
Part of these results were announced in a recent Letter, EPL {\bf 120}, 10006 (2017). 
\end{abstract}

\chapter{Fermions in rotation, complex Ginibre ensemble and $2d$ one component plasma}
\label{ch: rot_trap}
\begin{figure}
\centering
\includegraphics[width=0.4\textwidth]{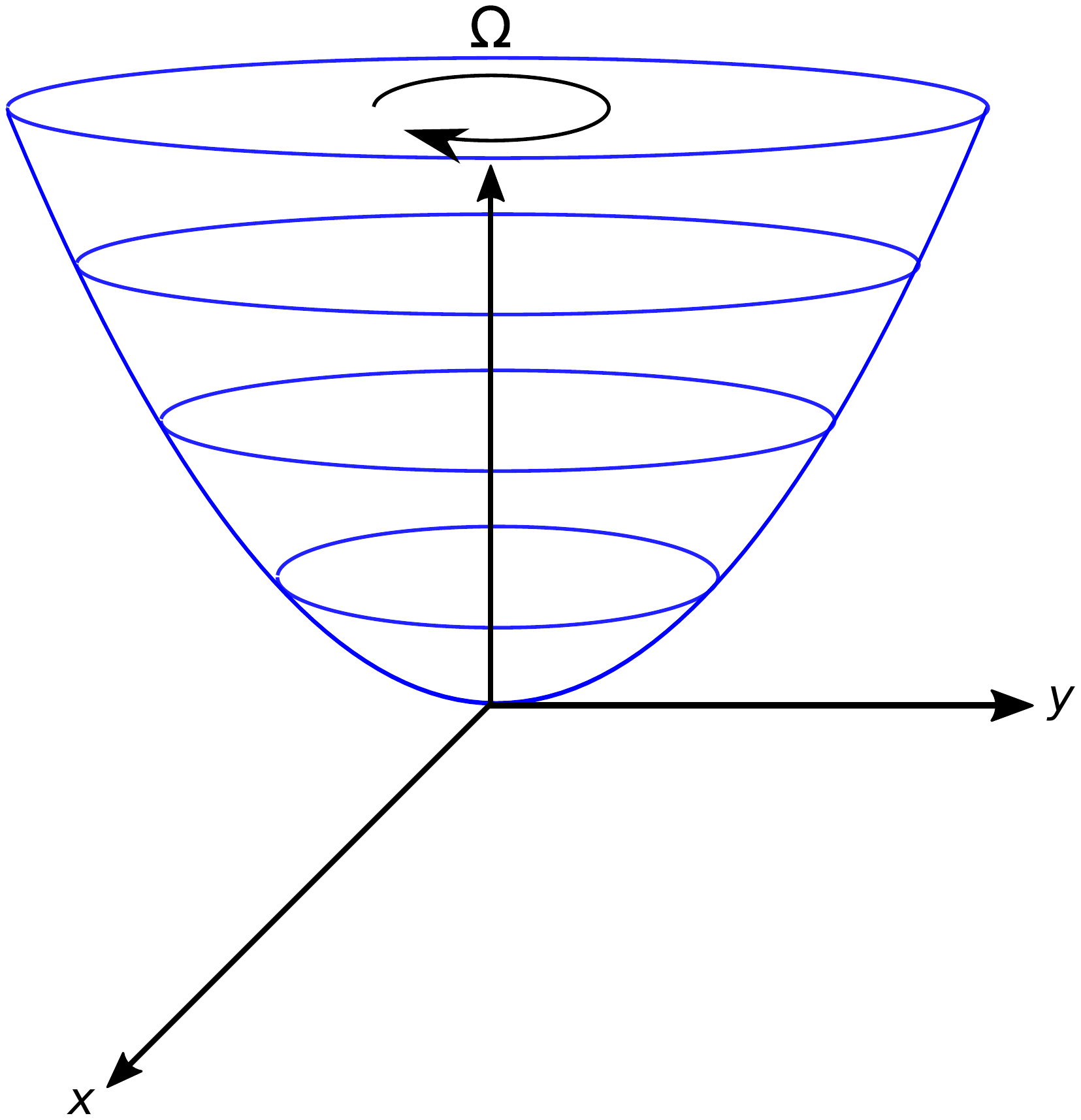}
\caption{Sketch of the two-dimensional rotating harmonic potential.}\label{Fig_rot_pot}
\end{figure}

In this chapter, we will first consider a system formed by $N$ non-interacting, spin-less, identical fermions of mass $m$ at zero temperature in a two-dimensional harmonic trap of frequency $\omega$, rotating at constant speed $\Omega$ (c.f. Fig. \ref{Fig_rot_pot}). For this trapped system, we will compute the bipartite entanglement entropy and the full counting statistics, i.e. the statistics of the number of particles, for a finite number of particles and in a finite domain of space. Note that powerful tools such as quantum and conformal field theory \cite{calabrese2004entanglement} have been developed to compute the entanglement entropy in the bulk, but the trapping potential often prevents to obtain these quantities close to the edge, where the translation invariance is explicitly broken. We developed alternative techniques based on an explicit mapping with the complex Ginibre ensemble.

In the second part of this chapter, we map this problem onto an equilibrium problem of classical statistical mechanics with long-range interaction: the two-dimensional one-component plasma \cite{forrester1998exact, serfaty2017systems}. In this system, particles of identical charges repel each other with the two-dimensional Coulomb repulsion and are confined by a potential $v(r)$. For a specific value of the temperature, both the typical \cite{chafai2014note} and atypical \cite{cunden2016large} fluctuations of the position of the particle the farthest away from the centre of the trap can be obtained. However, as noticed in \cite{cunden2016large} these two regimes do not match. We develop here the framework to solve this problem. 

We start by considering the problem of $N$ non-interacting fermions. The Hamiltonian of this system reads in the rotating frame \cite{landau1981statistical, leggett2006quantum}
\be\label{rot_ferm_H_1}
\hat {\cal H}_N=\sum_{i=1}^N \hat{H}_i=\sum_{i=1}^N  H(\hat{\bf x}_i,\hat{\bf p}_i)\;,\;\;H(\hat{\bf x},\hat{\bf p})=\frac{\hat {\bf p}^2}{2m}+\frac{1}{2}m\omega^2 \hat{\bf x}^2 -\Omega\hat{L}_z\;,
\ee
where the angular momentum operator reads $\hat{ L}_z= (\hat{\bf x}\wedge \hat{\bf p})\cdot {\bf u}_z$ and ${\bf u}_z^2=1$. Note that the first two terms of this single particle Hamiltonian in Eq. \eqref{rot_ferm_H_1} reproduce the two-dimensional harmonic oscillator, which was studied in chapter \ref{chap:fermions_intro}, while the last term accounts for the rotation of the system. This Hamiltonian can be rewritten as
\be\label{rot_ferm_H_2}
\hat H=\frac{\left(\hat {\bf p}-m\omega {\bf u}_z\wedge\hat{\bf x}\right)^2}{2m}+(\omega-\Omega)\hat{L}_z\;.
\ee
In this form, it appears clearly  that by rotating the system, one can introduce an artificial magnetic field acting on the system and model with cold atoms the properties of condensed matter. It is for instance used in practice to study the creation of vortices in Bose-Einstein condensates \cite{aftalion2005vortex}.  The Hamiltonian in Eq. \eqref{rot_ferm_H_1} can be diagonalised exactly for any value of $\omega$ and $\Omega$. Introducing the complex coordinates $z=x+\I y$ and $\bar z=x-\I y$, the Hamiltonian can be expressed as
\be
\hat H=-\frac{2\hbar^2}{m}\partial_z\partial_{\bar z}+\frac{1}{2}m\omega^2 z \bar{z}-\hbar\Omega(z\partial_z-\bar{z}\partial_{\bar z})\;,\;\;{\rm where}\;\;\partial_z=\frac{1}{2}\left(\partial_x-\I \partial_y\right)\;.
\ee
The wave-functions and energies of the Hamiltonian are labelled by two integers $n_1,\,n_2\in \mathbb{N}$ and read in this basis \cite{ho2000rapidly,aftalion2005vortex}
\begin{align}
&\phi_{n_1,n_2}(z,\bar z)=\frac{\displaystyle e^{-\alpha^2\frac{z\bar z}{2}}\partial_z^{n_1}\partial_{\bar z}^{n_2} e^{-\alpha^2 z\bar z}}{\alpha^{n_1+n_2-1}\sqrt{\pi n_1! n_2 !}}\;,\;\;\alpha=\sqrt{\frac{m\omega}{\hbar}}\;,\vspace{0.2cm}\\
&\;\;\epsilon_{n_1,n_2}=\hbar\omega(n_1+n_2+1)+\hbar\Omega(n_1-n_2)\;.\label{spect_rot}
\end{align}
Let us first discuss a few limiting cases of this Hamiltonian 
\begin{itemize}
\item For $\Omega>\omega$, the centrifugal force is stronger than the harmonic potential and the fermions are not confined by the potential. We will therefore not consider this case and restrict in the following to $\Omega<\omega$.
\item For $\Omega=\omega$, this problem can be exactly mapped onto the famous Landau problem where particles of charge $q$ are subjected to a uniform magnetic field ${\bf B}=B{\bf u}_z$, 
\be\label{H_LL}
\hat H_{\rm LL}=\frac{\left(2\hat {\bf p}- m\omega_c{\bf u}_z\wedge\hat{\bf x}\right)^2}{8m}\;,\;\;\omega_c=\frac{q B}{m}\;.
\ee
Introducing $\Omega=\omega$ in the spectrum in Eq. \eqref{spect_rot}, we see that the energies do not depend on the quantum number $n_2$ and are thus infinitely degenerate (see Fig. \ref{Fig_LLL}). 
\begin{figure}[h]
\centering
\includegraphics[width=0.6\textwidth]{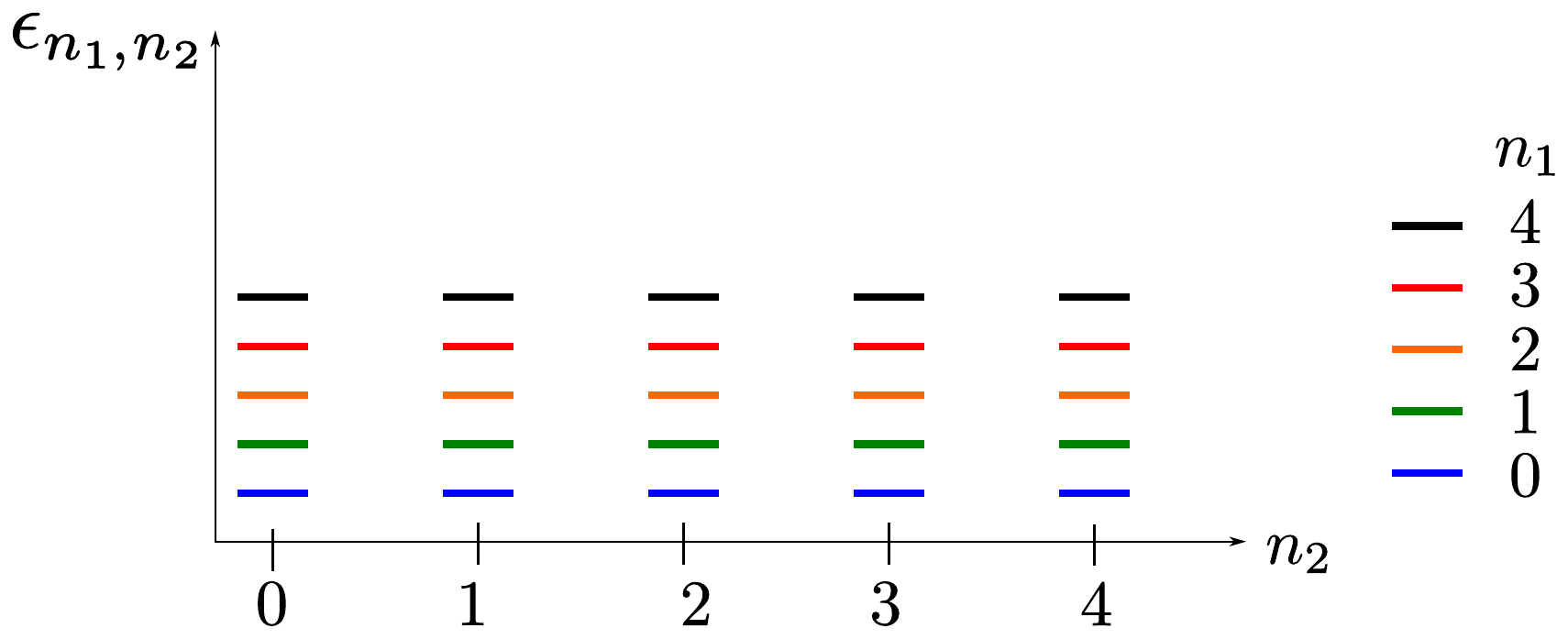}
\caption{Scheme of the single particle energy levels $\epsilon_{n_1,n_2}$ as a function of $n_1$ and $n_2$ in Eq. \eqref{spect_rot} corresponding to the Landau levels i.e. $\Omega=1$. The energies $\epsilon_{n_1,n_2}=\hbar\omega(2n_1+1)$ do not depend on $n_2$.}\label{Fig_LLL}
\end{figure}

\item For $\Omega=0$, the Hamiltonian reduces to the case of a standard two-dimensional harmonic potential with energy $\epsilon_{n_1,n_2}=\hbar\omega(n_1+n_2+1)$ (see Fig. \ref{Fig_oh2d}).

\begin{figure}[h]
\centering
\includegraphics[width=0.6\textwidth]{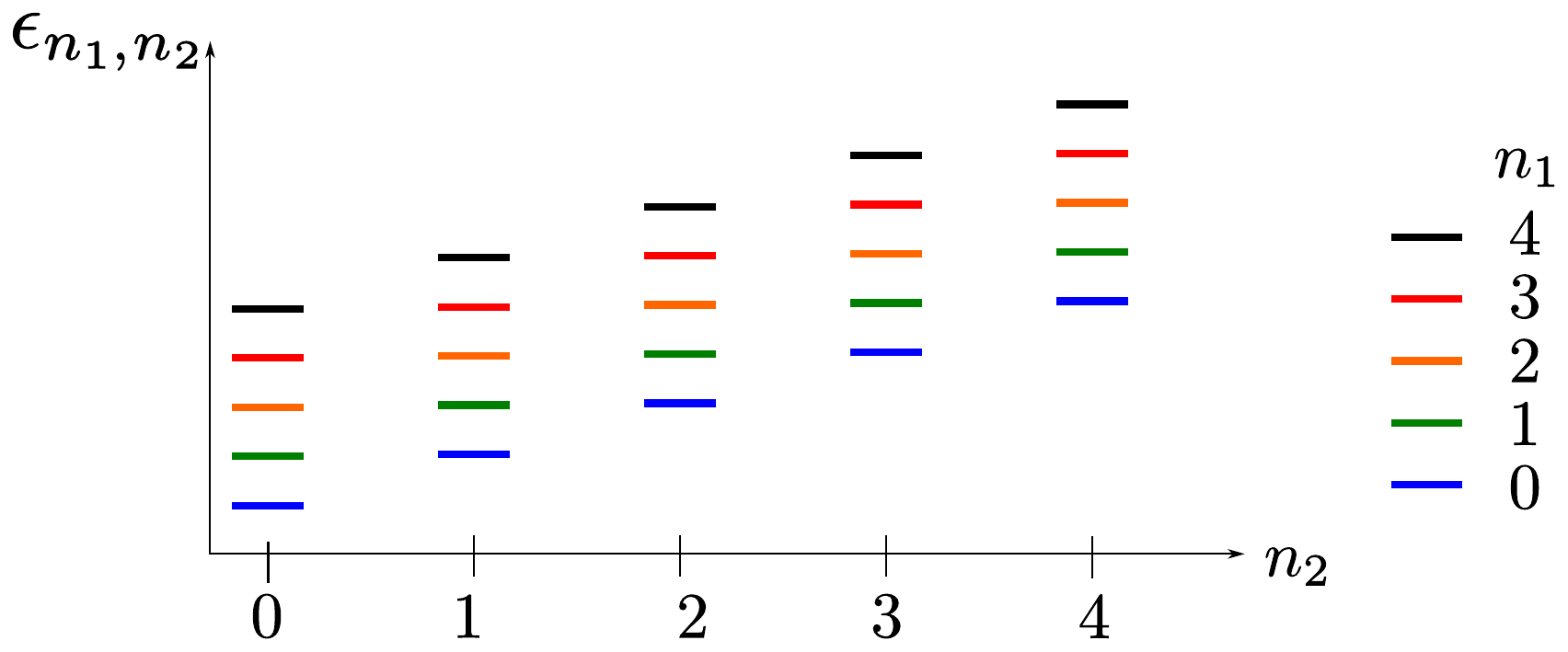}
\caption{Scheme of the single particle energy levels $\epsilon_{n_1,n_2}$ as a function of $n_1$ and $n_2$ in Eq. \eqref{spect_rot} corresponding to the $2d$ harmonic oscillator i.e. $\Omega=0$.}\label{Fig_oh2d}
\end{figure}

\end{itemize}
We set now and in the following $m=1$, $\hbar=1$ and $\omega=1$, which amounts to rescale the energy scales by $\hbar\omega$ and the length scales by $\alpha^{-1}$ with $\alpha=\sqrt{m\omega/\hbar}$. We will now consider the ground state of the many-body problem and obtain that for special choices of $\Omega$, this problem can be mapped exactly onto a random matrix ensemble: the complex Ginibre ensemble.

\section{Ground state probability and complex Ginibre ensemble}

\begin{figure}[h]
\centering
\includegraphics[width=0.6\textwidth]{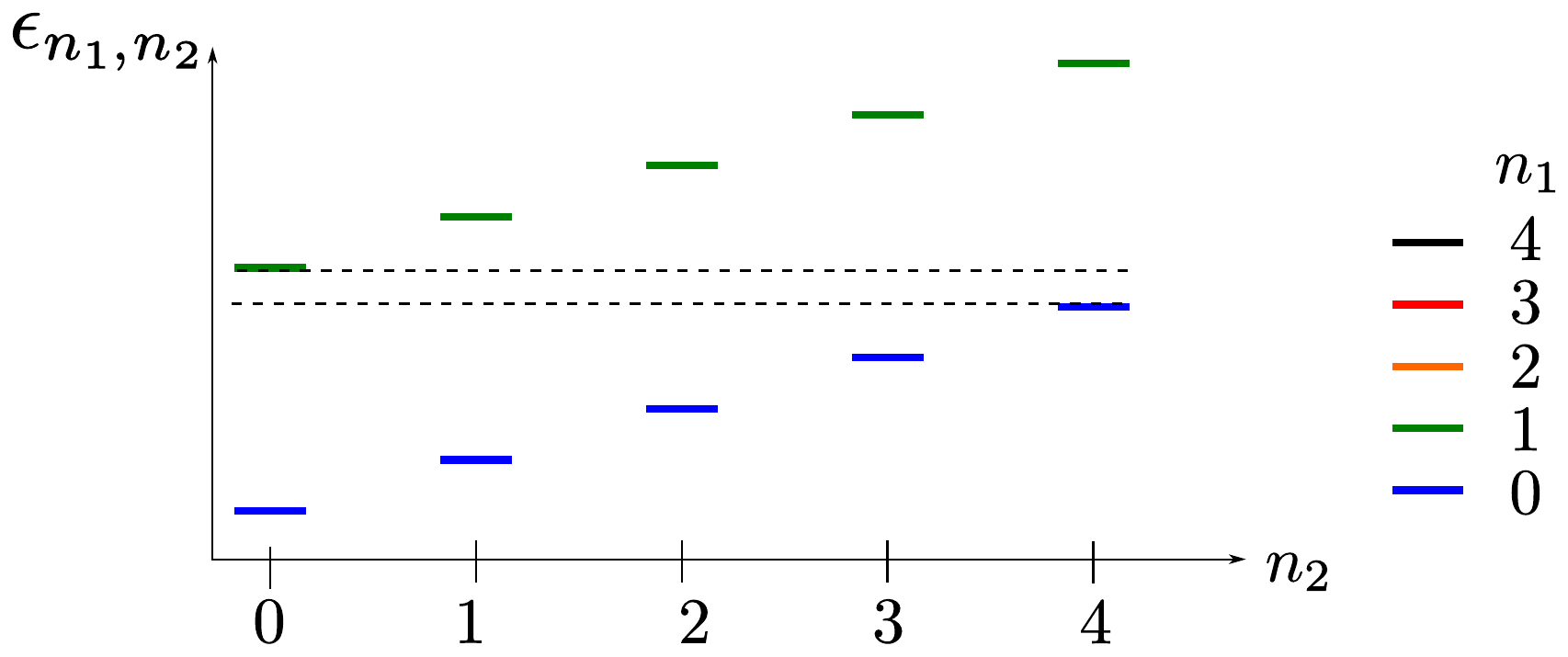}
\caption{Scheme of the single particle energy levels $\epsilon_{n_1,n_2}$ as a function of $n_1$ and $n_2$ in Eq. \eqref{spect_rot} corresponding to the condition in Eq. \eqref{cond_omega}.}\label{Fig_case_us}
\end{figure}

We now consider the problem for $N$ fermions at zero temperature. For general values of $\Omega$, it is rather complex to obtain the many-body wave function.
However, in the limit $\omega-\Omega=\delta\Omega\ll \omega$, the problem simplifies as we will now see. In this case the energies read
\be
\epsilon_{n_1,n_2}=1+2n_1+\delta\Omega(n_2-n_1)\;.
\ee
In particular, we see that (i) for $\delta\Omega>0$ the energies are non degenerate with respect to $n_2$ and (ii) for $\delta\Omega\ll 1$ the $N$ lowest energy states will lie in the state $n_1=0$ (see Fig. \ref{Fig_case_us}). This last condition requires
\be
\Delta\epsilon=\epsilon_{1,0}-\epsilon_{0,N-1}=2-(N-1)\delta\Omega>0\;,
\ee 
leading to the conditions for $\delta\Omega$ in the large $N$ limit
\be
\boxed{0<\delta\Omega=\frac{\omega-\Omega}{\omega}<\frac{2}{N}\;.}\label{cond_omega}
\ee

\noindent For this special set of conditions, the wave functions read
\be
\phi_{n_1=0,n_2=k}(z,\bar z)=\psi_k(z)=\frac{z^{k}}{\sqrt{\pi k!}}e^{-\frac{|z|^2}{2}}\;.\label{wf_psi_rot}
\ee
In this specific case, the ground state is unique and the occupied levels are ${\bf k}_{\rm GS}=\{0,1,\cdots,N-1\}$. The many-body wave function of the system is given by a single Slater determinant
\be
\Psi_0(z_1,\cdots,z_N)=\frac{1}{\sqrt{N!}}\det_{1\leq i,j\leq N}\psi_{j-1}(z_i)\;.
\ee
Note that if one considers the Landau problem $\Omega=\omega$, the single-particle wave-functions in the lowest Landau level are also given by Eq. \eqref{wf_psi_rot}. However in the latter the energy is degenerate with respect to $n_2=k$ and there is no reason to select the $N$-tuple ${\bf k}_{\rm GS}$ rather than any other $N$-tuple ${\bf k}=\{k_1,\cdots,k_N\}$. The many-body wave function will therefore read as an infinite superposition of Slater determinants corresponding to each $N$-tuple ${\bf k}$. 

For the specific choice of $\Omega$ given in Eq. \eqref{cond_omega} the associated joint probability of the complex positions $z_i$'s of the fermions can be computed exactly using the Vandermonde identity in Eq. \eqref{VdM} and reads
\be
\boxed{\begin{array}{rl}
 \left|\Psi_0(z_1,\cdots,z_N)\right|^2&\displaystyle=\frac{1}{N!}\det_{1\leq i,j\leq N}\psi_{j-1}(z_i)\det_{1\leq l,m\leq N}\overline{\psi}_{l-1}(z_m)\;,\vspace{0.1cm}\\
&\displaystyle =\frac{1}{Z_N}\prod_{i<j}|z_i-z_j|^2 \prod_{k=1}^N e^{-|z_k|^2}\;,\label{Gin_fermions}
\end{array}}
\ee
where $Z_N$ is a normalisation factor. We will now explain the connection between the joint PDF in Eq. \eqref{Gin_fermions} and the so-called complex Ginibre ensemble of random matrix theory.

\subsection{Complex Ginibre Ensemble}\label{sec_Gin}

\begin{figure}
\centering
\includegraphics[width=0.4\textwidth]{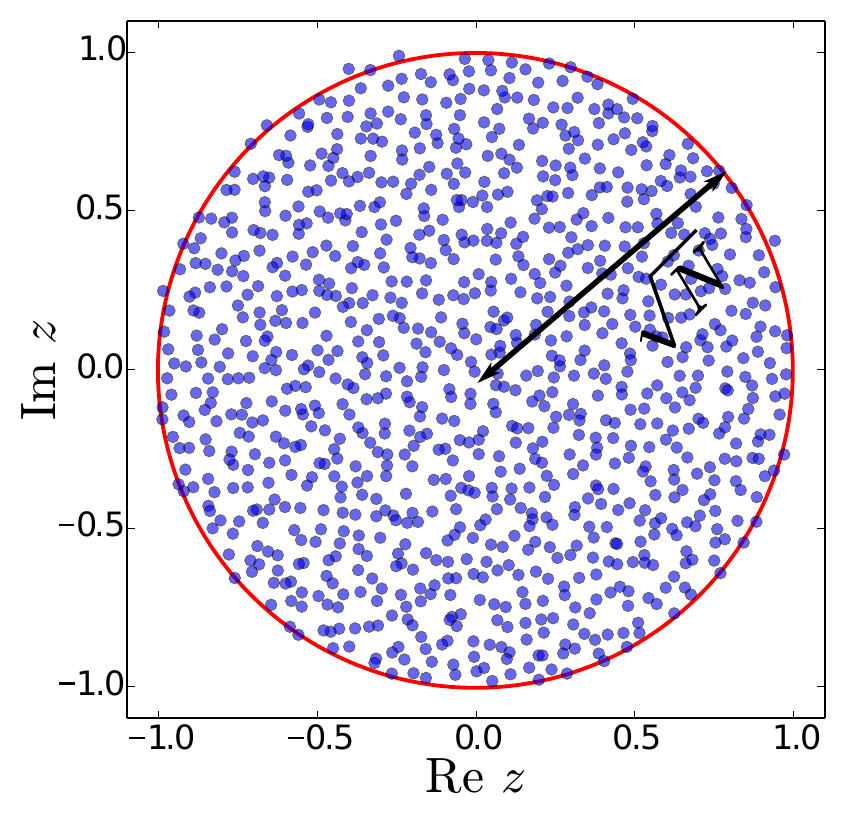}
\caption{Snapshot of the repartition of charges in the ground state. In the large $N$ limit, the density $\rho_N(z)\approx \left[\pi N\right]^{-1}$ is uniform within the disk of radius $\sqrt{N}$, while it is zero for $|z|>\sqrt{N}$ as seen in Eq. \eqref{Girko}.}\label{Fig_snp_Gin}
\end{figure}

\begin{figure}
\centering
\includegraphics[width=0.6\textwidth]{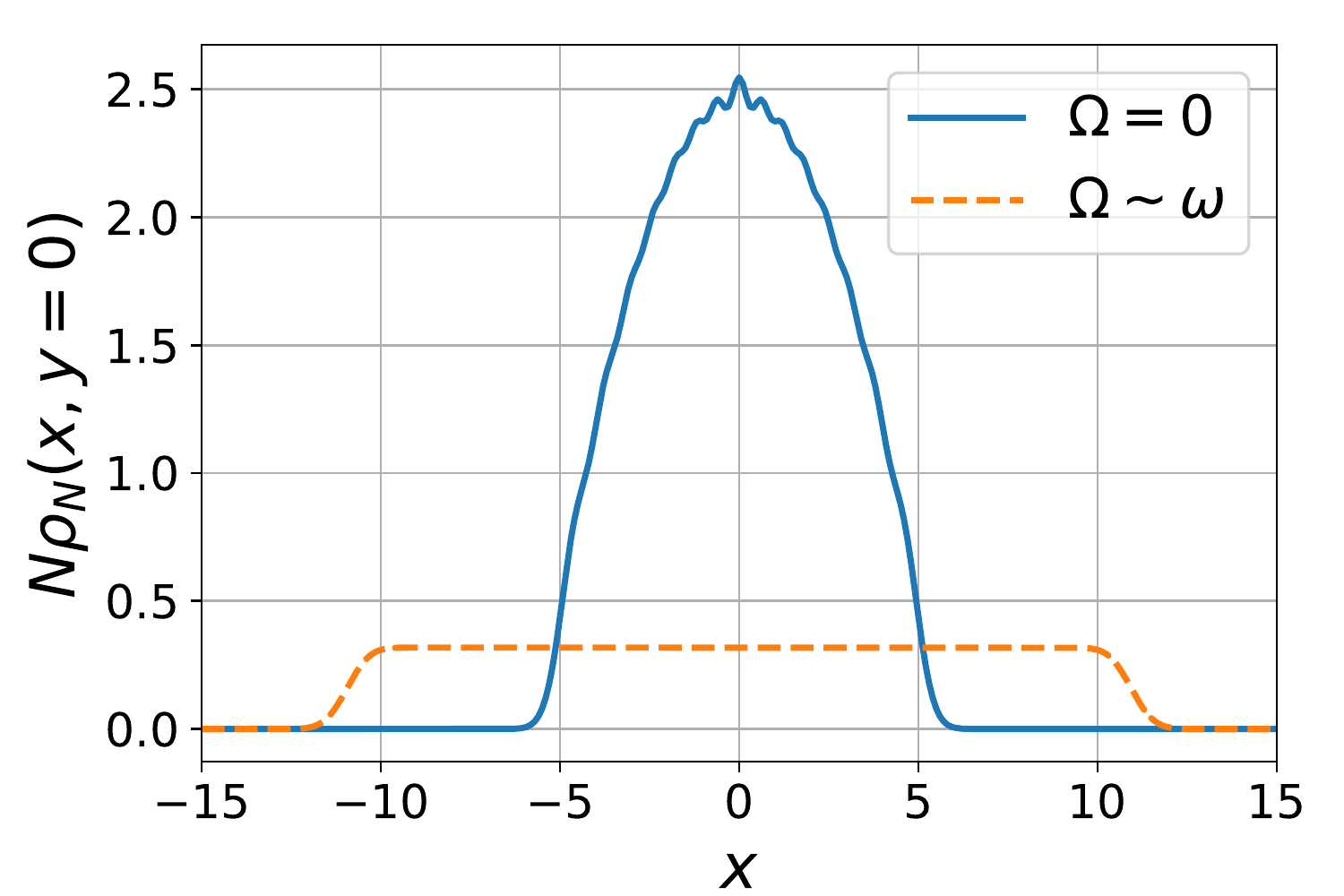}
\caption{Comparison between the exact rescaled density $N\rho_N(x,y=0)$ as a function of $x$ for $N=120$ fermions in the absence of rotation of the trap, i.e. $\Omega=0$, (in blue) and with rotation $\Omega\lesssim \omega$, c.f. Eq. \eqref{cond_omega}, (in dashed orange). In the latter, the density is nearly uniform in the bulk, extends farther and drops more abruptly at its edge.}\label{Fig_dens_comp}
\end{figure}

The complex Ginibre ensemble was introduced in 1965 by Jean Ginibre \cite{ginibre1965statistical} and consists of $N\times N$ matrices $G$ obtained by taking i.i.d. complex Gaussian entries
\be
g_{ij}\sim {\cal N}\left(0,\frac{1}{\sqrt{2N}}\right)+\I\; {\cal N}\left(0,\frac{1}{\sqrt{2N}}\right)\;.
\ee 
Note that this matrix $G$ can be factorised by unitary transformation to a tridiagonal matrix $G=U^{\dag}(\Lambda+\Delta)U$ where $U$ is unitary, $\Lambda={\rm diag}(z_1,\cdots,z_N)$ is the matrix of its complex eigenvalues and $\Delta$ is strictly upper triangular. The probability weight associated to this matrix is easily computed and reads \cite{akemann2011oxford}
\be
P(G)=\prod_{i,j}\sqrt{\frac{N}{\pi}} e^{-N |g_{ij}|^2}=\frac{e^{-N \Tr(\Lambda^{\dag}\Lambda+\Delta^{\dag}\Delta)}}{w_N^{\rm Gin}}\;,
\ee 
where $w_N^{\rm Gin}$ is a normalisation constant.
After integration over $U$ and $\Delta$, one obtains the distribution of its complex eigenvalues as \cite{akemann2011oxford} (c.f. chapter 18 there)
\be
P_{\rm joint}^{\rm Gin}(z_1,\cdots,z_N)=\frac{1}{Z_N^{\rm Gin}} \prod_{i<j}|z_i-z_j|^2 \prod_{i=1}^N e^{-N |z_i|^2 }\;.\label{P_joint_Gin}
\ee
This expression is very similar to the distribution of eigenvalues in the GUE \eqref{P_joint_GUE}, the only difference lying in the fact that the eigenvalues $z_i$'s are now complex. Remarkably, after a trivial rescaling $z\to \sqrt{N} z$ the joint PDF in Eqs. \eqref{Gin_fermions} and \eqref{P_joint_Gin} coincide exactly. From this exact mapping to the complex Ginibre ensemble, we obtain that the complex positions $z_i$'s of the fermions in rotation form a determinantal point process. The finite $N$ correlation kernel is obtained by rewriting the first line in Eq. \eqref{Gin_fermions} as a single determinant (using $\det(A)\det(B)=\det(A B)$), expressed in terms of the single particle wave-functions in Eq. \eqref{wf_psi_rot}
\begin{align}
K_N(z_1,z_2)&=\sum_{k=0}^{N-1} \overline{\psi}_{k}(z_1)\psi_k(z_2)=\frac{e^{-\frac{|z_1|^2+|z_2|^2}{2}}}{\pi}\sum_{k=0}^{N-1}\frac{(\bar z_1 z_2)^k}{k!}\nn\\
&=e^{-\frac{1}{2}(|z_1|^2+|z_2|^2-2\bar z_1 z_2)}\frac{\Gamma(N,\bar z_1 z_2)}{\pi \Gamma(N)}\;,
\end{align}
where $\Gamma(a,z)=\int_z^{\infty}t^{a-1}e^{-t}dt$ is the upper incomplete gamma function. 
The density associated to this system is given by
\be\label{rho_rot_n}
\rho_N(z)=\frac{1}{N}K_N(z,z)=\frac{\Gamma(N,|z|^2)}{\pi N!}\;.
\ee 
A snapshot of the repartition of fermions is given in Fig. \ref{Fig_snp_Gin} and a comparison with the density profile for the standard harmonic potential, i.e. $\Omega=0$ is given in Fig. \ref{Fig_dens_comp}. Note that the density extends on a larger scale for the rotating trap, as one naturally expects.

\begin{figure}
\centering
\includegraphics[width=0.6\textwidth]{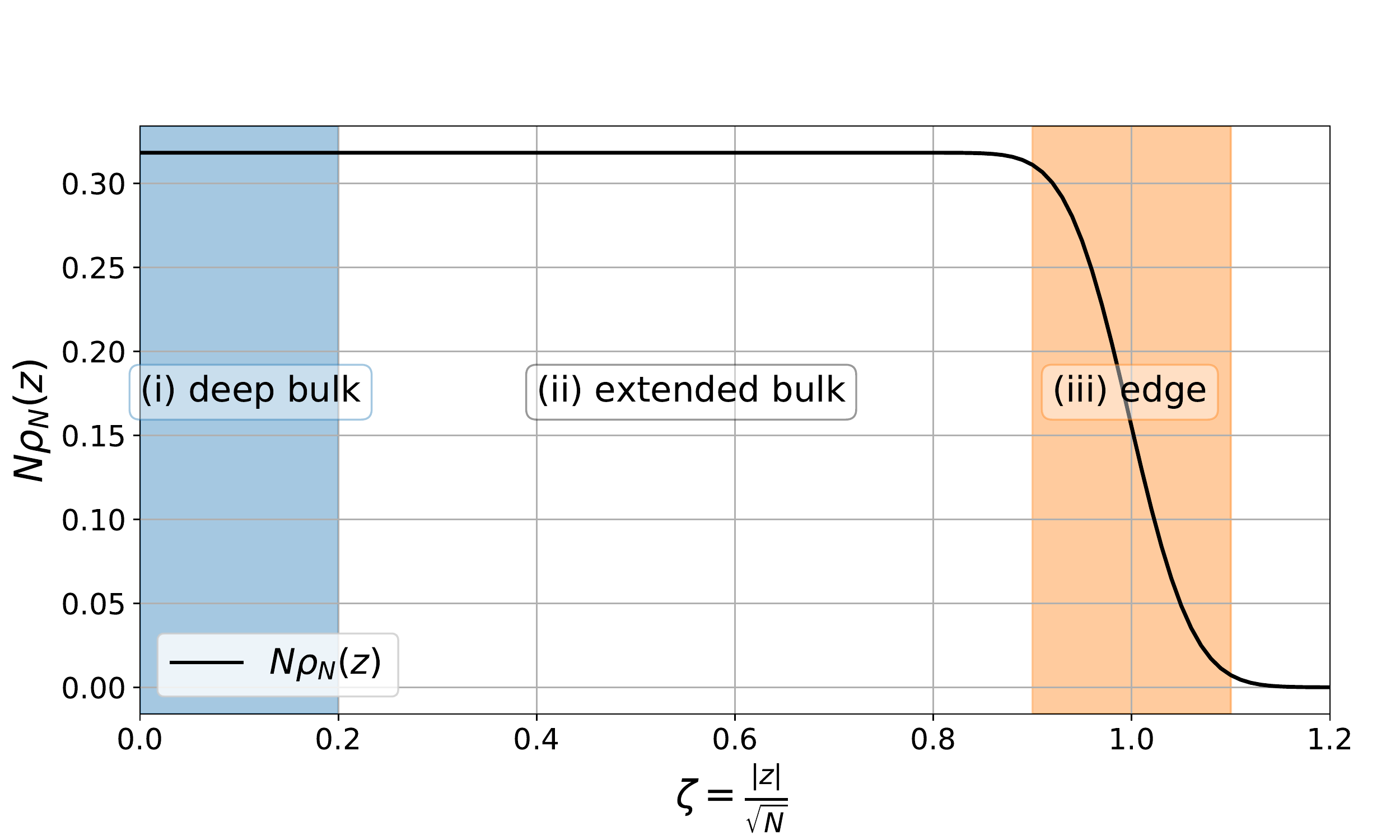}
\caption{Plot of the rescaled density $N\rho_N(z)$ as a function  of the rescaled variable $\zeta=\frac{|z|}{\sqrt{N}}$. One can identify three spatial regimes in this figure, (i) the deep bulk for $|z|=O(1)$ (shaded in blue), (ii) the extended bulk for $\zeta=\frac{|z|}{\sqrt{N}}=O(1)$ and (iii) the edge regime for $s=\sqrt{2}(|z|-\sqrt{N})=O(1)$ (which corresponds to the regime $|\zeta-1|\sim N^{-1/2}$ shaded in orange), where the density drops to zero.}\label{Fig_dens_Gin}
\end{figure}

We introduce three different rescaling of the complex position $z$ in order to analyse the limit $N\to \infty$ of this density
\begin{itemize}
\item In the limit $N\to \infty$ with $z=O(1)$, that we will refer to as ``deep bulk'' (see (i) in Fig. \ref{Fig_dens_Gin}) the rescaled average density is uniform over the whole complex plane
\be
\lim_{N\to\infty} N\rho_N\left(z\right)=\frac{1}{\pi}\;.
\ee
Note that in the deep bulk, the $p$-point correlation functions $R_p(z_1,\cdots, z_p)$ takes the simple form \cite{akemann2011oxford}
\be
R_p( z_1,\cdots, z_p)=\frac{N!}{(N-p)!}\sum_{i_1\neq i_2\neq \cdots\neq i_p}^N \moy{\prod_{l=1}^p\delta( z_k- z_{i_k})}=e^{-\sum_{k=1}^p |z_k|^2}\det_{1\leq i,j\leq p}\left(e^{-\bar{z_i}z_j}\right)\;.
\ee
\item In the limit $N\to \infty$ with $0<\zeta=|z|/\sqrt{N}<1$, that we will refer to as ``extended bulk'' (see (ii) in Fig. \ref{Fig_dens_Gin}), the rescaled density converges to the Girko's circular law \cite{girko1985circular}
\be
\lim_{N\to\infty}N \rho_N\left(z\right)=\rho_{\rm G}(\zeta)=\frac{1}{\pi}\Theta(1-\zeta)\;,\label{Girko}
\ee
with a uniform density over the disk of radius $\sqrt{N}$ (c.f. Figs. \ref{Fig_snp_Gin} and \ref{Fig_dens_comp}).
\item In the limit $N\to \infty$ with $s=\sqrt{2}(|z|-\sqrt{N})=O(1)$, that we will refer to as ``edge regime'' (see (iii) in Fig. \ref{Fig_dens_Gin}), the rescaled density reads \cite{forrester1999exact}
\be
\lim_{N\to\infty}N\rho_N\left(z\right)=\frac{1}{2\pi}\erfc\left(s\right)\;.
\ee
Note that while the behaviour of the density in Eq. \eqref{Girko} suggests a hard edge behaviour, the fact that the density is non-identically zero beyond the edge implies that this model does not belong to the same universality class. It does not belong either to the universality class of soft edges where the zero temperature density vanishes linearly at the edge in dimension $d=2$ as seen in Eq. \eqref{LDA_d}.

\end{itemize}

We will now focus for this Fermi gas -- which is of experimental relevance \cite{cooper2008rapidly} -- on a measurable observable \cite{cheuk2015quantum,haller2015single,parsons2015site}, the full counting statistics (FCS) and its connection to the entanglement entropy. While the latter is in general very hard to obtain for trapped gases (see however \cite{calabrese2015random} and the discussion in section \ref{FCS_EE}) we now show that it can be computed exactly for this Fermi gas.

\section{Full counting statistics and bipartite entanglement entropy} \label{FCS_EE_sec}

We want to characterise the quantum fluctuations in a circular domain ${\cal D}_r=\{|z|\leq r\}$ for this gas. In order to do so, we introduce two observables
\begin{itemize}
\item The full counting statistics (FCS), defined as the statistics of the observable
\be
\hat{N_r}=\sum_{i=1}^N \Theta(r-|\hat z_i|)\;.
\ee
FCS has attracted a lot of attention in mesoscopic physics \cite{levitov1993charge, lee1995universal, levitov1996electron}. 
Note that in the case of fermions, the recent development of Fermi quantum microscopes \cite{cheuk2015quantum,haller2015single,parsons2015site} allows to access to snapshots of the exact positions of the cold atoms in the gas and therefore to measure these FCS (c.f. Fig. \ref{Fig_qu_gas_micro}).
\item The R\'enyi bipartite entanglement entropy $S_q(N,r)$ of the domain ${\cal D}_r$ defined in terms of the reduced density matrix $\rho_r=\Tr_{\bar{\cal D}_r}[\rho]$ -- where the degrees of freedom of the complement $\bar{\cal D}_r=\{|z|\geq r\}$ of ${\cal D}_r$ have been traced out from the full density matrix $\rho$ -- as
\be
S_q(N,r)=\frac{1}{1-q}\ln \Tr\left[\rho_{r}^q\right]\;.
\ee
For $q\to 1$, the R\'enyi entropy coincides with the Von-Neumann entropy
\be
\lim_{q\to 1}S_q(N,r)=-\Tr\left[\rho_{r}\ln \rho_{r}\right]\;.
\ee
The entanglement entropy allows to characterise in particular the critical and topological phases of matter \cite{amico2008entanglement, calabrese2004entanglement}. This observable is in general rather difficult to measure experimentally \cite{amico2008entanglement} (see however e.g. \cite{islam2015measuring}).

\end{itemize}

It turns out that for non-interacting Fermi systems, these two quantities are actually related. One can indeed show that for a general system of non-interacting fermions, there exists an expression of the R\'enyi entanglement entropy as a series of the cumulants $\moy{N_{\cal D}}_c^p$ of order $p>2$  \cite{klich2009quantum, song2011entanglement, song2012bipartite},
\be
S_q({\cal D})=\sum_{p=2}^{\infty}\eta_{q,p}\moy{N_{\cal D}^p}_c\;,\label{S_q_cum_rot}
\ee
with in particular $\eta_{q,2}=\frac{\pi^2}{6q}(q+1)$. Note that there exists central limit theorems for a general determinantal point processes \cite{soshnikov2002gaussian}, which are therefore applicable to fermions at zero temperature. One could expect that in the large $N$ limit, the cumulants of order $p>2$ are subleading and that the entanglement entropy and the number variance $\var{N_{\cal D}}=\moy{N_{\cal D}^p}_c$ are always proportional to each other as was found in \cite{calabrese2015random}. We show that it is not always the case here and obtain an exact expression for the entanglement entropy close to the edge.

%
%
%
%

\subsection{Finite $N$ results}

In order to compute the FCS and entanglement entropy for the domain ${\cal D}_r=\{|z|\leq r\}$, we first define the overlap matrix
\begin{align}
\mathbb{A}_{kl}&=\int_{|z|\leq r} d^2 z\, \overline{\psi}_{k-1}(z)\psi_{l-1}(z)=\frac{1}{\pi \sqrt{\Gamma(k)\Gamma(l)}}\int_0^{2\pi}d\theta\, e^{\I \theta(k-l)}\int_0^{r}du\, u^{k+l-1}e^{-u^2}\;,\\
&=\frac{2\delta_{kl}}{\Gamma(k)}\int_0^{r}du\, u^{2k-1}e^{-u^2}=\delta_{kl} \frac{\gamma(k,r^2)}{\Gamma(k)}\;,
\end{align}
where we have used that $\int_0^{2\pi}d\theta e^{\I \theta(k-l)}=2\pi \delta_{kl}$ and $\gamma(a,z)=\int_0^z t^{a-1}e^{-t}dt$ is the lower incomplete gamma function. The overlap matrix is therefore diagonal for this particular choice of domain ${\cal D}_r$. Note that it would remain true for any choice of rotationally symmetric domain. In the following, we denote
\be
\lambda_k(r)= \frac{\gamma(k,r^2)}{\Gamma(k)}\;,\label{lambda_k_ferm}
\ee
the eigenvalues of the overlap matrix, which lie in $[0,1]$. From this overlap matrix, one can obtain exact expressions for finite $N$ and $r$ both for the entanglement entropy and the full counting statistics.

\begin{figure}
\centering
\includegraphics[width=0.6\textwidth]{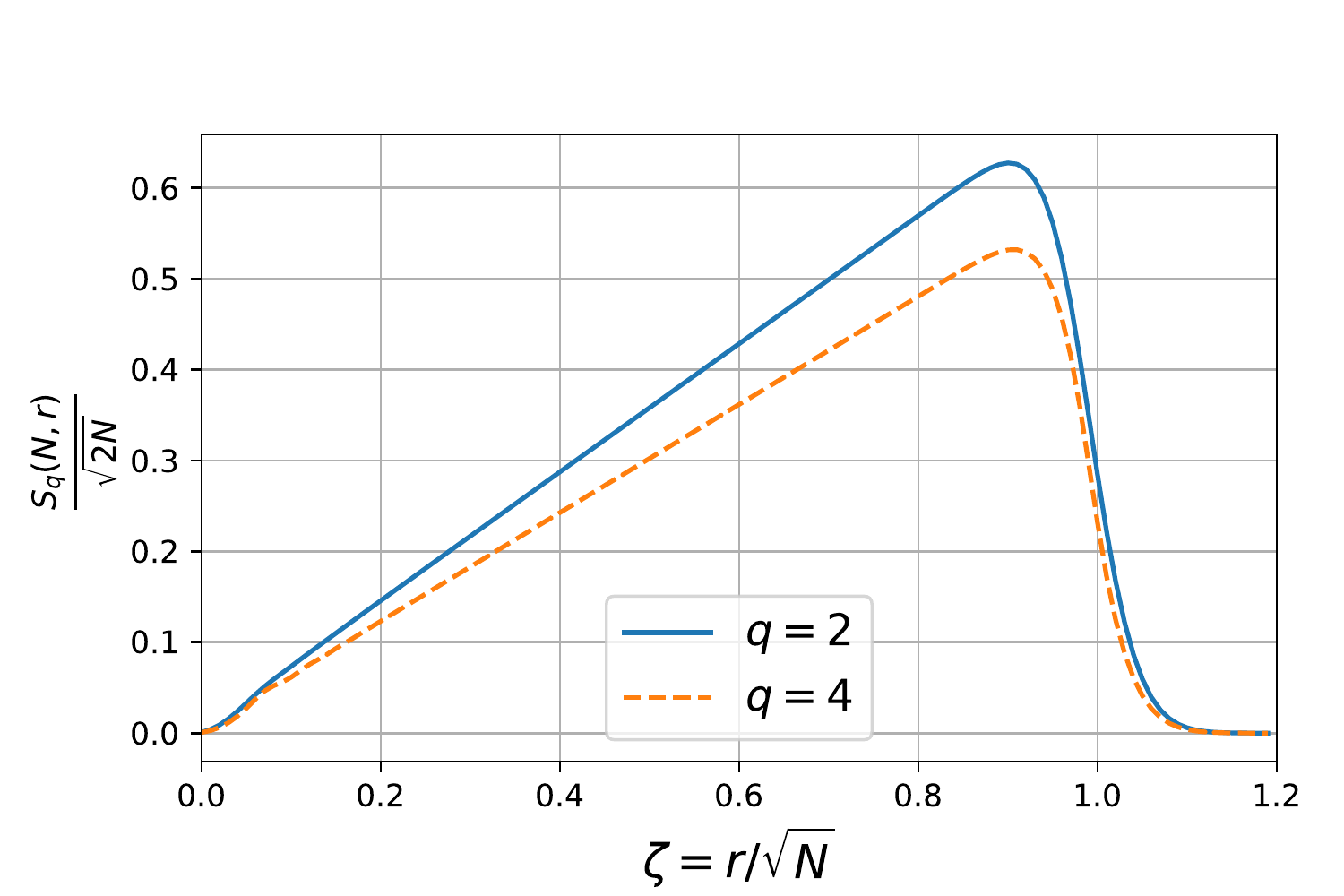}
\caption{Plot of the rescaled entanglement entropy $S_q(N,r)/\sqrt{2N}$ for $N=200$ and where $S_q(N,r)$ is given in Eq. \eqref{S_q_N} as a function of the rescaled radius $\zeta=r/\sqrt{N}$ for $q=2,4$ respectively in blue and orange.}\label{Fig_S_n}
\end{figure}

The entanglement entropy $S_q(N,r)$ is expressed in terms of this overlap matrix as \cite{klich2006lower}
\be
\boxed{S_q(N,r)=\frac{1}{1-q}\Tr[\ln(\mathbb{A}^q+(\mathbb{I}-\mathbb{A})^q)]=\frac{1}{1-q}\sum_{k=1}^N \ln\left[\lambda_k(r)^q+(1-\lambda_k)^q\right]\;.}\label{S_q_N}
\ee
This function is plotted in Fig. \ref{Fig_S_n} for $q=2$ and $q=4$ and $N=200$ fermions. It grows linearly in the bulk and vanishes rapidly at the edge for $r-\sqrt{N}=O(1)$.

We now consider the full counting statistics. We can also obtain an exact expression for the moment generating function (MGF) $\moy{e^{-\mu N_r}}$ of $N_r$ in terms of the overlap matrix as
\begin{align}
\moy{e^{-\mu N_r}}&=\int d^{2}z_1 \cdots\int d^{2}z_N e^{-\mu \sum_{i=1}^N \Theta(r-|z_i|)}|\Psi_0(z_1,\cdots,z_N)|^2\nn\\
&=\frac{1}{N!}\int d^{2}z_1 \cdots\int d^{2}z_N e^{-\mu \sum_{i=1}^N \Theta(r-|z_i|)}\det_{1\leq i,j\leq N}\psi_{j-1}(z_i)\det_{1\leq l,m\leq N}\overline{\psi}_{l-1}(z_m)\nn\\
&=\det_{1\leq i,j\leq N}\left(\mathbb{I}-(1-e^{-\mu})\mathbb{A}_{kl}\right)\nn\\
&=\prod_{k=1}^{N}\left(1-(1-e^{-\mu})\lambda_k(r)\right)\;,
\end{align}
where on the third line we have used the Cauchy-Binet-Andr{\'e}ief formula \eqref{Cauchy_Binet}. Using the definition of the moment generating function
\be
\boxed{\moy{e^{-\mu N_r}}=\sum_{k=0}^N e^{-\mu k}P_k(r)=\prod_{k=1}^N \left[\bar{\lambda}_k(r)+e^{-\mu}\lambda_k(r)\right]\;,}\label{MGF_ind_Gin}
\ee
where the $\bar{\lambda}_k(r)=1-\lambda_k(r)$ are the eigenvalues of $\bar{\mathbb{A}}=\mathbb{I}-\mathbb{A}$, one can extract the probability $P_k(r)=\Prob\left[N_r=k\right]$,
\be\label{explicit_PKN}
\boxed{P_k(r) =\prod_{l=1}^N \bar{\lambda}_{l}(r) e_k\left(\frac{\lambda_{1}(r)}{\bar{\lambda}_{1}(r)}, \cdots, \frac{\lambda_{N}(r)}{\bar{\lambda}_{N}(r)} \right)\;,\;\;{\rm with}\;\;e_k(x_1, \cdots, x_N) = \sum_{1\leq l_1 < \cdots < l_k \leq N} \prod_{m=1}^k x_{l_m}}
\ee
the elementary symmetric polynomial of $N$ variables and degree $k$. Finally, using the definition of the cumulant generating function,
\be\label{chi_mu_cum}
\chi_r(\mu)=\ln\moy{e^{-\mu N_r}}=\sum_{k=1}^N \ln\left[\bar{\lambda}_k(r)+e^{-\mu}\lambda_k(r)\right]=\sum_{p=1}^{\infty} \frac{(-\mu)^p}{p!}\moy{N_r^p}_c\;,
\ee
one can extract after a few manipulations (see the supplementary material of Article \ref{Art:rot} for details), the cumulants of arbitrary order $p$ as
\be\label{cumul_1}
\boxed{\moy{N_r^p}_c=-\sum_{k=1}^N \Li_{1-p}\left(-\frac{\lambda_k(r)}{\bar{\lambda}_k(r)}\right)=(-1)^{p+1}\sum_{k=1}^N \Li_{1-p}\left(-\frac{\bar{\lambda}_k(r)}{\lambda_k(r)}\right)\;,}
\ee
where $\Li_s(x)=\sum_{k=1}^\infty k^{-s} x^k$ is the polylogarithm function. Note that using $\bar{\lambda}_k(r)=1-\lambda_k(r)$, one can show that $\moy{N_r^p}_c$ is a polynomial of degree $p$ in the variable $\lambda_k(r)$. In particular, one can easily extract the average number of fermions $\moy{N_r}$ and the number variance $\var{N_r}$,
\begin{align}
&\moy{N_r}=\sum_{k=1}^N \frac{\gamma(k,r^2)}{\Gamma(k)}=\frac{\gamma(N+1,r^2)}{\Gamma(N)}+r^2 \frac{\Gamma(k,r^2)}{\Gamma(k)}\label{moy_n_Gin}\;,\\
&\var{N_r}=\sum_{k=1}^N \frac{\gamma(k,r^2)\Gamma(k,r^2)}{\Gamma(k)^2}\;.\label{var_n_Gin}
\end{align}

\begin{figure}
\centering
\includegraphics[width=0.6\textwidth]{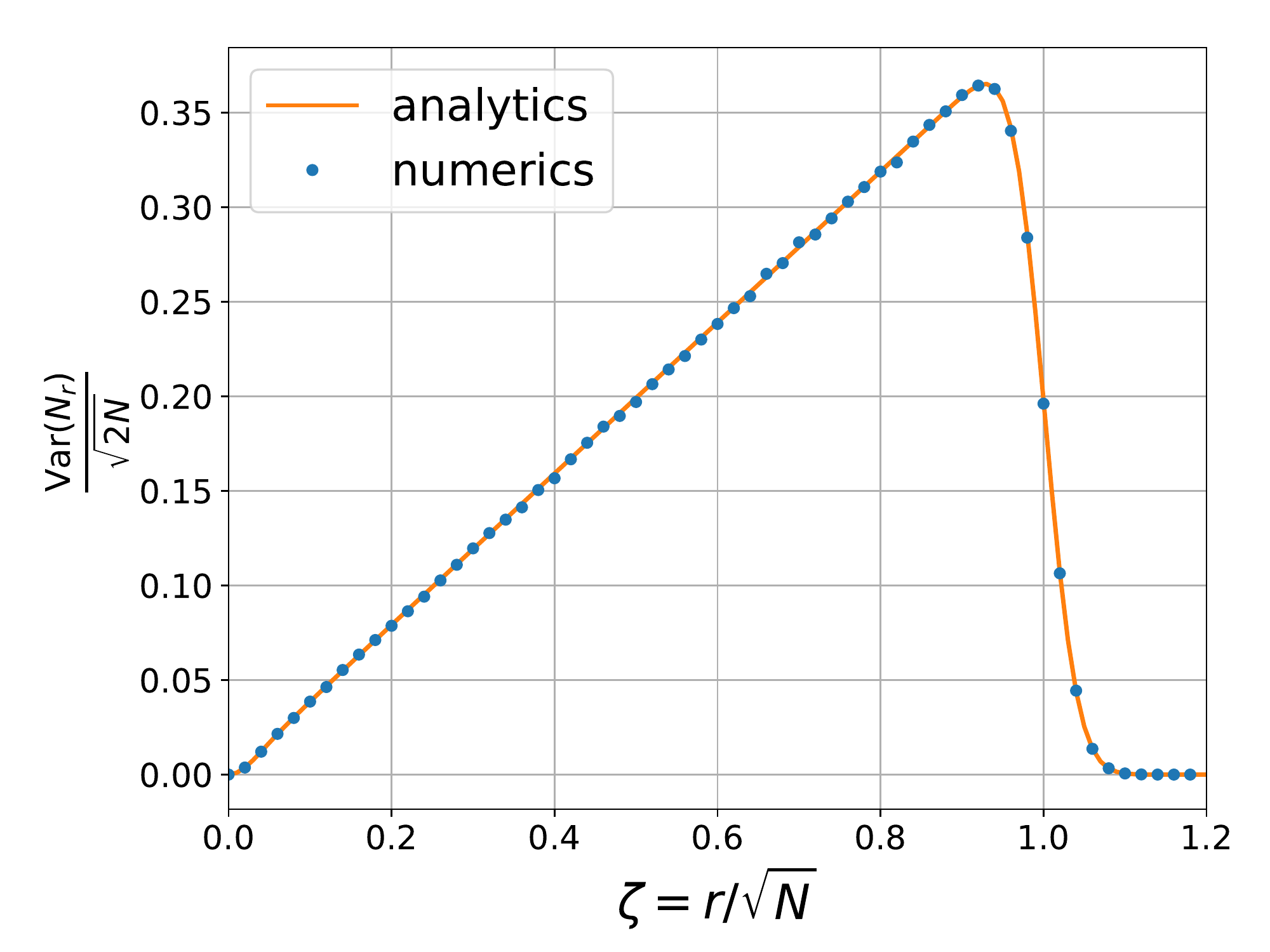}
\caption{Comparison between the rescaled variance $\var{N_r}/\sqrt{2N}$ for $N=200$ obtained by numerical simulation of $5. 10^{4}$ complex Ginibre matrices (blue dots) and the exact analytical formula in Eq. \eqref{var_n_Gin} (in orange) plotted as a function of the rescaled radius $\zeta=r/\sqrt{N}$. The comparison shows a perfect agreement.}\label{Fig_var_n}
\end{figure}

In Fig. \ref{Fig_var_n}, we show a plot of the rescaled variance $\var{N_r}/\sqrt{2N}$ as a function of the rescaled radius $\zeta=r/\sqrt{N}$ and a comparison with numerical results obtained using the exact mapping to the complex Ginibre ensemble.

We emphasise that the expressions for the FCS inside the disk of radius $r$ in Eq. \eqref{explicit_PKN}, its cumulants in Eq. \eqref{cumul_1} and the associated bipartite entanglement entropy in Eq. \eqref{S_q_N} are {\it exact} for any finite value of $N$ and $r$. We will now analyse the behaviour of the entanglement entropy and the full counting statistics in the large $N$ limit, separating our analysis in the different spatial regions represented in Fig. \ref{Fig_dens_Gin}: (i) the {\it deep bulk}, i.e. $r=O(1)$, (ii) the {\it extended bulk}, i.e. $0<\zeta=r/\sqrt{N}<1$ and (iii) the {\it edge regime}, i.e. $s=\sqrt{2}(r-\sqrt{N})=O(1)$.

\subsection{Results in the large $N$ limit}

\subsubsection{Results in the deep bulk}

In the deep bulk, taking the limit $N\to \infty$ while keeping $r=O(1)$, one can express the entanglement entropy as an infinite sum
\be
\boxed{S_q(N,r)\approx S_q^{\rm b}(r)=\frac{1}{1-q}\sum_{k=1}^{\infty}\ln\left[\left(\frac{\Gamma(k,r^2)}{\Gamma(k)}\right)^q+\left(\frac{\gamma(k,r^2)}{\Gamma(k)}\right)^q\right]\;.}\label{S_db}
\ee
Introducing the asymptotic expansion of the eigenvalues $\lambda_{k}(r)$ of the overlap matrix,
\be
\lambda_k(r)=\frac{\gamma(k,r^2)}{\Gamma(k)}\approx \begin{cases}
\displaystyle r^2 &\;,\;\;r\to 0\;.\\
&\\
\displaystyle \frac{1}{2}\erfc\left(\frac{k-r^2}{\sqrt{2}\,r}\right) &\;,\;\;r\to \infty\;,
\end{cases}\label{lambda_as_gin_db}
\ee
we obtain the asymptotic behaviours of the entanglement entropy
\be
S_q^{\rm b}(r)\to\begin{cases}
\displaystyle \frac{q}{q-1}r^2&\;,\;\;r\to 0\\
&\\
\displaystyle \frac{\alpha_q r}{\sqrt{\pi}}&\;,\;\;r\to\infty\;,
\end{cases}\;,\;q>1\;{\rm and}\;S_1^{\rm b}(r)\to\begin{cases}
\displaystyle -2r^2\ln r&\;,\;\;r\to 0\\
&\\
\displaystyle \frac{\alpha_1 r}{\sqrt{\pi}}&\;,\;\;r\to\infty\;,
\end{cases}\label{S_q_bulk_as}
\ee
where $\alpha_q$ is obtained after inserting in Eq. \eqref{S_db} the asymptotic behaviour of $\lambda_k(r)$ for $r\to \infty$ in the second line of Eq. \eqref{lambda_as_gin_db} and reads 
\be
\boxed{\alpha_q=\sqrt{2\pi}\int_{-\infty}^{\infty}\frac{ ds}{1-q}\ln\left[\frac{1}{2^q}\erfc(s)^q+\frac{1}{2^q}\erfc(-s)^q\right]\;.}\label{alpha_q}
\ee
Furthermore, the cumulants of arbitrary order $p$ are given by
\be
\boxed{\moy{N_r^p}_c\approx {\cal K}_p^{\rm b}(r)=-\sum_{k=1}^{\infty} \Li_{1-p}\left(-\frac{\gamma(k,r^2)}{\Gamma(k,r^2)}\right)\;.}\label{cum_finite_n}
\ee
Note that ${\cal K}_1^{\rm b}(r)=r^2$ for all values of $r$. Extracting the asymptotic behaviours of these cumulants, one obtains
\be\label{cum_db_as}
{\cal K}^{\rm b}_p(r)\approx \begin{cases}
r^2&\;,\;\;r\to 0\;,\\
&\\
\sqrt{2}\,r \, \kappa_p &\;,\;\;r\to \infty\;,
\end{cases}
\;,\;\;p\geq 2\;,
\ee
where the coefficient $\kappa_p$ is obtained following the same method as for $\alpha_q$. One can show that it is zero for odd values of $p$ while for even values of $p$, it reads
\be
\kappa_p=-\int_{-\infty}^{\infty}dx \Li_{1-p}\left(-\frac{\erfc(-x)}{\erfc(x)}\right)\;.\label{kappa_p}
\ee
In particular, one can show that $\kappa_2=(2\pi)^{-1/2}$. Note that the cumulants for $p=1,2$ were obtained in \cite{shirai2006large}. All the cumulants of order $p>1$ and the entropy are proportional to each other in the two asymptotic regimes $r\to 0$ and $r\to \infty$ but this is not the case for $r=O(1)$. We now detail the analysis starting by the limit $r\to 0$.

In the small $r$ limit all the cumulants are identical and one can then show that $N_r$ is Poisson distributed
\be
P_k(r)\approx \frac{r^{2k}}{k!}e^{-r^2}\;,\;\;r\to 0\;.
\ee
It is indeed a rare event to find a fermion in the disk of $r$ when the radius becomes small in comparison to the typical inter-particle distance in the deep bulk which is $O(1)$, hence it is not surprising to recover a Poisson distribution. If $N_{\cal D}$ is Poisson distributed with intensity $\lambda$, the behaviour of the associated entanglement entropy is given by \cite{calabrese2015random}
\be
S_q({\cal D})=\frac{q}{q-1}\lambda\;,\;\;q>1\;\;{\rm and}\;\;S_1({\cal D})=-\lambda\ln \lambda\;,
\ee  
hence the result for $r\to 0$ in Eq. \eqref{S_q_bulk_as}.

In the large $r$ limit, using the large $r$ asymptotic behaviour of the cumulants in Eq. \eqref{cum_db_as}, one can show that the rescaled variable
\be
n_r=\frac{N_r-\moy{N_r}}{\sqrt{\var{N_r}}}\approx \frac{\pi^{\frac{1}{4}}(N_r-r^2)}{\sqrt{r}}\;,\;\;r\to \infty\;,
\ee
has a normal distribution ${\cal N}(0,1)$ corresponding to the central limit theorem \cite{soshnikov2002gaussian, shirai2006large}. Indeed, after this rescaling, the cumulants of order $p\geq 3$ will be of order $O\left(r^{(2-p)/2}\right)$ and will therefore vanish in the limit $r\to \infty$. Furthermore, using the connection between this problem and the complex Ginibre ensemble, the atypical fluctuations of $N_r$ can be characterised \cite{shirai2006large}. In the limit $r\to \infty$, they follow a large deviation principle
\be
P_k(r)\approx \exp\left(-r^4 \Phi_{\rm b}\left(\frac{k-\moy{N_r}}{r^2}\right)\right)\;,\;\;r\to\infty\;,
\ee
where the large deviation rate function reads
\be
\Phi_{\rm b}(a)=\frac{1}{4}\left|2(1+a)^2 \ln (1+a)-a(3a+2)\right|\;.\label{Phi_b}
\ee
In particular $\Phi_{\rm b}(a)\approx |a|^3/6$ as $a\to 0$, which does not match with the Gaussian profile of the PDF in the typical regime of fluctuations.

\begin{figure}
\centering
\includegraphics[width=0.6\textwidth]{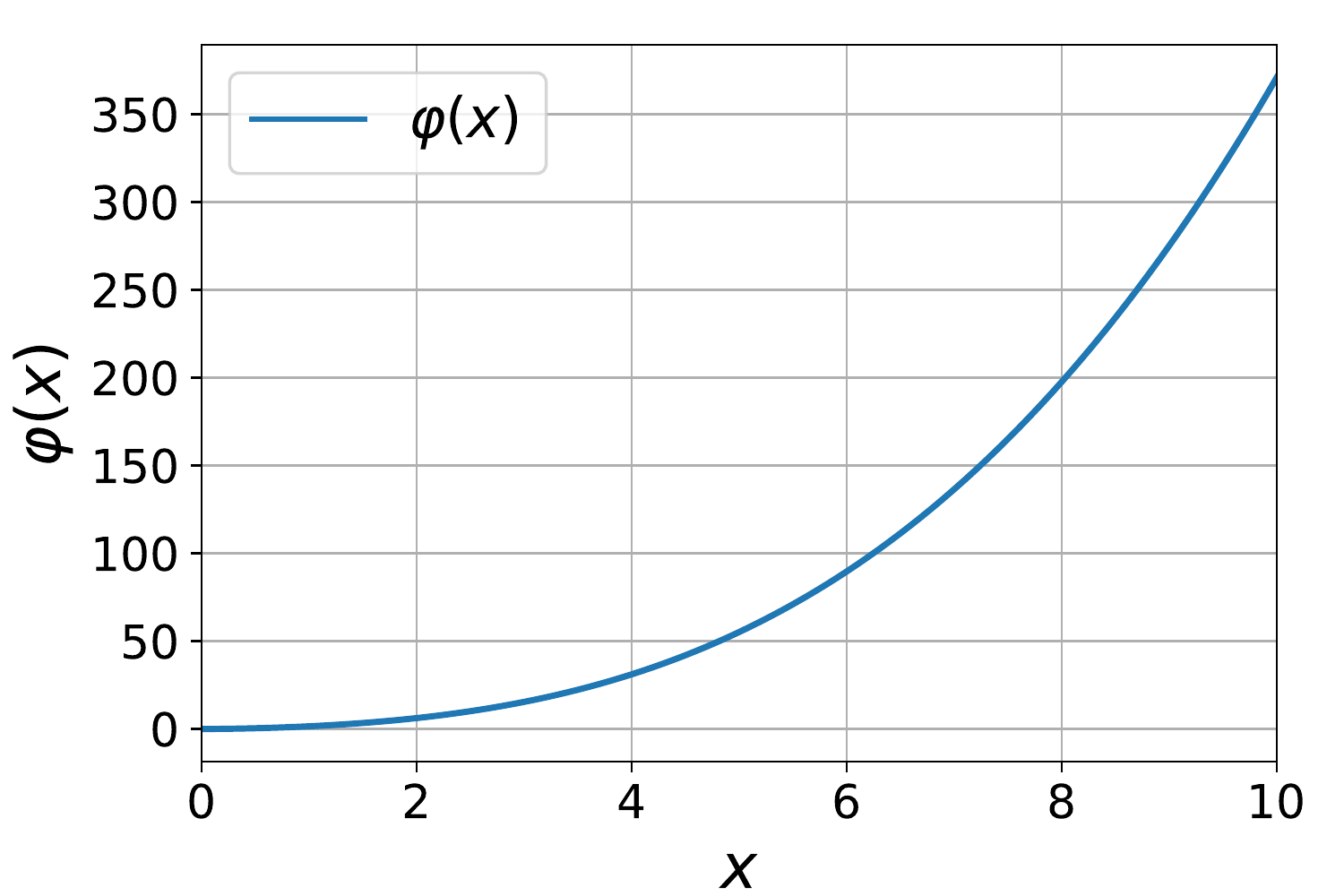}
\caption{Intermediate deviation function $\varphi(x)$ as a function of $x$. This function is quadratic for small values of $x$ and cubic for large $x$ as seen in Eq. \eqref{psi_asympt}, allowing respectively a smooth matching with the central Gaussian regime and the small $a$ cubic behaviour of the large deviation function $\Phi_{\rm b}(a)$ in Eq. \eqref{Phi_b}.}\label{Fig_varphi_FCS}
\end{figure}

In this case, we now show that there is an intermediate deviation regime for the fluctuations of $N_r$. It can be obtained by first computing the centred cumulant generating function
\be
\tilde \chi_r(\mu)=\chi_r(\mu)+\mu\moy{N_r}=\ln \moy{e^{-\mu N_r}}+\mu \moy{N_r}\approx \sqrt{2}\,r\chi(\mu)\;,\;\;r\to \infty\;,
\ee
where the scaling function $\chi(\mu)$ reads
\be
\boxed{\chi(\mu)=\int_0^{\infty}ds\ln\left[1+\sinh\left(\frac{\mu}{2}\right)^2 \erfc(s)\erfc(-s)\right]\;.}\label{chi_mu}
\ee
This scaling function is an even function of $\mu$ and has the asymptotic behaviours
\be\label{asympt_chi}
\chi(\mu)\approx \begin{cases}
\displaystyle\frac{\mu^2}{2\sqrt{2\pi}}&\;,\;\;\mu\to 0\;,\\
&\\
\displaystyle\frac{2}{3}|\mu|^{3/2}&\;,\;\;\mu\to \pm \infty\;.
\end{cases}
\ee
The PDF is then obtained in the intermediate regime $x=(k-\moy{N_r})/(\sqrt{2}r)=O(1)$ from this result by inverse Laplace transform
\be
P_k(r)=\int_{\cal C}\frac{d\mu}{2i\pi} e^{\sqrt{2} r \mu x+\mu \moy{N_r}}\moy{e^{-\mu N_r}}\approx\int_{\cal C}\frac{d\mu}{2i\pi} e^{\sqrt{2}r(\mu x+\chi(\mu))}\;,
\ee
where ${\cal C}$ is the Bromwich contour. Evaluating the integral by a saddle-point approximation, one obtains the scaling form
\be
P_k(r)\approx \exp\left(-\sqrt{2}\,r\varphi\left(\frac{k-\moy{N_r}}{\sqrt{2}\,r}\right)\right)\;,\;\;r\to \infty\;,
\ee
where the scaling function $\varphi(x)$ reads
\be
\boxed{\varphi(x)=-\min_{\mu \in \mathbb{R}}\left[x\mu+\chi(\mu)\right]\;.}\label{phi}
\ee
This expression is slightly formal but can be plotted quite simply as seen in Fig. \ref{Fig_varphi_FCS}. Furthermore, the asymptotic behaviours of $\varphi(x)$ can be extracted using Eq. \eqref{asympt_chi} and read
\be\label{psi_asympt}
\varphi(x)\approx \begin{cases}
\displaystyle\sqrt{\frac{\pi}{2}}x^2&\;,\;\;x\to 0\;,\\
&\\
\displaystyle\frac{1}{3}\,|x|^3&\;,\;\;x\to \infty\;.
\end{cases}
\ee

\begin{figure}
\centering
\includegraphics[width=0.6\textwidth]{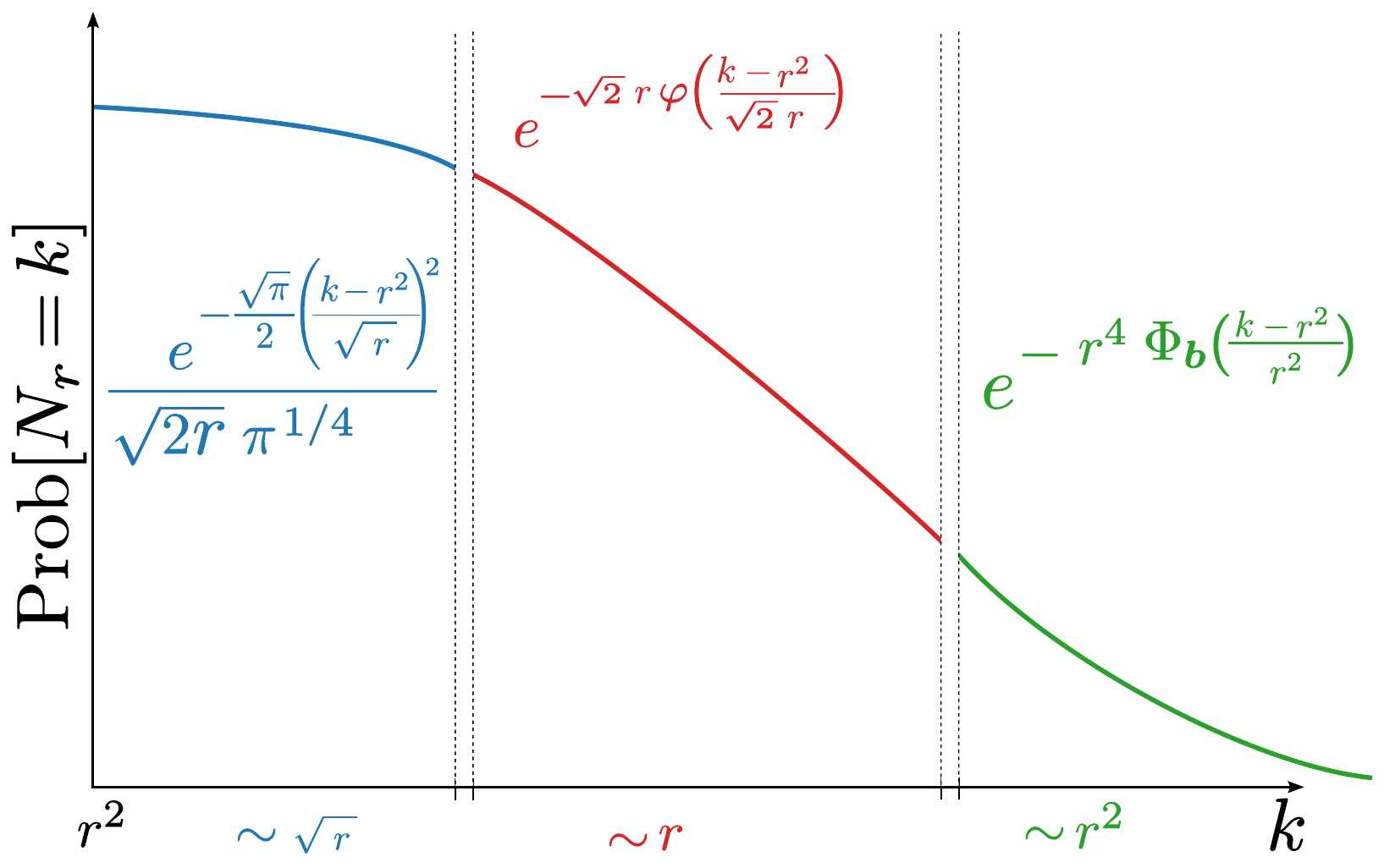}
\caption{Sketch of the regimes of typical (blue), intermediate (red) and large fluctuations (green) for the probability $\Prob[N_r=k]$ of the number $N_r$ of fermions in the disk of radius $r\gg 1$ in the limit $N\to\infty$ (see summary in Eq. \eqref{summary_P_nr_b}). Note that the distribution is symmetric around $\moy{N_r}=r^2$ and we represented only the part $k\geq \moy{N_r}$.}\label{Fig_P_nr_b}
\end{figure}

On the one hand, inserting $x=(k-r^2)/(\sqrt{2}\,r)\ll 1$ in the first line of Eq. \eqref{psi_asympt}, one shows that it smoothly matches with the Gaussian typical regime
\be
\sqrt{2}\,r\varphi\left(\frac{k-\moy{N_r}}{\sqrt{2}\,r}\right)\approx \sqrt{\pi}\left(\frac{k-\moy{N_r}}{\sqrt{2}\,r}\right)^2=\frac{1}{2}\left(\frac{\pi^{\frac{1}{4}}(k-r^2)}{\sqrt{r}}\right)^2\;.
\ee
On the other hand, inserting $x=(k-r^2)/(\sqrt{2}\,r)\gg 1$ in the second line of Eq. \eqref{psi_asympt}, one shows that it matches smoothly with the small $a=(k-r^2)/r^2$ behaviour of the rate function $r^4\Phi_{\rm b}(a)$,
\be
\sqrt{2}\,r\varphi\left(\frac{k-\moy{N_r}}{\sqrt{2}\,r}\right)\approx \frac{\sqrt{2}\,r}{3}\left|\frac{k-\moy{N_r}}{\sqrt{2}\,r}\right|^3=\frac{r^4}{6}\left|\frac{k-r^2}{r^2}\right|^3\;.
\ee
One can then summarise the fluctuations of $N_r$ in this deep bulk and in the asymptotic regime $r\to \infty$ as (see also Fig. \ref{Fig_P_nr_b})
\be
\boxed{
\Prob\left[N_r=k\right]\approx \begin{cases}
\displaystyle \frac{1}{\sqrt{2\sqrt{\pi}\,r}}\exp\left(-\frac{\sqrt{\pi}}{2}\left(\frac{k-r^2}{\sqrt{r}}\right)^2\right)&\;,\;\;|k-r^2|=O(\sqrt{r})\;,\\
&\\
\displaystyle e^{-\sqrt{2}\,r\varphi\left(\frac{k-r^2}{\sqrt{2}\,r}\right)}&\;,\;\;|k-r^2|=O(r)\;,\\
&\\
\displaystyle e^{-r^4 \Phi_{\rm b}\left(\frac{k-r^2}{r^2}\right)}&\;,\;\;|k-r^2|=O(r^2)\;.
\end{cases}}\label{summary_P_nr_b}
\ee

In the deep bulk, we are thus able to characterise in details the statistics of the number $N_r$ of fermions in the domain ${\cal D}_r=\{|z|\leq r\}$ in the two asymptotic limits $r\to 0$, where the typical fluctuations are Poissonian and $r\to \infty$ where the typical fluctuations are Gaussian, but the atypical fluctuations are non-trivial. In these two asymptotic regimes, the measurement of the cumulants allows to extract the exact form of the entanglement entropy. 
We will now consider the behaviour of the entropy and of the FCS in the extended bulk regime (c.f. (ii) in Fig. \ref{Fig_dens_Gin}).

\subsubsection{Results in the extended bulk}

In the extended bulk, for $0<\zeta=\frac{|z|}{\sqrt{N}}<1$, the results are similar as those obtained in the limit $r\to \infty$ of the deep bulk regime.
In the large $N$ limit, the eigenvalues of the overlap matrix $\lambda_k(r)$ take the scaling form
\be
\lambda_k(r)\approx  \frac{1}{2}\erfc\left(\frac{k-N \zeta^2}{\sqrt{2N}\,\zeta}\right)\;,
\ee
which matches smoothly with the limit $r\to \infty$ of the deep bulk result in the second line of Eq. \eqref{lambda_as_gin_db}. 

One can show that in this regime all the even cumulants are of the same order $\sim \sqrt{N}$ and grow linearly, while the odd cumulants of order $p>1$ vanish at order $O(\sqrt{N})$,
\be
\boxed{\moy{N_r}\approx N\zeta^2\;,\;\;{\rm and}\;\;\moy{N_r^p}_c\approx \sqrt{2N}\,\zeta\kappa_p\;,\;\;p\geq 2\;,}\label{cum_eb}
\ee
where $\kappa_p$ is given in Eq. \eqref{kappa_p}.
Inserting this behaviour in Eq. \eqref{S_q_cum_rot}, we obtain a linear growth of the entropy. One can indeed obtain in this regime the result for the entanglement entropy
\be
\boxed{S_q(N,r)\approx \sqrt{\frac{N}{\pi}} \alpha_q\zeta\;,\;\;N\to \infty\;,}\label{S_q_eb}
\ee
where $\alpha_q$ is given in Eq. \eqref{alpha_q}.

\begin{figure}
\centering
\includegraphics[width=0.6\textwidth]{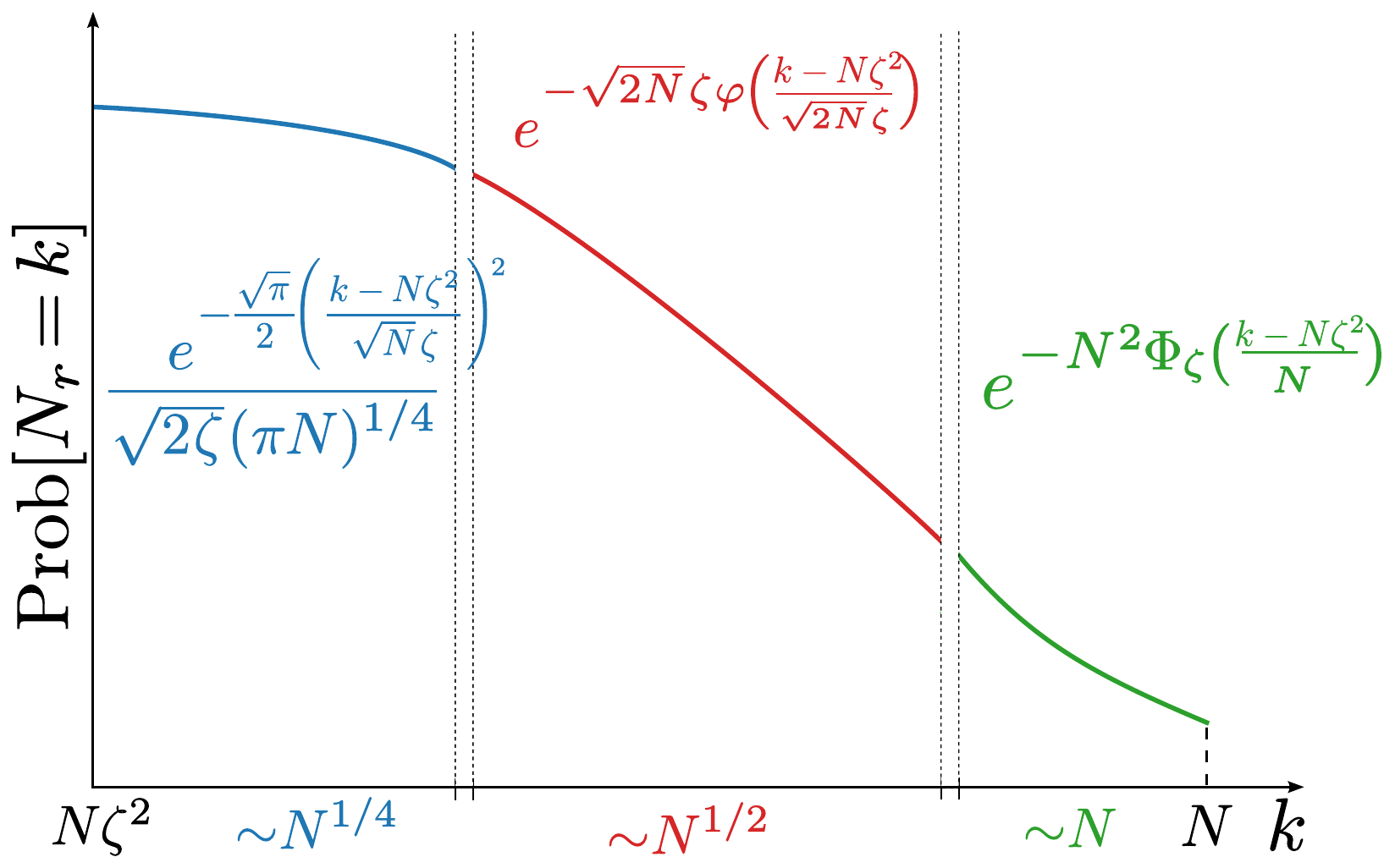}
\caption{Sketch of the regimes of typical (blue), intermediate (red) and large fluctuations (green) for the probability $\Prob[N_r=k]$ of the number $N_r$ of fermions in the disk of radius $r=\sqrt{N}\zeta$ in the limit $N\to\infty$ (see summary in Eq. \eqref{summary_N_r_eb}). Note that the distribution is symmetric around $\moy{N_r}=r^2$ and we represented only the part $k\geq \moy{N_r}$.}\label{Fig_P_nr}
\end{figure}

Finally, using the results for the cumulants we recover in this regime the three scales of fluctuations of $N_r$, summarised as (see also Fig. \ref{Fig_P_nr})
\be
\boxed{
\Prob\left[N_r=k\right]\approx \begin{cases}
\displaystyle \frac{1}{\sqrt{2\sqrt{N \pi }\,\zeta}}\exp\left(-\frac{\sqrt{\pi}}{2}\left(\frac{k-N \zeta^2}{\sqrt{N \zeta}}\right)^2\right)&\;,\;\;|k-N \zeta^2|=O(N^{1/4})\;,\\
&\\
\displaystyle e^{-\sqrt{2N}\,\zeta\varphi\left(\frac{k-N \zeta^2}{\sqrt{2N}\,\zeta}\right)}&\;,\;\;|k-N \zeta^2|=O(\sqrt{N})\;,\\
&\\
\displaystyle e^{-N^2 \Phi_{\zeta}\left(\frac{k-N \zeta^2}{N}\right)}&\;,\;\;|k-N \zeta^2|=O(N)\;,
\end{cases}}\label{summary_N_r_eb}
\ee
where the functions $\varphi(x)$ is given in Eq. \eqref{phi} (with $\chi(\mu)$ given in Eq. \eqref{chi_mu}) while $\Phi_{\zeta}(y)$ was also computed in this regime and reads \cite{allez2014index}
\be
\Phi_{\zeta}(y)=\zeta^4\Phi_{\rm b}\left(\frac{y}{\zeta^2}\right)=\frac{1}{4}\left|2(\zeta^2+y)^2 \ln \left(1+\frac{y}{\zeta^2}\right)-y(3y+2\zeta^2)\right|\;.\label{Phi_eb}
\ee
In the extended bulk of the Fermi gas, we are then able to characterise in details the statistics of the number of fermions and additionally to obtain from these fluctuations the behaviour of the bipartite entanglement entropy. Note that the cumulants were recently obtained in the analogous regime of extended bulk for a more general domain (compact K\"ahler manifold) in the context of quantum Hall states \cite{charles2018entanglement}. We now consider the behaviour close to the edge of the density (c.f. (iii) in Fig. \ref{Fig_dens_Gin}).

\subsubsection{Results at the edge}

Close to the edge, for $s=\sqrt{2}(r-\sqrt{N})=O(1)$, we can still use the asymptotic approximation
\be
\lambda_k(r)\approx  \frac{1}{2}\erfc\left(\frac{k-r^2}{\sqrt{2}\,r}\right)\approx\frac{1}{2}\erfc\left(\frac{k-N}{\sqrt{2N}\,}-s\right)\;.\label{lambda_edge}
\ee
For $|k-N|\gg \sqrt{N}$, the eigenvalues of the overlap matrix $\lambda_k(r)\to 1$. This can be understood recalling that
\be
\lambda_k(r)=\mathbb{A}_{kk}=\int_{|z|\leq r}d^2 z |\psi_{k-1}(z)|^2\;.
\ee
In this regime, the typical length scale of the wave-functions $\psi_{k-1}(z)$ in Eq. \eqref{wf_psi_rot} is $\sim \sqrt{k}$ and becomes small in comparison to the typical scale $r\sim \sqrt{N}$ of the domain over which we integrate. In the regime $x=(N-k)/\sqrt{2N}$, the eigenvalues are non-trivial and give the leading contributions to the fluctuations. 

We first consider the behaviour of the full counting statistics. Close to the edge, one can compute exactly the statistics of the number of fermions {\it outside of the disk of radius $r$}, i.e. $\overline{N_r}=N-N_r$. Inserting Eq. \eqref{lambda_edge} in the finite $N$ equation for the cumulants in Eq. \eqref{cum_finite_n} and replacing the discrete sum over $k$ by an integral over $x=(k-N)/\sqrt{2N}$, one obtains that all the cumulants of $\overline{N_r}$ are of the same order
\be
\moy{\overline{N_r}^p}_c=N\delta_{p,1}+(-1)^p\moy{N_r^p}_c\approx \sqrt{2N}\overline{\cal K}_p^{\rm e}(\sqrt{2}(r-\sqrt{N}))\;,\;\;p\geq 1\;,
\ee
where the scaling function $\overline{\cal K}_p^{\rm e}(s)$ reads
\be
\boxed{\overline{\cal K}_p^{\rm e}(s)=-\int_s^{\infty}dx \Li_{1-p}\left(-\frac{\erfc(x)}{\erfc(-x)}\right)\;.}\label{cum_edge}
\ee
This scaling function can be computed explicitly for $p=1,2$. Computing the asymptotic behaviours of these cumulant scaling functions, one obtains
\be
\overline{\cal K}_p^{\rm e}(s)\approx\begin{cases}
\displaystyle \kappa_p&\;,\;\;s\to -\infty\\
&\\
\displaystyle \frac{e^{-s^2}}{4\sqrt{\pi}s^2}&\;,\;\;s\to +\infty
\end{cases}\;,\;\;p>1\;\;{\rm and}\;\;
\overline{\cal K}_1^{\rm e}(s)\approx\begin{cases}
\displaystyle |s|&\;,\;\;s\to -\infty\\
&\\
\displaystyle\frac{e^{-s^2}}{4\sqrt{\pi}s^2}&\;,\;\;s\to +\infty\;.
\end{cases}\label{cum_as_edge}
\ee

In the asymptotic regime $s\to -\infty$ the cumulants $\moy{N_r^p}_c=\moy{\overline{N_r}^p}_c$ of even order $p$, match smoothly with the result in Eq. \eqref{cum_eb} for the cumulants of order $p>1$ of $N_r$ in the extended bulk. Note that for the average value of $\overline{N_r}$, inserting $r=\sqrt{N}+s/\sqrt{2}$ in the extended bulk result for the mean value of $\overline{N_r}$, it yields
\be
\moy{\overline{N_r}}=N-\moy{N_r}\approx N-\left(\sqrt{N}+s/\sqrt{2}\right)^2\approx -\sqrt{2N}s\;,
\ee
which matches exactly with the $s\to -\infty$ behaviour of $\overline{\cal K}_1^{\rm e}(s)$ in the first line of Eq. \eqref{cum_as_edge}. These results match exactly the extended bulk result for the statistics of $N_r$ and one recovers three regimes of fluctuations for $N_r$ at the edge by taking $\zeta=1$ in Eq. \eqref{summary_N_r_eb}.

Furthermore, in the regime $s\to +\infty$, all the cumulants become identical as seen in the second line of Eq. \eqref{cum_as_edge} and one obtains a Poisson distribution for $\overline{N_r}$, i.e.
\be
\boxed{\overline{P_k}(r)=\Prob\left[\overline{N_r}=k\right]\approx \frac{1}{k!}\left(\frac{e^{-s^2}}{4\sqrt{\pi}s^2}\right)^k \exp\left(-\frac{e^{-s^2}}{4\sqrt{\pi}s^2}\right)\;,\;\;s=\sqrt{2}(r-\sqrt{N})\to +\infty\;.}
\ee

Note that the regime $s=O(1)$ can be studied exactly by analysing the cumulant generating function $\overline{\chi_r}(\mu)=\ln\moy{e^{-\mu \overline{N_r}}}=\chi_r(-\mu)-\mu N$. In this regime, it can be obtained exactly and reads
\be
\overline{\chi_r}(\mu)=\chi_r(-\mu)-\mu N\approx \sqrt{2N}\, \Xi(\mu,\sqrt{2}(r-\sqrt{N}))\;,
\ee
where the scaling function $\Xi(\mu,s)$ is given by
\be
\boxed{\frame{\boxed{\Xi(\mu,s)=\int_s^{\infty}dx \ln\left[\frac{e^{-\mu}}{2}\erfc(x)+\frac{1}{2}\erfc(-x)\right]\;.}}}\label{Xi_edge}
\ee
We now analyse the behaviour of the entanglement entropy.

\begin{figure}
\centering
\includegraphics[width=0.6\textwidth]{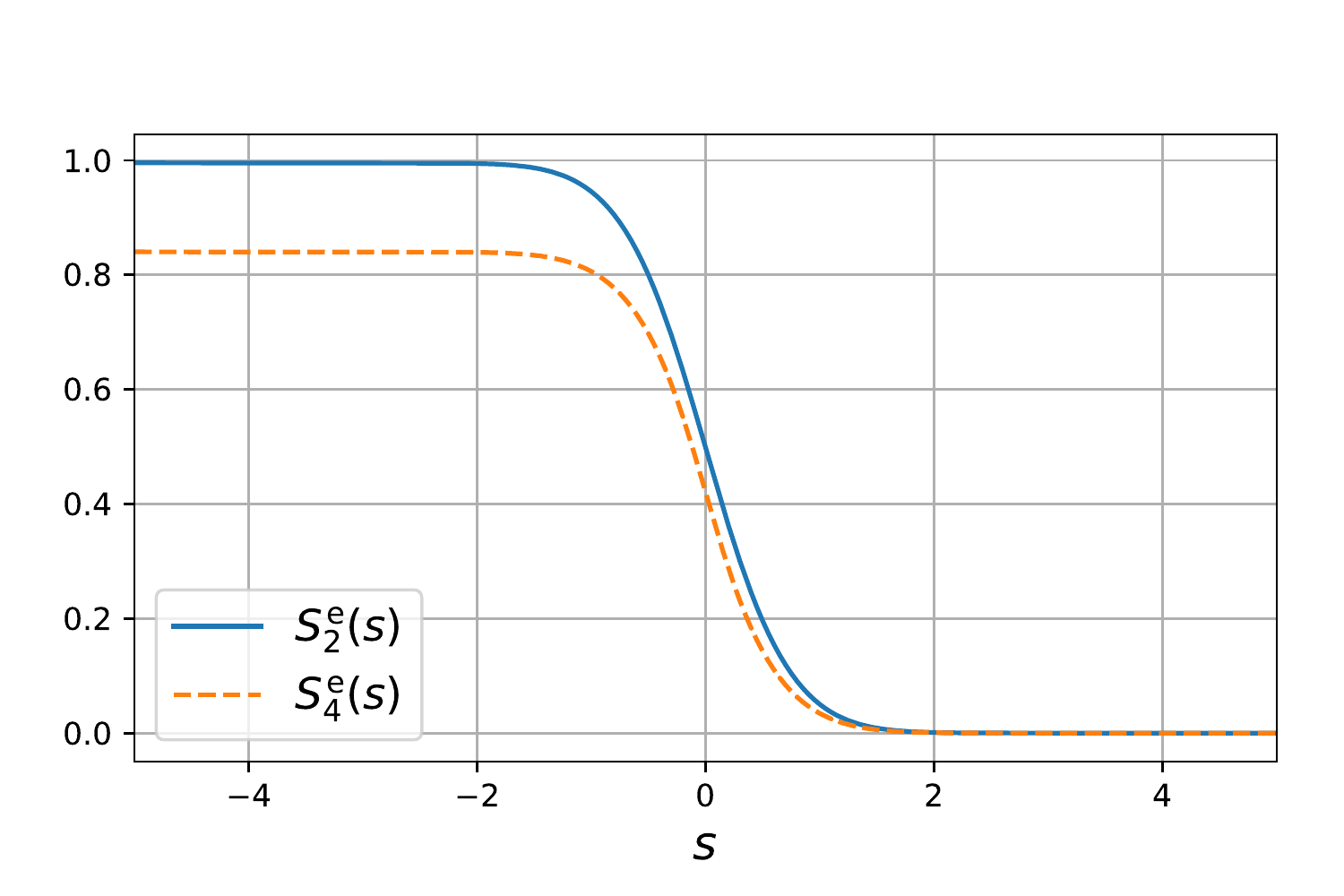}
\caption{Plot of the scaling function $S_q^{\rm e}(s)$ given in Eq. \eqref{S_q_e_eq} as a function of $s$ for $q=2,4$ respectively in blue and orange.}\label{Fig_S_e}
\end{figure}

For the entanglement entropy, inserting Eq. \eqref{lambda_edge} in the finite $N$ equation \eqref{S_q_N} and replacing the discrete sum over $k$ by an integral over $x=(k-N)/\sqrt{2N}$, one obtains
\be
S_q(N,r)\approx \sqrt{2N} S_{q}^{\rm e}(\sqrt{2}(r-\sqrt{N}))\;,
\ee
where the scaling function $S_{q}^{\rm e}(s)$ reads
\be
\boxed{S_{q}^{\rm e}(s)=\int_s^{\infty}\frac{ds}{1-q}\ln\left[\frac{1}{2^q}\erfc(s)^q+\frac{1}{2^q}\erfc(-s)^q\right]\;.}\label{S_q_e_eq}
\ee
Note that in the limit $s\to -\infty$, it is then trivial to obtain $S_{q}^{\rm e}(s)\to \alpha_q/\sqrt{2\pi}$, where $\alpha_q$ is given in Eq. \eqref{alpha_q}, therefore smoothly matching the extended bulk result. In the limit $s\to +\infty$, one can show that the scaling function reads
\be
S_{q}^{\rm e}(s)\approx \frac{q}{q-1}\frac{e^{-s^2}}{4\sqrt{\pi}s^2}\;,\;\;q>1\;\;{\rm and}\;\;S_1^{\rm e}(s)\approx \frac{e^{-s^2}}{4\sqrt{\pi}}\;.
\ee
In particular, for $s\to +\infty$, we recover the form of the entanglement entropy characteristic of a Poisson distribution \cite{calabrese2015random}. This scaling function is plotted in Fig. \ref{Fig_S_e} for $q=2$ and $q=4$. As expected, the entanglement entropy vanishes abruptly at the edge of the density. Note that close to the edge the scaling function for the cumulants in Eq. \eqref{cum_edge} (and in particular the variance) is not proportional to the scaling function for the entanglement entropy in Eq. \eqref{S_q_e_eq}, as the latter is a sum over all the contributions of all cumulants (which are all of the same order).

This result closes this section on the full counting statistics and the entanglement entropy. To summarise our findings, we have computed exactly for any finite value of $N$ the entanglement entropy and FCS of a gas of trapped fermions for a disk of radius $r$ around the centre of the trap. In the large $N$ limit, we have shown that the variance number $\var{N_r}$ and entanglement entropy are proportional to each other in the extended bulk $0<r/\sqrt{N}<1$, with a proportionality factor $\alpha_q$ computed exactly in Eq. \eqref{alpha_q}. This could prove useful to measure in a simple manner the entanglement entropy. However, this proportionality does not hold at the edge of the density.  

We will now highlight a connection between this problem of trapped fermions and the two-dimensional one-component plasma ($2d$ OCP), which is a gas of charged particles trapped in a potential $v(|z|)$.  


\section{Two-dimensional one-component plasma}

In this section, we show that the problem of non-interacting fermions in a rotating harmonic trap can be mapped exactly onto a different problem of classical statistical mechanics: the two-dimensional one-component plasma. In the model of the two-dimensional one-component plasma ($2d$ OCP), classical particles of same charge are confined by a potential. The joint probability of the complex positions $z_k=x_k+\I y_k$ of these charged particles in the confining potential $N V(z)/2$ at equilibrium at inverse temperature $\beta$ reads \cite{forrester1998exact} 
%
%
%
\be
P_{\rm joint}^{\rm OCP}(z_1,\cdots, z_N)=\frac{1}{Z_N^{\rm OCP}(\beta)}\prod_{i<j}|z_i-z_j|^{\beta}\prod_{k=1}^N e^{-\frac{N\beta}{2} V(z_k) }\;.\label{general_OCP}
\ee
Setting $\beta=2$ and $V(z)=|z|^2$ we recover the joint PDF of the eigenvalues in the complex Ginibre ensemble in Eq. \eqref{P_joint_Gin} and therefore after the change of variable $z\to z/\sqrt{N}$ the joint PDF of the fermions in Eq. \eqref{Gin_fermions}. For $\beta=2m$ with $m\in\mathbb{N}_+$ and $V(z)=|z|^2$, this joint probability corresponds to a Laughlin state which has applications in condensed matter and especially for the quantum hall effects \cite{ezawa}. Finally, we mention that setting $\beta=1$ or $\beta=4$, this joint PDF does {\it not} match the real and symplectic Ginibre ensembles, whose joint PDF of eigenvalues are more involved \cite{lehmann1991eigenvalue, ginibre1965statistical}.  
This two-dimensional Coulomb system has been studied in details both  in the mathematics and physics community \cite{cunden2017universality, cunden2016large, chafai2014note} (see \cite{serfaty2017systems, forrester1998exact, forrester2010log} for reviews on the subject). Note that there is no known connection between the model of plasma trapped by the potential $V(z)$ and the model of fermions in this same trapping potential, the case of the harmonic trapping $V(z)=|z|^2$ being an exception.

At the specific value of the inverse temperature $\beta=2$, the positions of the charges form a determinantal point process and the problem becomes exactly solvable. We can therefore generalise, though in a different context, the problem considered in the last section and obtain for a symmetric potential $V(z)=v(|z|)$ the full counting statistics of the charges, defined as
\be
N_r=\sum_{k=1}^N \Theta(r-|z_k|)\;.\label{index_ocp}
\ee
Note that the entanglement entropy could also be extended to this case, but for this classical system it does not refer to any physical property. Let us first characterise this system by computing the average density in the case of a rotationally symmetric potential $V(z)=v(|z|)$ that grows at infinity as $v(r)\gg 2\ln r$. To obtain this density, we use the method of the Coulomb gas.

\subsubsection{Two-dimensional Coulomb gas}

The equilibrium density can be obtained by first introducing the empirical density
\be
\hat\rho(z)=\frac{1}{N}\sum_{k=1}^N \delta(z-z_k)\;,
\ee 
and rewriting the energy of the Coulomb gas as
\begin{align}
E_N(z_1,\cdots,z_N)&=\frac{1}{N}\sum_{k=1}^N v(|z_k|)-\frac{2}{N^2}\sum_{i<j} \ln|z_i-z_j|\nn\\
&=\int d^2 z \hat \rho(z) v(|z|)-\iint d^2 z_1 d^2 z_2 \hat \rho(z_1)\hat \rho(z_2) \ln|z_1-z_2|\;.
\end{align}
In the large $N$ limit, one can replace the distribution $\hat \rho(z)$ by a continuous function $\rho(z)$ and the energy by a functional of this density 
\be
S[\rho]=\int d^2 z \rho(z) v(|z|)-\iint d^2 z_1 d^2 z_2  \rho(z_1)\rho(z_2) \ln|z_1-z_2|\;.
\ee
The equilibrium density $\rho_{\rm eq}(z)$ is the density that minimises this energy functional, under the constraint that it is normalised, and is therefore solution of the equation
%
%
%
%
%
%
%
%
%
%
%
%
%
%
%
\be\label{min_S_gin}
\left.\frac{\delta S}{\delta\rho}\right|_{\rho_{\rm eq}(z)}=v(|z|)- 2\int d^2 z_1 \rho_{\rm eq}(z_1)\ln|z-z_1|=0\;.
\ee
Note that this equation is only valid on the support of $\rho_{\rm eq}(z)$. In particular, for a potential $v(|z|)\gg 2\ln|z|$ as $|z|\to \infty$, Eq. \eqref{min_S_gin} cannot hold in the large $|z|$ limit and the density must have a finite edge $r_{\rm e}$. On the contrary, if $v(|z|)\sim 2\ln |z|$ as $|z|\to \infty$, the density will extend to the whole complex plane. The equation \eqref{min_S_gin} can be solved quite simply in this case using that the function $G(z)=\ln|z|$ is the Green's function of the two-dimensional Laplace operator
\be
\Delta G(z)=\delta(z)\;,\;\;{\rm with}\;\;\Delta=\frac{1}{r}\partial_r (r\partial_r ) +\frac{1}{r^2}\partial_\theta^2\;.
\ee
Applying the Laplace operator to Eq. \eqref{min_S_gin}, we obtain the value of the density
\be
\rho_{\rm eq}(z)=\frac{1}{4\pi|z|}\partial_r\left[r v'(r)\right]_{r=|z|}\;,\;\;r\leq r_{\rm e}\;.\label{eq_dens_2docp}
\ee
This density has a finite edge $r_{\rm e}$ which is obtained using the normalisation of the probability
\be
\int_{|z|\leq r_{\rm e}} d^2 z \rho_{\rm eq}(z)= \frac{r_{\rm e}v'(r_{\rm e})}{2}=1\;.\label{eq_edge}
\ee
Note that for $v(r)=r^2$, which coincides with the Ginibre ensemble, we recover the Girko's law with a uniform density $\rho_{\rm eq}(z)=\pi^{-1}$ and a finite edge for $r_{\rm e}=1$. For a general potential $v(r)$ such that $v(r)\gg 2\ln r$, one can clearly identify a bulk regime for $r<r_{\rm e}$ and an edge regime for $|r-r_{\rm e}|\ll 1$, where the density vanishes abruptly. We will now see how to compute the full counting statistics for this general potential. 

\subsection{Full counting statistics of the $2d$ OCP for $\beta=2$}

For the specific inverse temperature $\beta=2$, the complex positions $z_k$'s of the charged particles form a determinantal point process. In this case, 
we can first use the Vandermonde identity in Eq. \eqref{VdM} to rewrite the joint PDF as
\begin{align}
&P_{\rm joint}^{\rm OCP}(z_1,\cdots, z_N)=\frac{1}{N!}\det_{1\leq i,j\leq N} \phi_i(z_j)\det_{1\leq k,l\leq N} \overline{\phi_l}(z_k)\;,\\
&{\rm with}\;\;\phi_l(z)=\frac{z^{l-1}}{\sqrt{h_l}}e^{-\frac{N}{2}v(|z|)}\;\;{\rm and}\;\;h_l=\int d^2 z z^{2l-2}e^{-N v(|z|)}\;.\nn
\end{align}
Note the analogy between this form of the joint PDF and the first line of Eq. \eqref{Gin_fermions}. Using this representation, one can now compute exactly the moment generating function of the number $N_r$ of charges inside the disk of radius $r$ by using the Cauchy-Binet-Andr{\'e}ief formula in Eq. \eqref{Cauchy_Binet}
\begin{align}
\moy{e^{-\mu N_r}}=&\int d^2 z_1 \cdots \int d^2 z_N e^{-\mu \sum_{k=1}^N \Theta(r-|z_k|)}P_{\rm joint}^{\rm OCP}(z_1,\cdots, z_N)\\
&=\det_{1\leq i,j\leq N}\left(\int d^2 z  e^{-\mu\Theta(r-|z|)} \overline{\phi}_i(z)\phi_j(z)\right)\;.
\end{align}
Using the rotational invariance of $v(|z|)$ together with $\int_0^{2\pi}e^{\I \theta(k-l)}d\theta=2\pi \delta_{kl}$, we can simplify this result to obtain
\be
\moy{e^{-\mu N_r}}=\prod_{k=1}^{N}\left(1-(1-e^{-\mu})q_k(r)\right)\;\;{\rm with}\;\;q_k(r)=\frac{\displaystyle \int_0^{r}du\, u^{2k-1}e^{-N v(u)}}{\displaystyle \int_0^{\infty}du\, u^{2k-1}e^{-N v(u)}}\;.\label{MGF_2d_OCP}
\ee
Note finally that for $v(r)=r^2$, the connection is explicit with the case of fermions as $q_k(r)=\lambda_k(\sqrt{N} r)$ where $\lambda_k(r)$ is given in Eq. \eqref{lambda_k_ferm}. We will now show that the centred cumulant generating function defined as
\be\label{CGF_2d_OCP}
\tilde \chi_{r}(\mu)=\ln\moy{e^{-\mu (N_r-\moy{N_r})}}=\sum_{k=1}^N \left(\ln\left[1+(e^{-\mu}-1)q_k(r)\right]+\mu q_k(r)\right)
\ee
takes a universal scaling form.

In the large $N$ limit, the function $q_k(r)$ can be evaluated in the regime $x=\frac{k}{N}=O(1)$ using a saddle-point approximation as
\be\label{q_k}
q_k(r)=\frac{\displaystyle \int_0^{r}\frac{du}{u}e^{-N \varphi_x(u)}}{\displaystyle \int_0^{\infty}\frac{du}{u}e^{-N \varphi_x(u)}}\;,\;\;{\rm with}\;\;\varphi_x(u)=v(u)-2 x\ln u\;.
\ee
The rate function $\varphi_x(u)$ has a minimum $u^*(x)$ such that $u^*(x)v'[u^*(x)]=2x$. Note that for $x=1$, from Eq. \eqref{eq_edge} one recovers that $u^*(1)=r_{\rm e}$. Computing the second derivative of the rate function $\varphi_x(u)$, for $u=u^*(x)$, one obtains
\be
\left.\partial_u^2\varphi_x(u)\right|_{u=u^*(x)}=v''[u^*(x)]+\frac{2 x}{{u^*(x)}^2}=v''[u^*(x)]+\frac{v'[u^*(x)]}{u^*(x)}=\frac{\partial_u\left[u v'(u)\right]_{u=u^*(x)}}{u^*(x)}
\ee
where we used that $u^*(x)v'[u^*(x)]=2x$. It can then be expressed explicitly in terms of the equilibrium density $\rho_{\rm eq}(z)$ in Eq. \eqref{eq_dens_2docp} as $\left.\partial_u^2\varphi_x(u)\right|_{u=u^*(x)}=4\pi \rho_{\rm eq}\left[u^*(x)\right]$. Inserting in Eq. \eqref{q_k}, one finally obtains
\be
q_{k=N x}(r)\approx\frac{\displaystyle \int_0^{r}du e^{-\frac{N}{2}\left.\partial_u^2\varphi_x(u)\right|_{u=u^*(x)}(u-u^*(x))^2}}{\displaystyle \int_0^{\infty}du e^{-\frac{N}{2}\left.\partial_u^2\varphi_x(u)\right|_{u=u^*(x)}(u-u^*(x))^2 }}\approx\frac{1}{2}\erfc\left(\sqrt{2\pi N\rho_{\rm eq}\left[u^*(x)\right]}(u^*(x)-r)\right)\;.\label{q_k_sp}
\ee
In the bulk, i.e. for $r<r_{\rm e}=u^*(1)$, there exists a value $x_r=k_r/N$ such that $u^*(x_r)=r$. For values of $x=k/N$ close to $x_r$, one can use the Taylor series
\be
u^*(x)-r=\partial_{x}u^*(x)(x-x_r)+ O(x-x_r)^2=\frac{k-k_r}{2\pi N r \rho_{\rm eq}(r)}+O\left(\frac{k-k_r}{N}\right)^2\;.
\ee
Inserting \eqref{q_k_sp} in Eq. \eqref{CGF_2d_OCP} and replacing the discrete sum on $k$ by an integral over $u=(k-k_r)/(\sqrt{2\pi N\rho_{\rm eq}(r)} r)$, one finally obtains
\be
\boxed{\tilde \chi_{r}(\mu)=\ln\moy{e^{-\mu (N_r-\moy{N_r})}}\approx \sqrt{2\pi N\rho_{\rm eq}(r)}\,r\,\chi(\mu)\;,\;\;N\to \infty\;,}\label{CGF_bulk_2docp}
\ee
where the function $\chi(\mu)$ is given in Eq. \eqref{chi_mu}. This cumulant generating function is therefore universal and holds for general confining potentials. Note that it also appeared in very different contexts for the fluctuations of current of non-interacting Brownian walkers \cite{derrida2009current} and the displacement of tagged particles \cite{krapivsky2015tagged, sadhu2015large, cividini2017tagged}. This result implies in particular that for any potential $v(r)\gg 2\ln r$, the statistics of $N_r$ also has in the bulk (for $0<r<r_{\rm e}$) three regimes of fluctuations
\be
\boxed{
\Prob\left[N_r=k\right]\approx \begin{cases}
\displaystyle \frac{1}{\sqrt{2\pi\var{N_r}}}\exp\left(-\frac{(N_r-\moy{N_r})^2}{2\var{N_r}}\right)&\;,\;\;|k-\moy{N_r}|=O(N^{1/4})\;,\\
&\\
\displaystyle e^{-\sqrt{2\pi N\rho_{\rm eq}(r)}\,r\,\varphi\left(\frac{k-\moy{N_r}}{\sqrt{2\pi N\rho_{\rm eq}(r)}\,r}\right)}&\;,\;\;|k-\moy{N_r}|=O(\sqrt{N})\;,\\
&\\
\displaystyle e^{-N^2 \Lambda_{r}\left(\frac{k-\moy{N_r}}{N}\right)}&\;,\;\;|k-\moy{N_r}|=O(N)\;,
\end{cases}}\label{summary_FCS_2d_OCP}
\ee
where $\moy{N_r}=N r v'(r)/2$, $\var{N_r}\approx \sqrt{2N\rho_{\rm eq}(r)}\,r$ and where $\Lambda_{r}(x)\propto |x|^3$ as $x\to \infty$ as was obtained for fermions in the a rotating trap in Eq. \eqref{summary_N_r_eb} (see also Fig. \ref{Fig_P_nr}).

We close this section by mentioning two extensions of this result:

\begin{itemize}

\item One might extend this computation to any linear statistics for the charges inside the disk
\be\label{Lin_stat}
{\cal L}_r=\sum_{k=1}^N L(|z_k|)\Theta(r-|z_k|),
\ee
where $L(r)$ is a smooth function. Note that the fluctuations of ${\cal L}_r$ are two-fold: the number $N_r$ of charges involved in the linear statistics and the exact locations $z_k$'s of their positions. Using for instance $L(r)=r^2$, it corresponds to the moment of inertia, or in the context of fermions -- corresponding to the choice of potential $v(r)=r^2$ -- the potential energy of particles inside a disk of radius $r$. A similar type of {\it truncated linear statistics} was recently considered for Hermitian matrices \cite{Grabsch2017, Grabsch2017_2}. In the case of the $2d$ OCP, one can also compute exactly (at inverse temperature $\beta=2$), the moment generating function associated to this statistics which reads
\begin{align}
\moy{e^{-\mu {\cal L}_r}}&=\int d^2 z_1 \cdots \int d^2 z_N e^{-\mu \sum_{k=1}^N L(|z_k|) \Theta(r-|z_k|)}P_{\rm joint}^{\rm OCP}(z_1,\cdots, z_N)\\
&=\prod_{k=1}^{N}\left(1-Q_{k,L}(r,\mu)\right)\;\;{\rm with}\;\;Q_{k,L}(r,\mu)=\frac{\displaystyle \int_0^{r}du\, (1-e^{-\mu L(u)}) u^{2k-1}e^{-N v(u)}}{\displaystyle \int_0^{\infty}du\, u^{2k-1}e^{-N v(u)}}\;,\nn
\end{align}
where we used in the second line the Cauchy-Binet-Andr{\'e}ief formula \eqref{Cauchy_Binet}.
Evaluating $Q_{k,L}(r,\mu)$ for $x=k/N=O(1)$ with a saddle point approximation, which remains valid for $\mu=O(1)$, yields the simple relation 
\be
Q_{k=N x,L}(r,\mu)\approx(1-e^{-\mu L(u^*(x))}) q_{k=Nx}(r)\;,\;\;u^*(x)v'\left[u^*(x)\right]=2x\;,
\ee
where $q_{k=Nx}(r)$ is given in Eq. \eqref{q_k_sp}. This was the crucial point of our analysis and with this relation, everything follows through.
One can then show that in the large $N$ limit, the cumulants of ${\cal L}_r$ are governed by the same centred cumulant generating function
\be
\boxed{\frame{\boxed{\tilde \chi_{r,L}(\mu)=\ln\moy{e^{-\mu({\cal L}_r-\moy{{\cal L}_r})}}\approx \sqrt{2\pi N\rho_{\rm eq}(r)}\,r\,\chi\left(\mu L(r) \right)\;,\;\;N\to \infty\;.}}}\label{CGF_bulk_lin_2docp}
\ee
%
%
%
%
The result for the fluctuations of the FCS therefore extends for {\it any linear statistics} of the type of Eq. \eqref{Lin_stat}. It indicates that the fluctuations of the number $N_r$ of charges involved in the linear statistics are more prominent than the fluctuations of their positions. From this result, we expect for the fluctuations of any linear statistics ${\cal L}_r$ the same scenario 
as for $N_r$, with three regimes of fluctuations (c.f. Fig. \ref{Fig_P_nr}): (i) a typical Gaussian regime for fluctuations of order $|{\cal L}_r-\moy{{\cal L}_r}|=O(N^{1/4})$, (ii) an intermediate deviation regime given by $\varphi(x)$ in Eq. \eqref{phi} for fluctuations of order $|{\cal L}_r-\moy{{\cal L}_r}|=O(N^{1/2})$ and (iii) a large deviation regime for fluctuation of order $|{\cal L}_r-\moy{{\cal L}_r}|=O(N)$,  which vanishes cubically as ${\cal L}_r\to \moy{{\cal L}_r}$. 

\item We also mention that the full counting statistics can be obtained at the {\it edge} (and one should be able to extend it to any linear statistics of the form \eqref{Lin_stat}). Considering now the fluctuations of $\overline{N_r}=N-N_r$ close to the edge, with a similar analysis one obtains the cumulant generating function
\be
\overline{\chi_{r}}(\mu)=\ln\moy{e^{-\mu \overline{N_r}}}=\chi_r(-\mu)-\mu N=\sum_{k=1}^N \ln\left[e^{-\mu}(1-q_k(r))+q_k(r)\right]\label{chi_bar_r}
\ee
For $s=\sqrt{2N\pi \rho_{\rm eq}(r_{\rm e})}(r-r_{\rm e})$, and using the saddle-point approximation in Eq. \eqref{q_k_sp}, one can obtain by replacing the discrete sum on $k$ by an integral over $u=(N-k)/(\sqrt{2\pi N\rho_{\rm eq}(r_{\rm e})} r_{\rm e})$ the universal scaling form
\be
\boxed{\frame{\boxed{\overline{\chi_{r}}(\mu)=\ln\moy{e^{-\mu \overline{N_r}}}\approx \sqrt{2\pi N\rho_{\rm eq}(r_{\rm e})}\,r_{\rm e}\, \Xi\left[\mu,\sqrt{2N\pi \rho_{\rm eq}(r_{\rm e})}(r-r_{\rm e})\right]\;,}}}\label{CGF_edge_2docp}
\ee
where $\Xi(\mu,s)$ is given in Eq. \eqref{Xi_edge}. 

\end{itemize}

The results in Eqs. \eqref{CGF_bulk_2docp}, \eqref{CGF_bulk_lin_2docp}, \eqref{CGF_edge_2docp} together with the scaling functions in Eq. \eqref{chi_mu} and \eqref{Xi_edge} are the main results of this section and we will now analyse their implications. In particular, we will see that one can use these results to come back to our initial motivation and study in detail the extreme value statistics of the strongly correlated positions of the charged particles. 

\subsection{Order statistics for the complex Ginibre ensemble}
A direct application of our results on the full counting statistics  concerns the order statistics of the gas. It is defined by ordering the radii
\be
R_{1,N}=\max_{1\leq i\leq N} r_i\geq R_{2,N}\geq\cdots \geq R_{N,N}=\min_{1\leq i\leq N} r_i\;.
\ee
These ordered maxima are naturally linked to the full counting statistics by the exact relation
%
%
\be
\Prob\left[R_{k,N}\leq r\right]=\Prob\left[\overline{N_r}\leq k\right]=\sum_{l=0}^k \overline{P_l}(r)\;,
\ee
where $\overline{P_k}(r)=\Prob\left[\overline{N_r}=k\right]$. Indeed, if the particle with the $k^{\rm th}$ largest radius satisfies $R_k\leq r$, it yields trivially that there are $k$ or less particles outside of the disk of radius $r$. Using the results in Eq. \eqref{summary_FCS_2d_OCP} for the full counting statistics together with $\Prob\left[\overline{N_r}\leq k\right]=\Prob\left[\overline{N_r}\geq N-k\right]$, one can obtain the statistics of $R_{k,N}$ in the regime $N\to \infty$ with $\alpha=k/N=O(1)$. In particular, using the first line of Eq. \eqref{summary_FCS_2d_OCP}, one can show that the typical fluctuations of $R_{k,N}$ are Gaussian. This result is at variance with the typical distribution of the maxima close to the global maximum, i.e. in the regime $N\to \infty$ with $k=O(1)$ fixed, where one obtains \cite{rider2004order}
\be
\lim_{N\to \infty}\Prob\left[R_{k,N}< 1+a_N+s/b_N\right]=\frac{\Gamma(k,e^{-s})}{\Gamma(k)}\;.
\ee
This behaviour of the ordered maxima is identical to the behaviour obtained for i.i.d. random variables (see section \ref{sec_order_iid}). Next we consider another application of our results: the fluctuations of the global maximum $r_{\max}=R_{1,N}$ and in relation to the full counting statistics at the edge.

\subsection{Extreme value statistics: the case of the maximum}\label{r_max_Gin}

We will first introduce the problem of extreme value statistics for the $2d$ OCP and then show how the result in Eq. \eqref{Xi_edge} allows to solve a puzzle of matching between typical and large deviation regimes.
For the $2d$ OCP, and more particularly for the Ginibre ensemble, i.e. $v(r)=r^2$, the statistics of the radius $\displaystyle r_{\max}=\max_{1\leq i\leq N} |z_i|$
of the particle the farthest away from the centre of the trap was studied in details. For simplicity, we detail only the case of the Ginibre ensemble but these results can be extended to general rotationally symmetric potentials $V(z)=v(|z|)$ provided that $v(r)\gg  2\ln r$ as $r\to \infty$. 
For the particular value of inverse temperature $\beta=2$, we can use the determinantal structure of the process to obtain the CDF of $r_{\max}$ for a finite number $N$ of particles. It is obtained by integrating the joint PDF in Eq. \eqref{general_OCP} for all radii $r_i$'s over the interval $r_i\in[0,r]$
\be\label{r_max_Gin_prod}
\Prob\left[r_{\max}\leq r\right]=\prod_{k=1}^{N} q_k(r)\;,\;\;{\rm with}\;\;q_k(r)=\frac{\displaystyle \int_0^{r}du\, u^{2k-1}e^{-N v(u)}}{\displaystyle \int_0^{\infty}du\, u^{2k-1}e^{-N v(u)}}\;.
\ee
In the case of the Ginibre ensemble, this property can be obtained from Kostlan's theorem \cite{kostlan1992spectra}. From this formula, $r_{\max}$ can be interpreted as the maximum of a set $\{x_1,\cdots,x_N\}$ of independent but non-identically distributed random variables of individual CDFs $q_k(r)=\Prob\left[x_k\leq r\right]$. It is then possible to show that the typical fluctuations of $r_{\max}$ are given by a Gumbel distribution (as in the case of i.i.d. random variables). It was first obtained for the specific case of the complex Ginibre ensemble \cite{rider2003limit} 
\be
\lim_{N\to \infty}\Prob\left[r_{\max}\leq 1+a_N+\frac{s}{b_N}\right]=G_{\rm I}(s)=\exp\left(e^{-s}\right)\;,\label{r_max_Gin_typ}
\ee
where $a_N\sim \sqrt{\ln N/4N}$ and $b_N\sim \sqrt{4N\ln N}$. These results were then extended to a more general class of potentials \cite{chafai2014note}.
There are no exact result for the typical fluctuations for general values of $\beta$ but the Gumbel law was conjectured to hold \cite{chafai2014note, chafai2019simulating ,dubach}. The atypical fluctuations were also characterised for $v(r)=r^2$ \cite{cunden2016large} and the PDF was found to take a large deviation form both to the right $r\geq 1$ and to the left $r\leq 1$ of the typical fluctuations, with the scaling form \cite{cunden2016large}
\be
\partial_r \Prob\left[r_{\max}\leq r\right]=\begin{cases}
\displaystyle e^{-N^{2}\Psi_{-}^{\rm Gin}(r) }&\;,\;\;r\leq 1\;,\\
&\\
\displaystyle b_N G'\left(b_N(r-1-a_N)\right)&\;,\;\;|r-1-a_N|\sim b_N^{-1}\;,\\
&\\
\displaystyle e^{-N\Psi_{+}^{\rm Gin}(r) }&\;,\;\;r\geq 1\;.
\end{cases}
\ee
The expression of the right rate function is \cite{cunden2016large}
\be
\Psi_{+}^{\rm Gin}(r)=r^2-1-2\ln r\;,\;\;r\geq 1\;,
\ee
while the left rate function reads \cite{cunden2016large}
\be
\Psi_{-}^{\rm Gin}(r)=\frac{1}{4}(4r^2-3-r^4)-\ln r\;,\;\;r\leq 1\;.\label{r_max_Gin_LD}
\ee
The right large deviation function $\Psi_{+}^{\rm Gin}(r)\approx 2(r-1)^2$ is quadratic around its minimum at $r=1$. Inserting $r=1+a_N+s/b_N$, this yields
\be
N\Psi_{+}^{\rm Gin}\left(r=1+a_N+\frac{s}{b_N}\right)=2N\left(a_N+\frac{s}{b_N}\right)^2\approx\frac{1}{2}\ln N+s+O(\ln N)^{-1}\;,
\ee
where we used that $a_N\sim \sqrt{\ln N/4N}$ and $b_N\sim \sqrt{4N\ln N}$. This behaviour allows a smooth matching with the right exponential tail of the Gumbel distribution $-\ln G_{\rm I}'(s)\approx s$ for $s\to +\infty$. On the contrary, the behaviour of $\Psi_{-}^{\rm Gin}(r)\approx (4/3)(1-r)^3$ is cubic around its minimum. This behaviour cannot match with the super-exponential tail of the Gumbel distribution $-\ln G_{\rm I}'(s)\approx e^{-s}$ for $s\to -\infty$. This problem of matching is an indication that there may exist an {\it intermediate deviation regime} to the left of the Gumbel distribution, which we did not take into account in this first analysis. Note that these results on atypical fluctuations were extended to a more general potential \cite{cunden2017universality}, where in particular, the cubic behaviour of the left large deviation around its minimum was shown to be universal.

\begin{figure}
\centering
\includegraphics[width=0.6\textwidth]{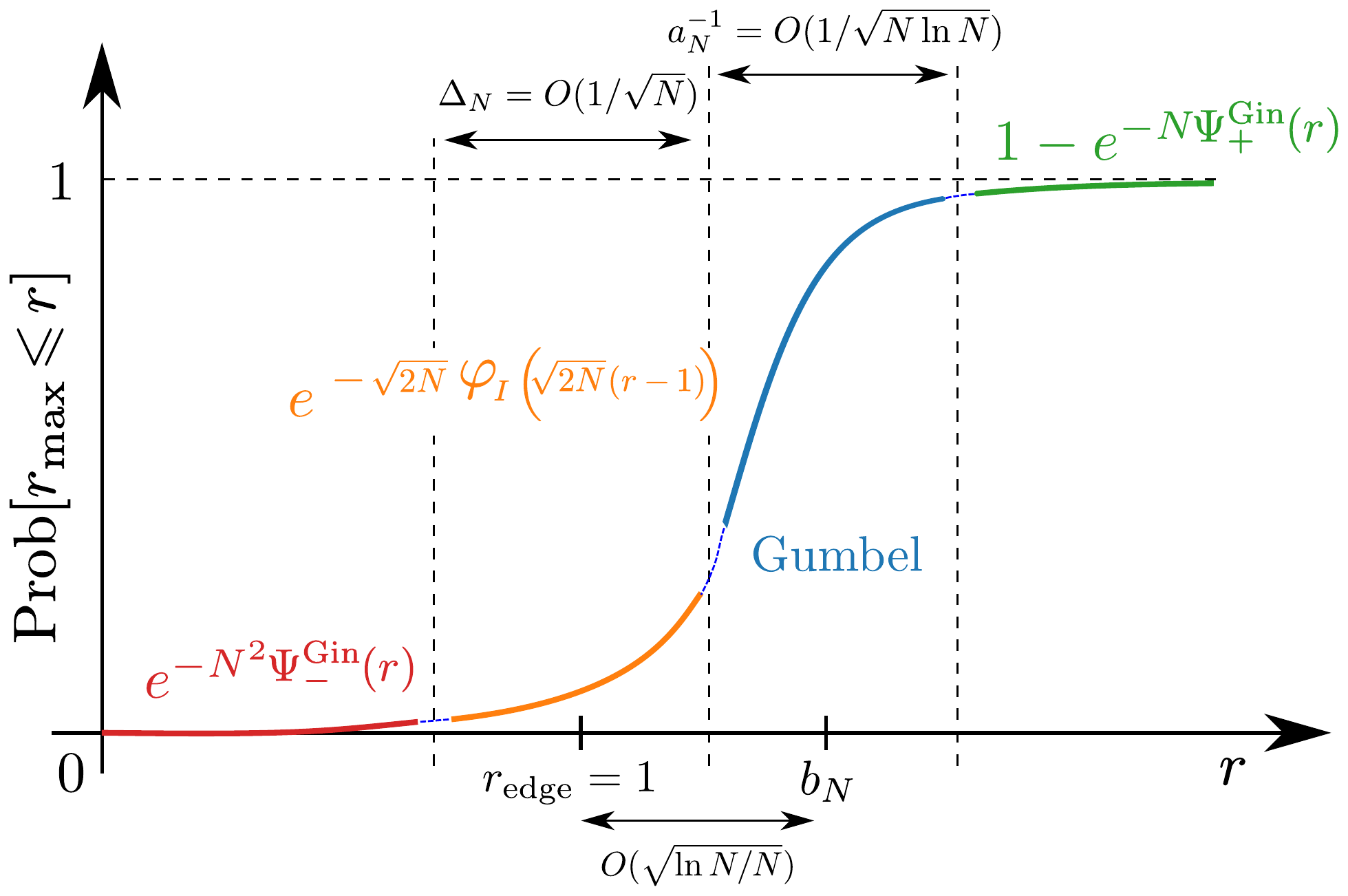}
\caption{Sketch of the typical Gumbel regime (blue), large deviation regime to the right (green), intermediate regime (orange) and large deviation regime to the left (red) for the CDF $\Prob[r_{\max}\leq r]$ of the radius $r_{\max}$ of the eigenvalue of largest modulus in the complex Ginibre ensemble.}\label{P_r_max_gin}
\end{figure}

To solve this puzzle of matching, we first use that the CDF $\Prob\left[r_{\max}\leq r\right]$ is the probability that there are no particles outside the disk of radius $r$,
\be
\Prob\left[r_{\max}\leq r\right]=\Prob\left[\overline{N_r}=0\right]\;.
\ee
This probability can be obtained by taking the limit $\mu\to \infty$ of Eq. \eqref{chi_bar_r}, which yields
\be
\Prob\left[r_{\max}\leq r\right]=\lim_{\mu \to\infty}\moy{e^{-\mu \overline{N_r}}}=\lim_{\mu \to\infty}\exp\left(\overline{ \chi_r}(\mu)\right)\;.
\ee
In the scaling regime close to the edge $s=\sqrt{2N}(r-1)=O(1)$, one can then use the scaling form for $\overline{ \chi_r}(\mu)$ given in Eq. \eqref{CGF_edge_2docp}. Taking the limit $\mu\to \infty$ in the scaling function $\Xi(\mu,s)$ given in Eq. \eqref{Xi_edge}, one finally obtains the intermediate deviation regime
\be
\Prob\left[r_{\max}\leq r\right]\approx \exp\left(-\sqrt{2N}\varphi_{I}\left(\sqrt{2N}(r-1)\right)\right)\;,\;\;|r-1|=O(N^{-1/2})\;,\label{scal_int_Gin}
\ee
where the intermediate deviation rate function $\varphi_{I}(s)$ reads
\be
\boxed{\varphi_{I}(s)=-\int_{s}^{\infty}dv \ln\frac{1}{2}\erfc(-v)\;.}\label{ind_dev_Gin}
\ee
The asymptotic behaviours of the function $\varphi_I(s)$ read
\be\label{int_dev_Gin_as}
\varphi_I(s)\approx\begin{cases}
\displaystyle  \frac{|s|^3}{3}&\;,\;\;s\to -\infty\;,\\
&\\
\displaystyle \frac{e^{-s^2}}{4\sqrt{\pi}s^2}&\;,\;\;s\to +\infty\;.
\end{cases}
\ee
Inserting $s=\sqrt{2N}(r-1)$ in $\varphi_I(s)$ and using the behaviour for $s\to -\infty$ in the first line of Eq. \eqref{int_dev_Gin_as}, one obtains
\be
\sqrt{2N}\varphi_I\left(\sqrt{2N}(r-1)\right)\approx \frac{4N^2}{3}|r-1|^3\;,\;\;(1-r)\gg N^{-1/2}\;,
\ee
which matches smoothly with the behaviour for $r\to 1$ of the left large deviation rate function $\Psi_{-}^{\rm Gin}(r)\approx \frac{4}{3}|r-1|^3$ given in Eq. \eqref{r_max_Gin_LD}.
Inserting  $s=\sqrt{2N}(a_N+u/b_N)$ in $\varphi_I(s)$, with $a_N\sim \sqrt{(\ln N-2\ln\ln N-\ln (2\pi))/(4N)}$ and $b_N\sim \sqrt{4N\ln N}$ in $\varphi_I(s)$ and using the behaviour for $s\to +\infty$ in the first line of Eq. \eqref{int_dev_Gin_as}, one obtains
\be
\sqrt{2N}\varphi_I\left(\sqrt{\frac{\ln N}{2}}+\frac{u}{\sqrt{2\ln N}}\right)\approx \frac{\sqrt{N}}{\sqrt{2\pi}\ln N}\exp\left(-\ln\left(\frac{\sqrt{N}}{\sqrt{2\pi}\ln N}\right)-u\right)=e^{-u}\;,
\ee
which matches smoothly with the Gumbel distribution $G_{I}(b_N(1-a_N-r))$ characterising the typical fluctuations. The complete description of the fluctuations of $r_{\max}$ in the complex Ginibre ensemble is therefore summarised as (see also Fig. \ref{P_r_max_gin})
\be\label{summary_Gin_fluc_rmax}
\boxed{
\partial_r \Prob\left[r_{\max}\leq r\right]=\begin{cases}
\displaystyle e^{-N^{2}\Psi_{-}^{\rm Gin}(r) }&\;,\;\;r\leq 1\;,\\
&\\
\displaystyle e^{-\sqrt{2N}\varphi_I\left(\sqrt{2N}(r-1)\right)}&\;,\;\;|r-1|\sim N^{-1/2}\;,\\
&\\
\displaystyle b_N G'\left(b_N(r-1-a_N)\right)&\;,\;\;|r-1-a_N|\sim b_N^{-1}\;,\\
&\\
\displaystyle e^{-N\Psi_{+}^{\rm Gin}(r) }&\;,\;\;r\geq 1\;.
\end{cases}}
\ee

We close this discussion on the fluctuations of $r_{\max}$ by two remarks. Form the universal result in Eq. \eqref{CGF_edge_2docp}, the scaling form for the regime of intermediate deviation in Eq. \eqref{scal_int_Gin} extends to an arbitrary potential $v(r)\gg 2\ln r$ for $r\to \infty$ 
\be
\Prob\left[r_{\max}\leq r\right]\approx \exp\left(-\sqrt{2\pi N\rho_{\rm e}(r_{\rm e})}\, r_{\rm e}\,\varphi_{I}\left(\sqrt{2\pi N\rho_{\rm eq}(r_{\rm e})}(r-r_{\rm e})\right)\right)\;,\;\;\sqrt{N}|r-r_{\rm e}|=O(1)\;.
\ee
(see Article \ref{Art:r_max} for potentials $v(r)\sim 2\ln r$ as $r\to \infty$ and potential with a finite hard edge such that $v(r>1)=+\infty$).

Furthermore, in the case of the real symplectic Ginibre ensemble $\beta=4$ -- whose joint PDF of eigenvalues is {\it not} obtained by insrting $\beta=4$ and $v(r)=r^2$ in Eq. \eqref{general_OCP} but has a more complicated structure \cite{ginibre1965statistical} -- the process of the complex positions $z_k$ is not determinantal but there exists a formula similar to Eq. \eqref{r_max_Gin_prod} for the probability of $r_{\max}$, which reads \cite{mehta2004random}
\be
\Prob\left[r_{\max}^{{\rm Gin},\beta=4}\leq r\right]=\prod_{k=1}^N \frac{\gamma(2k,2N r^2)}{\Gamma(2k)}\;.
\ee
The typical fluctuations of $r_{\max}$ are again characterised by a Gumbel distribution \cite{rider2004order}. One can then show, using the approach developed in this chapter that there exists also in this case an intermediate regime
\be
\boxed{\frame{\boxed{\Prob\left[r_{\max}^{{\rm Gin},\beta=4}\leq r\right]\approx \exp\left(-\sqrt{N}\varphi_{I}\left(2\sqrt{N}(r-1)\right)\right)\;,\;\;\sqrt{N}|r-1|=O(1)\;,}}}
\ee
whith the same scaling function $\varphi_{I}(s)$, given in Eq. \eqref{ind_dev_Gin}. 

We have seen in this section how our results on the fluctuations on the number of particles inside a disk can be used to compute explicitly the extreme value statistics for a gas of charged particles. We will now briefly summarise the main results obtained in this chapter.

\section{Summary of the results for fermions in rotation}

For better clarity, we first state the results for the model of non-interacting fermions and then for the two-dimensional one component plasma ($2d$ OCP).

\subsection{Results for fermions in rotation}

For a Fermi gas in a harmonic confining potential rotating at speed $\Omega$ satisfying the condition \eqref{cond_omega}, we have shown an exact mapping of the ground state many-body probability in Eq. \eqref{Gin_fermions} to the complex Ginibre ensemble. We have obtained exact formulae for any finite value of $N$ for (i) the bipartite entanglement entropy of the disk ${\cal D}_r=\{|z|\leq r\}$ \eqref{S_q_N} and (ii) the cumulants of arbitrary order $p$ of the number of particles $N_r$ inside this disk \eqref{cumul_1}. In the large $N$ limit, we have shown that entanglement entropy and cumulants are proportional to each other in the extended bulk, which should therefore allow a direct measurement of the entropy. This relation however breaks down at the edge of the density. Finally, we have also shown that the fluctuations of $N_r$ have a Gaussian typical regime, an intermediate and large deviation regime as summarised in Eq. \eqref{summary_N_r_eb} and Fig. \ref{Fig_P_nr}. This intermediate deviation regime allows to solve a puzzle of matching between typical and large deviation for the complex Ginibre ensemble.

\subsection{Results for the $2d$ OCP}

We first have shown that the problem of fermions considered in this chapter can also be mapped exactly onto the two-dimensional one component plasma for a confining potential $v(r)=r^2$ and at inverse temperature $\beta=2$ (values for which it is exactly solvable). We then have shown that the scaling function $\chi(\mu)$ given in Eqs. \eqref{chi_mu} of the centred cumulant generating function obtained in the case of fermions (i.e. $v(r)=r^2$) is universal with respect to the confining potential  \eqref{CGF_bulk_2docp} (for $v(r)\gg 2\ln r$ as $r\to \infty$). We realised that this universal scaling function actually extends beyond the full counting statistics and holds for any linear statistics ${\cal L}_r=\sum_{i=1}^N L(|z_i|)\Theta(|z_i|-r)$ within the disk of radius $r$ \eqref{CGF_bulk_lin_2docp}. A similar relation holds at the edge of the gas \eqref{CGF_bulk_2docp} where $\Xi(\mu,s)$ is given in Eq. \eqref{Xi_edge}. As an application of these edge results, we have obtained the regime of intermediate fluctuation of $r_{\max}$ in Eq. \eqref{ind_dev_Gin}, thus solving another puzzle of matching between typical and large deviations for the complex Ginibre ensemble (see summary of the four regime in \eqref{summary_Gin_fluc_rmax} and Fig. \ref{P_r_max_gin}).

\AddArticle{Art:rot}{Rotating trapped fermions in two dimensions and the complex Ginibre ensemble: Exact results for the entanglement entropy and number variance}

 \begin{center}
   {\large \textbf{Rotating trapped fermions in two dimensions and the complex Ginibre ensemble: Exact results for the entanglement entropy and number variance}}
 \end{center}


 \vspace{2cm}

 \noindent B. Lacroix-A-Chez-Toine, S.~N. Majumdar, G. Schehr,\\
 Phys. Rev. A {\bf 99} (2), 021602 (2019).\\

 \ding{43}
 \href{https://journals.aps.org/pra/abstract/10.1103/PhysRevA.99.021602}{https://journals.aps.org/pra/abstract/10.1103/PhysRevA.99.021602}

 \ding{43}
 \href{https://arxiv.org/abs/1809.05835}{https://arxiv.org/abs/1809.05835}

\begin{abstract}
We establish an exact mapping between the positions of $N$ noninteracting fermions in a $2d$ rotating harmonic trap in its ground-state 
and the eigenvalues of the $N \times N$ complex Ginibre ensemble of Random Matrix Theory (RMT). Using RMT techniques, we make precise 
predictions for the statistics of the positions of the fermions, both in the bulk as well as at the edge of the trapped Fermi gas. In addition, we
compute exactly, for any finite $N$, the R\'enyi entanglement entropy and the number variance of a disk of radius $r$ in the ground-state. 
We show that while these two quantities are proportional to each other in the (extended) bulk, this is no longer the case very close to the trap center nor 
at the edge. Near the edge, and for large $N$, we provide exact expressions for the scaling functions associated with these two observables.  
\end{abstract}

\AddArticle{Art:r_max}{Extremes of $2d$ Coulomb gas: universal intermediate deviation regime}

 \begin{center}
   {\large \textbf{Extremes of 2d Coulomb gas: universal intermediate deviation regime}}
 \end{center}


 \vspace{2cm}

 \noindent B. Lacroix-A-Chez-Toine, A. Grabsch, S.~N. Majumdar, G. Schehr,\\
 J. Stat. Mech {\bf 1}, 013203 (2018).\\

 \ding{43}
 \href{https://doi.org/10.1088/1742-5468/aa9bb2}{https://doi.org/10.1088/1742-5468/aa9bb2}

 \ding{43}
 \href{https://arxiv.org/abs/1710.06222}{https://arxiv.org/abs/1710.06222}

\begin{abstract}
In this paper, we study the extreme statistics in the complex Ginibre ensemble of $N \times N$ 
random matrices with complex Gaussian entries, but with no other symmetries. All the $N$ eigenvalues are
complex random variables and their joint distribution can be interpreted as a $2d$ Coulomb gas with a logarithmic
repulsion between any pair of particles and in presence of a confining harmonic potential $v(r) \propto r^2$. We study the statistics of the eigenvalue with the largest modulus $r_{\max}$
 in the complex plane. The typical and large fluctuations of $r_{\max}$ around its mean had been studied 
 before, and they match smoothly to the right of the mean. However, it remained a puzzle to understand why the large and typical fluctuations to the left of the mean did not match. In this paper, we show that there is indeed an intermediate fluctuation regime that interpolates smoothly between the large and the typical fluctuations to the left of the mean. Moreover, we compute explicitly this ``intermediate deviation function'' (IDF) and show that it is universal, i.e. independent of the confining potential $v(r)$ as long as it is spherically symmetric and increases faster than $\ln r^2$ for large $r$ with an unbounded support. If the confining potential $v(r)$ has a finite support, i.e. becomes infinite beyond a finite radius, we show via explicit computation that the corresponding IDF is different. Interestingly, in the borderline case where the confining potential grows very slowly as $v(r) \sim \ln r^2$ for $r \gg 1$ with an unbounded support, the intermediate regime disappears and there is a smooth matching between the central part and the left large deviation regime. 
\end{abstract}


\AddArticle{Art:FCSGin}{Intermediate deviation regime for the full eigenvalue statistics in the complex Ginibre ensemble}

 \begin{center}
   {\large \textbf{Intermediate deviation regime for the full eigenvalue statistics in the complex Ginibre ensemble}}
 \end{center}


 \vspace{2cm}

 \noindent B. Lacroix-A-Chez-Toine, J.~A.  Monroy Garz\'on, C.~S. Hidalgo Calva,\\
 I. P\'erez Castillo, A.  Kundu, S.~N. Majumdar, G. Schehr\\
arXiv preprint, arXiv: \textbf{1904.01813}, (2019).

 \ding{43}
 \href{https://arxiv.org/abs/1904.01813}{https://arxiv.org/abs/1904.01813}

\begin{abstract}
We study the Ginibre ensemble of $N \times N$  complex  random matrices and compute exactly, for any finite $N$, the full distribution as well as all the cumulants of the number $N_r$ of eigenvalues within a disk of radius $r$ centered at the origin. In the limit of large $N$, when the average density of eigenvalues becomes uniform over the unit disk, we show that for $0<r<1$ the fluctuations of $N_r$  around its mean value $\langle N_r \rangle \approx N r^2$ display three different regimes: (i) a typical Gaussian regime where the fluctuations are of order ${\cal O}(N^{1/4})$, (ii) an intermediate regime where $N_r - \langle N_r \rangle = {\cal O}(\sqrt{N})$, and (iii) a large deviation regime where $N_r - \langle N_r \rangle = {\cal O}({N})$. This intermediate behaviour (ii) had been overlooked in previous studies and we show here that it ensures a smooth matching between the typical  and the large deviation regimes. In addition, we demonstrate that this intermediate regime controls all the (centred) cumulants of $N_r$, which are all of order ${\cal O}(\sqrt{N})$, and we compute them explicitly. Our analytical results are corroborated by precise ``importance sampling'' Monte Carlo simulations.  
\end{abstract}


\part{Statistics of the gaps of random walks}
\label{Part:gaps}


\chapter{Order and gap statistics of i.i.d. random variables}\label{intro_iid}

In this part of the thesis, we study the {\it fluctuation theory} for sets of correlated random variables \cite{feller1968introduction, wendel1960order, port1962elementary}. These problems are not restricted to the value of the global maximum or minimum but extend to the order statistics and their associated times \cite{wendel1960order, port1962elementary, feller1968introduction, schehr2012universal, schehr2014exact, dassios1996sample, 10.2307/2959757, texier2000individual, hagendorf2008breaking}, the gap statistics \cite{Brunet_2009, Derrida2011,  ramola, battilana2017gap,schehr2012universal} or the statistics of records and their ages \cite{majumdar_ziff, PhysRevE.86.011119, Sabhapandit_2011, majumdar2012record, schehr2014exact, berkowitz_records_noise, wergen_borgner_krug, Godr_che_2017, Godr_che_2014}. While the fluctuations for independent and identically distributed (i.i.d.)~random variables have been well characterised, there is still a lot of activity for correlated sets of random variables \cite{pitman2018guide, revuz2013continuous}.

We start the second part of the thesis by recalling a few well-known results concerning the extreme value statistics of a set $\{x_1,x_2,\cdots,x_n\}$ of independent and identically distributed (i.i.d.)~random variables. To set the notations, these random variables $x_i$'s are drawn from the same probability distribution function (PDF) $p(x)$. We also define their cumulative distribution function (CDF) $q(x)=\int_{-\infty}^{x}p(x')dx'$ together with its complement $\bar q(x)=1-q(x)$. The joint PDF of these independent random variables is just the product of the individual probabilities
\be
P_{\rm joint}({\bf x})=P_{\rm joint}(x_1,\cdots,x_n)=\prod_{i=1}^{n}p(x_i)\;.\label{PDF_iid}
\ee


To characterise the extremes of this set, the most natural observable is simply the value of the maximum
\be\label{x_max}
x_{\max}=\max_{1\leq i\leq n} x_i\;.
\ee
We define its CDF as $Q_{n}(x)=\Prob\left[x_{\max}\leq x\right]$. The event ``$x_{\max}\leq x$'' is obviously equivalent to ``all $x_i\leq x$'', and therefore reads
\be
\Prob\left[x_{\max}\leq x\right]=\Prob\left[x_1\leq x,x_2\leq x,\cdots, x_N\leq x\right]=\int_{-\infty}^x d x_1\cdots \int_{-\infty}^x dx_N P_{\rm joint}({\bf x})\;.
\ee
In the case of i.i.d. random variables, the CDF of the maximum can be obtained exactly using Eq. \eqref{PDF_iid},
\be
Q_n(x)=\prod_{i=1}^N \int_{-\infty}^{x}p(x_i)dx_i=\left[q(x)\right]^n\;.\label{glob_max}
\ee
In the large $n$ limit, it is well-known that there are three universal classes, with corresponding scaling forms that only depends on the tail of the PDF $p(x)$ \cite{gumbel2012statistics}. 

\begin{figure}
\centering
\includegraphics[width=0.6\textwidth]{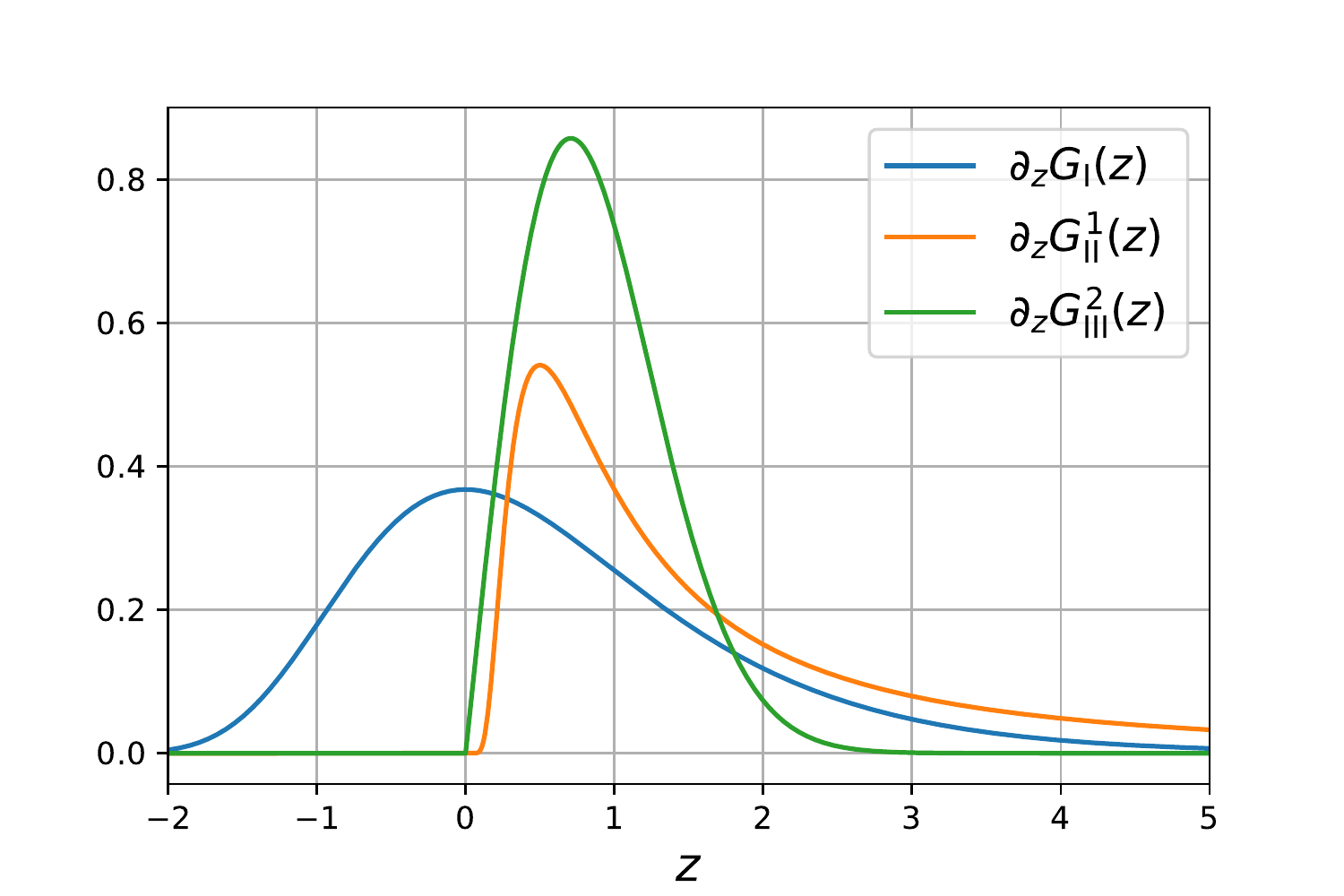}
\caption{Plot of the three universal PDF $G_a'(z)$ of the maximum corresponding to the Gumbel $a={\rm I}$ (blue), Fr\'echet $a={\rm II}$ (orange) and Weibull $a={\rm III}$ (green) universality classes.}\label{Fig_IID_max}
\end{figure}

\begin{itemize}
\item The Gumbel universality classes for PDF $p(x)$ such that $\forall \alpha\in \mathbb{R}$, $p(x)\ll x^{-\alpha}$ as $x\to \infty$. For this PDF
\be
\exists\; (a_n,b_n)\in \mathbb{R}^2\;,\;\;\lim_{n\to \infty} Q_n\left(a_n+b_n z\right)=G_{\rm I}(z)=\exp(-e^{-z})\;.
\ee
To obtain a probability of order $O(1)$ in the large $n$ limit,  one needs to ensure that there is $O(1)$ variable in the interval $[s,\infty)$, i.e. \cite{schehr2014exact}
\be
{\bar q}(a_n)=1-q(a_n)=n^{-1}.
\ee
On the other hand, the scale $b_n$ can be obtained as \cite{schehr2014exact}
\be
b_n=\frac{\int_{a_n}^{\infty}(x-a_n)p(x)dx}{\bar{q}(a_n)}\;,
\ee
which can be interpreted as the typical distance between $x$ and $a_n$ conditioned on $x>a_n$.
The PDF $G_{\rm I}'(z)$ associated to this class is plotted in blue in Fig. \ref{Fig_IID_max}.

\item The Fr\'echet universality class for PDF $p(x)$ such that $p(x)\sim x^{-\alpha-1}$ as $x\to \infty$ for $\alpha>0$. For this PDF
\be
\exists\;(a_n,b_n)\in \mathbb{R}^2\;,\;\;\lim_{n\to \infty} Q_n\left(a_n+b_n z\right)=G_{\rm II}^{\alpha}(z)=\Theta(z)\exp(-z^{-\alpha})\;.
\ee
For the Fr\'echet universality class, the coefficient $a_N=0$ is zero and the coefficient $b_N$ is evaluated by ensuring that there is $O(1)$ variables in the interval $[b_N, \infty)$, i.e. \cite{schehr2014exact}
\be
\bar{q}(b_N)=\int_{b_n}^{\infty}p(x)dx\sim \frac{1}{n} \Rightarrow b_n\sim n^{-\alpha}\;.
\ee
This universal PDF ${G_{\rm II}^{\alpha}}'(z)$ is plotted in orange in Fig. \ref{Fig_IID_max}.

\item The Weibull universality class for PDF $p(x)$ that have a finite edge in $x^*$ where $p(x)\sim (x^*-x)^{\alpha-1}$ with $\alpha>0$. For this PDF
\be
\exists\;(a_n,b_n)\in \mathbb{R}^2\;,\;\;\lim_{n\to \infty} Q_n\left(a_n+b_n z\right)= G_{\rm III}^{\alpha}(z)=\Theta(z)\exp(-z^{\alpha})\;.
\ee
For the Weibull universality class, the coefficient $a_N=x^*$ while the coefficient $b_N$ can be evaluated by ensuring that there is $O(1)$ variables in the interval $[x^*-b_n,b_n]$, i.e. \cite{schehr2014exact}
\be
\bar{q}(b_n)=\int_{x^{*}-b_n}^{b_n} p(x)dx \sim \frac{1}{n} \Rightarrow b_n\sim n^{-\alpha}\;.
\ee
This universal PDF ${G_{\rm III}^{\alpha}}'(z)$ is plotted in green in Fig. \ref{Fig_IID_max}.

\end{itemize}
These are the only three universality classes associated to the maximum for i.i.d.~random variables \cite{gumbel2012statistics}. 
These universality classes extend beyond the case of i.i.d.~random variables. For example, if one considers a set of weakly correlated random variables $\{x_1,x_2,\cdots,x_n\}$ with exponentially decreasing correlations,
\be
\moy{(x_i-\moy{x_i})(x_j-\moy{x_j})}\sim e^{-|i-j|/\zeta}\;,
\ee
one can define ``blocks'' of order $O(\zeta)$ random variables and for these block variables, the problem reduces to statistics of $O(N/\zeta)$ i.i.d.~random variables \cite{majumdar2014extreme}. If the random variables are independent but not identically distributed, the problem often reduces to the case of i.i.d.~random variables as seen in section \ref{r_max_Gin} for the Ginibre ensemble and section \ref{stat_zero_t} for fermions (but this is not always the case, c.f. Article \ref{Art:r_max}).

In many physical systems, it is important to characterise not only the maximum but also the ``near extreme events'', i.e. the events that happen in a close vicinity of the maximum/minimum \cite{perret2013near, sabhapandit2007density}. It is important for example in optimisation problems where one needs to know whether the maximum is isolated and can be clearly identified or not. We now extend the question of the maximum to the order statistics of i.i.d. random variables.

\section{Order statistics and occupation number for i.i.d. random variables}\label{sec_order_iid}

We define the order statistics by ordering the set of variables $x_i$'s as
\be
M_1=x_{\max}\geq M_2\geq \cdots\geq M_n=x_{\min}\;.
\ee
As we will see later on, it is useful to relate this order statistics to the occupation number, defined as
\be
N_+(x)=\sum_{i=1}^n \Theta(x_i-x)\;.\label{occ_num_iid}
\ee
Indeed, if $M_k$ is the $k^{\rm th}$ maximum, then there are exactly $k$ values in the set of $x_i$'s that are bigger than $M_k$. This leads to the useful identity\be\label{M_k_num_occ}
\Prob\left[M_{k}< x\right]=\Prob\left[N_+(x)< k\right]=\sum_{i=0}^{k-1} P_i(x)\;,
\ee
where $P_k(x)=\Prob\left[N_+(x)= k\right]$. This identity is valid for any set of random variables and not restricted to the i.i.d. case. In the latter, the probability $P_k(x)$ can be obtained quite simply. One just has to pick any $k$ random variables that will be larger than $x$, the remaining $n-k$ being smaller than $x$. It reads
\be\label{PDF_occ_num}
P_k(x)={{n}\choose{k}}{\bar q}(x)^k q(x)^{n-k}\;,
\ee
where the combinatorial factor ${{n}\choose{k}}$ counts the way to pick $k$ elements among $n$.
Note that if we set $k=1$, we recover that $\Prob\left[x_{\max}\leq x\right]=q(x)^n$ as in Eq. \eqref{glob_max}. 
For a general value of $k$, the CDF of the $k^{\rm th}$ maximum, resp. occupation number, reads
\be
\Prob\left[M_{k}< x\right]=\Prob\left[N_+(x)< k\right]=\sum_{j=0}^{k-1}{{n}\choose{j}}{\bar q}(x)^j q(x)^{n-j}\;.
\ee
A simple way to evaluate the typical scale of $M_{k}$ is to use the relation with the occupation number as 
\be
\bar{q}(\moy{M_k})=\int_{\moy{M_k}}^{\infty}p(x)dx\sim \frac{k}{N}\;.\label{M_k_typ}
\ee
This relation naturally distinguishes the maxima close to the global maximum, i.e. for $k=O(1)$ for which $\moy{M_k}\sim \moy{M_n}\sim a_n$ grows with $n$, or further in the ``bulk'', i.e. for $\alpha=k/n$ fixed, where $\moy{M_k}=O(1)$. Let us first consider the maxima $M_k$ for $k=O(1)$.

\subsection{Order statistics close to the global maximum}

\begin{figure}
\centering
\includegraphics[width=0.6\textwidth]{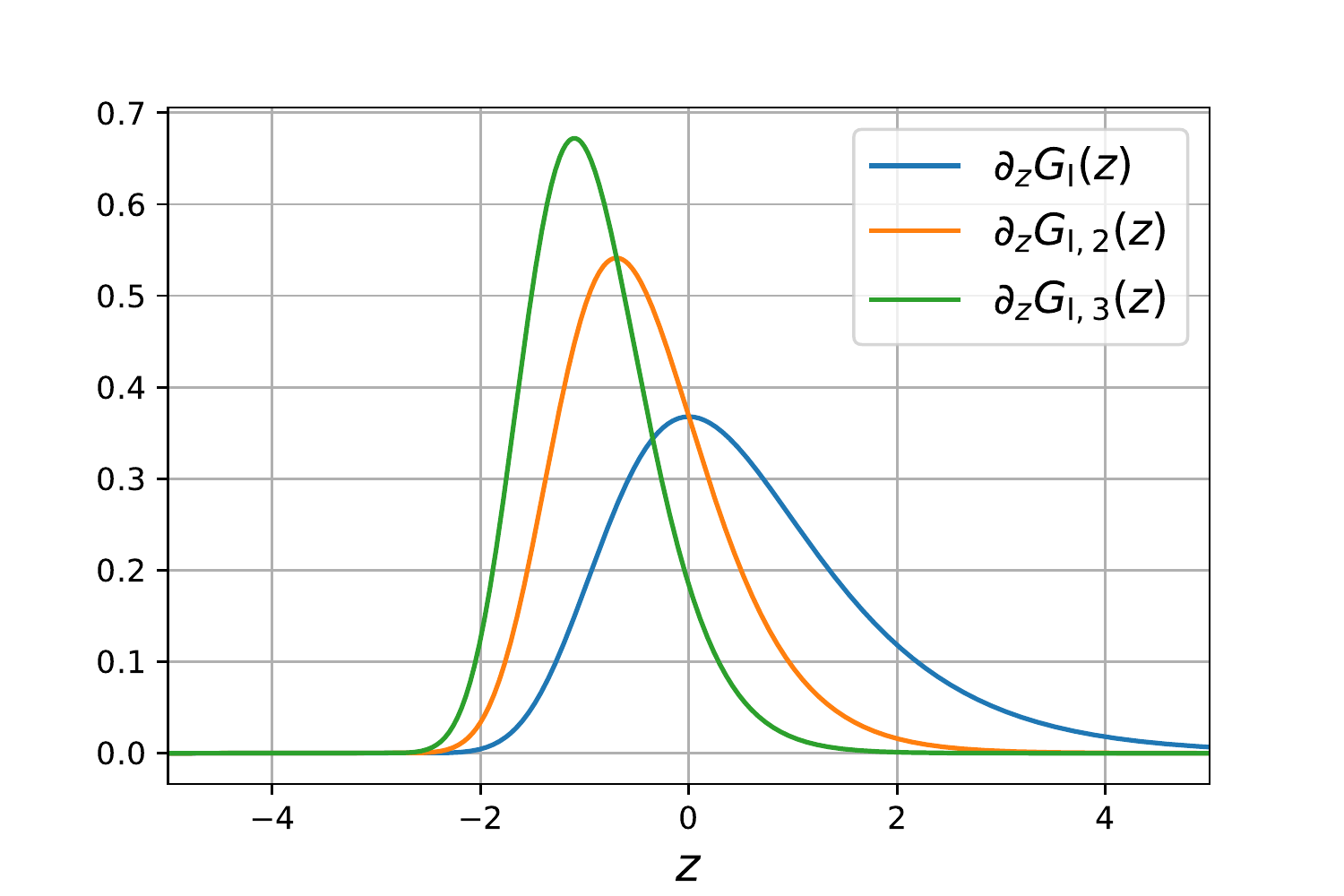}
\caption{Plot of the universal PDF $G_{{\rm I},k}'(z)$ of the $k^{\rm th}$ maximum corresponding to the Gumbel class for $k=1,2,3$ respectively in blue, orange and green.}\label{Fig_IID_kmax}
\end{figure}

Close to the global maximum, imposing the rescaling $x=a_n+b_n z$, (where the coefficients $a_n$ and $b_n$ are the same as for the global maximum), such that $n\bar q(x=a_n+b_n z)=O(1)$, and taking the large $n$ limit it can be shown that
\be
\lim_{n\to \infty}\Prob\left[M_{k}< a_n+b_n z\right]=G_{a,k}(z)=G_a(z)\sum_{i=0}^{k-1}\frac{[-\ln G_a(z)]^j}{j!}=\frac{\Gamma(k,-\ln G_a(z))}{\Gamma(k)}\;,
\ee
where $\Gamma(a,z)=\int_z^{\infty} t^{a-1}e^{-t}$ is the lower incomplete gamma function. The order statistics for $k=O(1)$ fall into the same three universal classes as for the value of the maximum, \cite{feller1968introduction,schehr2014exact}. For $k=1$, as $\Gamma(1,z)=e^{-z}$, one recovers the result for the global maximum. The PDF $\partial_z G_{{\rm I},k}(z)$, corresponding to the Gumbel class, are plotted in Fig. \ref{Fig_IID_kmax}.

 We will now consider the order statistics far from the global maximum, i.e. in the ``bulk'' of maxima where $k/n=O(1)$.

\subsection{Order statistics in the bulk}

Deep in the bulk, for $\alpha=k/n=O(1)$, we will now see that the statistics become even more universal: the results become identical for all classes of universality, i.e. Gumbel, Weibull and Fr\'echet universality classes \cite{feller1968introduction}. In the regime of large $n$ with fixed $\alpha=k/n$, one can use that the occupation number $N_+(x)$ defined in Eq. \eqref{occ_num_iid} is the sum of i.i.d. random variables
\be
N_+(x)=\sum_{i=1}^N n_i(x)\;,\;\;{\rm with}\;\;n_i(x)=\Theta(x_i-x)\;.
\ee
The mean value of $n_i(x)$ is simply $\moy{n_i(x)}=\bar{q}(x)=\int_x^{\infty}p(x')dx'$ and its variance is
\be
\var{n_i}=\moy{n_i^2(x)}-\moy{n_i(x)}^2=\moy{n_i(x)}(1-\moy{n_i(x)})=\bar{q}(x)q(x)\;,
\ee
where we used that $n_i(x)^2=n_i(x)$ and $q(x)=1-\bar{q}(x)$. This variance is finite as the variables $n_i=0,1$ are finite and it is identical for all the i.i.d. random variables $n_i$'s. In the regime of large $n$ and for $x=O(1)$, one therefore obtains from the central limit theorem that $N_+(x)$ has a Gaussian distribution
\be
\Prob\left[N_+(x)<k\right]\approx \frac{1}{2}\erfc\left(\sqrt{\frac{n}{2q(x)\bar{q}(x)}}(\bar{q}(x)-k)\right)\;.
\ee
Using Eq. \eqref{M_k_num_occ}, it is then simple to prove that in the regime $\alpha=k/n=O(1)$, the CDF of $M_k$ reads
%
%
%
\cite{feller1968introduction}
\be
\lim_{n,k\to\infty}\Prob\left[M_{k}\leq \xi_\alpha+\frac{z}{\sqrt{n}}\right]=\frac{1}{2}\erfc\left(\frac{p(\xi_\alpha)z}{\sqrt{2\pi\alpha(1-\alpha)}}\right)\;,
\ee
where $\xi_{\alpha}$ is the mean value of $M_{k}$. From Eq. \eqref{M_k_typ}, one obtains that ${\bar q}(\xi_\alpha)=\alpha$. Note that in this ``bulk'' regime the results are even more universal as here there is a single universality class with normal distribution, valid for any PDF $p(x)$.

To describe the ``near extreme statistics'', and in particular the crowding of these maxima it is important not only to obtain the statistics of the $k^{\rm th}$ maxima but also to consider the statistics of the gaps between them. In this vein, the description of the energy levels of disordered system by i.i.d. random variables has known a huge success since its first introduction by Derrida \cite{derrida1981random}. In particular, for this model, the maxima and gaps represent	 the (low-lying) excitations of the ground state and the corresponding spectral gaps. Note that one can simply connect this problem to the models of non-interacting fermions studied in the previous part of this thesis but in a disordered environment \cite{schawe2018ground}.

\section{Gap statistics of i.i.d. random variables}

We close this introduction on i.i.d. random variables by mentioning results on the gaps between consecutive maxima defined as
\be
d_k=M_k-M_{k+1}\;,\;\;k=1,\cdots,n-1\;.
\ee
The distribution of $d_{k}$ partly encodes the correlations between the maxima, which are not independent even for i.i.d. random variables as $M_k\geq M_{k+1}$.  
 For finite $n$, the CDF of these gaps can be obtained  quite simply as
\be
\Prob\left[d_k\leq \delta\right]=\frac{n!}{(k-1)!(n-k-1)!}\Theta(\delta)\int_{-\infty}^{\infty}dx\int_{x-\delta}^{x}dy p(x)p(y)\bar q(x)^{k-1}q(y)^{n-k-1}\;.\label{C_gap_iid_1}
\ee
Note that by changing variables from $x\to u=q(x)$ and $y\to v=\bar{q}(y)$, we obtain
\be
\Prob\left[d_k\leq \delta\right]=\frac{n!}{(k-1)!(n-k-1)!}\Theta(\delta) \int_{0}^{1} du\int_{u}^{\bar q(\xi_u-\delta)}dv (1-v)^{k-1}u^{n-k-1}\;,\label{C_gap_iid_2}
\ee
where ${\bar q}(\xi_u)=u$. As for the order statistics, the behaviour of this gap statistics will be quite different for gaps close to the global maximum, i.e. taking $n\to\infty$ with fixed $k$, or in the bulk, i.e. taking $n\to\infty$ with $\alpha=k/n$ fixed.

\subsection{Gap statistics close to the global maximum}

This regime was analysed in \cite{schehr2014exact}, where the authors found that the gap CDF falls into the same Gumbel, Fr\'echet and Weibull universality classes denoted respectively as $a={\rm I},\;{\rm II}\;,{\rm III}$. Rescaling the CDF by the typical scale of the PDF of the global maximum $b_n$, it reads
\be
\lim_{n\to \infty}\Prob\left[d_k\leq b_n\delta\right]=G_a^{\rm gap}(\delta)=\frac{\Theta(\delta)}{(k-1)!}\int_{-\infty}^{\infty} dx G'_a(x)\left[-\ln G_a(x)\right]^{k-1}\left[1-\frac{G_a(x-\delta)}{G_a(x)}\right]\;.
\ee
Note that in the case of the Gumbel universality class, the gap distribution is simply an exponential distribution
\be
\lim_{n\to \infty}\Prob\left[d_k\leq b_n\delta\right]=1-e^{-k\,\delta}\;.
\ee
We will now consider the bulk limit of these gap distribution.

\subsection{Gap statistics in the bulk}

In the regime of large $n$, keeping $\alpha=k/n$ fixed, we expect that the gaps will scale as $n d_{k}=O(1)$. Introducing this rescaling in the CDF, we obtain
\be
\Prob\left[d_k\leq \frac{\delta}{n}\right]\approx \sqrt{\frac{n\alpha(1-\alpha)}{2\pi}}\Theta(\delta)\int_{0}^{\delta}du \int_{-\infty}^{\infty}dx \frac{p(x)^2}{q(x)\bar q(x)}e^{n\left[\alpha \ln\left(\frac{\bar q(x)}{\alpha}\right)+(1-\alpha) \ln\left(\frac{q(x)}{1-\alpha}\right)\right]}e^{-\frac{\alpha p(x)}{{\bar q(x)}}u}\;.
\ee
The integral over $x$ can be evaluated using a saddle-point approximation, where the saddle is at $x=\xi_\alpha$ such that ${\bar q}(\xi_\alpha)=\alpha$. This yields the universal scaling form
\be
\lim_{n,k\to \infty}\Prob\left[d_k\leq \frac{\delta}{n}\right]\approx 1-e^{-p(\xi_\alpha)\,\delta}\;.
\ee
As for the maximum, in the bulk we recover that there is a single universality class with exponential distribution instead of the three classes close to the global maximum.

We have seen that in the case of independent and identically distributed random variables, many extreme value observables can be obtained. However, in physical systems, the degrees of freedom are often correlated. It is therefore essential to consider the effects of correlations on the extreme value statistics. This is unfortunately a very difficult task for an arbitrary correlated system and one needs to consider simple models that can be solved to progress. As a step in this direction, we consider in the following both the order and the gap statistics for a strongly correlated system: the positions of a random walker. This ubiquitous model has been extensively studied as we will see in the next section and constitutes an ideal toy model for testing the effects of correlations.

\chapter{Introduction to the extreme value statistics of random walks and Brownian motion}
\label{ch:gaps_intro}
In this chapter we will consider one-dimensional diffusion processes, either continuous in time, i.e. Brownian motion, or discrete in time, i.e. random walks. Let us start with the former, which will turn out to be easier to study.

\section{Extreme value properties of the Brownian motion}
The Brownian motion $x(t)$, also called Wiener process $x(\tau)=W_{\tau}+x_0$, can be defined through the Langevin equation
\be\label{BM_def}
\frac{dx}{dt}=\xi(t)\;,\;\;{\rm with}\;\;x(0)=x_0
\ee
where $\xi(t)$ is a Gaussian white noise with $\moy{\xi(t)}=0$ and $\moy{\xi(t)\xi(t')}=2D\delta(t-t')$.
We fix now and for the rest of the manuscript the value of the diffusion constant to $D=1/2$. This process is Markovian and its propagator $G(x,t|x_0,t_0)=\partial_x\Prob\left[x(t)\leq x|x(t_0)=x_0\right]$ is solution of the Fokker-Planck equation
\be\label{FP_MP}
\partial_t G(x,t|x_0,t_0)=D\partial_x^2 G(x,t|x_0,t_0) \;,\;\;{\rm with}\;\;G(x,t_0|x_0,t_0)=\delta(x-x_0)\;.
\ee
This equation is well-known in physics as the free one-dimensional diffusion equation, with a solution that is 
translation invariant in time and space $G(x,t|x_0,t_0)=\Prob\left[x(t-t_0)=x-x_0|x(0)=0\right]=G(x-x_0,t-t_0)$, given by
\be\label{prop_BM}
G(z,t)=\frac{1}{\sqrt{2\pi t}}\exp\left(-\frac{z^2}{2 t}\right)\;.
\ee
This Markov propagator is symmetric in space $G(-z,t)=G(z,t)$. Adding a drift term $+v$ in the right hand side of Eq. \eqref{BM_def} would break this spatial symmetry and the propagator would be simply obtained as $G(z-v\, t,t)$.
We will now show how the Feynman-Kac formalism allows to compute general functionals of the Brownian motion.



\subsection{Feynman-Kac formalism}

We now consider a Brownian functional $O=\displaystyle\int_0^{t}d\tau {\cal O}[x(\tau)]$ that depends explicitly on the path realisation of the Brownian motion. To compute the statistics of this observable, we will integrate over all the possible paths of the Brownian with a Gaussian weight for each trajectory on the time interval $[0,t]$
\be
P[x(\tau)]\propto \exp\left(-\frac{1}{2}\int_0^{t}d\tau \left(\frac{dx}{d\tau}\right)^{2}\right)\;.
\ee
Therefore, the PDF of $O$ for the ensemble of Brownian trajectories starting from position $x(0)=x_0$ is obtained as
\be
P(O;x_0)=\frac{1}{Z(x_0)}\int_{-\infty}^{\infty}dx \int_{x(0)=x_0}^{x(t)=x} {\cal D}x(\tau)e^{- \frac{1}{2}\int_0^{t}d\tau \left(\frac{dx}{d\tau}\right)^{2}}\delta\left[O-\int_0^{t}d\tau {\cal O}[x(\tau)]\right]\;,
\ee
where $Z(x_0)$ plays the role of a partition function for the ensemble of trajectories
\be
Z(x_0)=\int_{-\infty}^{\infty}dx \int_{x(0)=x_0}^{x(t)=x} {\cal D}x(\tau)e^{- \frac{1}{2}\int_0^{t}d\tau\left(\frac{dx}{d\tau}\right)^{2}}\;.
\ee
For a positive random variable $O$, it will be convenient to consider the Moment Generating Function (MGF) $\mathbb{E}_{x_0}\left[e^{-\mu O}\right]$ of the observable $O$ instead of its PDF, where
\be
\mathbb{E}_{x_0}\left[e^{-\mu O}\right]=Q(\mu;x_0,t)=\int_{-\infty}^{\infty}G(\mu;x_0,x,t)dx\;.
\ee
Using the path integral formalism of quantum mechanics introduced by Feynman, the function $G(\mu;x_0,x,t)$ can be interpreted as a quantum Euclidean propagator between the positions $x_0$ at time $0$ and $x$ at time $t$,
\be
G(\mu;x_0,x,t)=\langle x|e^{-\hat{H}t}|x_0\rangle=\int_{x(0)=x_0}^{x(t)=x} \frac{{\cal D}x(\tau)}{Z(x_0)}e^{-\int_0^{t}d\tau \left[\frac{1}{2}\left(\frac{dx}{d\tau}\right)^{2}+\mu {\cal O}[x(\tau)]\right]}\;,
\ee
where the effective Hamiltonian of the quantum system reads
\be
\hat H=\displaystyle\frac{\hat{p}^2}{2}+\mu {\cal O}(\hat{x})\;.
\ee
This propagator is solution of the partial differential equation $-\partial_t G=\hat{H} G$, i.e.
\be\label{f_diff}
\partial_{t}G=\frac{1}{2}\partial_{x}^2 G-\mu {\cal O}(x)G\;,\;\;{\rm with}\;\;G(\mu;x_0,x,0)=\delta(x-x_0)\;.
\ee
In the absence of a potential ${\cal O}(x)=0$, we recover that the Markov propagator is the free Euclidean quantum propagator in Eq. \eqref{prop_BM}. Note that the Euclidean propagator was already introduced in a different context in section \ref{univ_prop_1d}.

To compute a probability integrated over the final position $x$, one can instead directly solve the backward diffusion equation \cite{majumdar2007brownian}
\be\label{b_diff}
\partial_{t}Q=\frac{1}{2}\partial_{x_0}^2 Q-\mu {\cal O}(x_0)Q\;,\;\;{\rm with}\;\;Q(\mu;x_0,t=0)=1\;.
\ee
Note that the difference between Eqs. \eqref{f_diff} and \eqref{b_diff} only lies in the initial condition. We will now apply this formalism to obtain extreme value observables of the Brownian motion.

\subsection{Survival probability and maximum of the Brownian motion}

\begin{figure}
\centering
\includegraphics[width=0.45\textwidth]{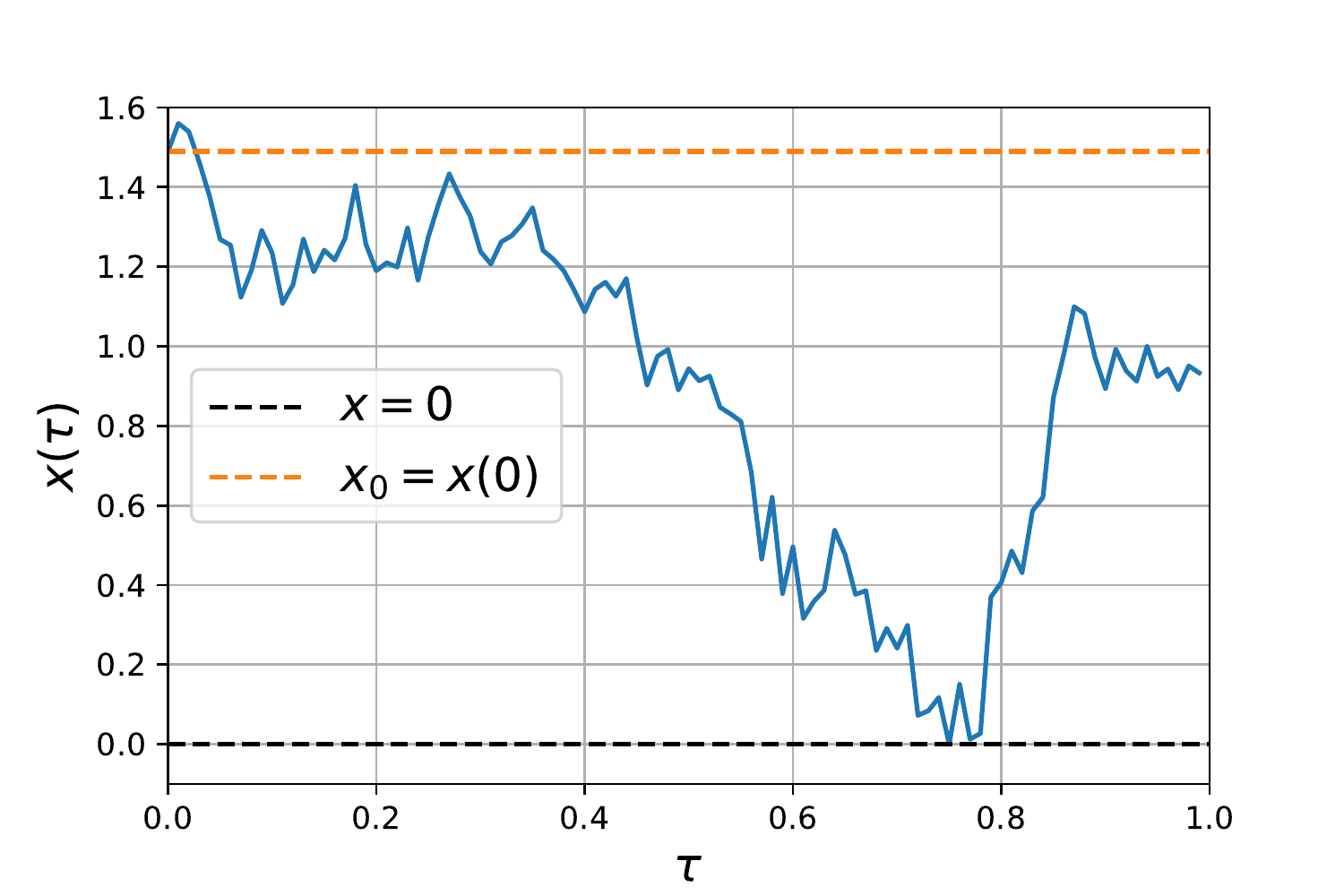}
\includegraphics[width=0.45\textwidth]{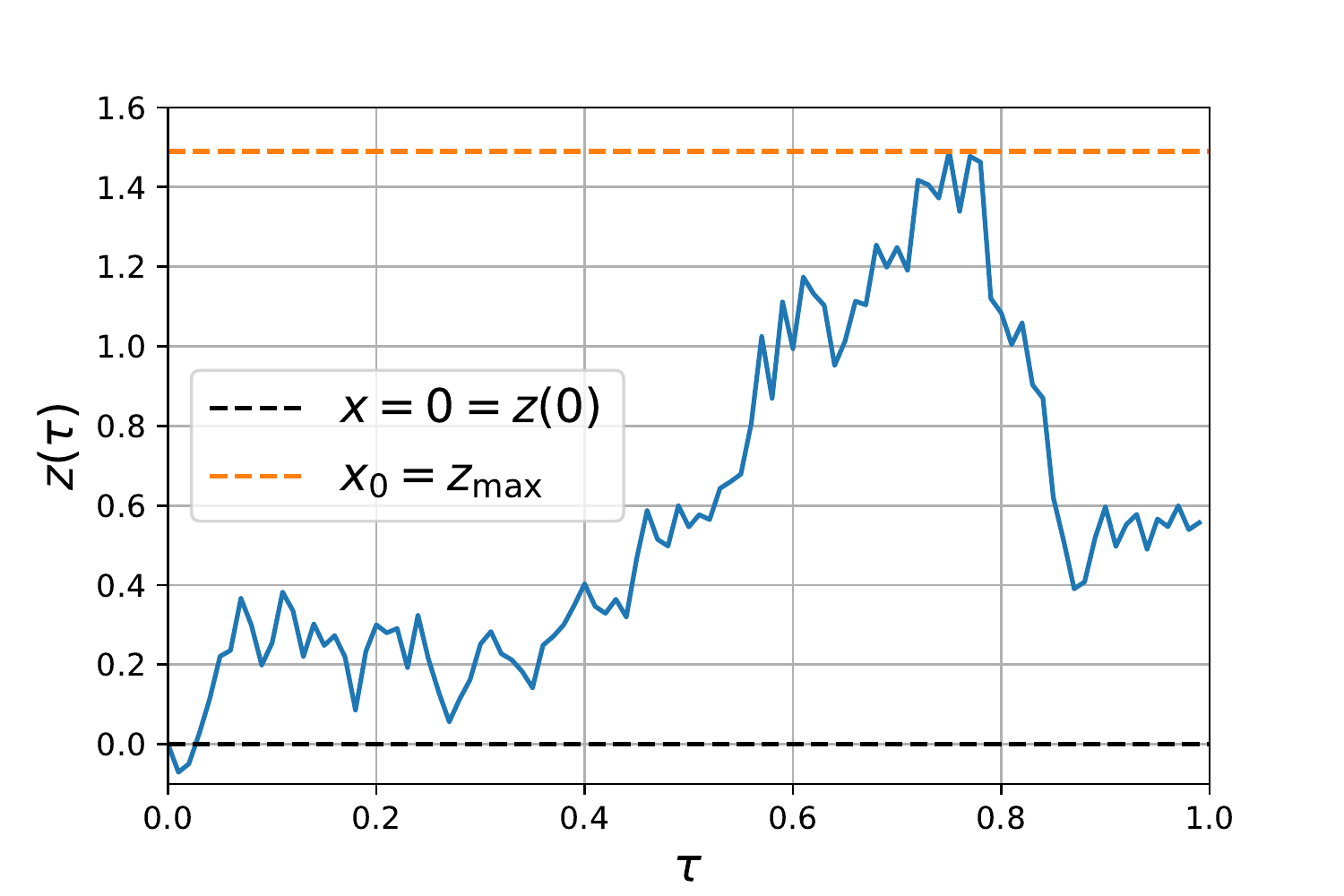}
\caption{In the Brownian trajectory $x(\tau)$ on the left panel, the trajectory starts from position $x(0)=x_0$ (indicated in dashed orange) and survives up to time $t=1$ (with $x=0$ indicated in dashed black). On the right panel, the Brownian trajectory built as $z(\tau)=x_{0}-x(\tau)$ starts from $z(0)=0$ and is less than its maximum $z_{\max}=x_0$.}\label{Fig_surv_max}
\end{figure}

Let us first compute the distribution of the survival probability defined as
\be
Q(x_0,t)=\Prob\left[\lbrace x(\tau)\geq 0\;,\;\;\tau\in [0,t]\rbrace|x(0)=x_0\right]\;.
\ee
Using the spatial translation and symmetry of the Brownian motion and defining $z(\tau)=x_0-x(\tau)$ for $\tau\in[0,t]$ (c.f. Fig. \ref{Fig_surv_max}), we obtain that this probability can also be expressed as
\be
Q(x_0,t)=\Prob\left[z_{\max}=\max_{\tau\in [0,t]}z(\tau)\leq x_0|z(0)=0\right]\;.
\ee

To compute this survival probability, we must discard all the trajectories that become negative on the time interval $[0,t]$. We define the hard-wall potential ${\cal O}(x)=0$ if $x\geq 0$ and ${\cal O}(x)=+\infty$ if $x<0$. For this specific choice, we have
\be
\mathbb{E}_{x_0}\left[e^{-O}\right]=Q(x_0,t).
\ee
It is clear that $Q(x_0\leq 0,t)=0$ as the Brownian trajectories are continuous. For $x_0>0$, $Q(x_0,t)$ is solution of the free diffusion equation on the positive half-space with Dirichlet boundary conditions. To solve this diffusion equation, we introduce the Laplace transform $\tilde Q(x_0;s)=\int_0^{\infty} Q(\mu;x_0,t)e^{-s t}dt$. This yields
\be
\frac{1}{2}\partial_{x_0}^2 \tilde Q(x_0;s)-s \tilde Q(x_0;s)=-1\;,\;\;{\rm for}\;\;x_0>0\;,
\ee
where we used the initial condition $Q(x_0,0)=\Theta(x_0)$. Solving this equation, ensuring the continuity in $x=0$ such that $\tilde Q(x_0=0,s)=0$, we find
\be
\tilde Q(x_0,s)=\frac{1-e^{-\sqrt{2s}x_0}}{s}\;,\;\;x_0\geq 0\;.
\ee 
Using Eq. \eqref{LT_diff_prop_int} in the table of Appendix \ref{LT} to invert the Laplace transform from $s$ to $t$, we obtain
\be\label{surv_BM}
Q(x_0,t)=\Theta(x_0)\erf\left(\frac{x_0}{\sqrt{2t}}\right)\;.
\ee
The PDF $\partial_{x_0} Q(x_0,t)$ of the maximum of the Brownian motion is simply given by a half Gaussian. Using its symmetry in space, the CDF of its minimum is $\Prob\left[z_{\min}\geq x_0|z(0)=0\right]=Q(-x_0,t)$. We can extract from this result the mean value $\moy{x_{\max}}_t$ of the maximum of the Brownian motion on the time interval $[0,t]$ as
\be\label{x_max_mean_BM}
\moy{x_{\max}}_t=\int_0^{\infty}x_0\partial_{x_0}Q(x_0,t)dx_0=\sqrt{\frac{2t}{\pi}}\;.
\ee
Coming back to the survival probability, we see that it behaves in the long time limit $t\to \infty$ as $Q(x_0,t)\approx \frac{2}{\sqrt{\pi t}}$. 
We will now see how to use the PDF $\partial_{x_0} Q(x_0,t)$ of the maximum to obtain another extreme value observable: the time to reach this maximum.

\subsection{Time to reach the maximum and arcsine laws}

\begin{figure}
\centering
\includegraphics[width=0.45\textwidth]{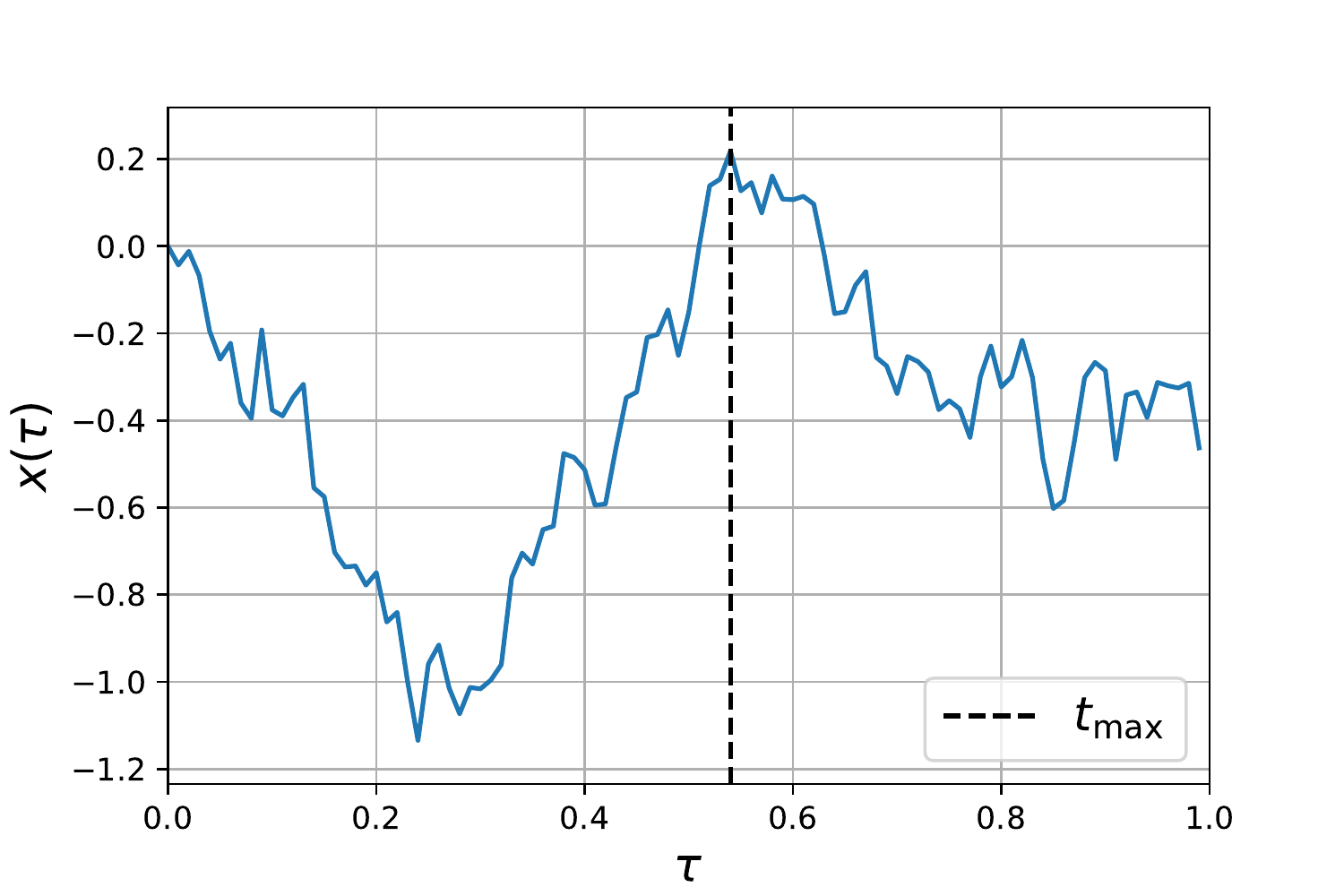}
\includegraphics[width=0.45\textwidth]{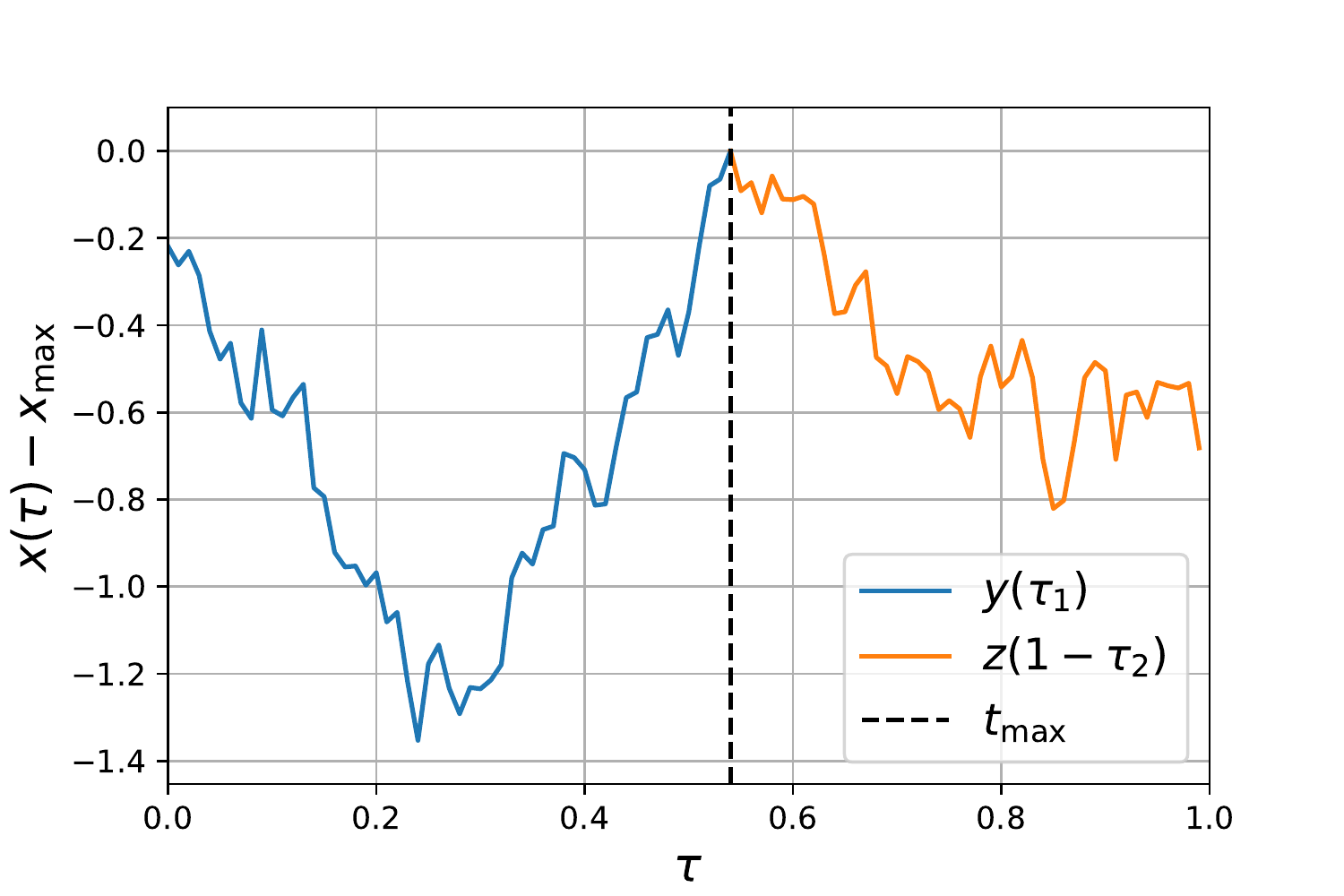}
\caption{In the Brownian trajectory $x(\tau)$ on the left panel, the trajectory starts from position $x(0)=0$ and reach its maximum $x_{\max}$ at time $t_{\max}$. On the right panel, the trajectory is decomposed into a blue path starting from $y(0)=x(0)-x_{\max}$ and reaching its maximum $y_{\max}=0$ at its endpoint at time $t_{\max}$ and an independent orange path (using the Markovian property of Brownian motion), starting from $z(0)=x(1)-x_{\max}$ and reaching its maximum $z_{\max}=0$ at its endpoint at time $1-t_{\max}$.}\label{Fig_t_reach}
\end{figure}

At variance with the case of i.i.d.~random variables, there is a clear notion of time for both the Brownian motion and the random walk. The observables of these stochastic processes are not time invariant and the question of the time to reach the maximum is therefore non-trivial. The distribution of the time $t_{\max}$ at which the maximum is reached, i.e. $x(t_{\max})=x_{\max}$, can be obtained by using a path decomposition. We consider that the maximum is reached at time $t_{\max}=\tau$ and separate the Brownian motion in two independent parts (using the Markovian property of Brownian motion) by redefining $y(\tau_1)=x(\tau_1)-x_{\max}\leq 0$ for $\tau_1\in [0,\tau]$ and $z(\tau_2)=x(1-\tau_2)-x_{\max}\leq 0$ for $\tau_2\in [0,1-\tau]$ (c.f. Fig \ref{Fig_t_reach}). The maximum for the two Brownian motions is reached respectively for $y_{\max}=y(0)=0$ and $z_{\max}=z(0)=0$. Therefore, the PDF of the time to reach the maximum for the Brownian $x(\tau)$ is given by
\be\label{arcsine}
P_{\arcsin}(\tau)=\left.\partial_{z}Q(z,1-\tau)\right|_{z=0}\left.\partial_{y}Q(y,\tau)\right|_{y=0}=\frac{1}{\pi\sqrt{\tau(1-\tau)}}\;.
\ee

\begin{figure}
\centering
\includegraphics[width=0.32\textwidth]{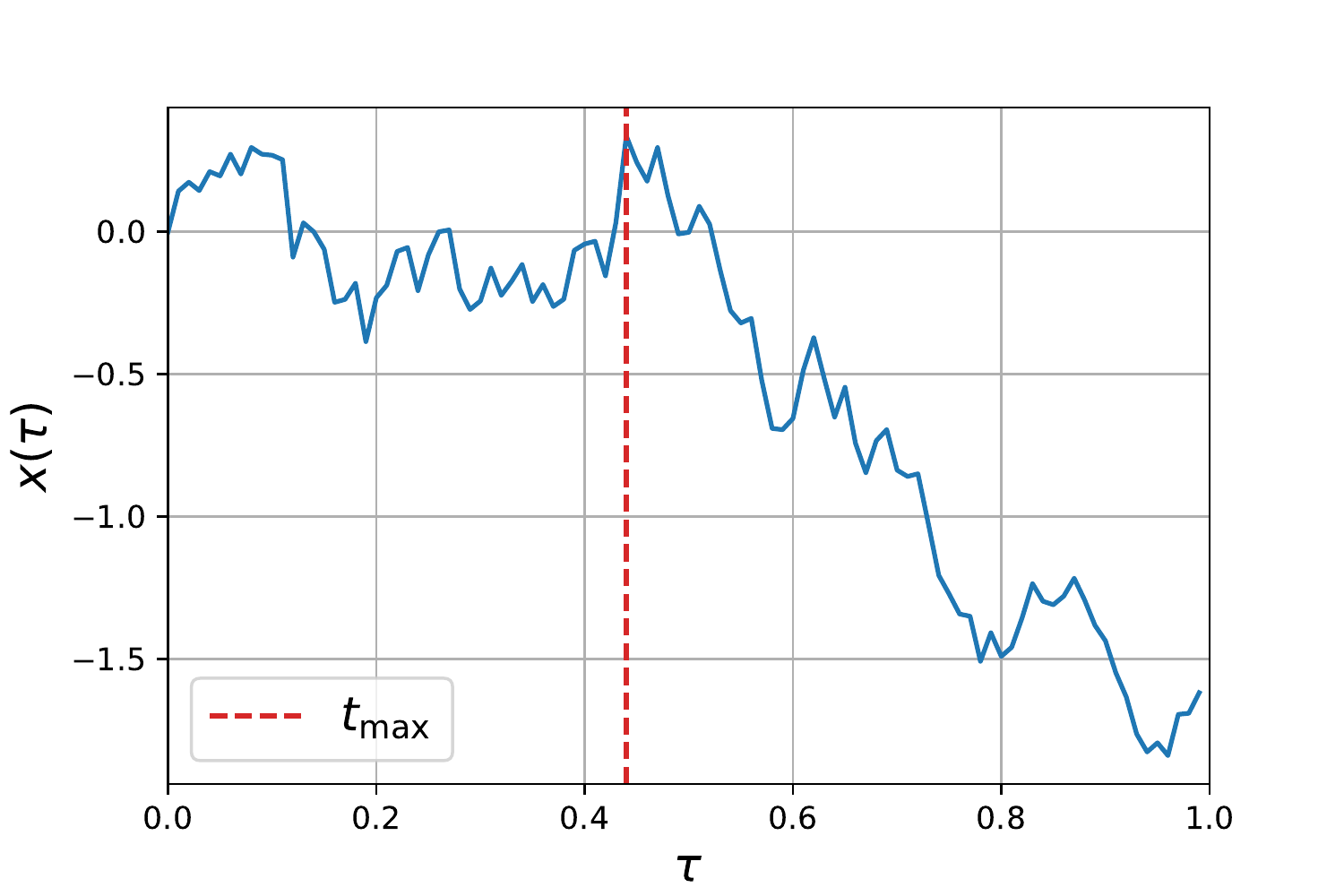}
\includegraphics[width=0.32\textwidth]{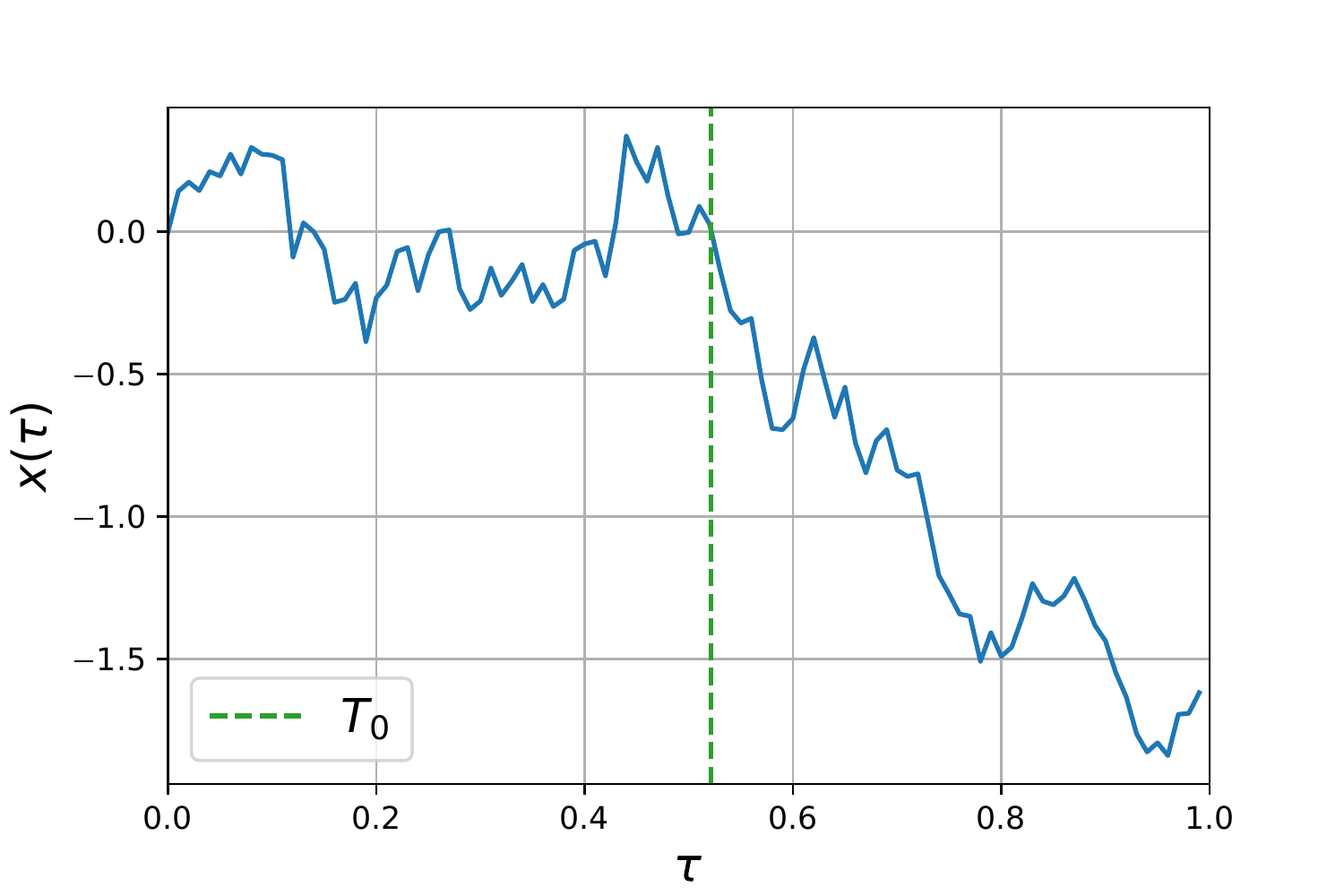}
\includegraphics[width=0.32\textwidth]{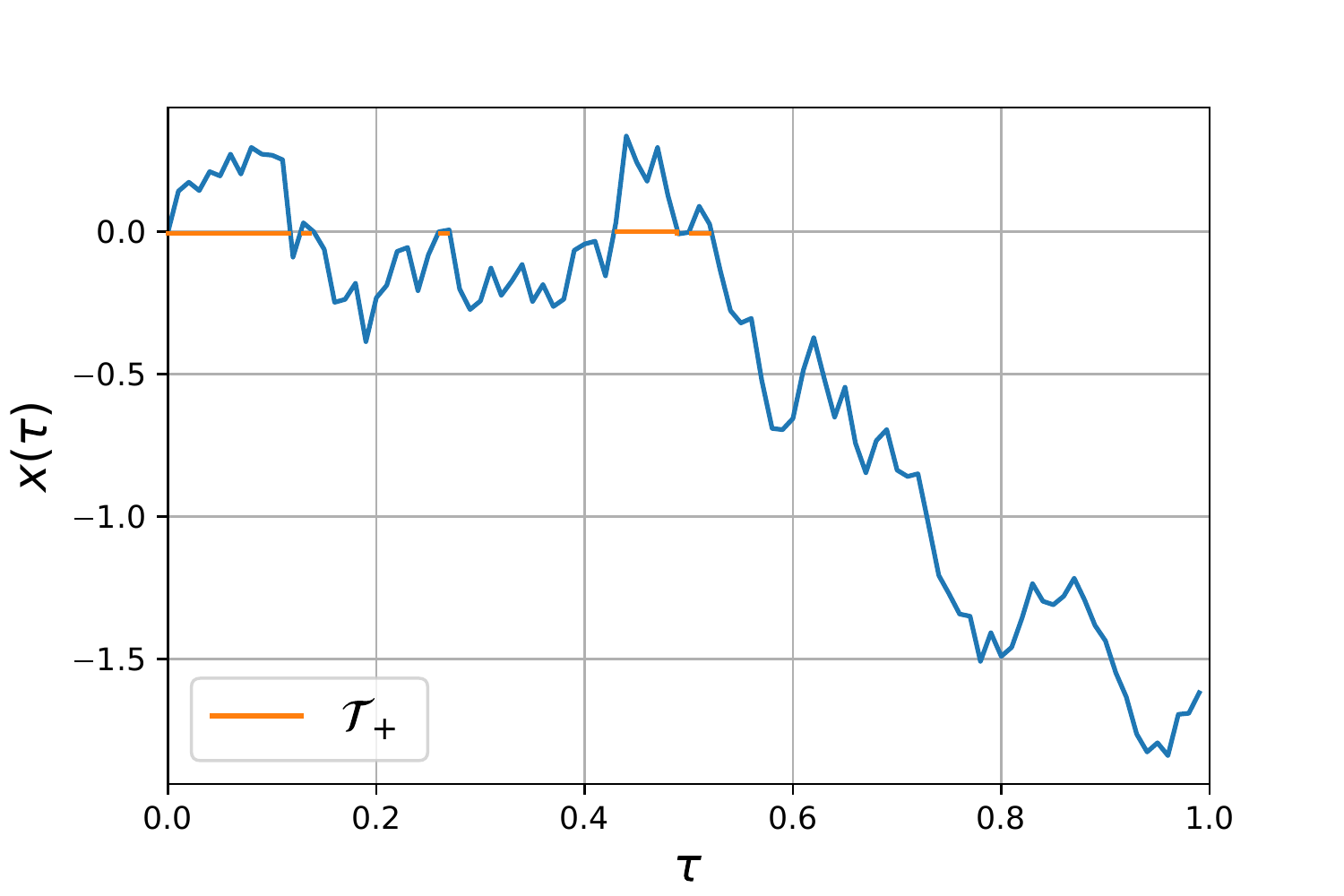}
\caption{For a given trajectory, definition of the time $t_{\max}$ to reach the global maximum (left), the time $T_0$ of the last crossing of $0$ (centre) and the time ${\cal T}_+={\cal T}_+(0)$ spent by the trajectory above $0$ (right).}\label{Fig_3_arcsine}
\end{figure}

Remarkably, this arcsine law is the common PDF for three observables of the Brownian motion \cite{pitman2018guide, revuz2013continuous}: (see Fig. \ref{Fig_3_arcsine})
\begin{itemize}
\item For the time $t_{\max}$ to reach of the maximum (or minimum) of Brownian motion
\item For the time $T_0$ of the last crossing of the Brownian motion $x(\tau)$ to $x(0)=0$,
\be
T_0=\sup_{\tau\in[0,1]}\left[x(\tau)=0 |x(0)=0\right]
\ee 
\item For the occupation time ${\cal T}_{\pm}(0)$ of the Brownian above (below) $z=x(0)=0$,
\be\label{occ_time}
{\cal T}_+(z)=\int_0^{1}d\tau \Theta(x(\tau)-z)\;.
\ee
\end{itemize}

The laws of these three quantities have been recently considered for non-Markovian processes e.g. fractional Brownian motion \cite{sadhu2018generalized}.

The minimum and maximum (together with their time of reach) give already a few informations on the extreme statistics. However, for many physical processes, this gives only a partial information on the extreme events of the system. To obtain a clearer description of the system one should also consider the near-extreme events \cite{perret2013near,sabhapandit2007density}. In the case of the Brownian motion, this can be partly considered  by computing the statistics of the quantiles of Brownian motion $q(\alpha)$ \cite{yor1995distribution}, which we focus on in the next section.

\subsection{Occupation time of the Brownian}

\begin{figure}
\centering
\includegraphics[width=0.6\textwidth]{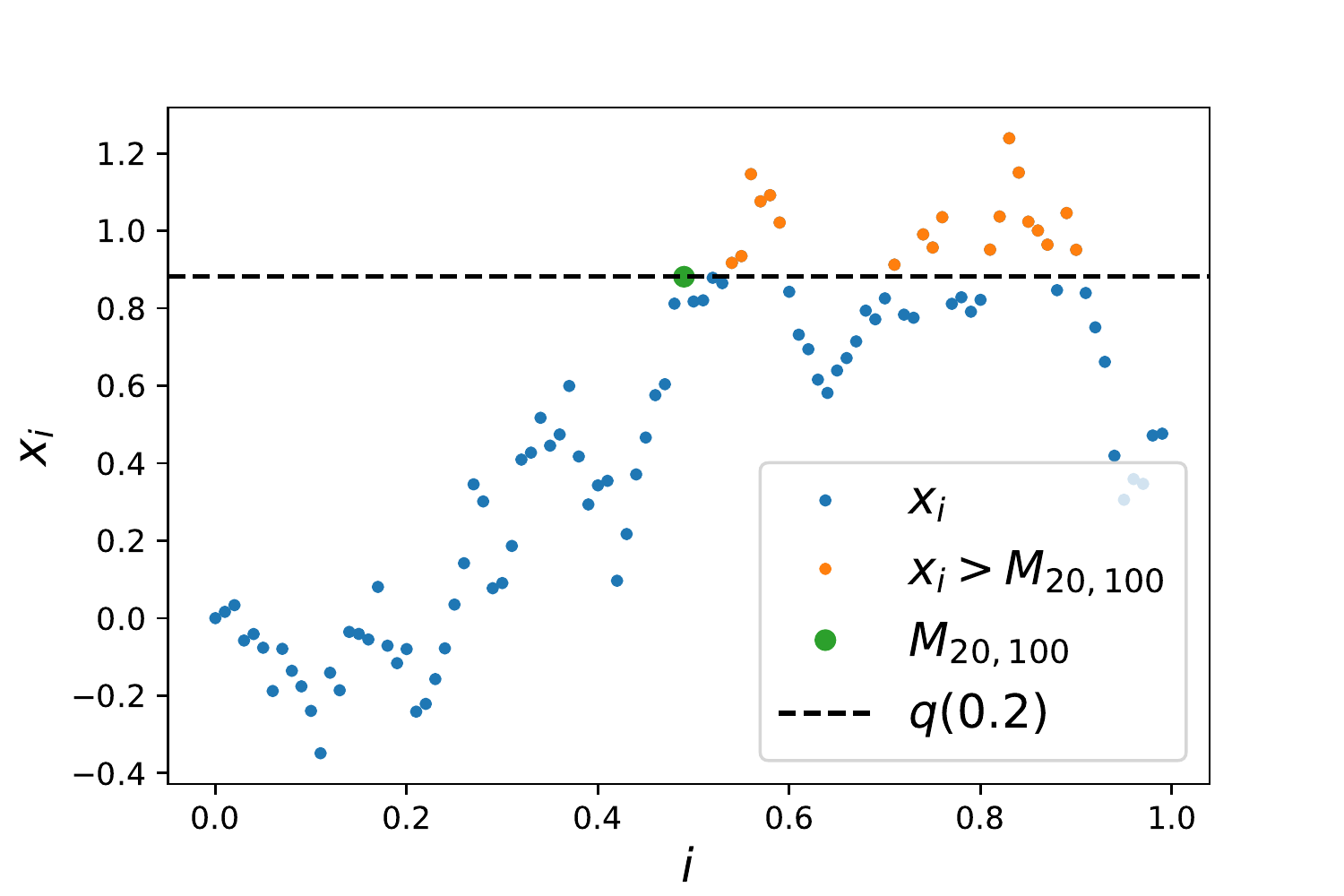}
\caption{Discrete random walk of $n=100$ steps. There are exactly $20$ positions of the walk above $M_{20,100}$. In the large $n$ limit, this walk becomes a continuous path with a fraction $\alpha=0.2$ of the trajectory (in orange) lying above $q(0.2)$. }\label{Fig_walk_q_alpha}
\end{figure}

As stated before, we can define an observable that will generalise the case of the maximum (and is an analogous of the $k^{\rm th}$ maximum for the continuous process): the quantile of Brownian motion, defined as
\be\label{def_quant}
q(\alpha)=\inf\lbrace z\;{\rm such\;that}\; {\cal T}_+(z)=\int_0^{1}d\tau \Theta(x(\tau)-z) \geq \alpha\rbrace\;.
\ee
Note that it is also conveniently defined by considering first a discrete version $x_i$ of this process with $n$ steps and ordering its positions as $M_{1,n}\leq M_{2,n}\leq \cdots\leq M_{n+1,n}$. There are exactly $k$ positions in this discrete process lying above the $k^{\rm th}$ maximum $M_{k,n}$ (c.f. Fig. \ref{Fig_walk_q_alpha}). In the large $n$ limit with $\alpha=k/n$ fixed, it converges naturally to $q(\alpha)$, with a fraction $\alpha$ of the full trajectory lying above $q(\alpha)$.
We may relate easily this observable to the occupation time ${\cal T}_+(z)$ via the identity
\be
\Prob\left[{\cal T}_+(z)\leq \alpha\right]=\Prob\left[q(\alpha)\leq z\right]\;.
\ee 
We will now compute the PDF of this occupation time \eqref{occ_time}. We introduce its Moment Generating Function (MGF),
\be
\mathbb{E}_{x_0}\left[e^{-\mu {\cal T}_+(z)}\right]=Q(\mu;x_0,z,t)\;,
\ee
and will use the Feynman-Kac formalism to derive $Q(\mu;x_0,z,t)$.
This function is solution of the diffusion equation
\be
\partial_t Q=\frac{1}{2}\partial_{x_0}^2 Q-\mu \Theta(x_0-z)Q\;,\;\;{\rm with}\;\;Q(\mu;x_0,z,t=0)=1\;.
\ee
In this case also, the equation can be solved by introducing the Laplace transform $\tilde Q(\mu;x_,z,s)=\int_0^{\infty}Q(\mu;x_0,z,t)e^{-st}dt$, which is solution of
\be
\frac{1}{2}\partial_{x_0}^2 \tilde Q-\left[s+\mu \Theta(x_0-z)\right]\tilde Q=-1\;.
\ee
Solving this equation and setting $x_0=0$, we obtain
\be
 \tilde Q(\mu;x_0=0,z,s)=\begin{cases}
\displaystyle \frac{e^{\sqrt{2(s+\mu)}z}}{\sqrt{s(s+\mu)}}+\frac{1-e^{\sqrt{2(s+\mu)}z}}{s+\mu} &\;,\;\;z\leq 0\;\\
&\\
\displaystyle  \frac{e^{-\sqrt{2s}z}}{\sqrt{s(s+\mu)}}+\frac{1-e^{-\sqrt{2s}z}}{s}&\;,\;\;z>0\;.
\end{cases}
\ee
Note the symmetry $ \tilde Q(\mu;x_0=0,-z,s)= \tilde Q(-\mu;x_0=0,z,s+\mu)$. Using Eqs. \eqref{LT_diff_prop} and \eqref{LT_diff_prop_int} in the table of Appendix \ref{LT}, we invert the Laplace transform from $s$ to $t$ and obtain the MGF of the occupation time ${\cal T}_+(z)$,
\be
\mathbb{E}_{x_0=0}\left[e^{-\mu {\cal T}_+(z)}\right]=
\begin{cases}
\displaystyle \erf\left(\frac{z}{\sqrt{2t}}\right)+\int_0^{t}d\tau \frac{e^{-\frac{z^2}{2\tau}}e^{-\mu (t-\tau)}}{\pi\sqrt{\tau(t-\tau)}}&\;,\;\;z\geq 0\\
&\\
\displaystyle e^{-\mu}\erf\left(-\frac{z}{\sqrt{2t}}\right)+\int_0^{t}d\tau \frac{e^{-\frac{z^2}{2\tau}}e^{-\mu \tau}}{\pi\sqrt{\tau(t-\tau)}}&\;,\;\;z<0\;.
\end{cases}
\ee
Note that taking the limit $\mu\to \infty$, this expression yields back the result for the survival probability $Q(z,t)=\lim_{\mu \to \infty} \mathbb{E}_{x_0=0}\left[e^{-\mu {\cal T}_+(z)}\right]$ in Eq. \eqref{surv_BM}. To obtain the PDF from this expression, we invert the Laplace transform from $\mu$ to $\tau$ using Eq. \eqref{LT_dirac} in the table of Appendix \ref{LT}
\be\label{occ_time_PDF_BM}
P_{ {\cal T}_+}(\tau;z)=\begin{cases}
\displaystyle \delta(\tau)\erf\left(\frac{z}{\sqrt{2t}}\right)+\frac{e^{-\frac{z^2}{2(t-\tau)}}}{\pi\sqrt{\tau(t-\tau)}}&\;,\;\;z\geq 0\\
&\\
\displaystyle \delta(t-\tau)\erf\left(-\frac{z}{\sqrt{2t}}\right)+\frac{e^{-\frac{z^2}{2\tau}}}{\pi\sqrt{\tau(t-\tau)}}&\;,\;\;z<0\;.
\end{cases}
\ee
Taking $z=0$ in this expression, it corresponds to the occupation time on the positive half space $\mathbb{R_+}$ and one recovers the arcsine law $P_{ {\cal T}_+}(\tau;z=0)=\frac{1}{t}P_{\arcsin}\left(\frac{\tau}{t}\right)$ where $P_{\arcsin}(\tau)$ is given in Eq. \eqref{arcsine}. The Dirac delta term in \eqref{occ_time_PDF_BM} can be explained as such: for $z>0$, the probability for $z$ to be larger than the maximum of the Brownian is positive $Q(z,t)=\erf\left(\frac{z}{\sqrt{2t}}\right)>0$. If this is the case, then the full trajectory remains below $z$, i.e. $x(\tau)<z$  for $\tau \in [0,t]$ and ${\cal T}_+(z)=0$. Integrating this function with respect to $\tau$ for $\tau\in[0,\alpha]$, we obtain the CDF of the quantiles of Brownian motion \cite{yor1995distribution,dassios1995distribution, embrechts1995proof}
\be\label{CDF_quantiles}
\Prob\left[{\cal T}_+(z)\leq \alpha\right]=\Prob\left[q(\alpha)\leq z\right]=\begin{cases}
\displaystyle \int_0^{t-\alpha}\frac{e^{-\frac{z^2}{2\tau}}}{\pi\sqrt{\tau(t-\tau)}}d\tau&\;,\;\;z\geq 0\\
&\\
\displaystyle \int_0^{\alpha}\frac{e^{-\frac{z^2}{2\tau}}}{\pi\sqrt{\tau(t-\tau)}}d\tau&\;,\;\;z<0\;.
\end{cases}
\ee

 This concludes our introduction on the Brownian motion. We will now introduce its discrete time counterpart: the random walk.

\section{Extreme value properties of random walks}

We define the one-dimensional random walk by the initial position $x_0$ and the recursion relation
\be\label{walk}
x_i=x_{i-1}+\eta_i\;,\;\;{\rm for}\;\;i=1,\cdots,n\;,
\ee
where the random variables $\eta_i$'s are independent and identically distributed (i.i.d.), with common PDF $f(\eta)$. The random walks are ubiquitous models that have been studied for more than a century \cite{RevModPhys.15.1}, with applications in many different fields of research such as mathematics, physics, biology, ecology, finance, ... Many exact results have been obtained for this discrete process, and we will state the most important ones to study the extreme value observables of this process. We will restrict our study to symmetric random walks $f(\eta)=f(-\eta)$ (see \cite{majumdar2012record,Mounaix_2018} for cases with drift). The random walk defined in Eq. \eqref{walk} is a Markov chain, with a propagator that is translation invariant in time and space
\be\label{M_gen_RW}
P_{i,j}(x,y)=\Prob\left[x_i=x|x_j=y\right]=\Prob\left[x_{i-j}=x-y|x_0=0\right]=P_{i-j}(x-y).
\ee 
This propagator is solution of the forward integral equation
\be\label{fw_prop_eq}
P_{n+1}(x)=\int_{-\infty}^{\infty}f(x-x')P_n(x')dx'\;,
\ee
which states that the probability to arrive at position $x$ at step $n+1$ is the probability to start from any position $x'$ at step $n$ and make a jump of length $x-x'$ on the last step. This equation can be trivially solved by Fourier transform, yielding
\be\label{prop_RW}
P_n(x)=\int_{-\infty}^{\infty}\frac{dk}{2\pi}\left[\hat f(k)\right]^n e^{-\I kx}\;,
\ee 
where the function $\hat f(k)=\int_{-\infty}^{\infty}f(\eta)e^{\I k\eta}d\eta$ is the Fourier transform of $f(\eta)$.

\subsection{Finite variance vs L\'evy flights}

Taking the large $n$ limit in the expression of the propagator in Eq. \eqref{prop_RW}, the expression will depend on the small $k$ behaviour of $\hat f(k)$. Indeed, the propagator between position $x_n=\sum_{k=1}^n \eta_k+x_0$ and $x_0$ is simply the PDF of $\sum_{k=1}^n \eta_k$, which is a sum of i.i.d. random variables. If these symmetric variables have a finite variance $\sigma^2$, the Fourier transform of the jump PDF reads
\be\label{FT_finite_variance}
\hat{f}(k)=\int_{-\infty}^{\infty}f(\eta)e^{\I k\eta}d\eta\approx 1-\frac{(\sigma k)^2}{2}+o(k^2)\;,
\ee 
and the central limit theorem applies. In this case, the full process converges towards the Brownian motion and in particular for its propagator,
\be
P_n(x)\approx \frac{1}{\sigma\sqrt{n}}G\left(\frac{x}{\sigma\sqrt{n}},1\right)\;,
\ee
where $G(z,t)$ is the propagator of the Brownian motion in Eq. \eqref{prop_BM}.

However, if the jump distribution is heavy-tailed $f(\eta)\approx |\eta|^{-1-\mu}$ for $|\eta|\gg 1$ and $0<\mu<2$, the jump length has infinite variance $\sigma^2=\infty$ and therefore the central limit theorem does not apply. Using the behaviour of the Fourier transform of the jump PDF,
\be\label{FT_mu}
\hat{f}(k)=\int_{-\infty}^{\infty}f(\eta)e^{\I k\eta}d\eta\approx 1-|a_{\mu} k|^{\mu}+o(k^\mu)\;,
\ee 
one can show instead the convergence of the process towards a L\'evy flight of index $\mu$,
\be\label{levy_scal}
P_n(x)\approx \frac{1}{a_{\mu}(n\tau)^{1/\mu}}{\cal L}_{\mu}\left( \frac{x}{a_{\mu}(n\tau)^{1/\mu}}\right)\;,\;\;{\rm with}\;\;{\cal L}_{\mu}(z)=\int_{-\infty}^{\infty}\frac{dq}{2\pi}e^{-\I q z- |q|^{\mu}}\;,
\ee
where ${\cal L}_{\mu}(z)$ is called the L\'evy stable law of index $\mu$. Note that setting $\mu=2$ in this expression and using $\sigma=\sqrt{2}a_2$, we recover the propagator of the Brownian motion ${\cal L}_{\mu=2}(z)=G(z/\sqrt{2},1)$. This distinction between finite variance and L\'evy flight (for which $0<\mu<2$) will be essential to analyse the emerging behaviours for walks with a large number of steps. We will now consider the different extreme value observables for the random walk and analyse how they differ or not from what we obtained for the Brownian motion in the previous section.

\begin{figure}
\centering
\includegraphics[width=0.6\textwidth]{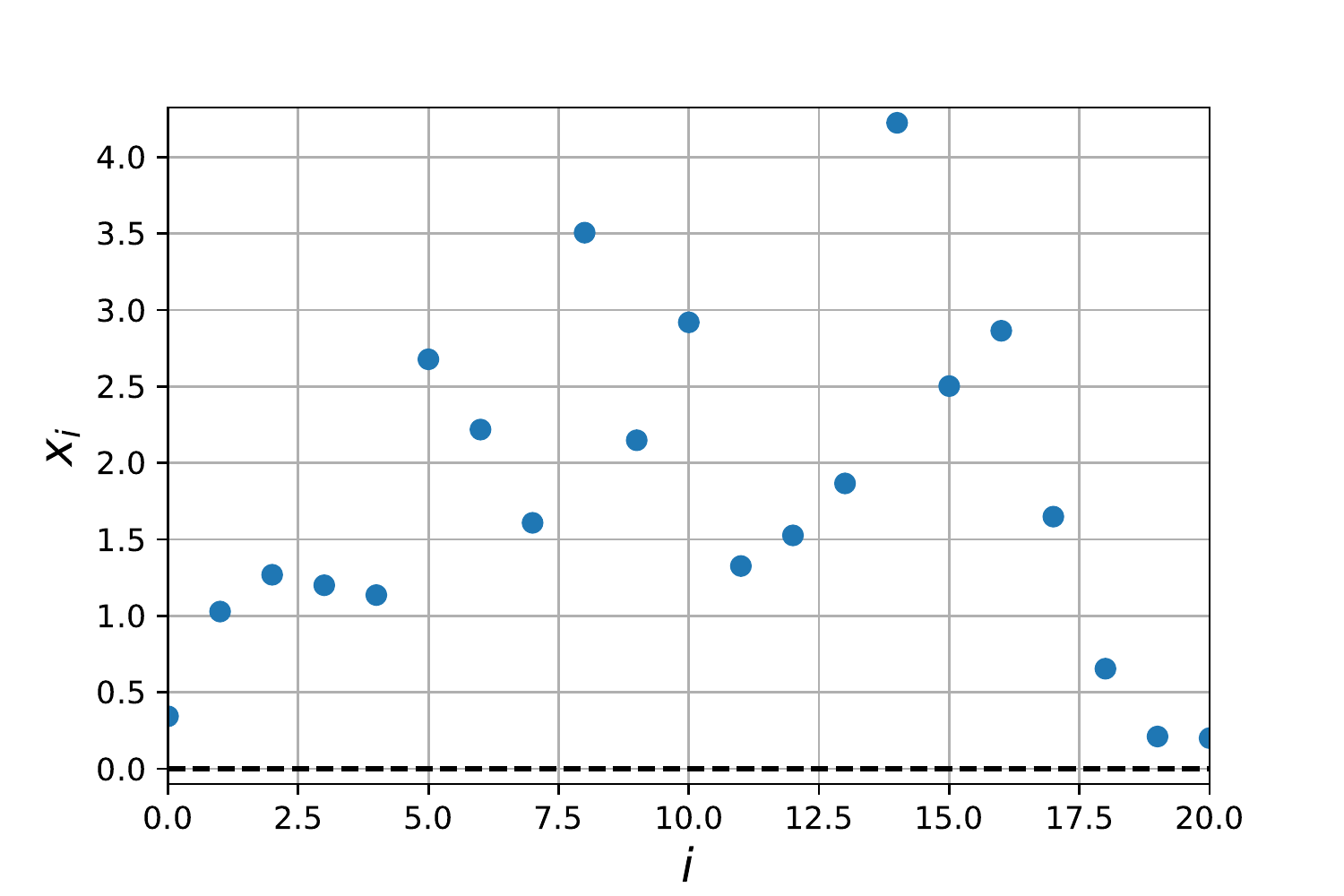}
\caption{Example of a random walk starting at a position $x_0>0$ which survives up to $n=20$ steps. All the positions of the walk are positive.}\label{Fig_surv_walk}
\end{figure}

\subsection{Survival probability and PDF of the maximum}\label{max_k_surv}

As in the case of the Brownian motion, we start by considering the survival probability $Q_n(x_0)$, defined as the probability that a random walk starting at $x_0\geq 0$ remains positive afterwards (c.f. Fig \ref{Fig_surv_walk})
\be\label{survival}
Q_n(x_0)=\Prob\left[x_1\geq 0,\;x_2 \geq 0,\;\cdots\;,\;x_n\geq 0|x_0\right]\;.
\ee
Note that $Q_{n=0}(x_0)=1$ for $x_0>0$. It is convenient to write for this survival probability a backward recursion, depending on the initial position $x_0$. This backward equation is therefore quite different from the forward equation that we obtained for the propagator in Eq. \eqref{fw_prop_eq}. It reads
\be\label{bw_surv_eq}
Q_{n+1}(x_0)=\int_0^{\infty}Q_n(x)f(x-x_0)dx\;,\;\;x_0\geq 0\;.
\ee
Considering only the initial step, the walk survives if it starts from a position $x_0\geq 0$, makes an initial jump of length $\eta_1=x-x_0$, arriving at a position $x_1=x\geq 0$, and survives for the remaining $n$ steps. At variance with the propagator equation, the domain of integration in Eq. \eqref{bw_surv_eq} is restricted to the positive half-space. These Wiener-Hopf type of equations are much more difficult to study as they cannot be solved simply by Fourier or Laplace transform. Similarly to the case of the Brownian motion, we introduce a path transformation of the random walk $z_k=x_0-x_k$ and using the space invariance and symmetry of the Markov propagator in Eq. \eqref{M_gen_RW}, allowing to make a clear connection between the survival probability and the CDF of the maximum $z_{\max}=\displaystyle \max_{1\leq k\leq n}z_k$ of the walk, \cite{majumdar2010universal}
\be
Q_n(x_0)=\Prob\left[z_1\leq x_0,\;z_2 \leq x_0,\;\cdots\;,\;z_n\leq x_0|0\right]=\Prob\left[z_{\max}\leq x_0|z_0=0\right]\;.
\ee
There exists an exact formula for the Laplace transform of the generating function of $Q_n(x_0)$, valid for any jump PDF known as the Pollaczek-Spitzer formula \cite{pollaczek1952fonctions,spitzer1956combinatorial,spitzer1957wiener}. It reads
\be
\int_0^{\infty}\left[\sum_{n=0}^{\infty}s^n Q_n(x_0)\right]e^{-p x_0}dx_0=\frac{1}{p\sqrt{1-s}}\exp\left[-\frac{p}{\pi}\int_0^{\infty}\frac{\ln[1-s\hat{f}(q)]}{p^2+q^2}dq\right]\;.\label{PS}
\ee
This formula is still quite hard to analyse and the generating and Laplace transforms cannot be inverted in general. 

As a first application of this formula, we may extract the mean value of the maximum $\moy{x_{\max}}_n$ of the walk of $n$ steps which will be useful in the following. The generating function for this maximum can be obtained by multiplying Eq. \eqref{PS} by $p$, deriving with respect to $p$ and finally taking $p=0$
\be
-\int_0^{\infty}\partial_{p}\left[p\sum_{n=0}^{\infty}s^n Q_n(x_0)e^{-p x_0}\right]_{p=0}dx_0=\int_0^{\infty}\sum_{n=0}^{\infty}s^n x_0 Q'_n(x_0)dx_0
=\sum_{n=0}^{\infty}s^n \moy{x_{\max}}_n
\ee
Inverting this generating function for large $n$, one obtains \cite{comtet2005precise}
\be\label{mean_max_mu}
\frac{\moy{x_{\max}}_n}{a_{\mu}}\approx\frac{\mu}{\pi}\Gamma\left(1-\frac{1}{\mu}\right)n^{1/\mu}\;,
\ee
with in particular for finite variance given by $\sigma^2=2a_{2}^2<\infty$ (corresponding to $\mu=2$),
\be\label{mean_max_fv}
\frac{\moy{x_{\max}}_n}{\sigma}\approx \sqrt{\frac{2n}{\pi}}\;.
\ee
Thus, one recovers the result obtained for the Brownian motion in Eq. \eqref{x_max_mean_BM}. Analysing in detail Eq. \eqref{PS} for a finite variance jump PDF we actually recover the result for the full CDF of the maximum of the Brownian motion
\be
Q_n(x_0)\approx \erf\left(\frac{x_0}{\sigma\sqrt{2n}}\right)\;,\;\;{\rm for}\;\;x_0\geq 0\;.
\ee
As a third application of this Pollaczek-Spitzer formula, we will see that for $x_0=0$, the survival probability $Q_n(x_0=0)$ is universal for any finite value of $n$.

\subsection{Sparre Andersen formula}

The survival probability $Q_n(0)$ of the walk starting from $x_0=0$ was computed explicitly by Sparre Andersen \cite{andersen1955fluctuations}. It can be obtained from the Pollaczek-Spitzer formula \eqref{PS} by multiplying by $p$ and changing the variable of integration $x_0\to z=p x_0$. This yields
\be
\int_0^{\infty}\left[\sum_{n=0}^{\infty}s^n Q_n\left(\frac{x_0}{p}\right)\right]e^{-z}dz=\frac{1}{\sqrt{1-s}}\exp\left[-\frac{p}{\pi}\int_0^{\infty}\frac{\ln[1-s\hat{f}(q)]}{p^2+q^2}dq\right]\;.\label{PS_1}
\ee
Taking finally the $p\to \infty$ limit in Eq. \eqref{PS_1}, we obtain the generating function (GF) of the survival probability for $x_0=0$, \cite{majumdar2010universal}
\be
\sum_{n=0}^{\infty}s^n Q_n\left(0\right)=\frac{1}{\sqrt{1-s}}\;.
\ee
From this equation, we extract the expression of $Q_n(0)$ for all values of $n$,
\be\label{SA}
Q_n(0)=\Prob\left[x_1\geq 0,\;x_2 \geq 0,\;\cdots\;,\;x_n\geq 0|x_0=0\right]={{2n}\choose{n}}2^{-2n}\;.
\ee
Remarkably this formula is universal and holds for {\it any} jump PDF $f(\eta)$. Taking the large $n$ limit, we obtain the decay of the probability $Q_n(0)\approx (\pi n)^{-1/2}$. As a nice application of this formula, we can extract the time to reach the maximum of the random walk.

\subsection{Time to reach the maximum and occupation time}

Let us consider a walk with a maximum $\displaystyle x_{\max, n}=\max_{0\leq k\leq n} x_k$ that is reached at step $l$, i.e. $x_{\max}=x_l$. We obtain this probability using a similar decomposition as in the case of the Brownian motion (c.f. Fig. \ref{Fig_t_reach}). Using the Markov property of the process, we decompose the walk in two independent walks. A first walk defined as $y_j=x_l-x_{l-j}\leq 0$ (as $x_{\max, n}=x_l$) for $j=1,\cdots,l$ starting at $y_0=0$, and a second walk defined as $z_k=x_l-x_{l+k}\leq 0$ for $k=1,\cdots,n-l$, starting also at $z_0=0$. Each walk starts from $0$ and stays negative afterwards. Using the Sparre Andersen formula together with the independence of the two walks, the probability to reach the maximum $x_{\max}$ at step $l$ reads \cite{feller1968introduction}
\be
P_{n,l}=\Prob\left[x_{\max, n}=x_l\right]=Q_{l}(0)Q_{n-l}(0)={{2l}\choose{l}}{{2(n-l)}\choose{n-l}}2^{-2n}\;.
\ee 
It turns out that, as for the Brownian motion, this law also describes another random variable characteristic
of the random walk: the occupation time $N_{+}(x=0)$.  For a discrete process, the occupation time $N_{+}(x)$ is the counting process of the number of steps taken by the random walker above the level $x$,
\be\label{occ}
N_{+}(x)=\sum_{k=0}^n \Theta(x_k-x)\;,
\ee
where $\Theta(x)$ is the Heaviside step-function. In the specific case where $x=x_0=0$, the probability $\Prob\left[N_{+}(0)=l\right]=P_{n,l}$ \cite{feller1968introduction}. Taking the large $n$ limit with $\tau=l/n$ fixed for the probability $P_{n,l}$, we recover the arcsine law
\be
 P_{n,l}\approx \frac{1}{n}P_{\arcsin}\left(\frac{l}{n}\right)\;,\;\;{\rm with}\;\;P_{\arcsin}(\tau)=\frac{1}{\pi\sqrt{\tau(1-\tau)}}\;.
\ee
Note that initially, we only computed the arcsine law for Brownian motion but we see using the Sparre-Andersen formula that it remains valid for L\'evy flights. This result closes this introduction on the random walks and their global maximum. We will turn in the next chapter to the order statistics of the random walks, i.e. the statistics of the ordered maxima (first, second, third...).

%
%
%



\begin{thebibliography}{100}

\bibitem{andersen1955fluctuations}
E.~Sparre Andersen.
\newblock On the fluctuations of sums of random variables ii.
\newblock {\em Mathematica Scandinavica}, pages 195--223, 1955.

\bibitem{andreief1883note}
C.~Andr{\'e}ief.
\newblock Note sur une relation les int{\'e}grales d{\'e}finies des produits
  des fonctions.
\newblock {\em M{\'e}m. de la Soc. Sci. Bordeaux}, 2(1):1--14, 1883.

\bibitem{andrews1976singular}
M.~Andrews.
\newblock Singular potentials in one dimension.
\newblock {\em American Journal of Physics}, 44(11):1064--1066, 1976.

\bibitem{bachelier1900theorie}
L.~Bachelier.
\newblock Th{\'e}orie de la sp{\'e}culation.
\newblock In {\em Annales scientifiques de l'{\'E}cole normale sup{\'e}rieure},
  volume~17, pages 21--86, 1900.

\bibitem{pathria}
R.~K. Pathria; P.~D. Beale.
\newblock {\em Statistical Mechanics (Third Edition)}.
\newblock Academic Press, 2011.

\bibitem{beenakker1997random}
C.~W.~J. Beenakker.
\newblock Random-matrix theory of quantum transport.
\newblock {\em Reviews of modern physics}, 69(3):731, 1997.

\bibitem{PhysRevLett.78.4737}
P.~W. Brouwer; K. M. Frahm; C. W.~J. Beenakker.
\newblock Quantum mechanical time-delay matrix in chaotic scattering.
\newblock {\em Phys. Rev. Lett.}, 78:4737--4740, Jun 1997.

\bibitem{berkowitz_records_noise}
Y.~Edery; A.~B. Kostinski; S.~N. Majumdar;~B. Berkowitz.
\newblock Record-breaking statistics for random walks in the presence of
  measurement error and noise.
\newblock {\em Phys. Rev. Lett.}, 110:180602, May 2013.

\bibitem{vivo2007large}
P.~Vivo; S.~N. Majumdar;~O. Bohigas.
\newblock Large deviations of the maximum eigenvalue in wishart random
  matrices.
\newblock {\em Journal of Physics A: Mathematical and Theoretical},
  40(16):4317, 2007.

\bibitem{PhysRevLett.101.216809}
P.~Vivo; S.~N. Majumdar;~O. Bohigas.
\newblock Distributions of conductance and shot noise and associated phase
  transitions.
\newblock {\em Phys. Rev. Lett.}, 101:216809, Nov 2008.

\bibitem{PhysRevB.81.104202}
P.~Vivo; S.~N. Majumdar;~O. Bohigas.
\newblock Probability distributions of linear statistics in chaotic cavities
  and associated phase transitions.
\newblock {\em Phys. Rev. B}, 81:104202, Mar 2010.

\bibitem{bornemann2010numerical}
F.~Bornemann.
\newblock On the numerical evaluation of fredholm determinants.
\newblock {\em Mathematics of Computation}, 79(270):871--915, 2010.

\bibitem{bornemann2011scaling}
F.~Bornemann.
\newblock On the scaling limits of determinantal point processes with kernels
  induced by sturm-liouville operators.
\newblock {\em arXiv preprint arXiv:1104.0153}, 2011.

\bibitem{dubail2017conformal}
J.~Dubail; J.~M. St{\'e}phan; J. Viti;~P. Calabrese.
\newblock Conformal field theory for inhomogeneous one-dimensional quantum
  systems: the example of non-interacting fermi gases.
\newblock {\em SciPost Physics}, 2(1):002, 2017.

\bibitem{calabrese2004entanglement}
P.~Calabrese;~J. Cardy.
\newblock Entanglement entropy and quantum field theory.
\newblock {\em Journal of Statistical Mechanics: Theory and Experiment},
  2004(06):P06002, 2004.

\bibitem{calabrese2006time}
P.~Calabrese;~J. Cardy.
\newblock Time dependence of correlation functions following a quantum quench.
\newblock {\em Physical review letters}, 96(13):136801, 2006.

\bibitem{calabrese2007quantum}
P.~Calabrese;~J. Cardy.
\newblock Quantum quenches in extended systems.
\newblock {\em Journal of Statistical Mechanics: Theory and Experiment},
  2007(06):P06008, 2007.

\bibitem{case2008wigner}
W.~B. Case.
\newblock Wigner functions and weyl transforms for pedestrians.
\newblock {\em American Journal of Physics}, 76(10):937--946, 2008.

\bibitem{castin2006basic}
Y.~Castin.
\newblock Basic theory tools for degenerate fermi gases.
\newblock {\em arXiv preprint cond-mat/0612613}, 2006.

\bibitem{RevModPhys.15.1}
S.~Chandrasekhar.
\newblock Stochastic problems in physics and astronomy.
\newblock {\em Rev. Mod. Phys.}, 15:1--89, Jan 1943.

\bibitem{chaumont1999path}
L.~Chaumont.
\newblock A path transformation and its applications to fluctuation theory.
\newblock {\em Journal of the London Mathematical Society}, 59(2):729--741,
  1999.

\bibitem{ho2000rapidly}
T.~L. Ho; C.~V. Ciobanu.
\newblock Rapidly rotating fermi gases.
\newblock {\em Physical review letters}, 85(22):4648, 2000.

\bibitem{bertin2006generalized}
E.~Bertin;~M. Clusel.
\newblock Generalized extreme value statistics and sum of correlated variables.
\newblock {\em Journal of Physics A: Mathematical and General}, 39(24):7607,
  2006.

\bibitem{PhysRevLett.82.4220}
C.~Texier;~A. Comtet.
\newblock Universality of the wigner time delay distribution for
  one-dimensional random potentials.
\newblock {\em Phys. Rev. Lett.}, 82:4220--4223, May 1999.

\bibitem{majumdar2005airy}
S.~N. Majumdar;~A. Comtet.
\newblock Airy distribution function: from the area under a brownian excursion
  to the maximal height of fluctuating interfaces.
\newblock {\em Journal of Statistical Physics}, 119(3-4):777--826, 2005.

\bibitem{cooper2008rapidly}
N.~R. Cooper.
\newblock Rapidly rotating atomic gases.
\newblock {\em Advances in Physics}, 57(6):539--616, 2008.

\bibitem{aftalion2005vortex}
A.~Aftalion; X. Blanc;~J. Dalibard.
\newblock Vortex patterns in a fast rotating bose-einstein condensate.
\newblock {\em Physical Review A}, 71(2):023611, 2005.

\bibitem{dassios1995distribution}
A.. Dassios.
\newblock The distribution of the quantile of a brownian motion with drift and
  the pricing of related path-dependent options.
\newblock {\em The Annals of Applied Probability}, pages 389--398, 1995.

\bibitem{dassios1996sample}
A.~Dassios.
\newblock Sample quantiles of stochastic processes with stationary and
  independent increments.
\newblock {\em The Annals of Applied Probability}, pages 1041--1043, 1996.

\bibitem{fridman2012measuring}
M.~Fridman; R. Pugatch; M. Nixon; A. Friesem;~N. Davidson.
\newblock Measuring maximal eigenvalue distribution of wishart random matrices
  with coupled lasers.
\newblock {\em Physical Review E}, 85(2):020101, 2012.

\bibitem{derrida1981random}
B.~Derrida.
\newblock Random-energy model: An exactly solvable model of disordered systems.
\newblock {\em Physical Review B}, 24(5):2613, 1981.

\bibitem{Brunet_2009}
E.~Brunet;~B. Derrida.
\newblock Statistics at the tip of a branching random walk and the delay of
  traveling waves.
\newblock {\em EPL (Europhysics Letters)}, 87(6):60010, sep 2009.

\bibitem{Derrida2011}
E.~Brunet;~B. Derrida.
\newblock A branching random walk seen from the tip.
\newblock {\em Journal of Statistical Physics}, 143(3):420, Apr 2011.

\bibitem{sadhu2015large}
T.~Sadhu;~B. Derrida.
\newblock Large deviation function of a tracer position in single file
  diffusion.
\newblock {\em Journal of Statistical Mechanics: Theory and Experiment},
  2015(9):P09008, 2015.

\bibitem{carpentier2001glass}
D.~Carpentier; P.~Le Doussal.
\newblock Glass transition of a particle in a random potential, front selection
  in nonlinear renormalization group, and entropic phenomena in liouville and
  sinh-gordon models.
\newblock {\em Physical review E}, 63(2):026110, 2001.

\bibitem{dubach}
G.~Dubach.
\newblock Powers of ginibre eigenvalues.
\newblock {\em arXiv preprint arXiv:1711.03151v2}, 2017.

\bibitem{dumitriu2003eigenvalue}
I.~Dumitriu.
\newblock {\em Eigenvalue statistics for beta-ensembles}.
\newblock PhD thesis, Massachusetts Institute of Technology, 2003.

\bibitem{bartel1985extended}
J~Bartel; M. Brack;~M. Durand.
\newblock Extended thomas-fermi theory at finite temperature.
\newblock {\em Nuclear Physics A}, 445(2):263--303, 1985.

\bibitem{dyson1962statistical}
F.~J. Dyson.
\newblock Statistical theory of the energy levels of complex systems. i.
\newblock {\em Journal of Mathematical Physics}, 3(1):140--156, 1962.

\bibitem{dyson1962statistical2}
F.~J. Dyson.
\newblock Statistical theory of the energy levels of complex systems. ii.
\newblock {\em Journal of Mathematical Physics}, 3(1):157--165, 1962.

\bibitem{dyson1962statistical3}
F.~J. Dyson.
\newblock Statistical theory of the energy levels of complex systems. iii.
\newblock {\em Journal of Mathematical Physics}, 3(1):166--175, 1962.

\bibitem{dyson1962threefold}
F.~J. Dyson.
\newblock The threefold way. algebraic structure of symmetry groups and
  ensembles in quantum mechanics.
\newblock {\em Journal of Mathematical Physics}, 3(6):1199--1215, 1962.

\bibitem{dyson1976fredholm}
F.~J. Dyson.
\newblock Fredholm determinants and inverse scattering problems.
\newblock {\em Communications in Mathematical Physics}, 47(2):171--183, 1976.

\bibitem{edelman1989eigenvalues}
A.~Edelman.
\newblock Eigenvalues and condition numbers of random matrices. phd thesis.
\newblock {\em Department of Mathematics, Massachusetts Institute of
  Technology, Cambridge, MA}, 1989.

\bibitem{eisler2013universality}
V.~Eisler.
\newblock Universality in the full counting statistics of trapped fermions.
\newblock {\em Physical review letters}, 111(8):080402, 2013.

\bibitem{charles2018entanglement}
L.~Charles;~B. Estienne.
\newblock Entanglement entropy and berezin-toeplitz operators.
\newblock {\em arXiv preprint arXiv:1803.03149}, 2018.

\bibitem{ezawa}
Z.~F. Ezawa.
\newblock {\em Quantum Hall Effects: Field Theoretical Approach and Related
  Topics}.
\newblock World Scientific, 2008.

\bibitem{feller1968introduction}
W.~Feller.
\newblock {\em An introduction to probability theory and its applications},
  volume~1.
\newblock Wiley New York, 1968.

\bibitem{chafai2019simulating}
D.~Chafa{\"\i};~G. Ferr{\'e}.
\newblock Simulating coulomb and log-gases with hybrid monte carlo algorithms.
\newblock {\em Journal of Statistical Physics}, 174(3):692--714, 2019.

\bibitem{Forrester1989}
P.~J. Forrester.
\newblock Vicious random walkers in the limit of a large number of walkers.
\newblock {\em Journal of Statistical Physics}, 56(5):767--782, Sep 1989.

\bibitem{forrester1998exact}
P.~J. Forrester.
\newblock Exact results for two-dimensional coulomb systems.
\newblock {\em Physics reports}, 301(1-3):235--270, 1998.

\bibitem{forrester2010log}
P.~J. Forrester.
\newblock {\em Log-gases and random matrices (LMS-34)}.
\newblock Princeton University Press, 2010.

\bibitem{forrester1993spectrum}
P.J. Forrester.
\newblock The spectrum edge of random matrix ensembles.
\newblock {\em Nuclear Physics B}, 402(3):709--728, 1993.

\bibitem{akemann2011oxford}
G.~Akemann; J. Baik; P.~Di Francesco.
\newblock {\em The Oxford handbook of random matrix theory}.
\newblock Oxford University Press, 2011.

\bibitem{ferrari2013spatial}
P.~Ferrari;~R. Frings.
\newblock On the spatial persistence for airy processes.
\newblock {\em Journal of Statistical Mechanics: Theory and Experiment},
  2013(02):P02001, 2013.

\bibitem{perfetto2017ballistic}
G.~Perfetto;~A. Gambassi.
\newblock Ballistic front dynamics after joining two semi-infinite quantum
  ising chains.
\newblock {\em Physical Review E}, 96(1):012138, 2017.

\bibitem{derrida2009current}
B.~Derrida;~A. Gerschenfeld.
\newblock Current fluctuations in one dimensional diffusive systems with a step
  initial density profile.
\newblock {\em Journal of Statistical Physics}, 137(5-6):978, 2009.

\bibitem{ginibre1965statistical}
J.~Ginibre.
\newblock Statistical ensembles of complex, quaternion, and real matrices.
\newblock {\em Journal of Mathematical Physics}, 6(3):440--449, 1965.

\bibitem{girardeau1960relationship}
M.~Girardeau.
\newblock Relationship between systems of impenetrable bosons and fermions in
  one dimension.
\newblock {\em Journal of Mathematical Physics}, 1(6):516--523, 1960.

\bibitem{girardeau1965permutation}
M.~D. Girardeau.
\newblock Permutation symmetry of many-particle wave functions.
\newblock {\em Physical Review}, 139(2B):B500, 1965.

\bibitem{girko1985circular}
V.~L. Girko.
\newblock Circular law.
\newblock {\em Theory of Probability \& Its Applications}, 29(4):694--706,
  1985.

\bibitem{islam2015measuring}
R.~Islam; R. Ma; P.~R. Preiss; M.~E. Tai; A. Lukin; M. Rispoli;~M. Greiner.
\newblock Measuring entanglement entropy in a quantum many-body system.
\newblock {\em Nature}, 528(7580):77, 2015.

\bibitem{arous2001aging}
G.~Ben Arous; A. Dembo;~A. Guionnet.
\newblock Aging of spherical spin glasses.
\newblock {\em Probability theory and related fields}, 120(1):1--67, 2001.

\bibitem{gumbel2012statistics}
E.~J. Gumbel.
\newblock {\em Statistics of extremes}.
\newblock Courier Corporation, 2012.

\bibitem{NAGAO200329}
T.~Nagao; M. Katori;~T. Hideki.
\newblock Dynamical correlations among vicious random walkers.
\newblock {\em Physics Letters A}, 307(1):29 -- 35, 2003.

\bibitem{forrester1999exact}
P.~J. Forrester;~G. Honner.
\newblock Exact statistical properties of the zeros of complex random
  polynomials.
\newblock {\em Journal of Physics A: Mathematical and General}, 32(16):2961,
  1999.

\bibitem{forrester1994complex}
P.~J. Forrester; T.~D. Hughes.
\newblock Complex wishart matrices and conductance in mesoscopic systems: exact
  results.
\newblock {\em Journal of Mathematical Physics}, 35(12):6736--6747, 1994.

\bibitem{song2011entanglement}
H.~F. Song; C. Flindt; S. Rachel; I. Klich; K.~Le Hur.
\newblock Entanglement entropy from charge statistics: Exact relations for
  noninteracting many-body systems.
\newblock {\em Physical Review B}, 83(16):161408, 2011.

\bibitem{song2012bipartite}
H.~F. Song; S. Rachel; C. Flindt; I. Klich; N. Laflorencie; K.~Le Hur.
\newblock Bipartite fluctuations as a probe of many-body entanglement.
\newblock {\em Physical Review B}, 85(3):035409, 2012.

\bibitem{regal2003tuning}
C.~A. Regal; C. Ticknor; J.~L. Bohn; D.~S. Jin.
\newblock Tuning p-wave interactions in an ultracold fermi gas of atoms.
\newblock {\em Physical review letters}, 90(5):053201, 2003.

\bibitem{baik1999distribution}
J.~Baik; P. Deift;~K. Johansson.
\newblock On the distribution of the length of the longest increasing
  subsequence of random permutations.
\newblock {\em Journal of the American Mathematical Society}, 12(4):1119--1178,
  1999.

\bibitem{johansson2000shape}
K.~Johansson.
\newblock Shape fluctuations and random matrices.
\newblock {\em Communications in mathematical physics}, 209(2):437--476, 2000.

\bibitem{johansson2005random}
K.~Johansson.
\newblock Random matrices and determinantal processes.
\newblock {\em arXiv preprint math-ph/0510038}, 2005.

\bibitem{johansson2007gumbel}
K.~Johansson.
\newblock From gumbel to tracy-widom.
\newblock {\em Probability theory and related fields}, 138(1-2):75--112, 2007.

\bibitem{klich2006lower}
I.~Klich.
\newblock Lower entropy bounds and particle number fluctuations in a fermi sea.
\newblock {\em Journal of Physics A: Mathematical and General}, 39(4):L85,
  2006.

\bibitem{dumitriu2008distributions}
I.~Dumitriu;~P. Koev.
\newblock Distributions of the extreme eigenvaluesof beta--jacobi random
  matrices.
\newblock {\em SIAM Journal on Matrix Analysis and Applications}, 30(1):1--6,
  2008.

\bibitem{kostlan1992spectra}
E.~Kostlan.
\newblock On the spectra of gaussian matrices.
\newblock {\em Linear algebra and its applications}, 162:385--388, 1992.

\bibitem{wergen_borgner_krug}
G.~Wergen; M. Bogner;~J. Krug.
\newblock Record statistics for biased random walks, with an application to
  financial data.
\newblock {\em Phys. Rev. E}, 83:051109, May 2011.

\bibitem{gohberg2012traces}
I.~Gohberg; S. Goldberg;~N. Krupnik.
\newblock {\em Traces and determinants of linear operators}, volume 116.
\newblock Birkh{\"a}user, 2012.

\bibitem{haller2015single}
E.~Haller; J. Hudson; A. Kelly; D.~A. Cotta; B. Peaudecerf; G.~D. Bruce;~S.
  Kuhr.
\newblock Single-atom imaging of fermions in a quantum-gas microscope.
\newblock {\em Nature Physics}, 11(9):738, 2015.

\bibitem{kuhr2016quantum}
S.~Kuhr.
\newblock Quantum-gas microscopes: a new tool for cold-atom quantum simulators.
\newblock {\em National Science Review}, 3(2):170--172, 2016.

\bibitem{Jayannavar1989}
A.~M. Jayannavar; G. V. Vijayagovindan;~N. Kumar.
\newblock Energy dispersive backscattering of electrons from surface resonances
  of a disordered medium and 1/f noise.
\newblock {\em Zeitschrift f{\"u}r Physik B Condensed Matter}, 75(1):77--79,
  Mar 1989.

\bibitem{cividini2017tagged}
J.~Cividini;~A. Kundu.
\newblock Tagged particle in single-file diffusion with arbitrary initial
  conditions.
\newblock {\em Journal of Statistical Mechanics: Theory and Experiment},
  2017(8):083203, 2017.

\bibitem{costin1995gaussian}
O.~Costin; J.~L. Lebowitz.
\newblock Gaussian fluctuation in random matrices.
\newblock {\em Physical Review Letters}, 75(1):69, 1995.

\bibitem{leggett2006quantum}
A.~J. Leggett.
\newblock {\em Quantum liquids: Bose condensation and Cooper pairing in
  condensed-matter systems}.
\newblock Oxford university press, 2006.

\bibitem{levitov1993charge}
L.~S. Levitov; G.~B. Lesovik.
\newblock Charge distribution in quantum shot noise.
\newblock {\em JETP LETTERS C/C OF PIS'MA V ZHURNAL EKSPERIMENTAL'NOI
  TEORETICHESKOI FIZIKI}, 58:230--230, 1993.

\bibitem{levitov1996electron}
L.~S. Levitov; H. Lee; G.~B. Lesovik.
\newblock Electron counting statistics and coherent states of electric current.
\newblock {\em Journal of Mathematical Physics}, 37(10):4845--4866, 1996.

\bibitem{klich2009quantum}
I.~Klich;~L. Levitov.
\newblock Quantum noise as an entanglement meter.
\newblock {\em Physical review letters}, 102(10):100502, 2009.

\bibitem{PhysRev.130.1605}
E.~H. Lieb;~W. Liniger.
\newblock Exact analysis of an interacting bose gas. i. the general solution
  and the ground state.
\newblock {\em Phys. Rev.}, 130:1605--1616, May 1963.

\bibitem{moshinsky1996contemporary}
Y.F.~Smirnov M.~Moshinsky.
\newblock Contemporary concepts in physics (volume 9): The harmonic oscillator
  in modern physics.
\newblock {\em Editions Harwood Academic Publishers, Amsterdam}, 1996.

\bibitem{macchi1975coincidence}
O.~Macchi.
\newblock The coincidence approach to stochastic point processes.
\newblock {\em Advances in Applied Probability}, 7(1):83--122, 1975.

\bibitem{mahan2013many}
G.~D. Mahan.
\newblock {\em Many-particle physics}.
\newblock Springer Science \& Business Media, 2013.

\bibitem{comtet2005precise}
A.~Comtet; S.~N. Majumdar.
\newblock Precise asymptotics for a random walker’s maximum.
\newblock {\em Journal of Statistical Mechanics: Theory and Experiment},
  2005(06):P06013, 2005.

\bibitem{nadal2009nonintersecting}
C.~Nadal; S.~N. Majumdar.
\newblock Nonintersecting brownian interfaces and wishart random matrices.
\newblock {\em Physical Review E}, 79(6):061117, 2009.

\bibitem{dean2001extreme}
D.~S. Dean; S.~N. Majumdar.
\newblock Extreme-value statistics of hierarchically correlated variables
  deviation from gumbel statistics and anomalous persistence.
\newblock {\em Physical Review E}, 64(4):046121, 2001.

\bibitem{dean2006large}
D.~S. Dean; S.~N. Majumdar.
\newblock Large deviations of extreme eigenvalues of random matrices.
\newblock {\em Physical review letters}, 97(16):160201, 2006.

\bibitem{dean2008extreme}
D.~S. Dean; S.~N. Majumdar.
\newblock Extreme value statistics of eigenvalues of gaussian random matrices.
\newblock {\em Physical Review E}, 77(4):041108, 2008.

\bibitem{schehr2012universal}
G.~Schehr; S.~N. Majumdar.
\newblock Universal order statistics of random walks.
\newblock {\em Physical review letters}, 108(4):040601, 2012.

\bibitem{schehr2014exact}
G.~Schehr; S.~N. Majumdar.
\newblock Exact record and order statistics of random walks via first-passage
  ideas.
\newblock In {\em First-Passage Phenomena and Their Applications}, pages
  226--251. World Scientific, 2014.

\bibitem{calabrese2015random}
P.~Calabrese; P. Le Doussal; S.~N. Majumdar.
\newblock Random matrices and entanglement entropy of trapped fermi gases.
\newblock {\em Physical Review A}, 91(1):012303, 2015.

\bibitem{krapivsky2000traveling}
P.~L. Krapivsky; S.~N. Majumdar.
\newblock Traveling waves, front selection, and exact nontrivial exponents in a
  random fragmentation problem.
\newblock {\em Physical review letters}, 85(26):5492, 2000.

\bibitem{majumdar2007brownian}
S.~N. Majumdar.
\newblock Brownian functionals in physics and computer science.
\newblock In {\em The Legacy Of Albert Einstein: A Collection of Essays in
  Celebration of the Year of Physics}, pages 93--129. World Scientific, 2007.

\bibitem{majumdar2007course}
S.~N. Majumdar.
\newblock Course 4 random matrices, the ulam problem, directed polymers \&
  growth models, and sequence matching.
\newblock {\em Les Houches}, 85:179--216, 2007.

\bibitem{majumdar2010universal}
S.~N. Majumdar.
\newblock Universal first-passage properties of discrete-time random walks and
  l{\'e}vy flights on a line: Statistics of the global maximum and records.
\newblock {\em Physica A: Statistical Mechanics and its Applications},
  389(20):4299--4316, 2010.

\bibitem{sabhapandit2007density}
S.~Sabhapandit; S.~N. Majumdar.
\newblock Density of near-extreme events.
\newblock {\em Physical review letters}, 98(14):140201, 2007.

\bibitem{krapivsky2018quantum}
P.~L. Krapivsky; J.~M. Luck;~K. Mallick.
\newblock Quantum return probability of a system of n non-interacting lattice
  fermions.
\newblock {\em Journal of Statistical Mechanics: Theory and Experiment},
  2018(2):023104, 2018.

\bibitem{krapivsky2019return}
P.~L. Krapivsky; J.~M. Luck;~K. Mallick.
\newblock Return probability of n fermions released from a 1d confining
  potential.
\newblock {\em Journal of Statistical Mechanics: Theory and Experiment},
  2019(2):023103, 2019.

\bibitem{martin1998final}
A.~Martin-L{\"o}f.
\newblock The final size of a nearly critical epidemic, and the first passage
  time of a wiener process to a parabolic barrier.
\newblock {\em Journal of Applied probability}, 35(3):671--682, 1998.

\bibitem{kohn1998edge}
W.~Kohn; A.~E. Mattsson.
\newblock Edge electron gas.
\newblock {\em Physical review letters}, 81(16):3487, 1998.

\bibitem{mehta2004random}
M.~L. Mehta.
\newblock {\em Random matrices}, volume 142.
\newblock Elsevier, 2004.

\bibitem{bouchaud1997universality}
J.~P. Bouchaud;~M. M{\'e}zard.
\newblock Universality classes for extreme-value statistics.
\newblock {\em Journal of Physics A: Mathematical and General}, 30(23):7997,
  1997.

\bibitem{grava2016tracy}
T.~Grava; A. Its; A. Kapaev;~F. Mezzadri.
\newblock On the tracy-widom $\beta $ distribution for $\beta= 6$.
\newblock {\em arXiv preprint arXiv:1607.01351}, 2016.

\bibitem{parsons2015site}
M.~F. Parsons; F. Huber; A. Mazurenko; C.~S. Chiu; W. Setiawan; K.
  Wooley-Brown; S.~Blatt; M.Greiner.
\newblock Site-resolved imaging of fermionic li 6 in an optical lattice.
\newblock {\em Physical review letters}, 114(21):213002, 2015.

\bibitem{makri1988correct}
N.~Makri; W.~H. Miller.
\newblock Correct short time propagator for feynman path integration by power
  series expansion in $\delta$t.
\newblock {\em Chemical physics letters}, 151(1-2):1--8, 1988.

\bibitem{vignolo2001one}
P.~Vignolo;~A. Minguzzi.
\newblock One-dimensional non-interacting fermions in harmonic confinement:
  equilibrium and dynamical properties.
\newblock {\em Journal of Physics B: Atomic, Molecular and Optical Physics},
  34(23):4653, 2001.

\bibitem{le2003exact}
P.~Le Doussal;~C. Monthus.
\newblock Exact solutions for the statistics of extrema of some random 1d
  landscapes, application to the equilibrium and the dynamics of the toy model.
\newblock {\em Physica A: Statistical Mechanics and its Applications},
  317(1-2):140--198, 2003.

\bibitem{kulkarni2018quantum}
M.~Kulkarni; G. Mandal;~T. Morita.
\newblock Quantum quench and thermalization of one-dimensional fermi gas via
  phase-space hydrodynamics.
\newblock {\em Physical Review A}, 98(4):043610, 2018.

\bibitem{hueck2018two}
K.~Hueck; N. Luick; L. Sobirey; J. Siegl; T. Lompe;~H. Moritz.
\newblock Two-dimensional homogeneous fermi gases.
\newblock {\em Physical review letters}, 120(6):060402, 2018.

\bibitem{embrechts1995proof}
P.~Embrechts; L.~C.~G.~Rogers; M.Yor.
\newblock A proof of dassios' representation of the $\alpha $-quantile of
  brownian motion with drift.
\newblock {\em The Annals of Applied Probability}, 5(3):757--767, 1995.

\bibitem{bloch2012quantum}
I.~Bloch; J. Dalibard;~S. Nascimbene.
\newblock Quantum simulations with ultracold quantum gases.
\newblock {\em Nature Physics}, 8(4):267, 2012.

\bibitem{luck1991low}
J.~M. Luck; M. Funke; Th.~M. Nieuwenhuizen.
\newblock Low-temperature thermodynamics of random-field ising chains: exact
  results.
\newblock {\em Journal of Physics A: Mathematical and General}, 24(17):4155,
  1991.

\bibitem{ott2016single}
H.~Ott.
\newblock Single atom detection in ultracold quantum gases: a review of current
  progress.
\newblock {\em Reports on Progress in Physics}, 79(5):054401, 2016.

\bibitem{majumdar2014extreme}
S.~N. Majumdar;~A. Pal.
\newblock Extreme value statistics of correlated random variables.
\newblock {\em arXiv preprint arXiv:1406.6768}, 2014.

\bibitem{marvcenko1967distribution}
V.~A. Mar{\v{c}}enko; L.~A. Pastur.
\newblock Distribution of eigenvalues for some sets of random matrices.
\newblock {\em Sbornik: Mathematics}, 1(4):457--483, 1967.

\bibitem{Pearson1905}
K.~Pearson.
\newblock The problem of the random walk.
\newblock {\em Nature}, 72(1867):342, 1905.

\bibitem{edelman2016beyond}
A.~Edelman; A. Guionnet;~S. P{\'e}ch{\'e}.
\newblock Beyond universality in random matrix theory.
\newblock {\em The Annals of Applied Probability}, 26(3):1659--1697, 2016.

\bibitem{chafai2014note}
D.~Chafa{\"\i};~S. P{\'e}ch{\'e}.
\newblock A note on the second order universality at the edge of coulomb gases
  on the plane.
\newblock {\em Journal of Statistical Physics}, 156(2):368--383, 2014.

\bibitem{pollaczek1952fonctions}
F.~Pollaczek.
\newblock Fonctions caract\'eristiques de certaines r\'epartitions d\'efinies
  au moyen de la notion d'ordre-application a la th\'eorie des attentes.
\newblock {\em Comptes rendus hebdomadaires des s\'eances de l'academie des
  sciences}, 234(24):2334--2336, 1952.

\bibitem{pollaczek1975order}
F.~Pollaczek.
\newblock Order statistics of partial sums of mutually independent random
  variables.
\newblock {\em Journal of Applied Probability}, 12(2):390--395, 1975.

\bibitem{port1962elementary}
S.~C. Port.
\newblock An elementary probability approach to fluctuation theory.
\newblock Technical report, NORTHWESTERN UNIV EVANSTON IL, 1962.

\bibitem{biroli2007extreme}
G.~Biroli; J.~P.~Bouchaud;M. Potters.
\newblock Extreme value problems in random matrix theory and other disordered
  systems.
\newblock {\em Journal of Statistical Mechanics: Theory and Experiment},
  2007(07):P07019, 2007.

\bibitem{amir2011probability}
G.~Amir; I. Corwin;~J. Quastel.
\newblock Probability distribution of the free energy of the continuum directed
  random polymer in 1+ 1 dimensions.
\newblock {\em Communications on pure and applied mathematics}, 64(4):466--537,
  2011.

\bibitem{eisler2013full}
V.~Eisler;~Z. R{\'a}cz.
\newblock Full counting statistics in a propagating quantum front and random
  matrix spectra.
\newblock {\em Physical review letters}, 110(6):060602, 2013.

\bibitem{PhysRevLett.101.150601}
G.~Schehr; S. N. Majumdar; A. Comtet;~J. Randon-Furling.
\newblock Exact distribution of the maximal height of $p$ vicious walkers.
\newblock {\em Phys. Rev. Lett.}, 101:150601, Oct 2008.

\bibitem{landau1981statistical}
L.~D. Landau; E.~M. Lifshitz; L.~E. Reichl.
\newblock Statistical physics, part 1.
\newblock {\em Physics Today}, 34:74, 1981.

\bibitem{rider2003limit}
B.~Rider.
\newblock A limit theorem at the edge of a non-hermitian random matrix
  ensemble.
\newblock {\em Journal of Physics A: Mathematical and General}, 36(12):3401,
  2003.

\bibitem{rider2004order}
B.~Rider.
\newblock Order statistics and ginibre's ensembles.
\newblock {\em Journal of statistical physics}, 114(3-4):1139--1148, 2004.

\bibitem{butts1997trapped}
D.~A. Butts; D.~S. Rokhsar.
\newblock Trapped fermi gases.
\newblock {\em Physical Review A}, 55(6):4346, 1997.

\bibitem{calabrese2010free}
P.~Calabrese; P. Le Doussal;~A. Rosso.
\newblock Free-energy distribution of the directed polymer at high temperature.
\newblock {\em EPL (Europhysics Letters)}, 90(2):20002, 2010.

\bibitem{texier2017physique}
C.~Texier;~G. Roux.
\newblock {\em Physique statistique: des processus {\'e}l{\'e}mentaires aux
  ph{\'e}nom{\`e}nes collectifs}.
\newblock Dunod, 2017.

\bibitem{Sabhapandit_2011}
S.~Sabhapandit.
\newblock Record statistics of continuous time random walk.
\newblock {\em EPL (Europhysics Letters)}, 94(2):20003, apr 2011.

\bibitem{krapivsky2015tagged}
P.~L. Krapivsky; K. Mallick;~T. Sadhu.
\newblock Tagged particle in single-file diffusion.
\newblock {\em Journal of Statistical Physics}, 160(4):885--925, 2015.

\bibitem{takeuchi2010universal}
K.~A. Takeuchi;~M. Sano.
\newblock Universal fluctuations of growing interfaces: evidence in turbulent
  liquid crystals.
\newblock {\em Physical review letters}, 104(23):230601, 2010.

\bibitem{jimbo1980density}
M.~Jimbo; T. Miwa; Y. M{\^o}ri;~M. Sato.
\newblock Density matrix of an impenetrable bose gas and the fifth painlev{\'e}
  transcendent.
\newblock {\em Physica D: Nonlinear Phenomena}, 1(1):80--158, 1980.

\bibitem{dhar2017exact}
A.~Dhar; A. Kundu; S.~N. Majumdar; S. Sabhapandit;~G. Schehr.
\newblock Exact extremal statistics in the classical 1d coulomb gas.
\newblock {\em Physical review letters}, 119(6):060601, 2017.

\bibitem{dhar2018extreme}
A.~Dhar; A. Kundu; S. N. Majumdar; S. Sabhapandit;~G. Schehr.
\newblock Extreme statistics and index distribution in the classical 1d coulomb
  gas.
\newblock {\em Journal of Physics A: Mathematical and Theoretical},
  51(29):295001, 2018.

\bibitem{perret2013near}
A.~Perret; A. Comtet; S.~N. Majumdar;~G. Schehr.
\newblock Near-extreme statistics of brownian motion.
\newblock {\em Physical review letters}, 111(24):240601, 2013.

\bibitem{Godr_che_2014}
C.~Godr{\`{e}}che; S. N. Majumdar;~G. Schehr.
\newblock Universal statistics of longest lasting records of random walks and
  l{\'{e}}vy flights.
\newblock {\em Journal of Physics A: Mathematical and Theoretical},
  47(25):255001, jun 2014.

\bibitem{Godr_che_2017}
C.~Godr{\`{e}}che; S. N. Majumdar;~G. Schehr.
\newblock Record statistics of a strongly correlated time series: random walks
  and l{\'{e}}vy flights.
\newblock {\em Journal of Physics A: Mathematical and Theoretical},
  50(33):333001, jul 2017.

\bibitem{dean2015finite}
D.~S. Dean; P. Le Doussal; S. N. Majumdar;~G. Schehr.
\newblock Finite-temperature free fermions and the kardar-parisi-zhang equation
  at finite time.
\newblock {\em Physical review letters}, 114(11):110402, 2015.

\bibitem{dean2015universal}
D.~S. Dean; P. Le Doussal; S. N. Majumdar;~G. Schehr.
\newblock Universal ground-state properties of free fermions in a d-dimensional
  trap.
\newblock {\em EPL (Europhysics Letters)}, 112(6):60001, 2015.

\bibitem{dean2016noninteracting}
D.~S. Dean; P. Le Doussal; S. N. Majumdar;~G. Schehr.
\newblock Noninteracting fermions at finite temperature in a d-dimensional
  trap: Universal correlations.
\newblock {\em Physical Review A}, 94(6):063622, 2016.

\bibitem{dean2017statistics}
D.~S. Dean; P. Le Doussal; S. N. Majumdar;~G. Schehr.
\newblock Statistics of the maximal distance and momentum in a trapped fermi
  gas at low temperature.
\newblock {\em Journal of Statistical Mechanics: Theory and Experiment},
  2017(6):063301, 2017.

\bibitem{dean2018wigner}
D.~S. Dean; P. Le Doussal; S. N. Majumdar;~G. Schehr.
\newblock Wigner function of noninteracting trapped fermions.
\newblock {\em Physical Review A}, 97(6):063614, 2018.

\bibitem{dean2019nonequilibrium}
D.~S. Dean; P. Le Doussal; S. N. Majumdar;~G. Schehr.
\newblock Nonequilibrium dynamics of noninteracting fermions in a trap.
\newblock {\em arXiv preprint arXiv:1902.02594}, 2019.

\bibitem{Dean_2019}
D.~S. Dean; P. Le Doussal; S. N. Majumdar;~G. Schehr.
\newblock Noninteracting fermions in a trap and random matrix theory.
\newblock {\em Journal of Physics A: Mathematical and Theoretical},
  52(14):144006, mar 2019.

\bibitem{PhysRevE.86.011119}
G.~Wergen; S. N. Majumdar;~G. Schehr.
\newblock Record statistics for multiple random walks.
\newblock {\em Phys. Rev. E}, 86:011119, Jul 2012.

\bibitem{schawe2018ground}
H.~Schawe; A. Hartmann; S.~N. Majumdar;~G. Schehr.
\newblock Ground-state energy of noninteracting fermions with a random energy
  spectrum.
\newblock {\em EPL (Europhysics Letters)}, 124(4):40005, 2018.

\bibitem{baik2012joint}
J.~Baik; K. Liechty;~G. Schehr.
\newblock On the joint distribution of the maximum and its position of the
  airy2 process minus a parabola.
\newblock {\em Journal of Mathematical Physics}, 53(8):083303, 2012.

\bibitem{grela2017kinetic}
J.~Grela; S. N. Majumdar;~G. Schehr.
\newblock Kinetic energy of a trapped fermi gas at finite temperature.
\newblock {\em Physical review letters}, 119(13):130601, 2017.

\bibitem{ramola}
K.~Ramola; S.~N. Majumdar;~G. Schehr.
\newblock Universal order and gap statistics of critical branching brownian
  motion.
\newblock {\em Physical Review Letters}, 112:210602, 2014.
\newblock 5 pages, 3 Figures.

\bibitem{ramola2}
K.~Ramola; S. N. Majumdar;~G. Schehr.
\newblock Branching brownian motion conditioned on particle numbers.
\newblock {\em {Chaos, Solitons and Fractals}}, 74:79, 2015.
\newblock 19 pages, 5 figures.

\bibitem{battilana2017gap}
M.~Battilana; S.~N. Majumdar;~G. Schehr.
\newblock Gap statistics for random walks with gamma distributed jumps.
\newblock {\em arXiv preprint arXiv:1711.08744}, 2017.

\bibitem{FORRESTER2011500}
P.~J. Forrester; S. N. Majumdar;~G. Schehr.
\newblock Non-intersecting brownian walkers and yang–mills theory on the
  sphere.
\newblock {\em Nuclear Physics B}, 844(3):500 -- 526, 2011.

\bibitem{le2016exact}
P.~Le Doussal; S. N. Majumdar; A. Rosso;~G. Schehr.
\newblock Exact short-time height distribution in the one-dimensional
  kardar-parisi-zhang equation and edge fermions at high temperature.
\newblock {\em Physical review letters}, 117(7):070403, 2016.

\bibitem{le2017periodic}
P.~Le Doussal; S. N. Majumdar;~G. Schehr.
\newblock Periodic airy process and equilibrium dynamics of edge fermions in a
  trap.
\newblock {\em Annals of Physics}, 383:312--345, 2017.

\bibitem{le2018multicritical}
P.~Le Doussal; S. N. Majumdar;~G. Schehr.
\newblock Multicritical edge statistics for the momenta of fermions in
  nonharmonic traps.
\newblock {\em Physical review letters}, 121(3):030603, 2018.

\bibitem{Mounaix_2018}
P.~Mounaix; S.~N. Majumdar;~G. Schehr.
\newblock Asymptotics for the expected maximum of random walks and l{\'{e}}vy
  flights with a constant drift.
\newblock {\em Journal of Statistical Mechanics: Theory and Experiment},
  2018(8):083201, aug 2018.

\bibitem{majumdar2014top}
S.~N. Majumdar;~G. Schehr.
\newblock Top eigenvalue of a random matrix: large deviations and third order
  phase transition.
\newblock {\em Journal of Statistical Mechanics: Theory and Experiment},
  2014(1):P01012, 2014.

\bibitem{serfaty2017systems}
S.~Serfaty.
\newblock Systems of points with coulomb interactions.
\newblock {\em arXiv preprint arXiv:1712.04095}, 2017.

\bibitem{shirai2006large}
T.~Shirai.
\newblock Large deviations for the fermion point process associated with the
  exponential kernel.
\newblock {\em Journal of statistical physics}, 123(3):615--629, 2006.

\bibitem{duenez2010lowest}
E.~Due{\~n}ez; D.~K. Huynh; J.~P. Keating; S.~J. Miller; N.~C. Snaith.
\newblock The lowest eigenvalue of jacobi random matrix ensembles and
  painlev{\'e} vi.
\newblock {\em Journal of Physics A: Mathematical and Theoretical},
  43(40):405204, 2010.

\bibitem{vallee2010airy}
O.~Vall{\'e}e;~M. Soares.
\newblock {\em Airy functions and applications to physics}.
\newblock World Scientific Publishing Company, 2010.

\bibitem{lehmann1991eigenvalue}
N.~Lehmann; H.-J. Sommers.
\newblock Eigenvalue statistics of random real matrices.
\newblock {\em Physical review letters}, 67(8):941, 1991.

\bibitem{soshnikov2002gaussian}
A.~Soshnikov.
\newblock Gaussian limit for determinantal random point fields.
\newblock {\em Annals of probability}, pages 171--187, 2002.

\bibitem{spitzer1956combinatorial}
F.~Spitzer.
\newblock A combinatorial lemma and its application to probability theory.
\newblock {\em Transactions of the American Mathematical Society},
  82(2):323--339, 1956.

\bibitem{spitzer1957wiener}
F.~Spitzer.
\newblock The wiener-hopf equation whose kernel is a probability density.
\newblock {\em Duke Mathematical Journal}, 24(3):327--343, 1957.

\bibitem{takeuchi2011growing}
K.~Takeuchi; M. Sano; T. Sasamoto;~H. Spohn.
\newblock Growing interfaces uncover universal fluctuations behind scale
  invariance.
\newblock {\em Scientific reports}, 1:34, 2011.

\bibitem{prahofer2000universal}
M.~Pr{\"a}hofer;~H. Spohn.
\newblock Universal distributions for growth processes in 1+ 1 dimensions and
  random matrices.
\newblock {\em Physical review letters}, 84(21):4882, 2000.

\bibitem{sasamoto2010one}
T.~Sasamoto;~H. Spohn.
\newblock One-dimensional kardar-parisi-zhang equation: an exact solution and
  its universality.
\newblock {\em Physical review letters}, 104(23):230602, 2010.

\bibitem{stephan2019free}
J.~M. St{\'e}phan.
\newblock Free fermions at the edge of interacting systems.
\newblock {\em arXiv preprint arXiv:1901.02770}, 2019.

\bibitem{scardicchio2009statistical}
A.~Scardicchio; C.~E.~Zachary; S.Torquato.
\newblock Statistical properties of determinantal point processes in
  high-dimensional euclidean spaces.
\newblock {\em Physical Review E}, 79(4):041108, 2009.

\bibitem{giorgini2008theory}
S.~Giorgini; L.~P. Pitaevskii;~S. Stringari.
\newblock Theory of ultracold atomic fermi gases.
\newblock {\em Reviews of Modern Physics}, 80(4):1215, 2008.

\bibitem{sutherland2004beautiful}
B.~Sutherland.
\newblock {\em Beautiful models: 70 years of exactly solved quantum many-body
  problems}.
\newblock World Scientific Publishing Company, 2004.

\bibitem{10.2307/2959757}
L.~Tak\'acs.
\newblock Random walk processes and their applications in order statistics.
\newblock {\em The Annals of Applied Probability}, 2(2):435--459, 1992.

\bibitem{TAKEUCHI201877}
K.~A. Takeuchi.
\newblock An appetizer to modern developments on the kardar–parisi–zhang
  universality class.
\newblock {\em Physica A: Statistical Mechanics and its Applications}, 504:77
  -- 105, 2018.
\newblock Lecture Notes of the 14th International Summer School on Fundamental
  Problems in Statistical Physics.

\bibitem{Grabsch2017_2}
A.~Grabsch; S.~N. Majumdar;~C. Texier.
\newblock Truncated linear statistics associated with the eigenvalues of random
  matrices ii. partial sums over proper time delays for chaotic quantum dots.
\newblock {\em Journal of Statistical Physics}, 167(6):1452--1488, Jun 2017.

\bibitem{Grabsch2017}
A.~Grabsch; S.~N. Majumdar;~C. Texier.
\newblock Truncated linear statistics associated with the top eigenvalues of
  random matrices.
\newblock {\em Journal of Statistical Physics}, 167(2):234--259, Apr 2017.

\bibitem{grabsch2018fluctuations}
A.~Grabsch; S. N. Majumdar; G. Schehr;~C. Texier.
\newblock Fluctuations of observables for free fermions in a harmonic trap at
  finite temperature.
\newblock {\em SciPost Physics}, 4(3):014, 2018.

\bibitem{texier2000individual}
C.~Texier.
\newblock Individual energy level distributions for one-dimensional diagonal
  and off-diagonal disorder.
\newblock {\em Journal of Physics A: Mathematical and General}, 33(35):6095,
  2000.

\bibitem{hagendorf2008breaking}
C.~Hagendorf;~C. Texier.
\newblock Breaking supersymmetry in a one-dimensional random hamiltonian.
\newblock {\em Journal of Physics A: Mathematical and Theoretical},
  41(40):405302, 2008.

\bibitem{giraud2018correlations}
O.~Giraud; A. Grabsch;~C. Texier.
\newblock Correlations of occupation numbers in the canonical ensemble and
  application to a bose-einstein condensate in a one-dimensional harmonic trap.
\newblock {\em Physical Review A}, 97(5):053615, 2018.

\bibitem{vignolo2000exact}
P.~Vignolo; A. Minguzzi; M.~P. Tosi.
\newblock Exact particle and kinetic-energy densities for one-dimensional
  confined gases of noninteracting fermions.
\newblock {\em Physical review letters}, 85(14):2850, 2000.

\bibitem{vignolo2002degenerate}
P.~Vignolo; A. Minguzzi; M.~P. Tosi.
\newblock Degenerate gases under harmonic confinement in one dimension:
  Rigorous results in the impenetrable-bosons/spin-polarized-fermions limit.
\newblock {\em International Journal of Modern Physics B}, 16(16):2161--2184,
  2002.

\bibitem{PhysRevA.63.033601}
M.~D. Girardeau; E. M. Wright; J.~M. Triscari.
\newblock Ground-state properties of a one-dimensional system of hard-core
  bosons in a harmonic trap.
\newblock {\em Phys. Rev. A}, 63:033601, Feb 2001.

\bibitem{amico2008entanglement}
L.~Amico; R. Fazio; A. Osterloh;~V. Vedral.
\newblock Entanglement in many-body systems.
\newblock {\em Reviews of modern physics}, 80(2):517, 2008.

\bibitem{garcia2000chiral}
A.~M. Garcia-Garcia; J.~M. Verbaarschot.
\newblock Chiral random matrix model for critical statistics.
\newblock {\em Nuclear Physics B}, 586(3):668--685, 2000.

\bibitem{majumdar2009large}
S.~N. Majumdar;~M. Vergassola.
\newblock Large deviations of the maximum eigenvalue for wishart and gaussian
  random matrices.
\newblock {\em Physical review letters}, 102(6):060601, 2009.

\bibitem{calabrese2011entanglement}
P.~Calabrese; M. Mintchev;~E. Vicari.
\newblock Entanglement entropy of one-dimensional gases.
\newblock {\em Physical review letters}, 107(2):020601, 2011.

\bibitem{calabrese2011entanglement2}
P.~Calabrese; M. Mintchev;~E. Vicari.
\newblock The entanglement entropy of one-dimensional systems in continuous and
  homogeneous space.
\newblock {\em Journal of Statistical Mechanics: Theory and Experiment},
  2011(09):P09028, 2011.

\bibitem{bloemendal2013limits}
A.~Bloemendal;~B. Vir{\'a}g.
\newblock Limits of spiked random matrices i.
\newblock {\em Probability Theory and Related Fields}, 156(3-4):795--825, 2013.

\bibitem{hough2006determinantal}
J.~B. Hough; M. Krishnapur; Y. Peres;~B. Vir{\'a}g.
\newblock Determinantal processes and independence.
\newblock {\em Probability surveys}, 3:206--229, 2006.

\bibitem{cunden2016large}
F.~D. Cunden; F. Mezzadri;~P. Vivo.
\newblock Large deviations of radial statistics in the two-dimensional
  one-component plasma.
\newblock {\em Journal of Statistical Physics}, 164(5):1062--1081, 2016.

\bibitem{cunden2017universality}
F.~D. Cunden; P. Facchi; M. Ligab{\`o};~P. Vivo.
\newblock Universality of the third-order phase transition in the constrained
  coulomb gas.
\newblock {\em Journal of Statistical Mechanics: Theory and Experiment},
  2017(5):053303, 2017.

\bibitem{livan2018introduction}
G.~Livan; M. Novaes;~P. Vivo.
\newblock {\em Introduction to Random Matrices: Theory and Practice},
  volume~26.
\newblock Springer, 2018.

\bibitem{marino2014phase}
R.~Marino; S.~N. Majumdar; G. Schehr;~P. Vivo.
\newblock Phase transitions and edge scaling of number variance in gaussian
  random matrices.
\newblock {\em Physical review letters}, 112(25):254101, 2014.

\bibitem{marino2016number}
R.~Marino; S. N. Majumdar; G. Schehr;~P. Vivo.
\newblock Number statistics for $\beta$-ensembles of random matrices:
  applications to trapped fermions at zero temperature.
\newblock {\em Physical Review E}, 94(3):032115, 2016.

\bibitem{wadia1980n}
S.~R. Wadia.
\newblock N=$\infty$ phase transition in a class of exactly soluble model
  lattice gauge theories.
\newblock {\em Physics Letters B}, 93(4):403--410, 1980.

\bibitem{allez2014index}
R.~Allez; J. Touboul;~G. Wainrib.
\newblock Index distribution of the ginibre ensemble.
\newblock {\em Journal of Physics A: Mathematical and Theoretical},
  47(4):042001, 2014.

\bibitem{wendel1960order}
J.~G. Wendel.
\newblock Order statistics of partial sums.
\newblock {\em The Annals of Mathematical Statistics}, 31(4):1034--1044, 1960.

\bibitem{majumdar2012record}
S.~N. Majumdar; G. Schehr;~G. Wergen.
\newblock Record statistics and persistence for a random walk with a drift.
\newblock {\em Journal of Physics A: Mathematical and Theoretical},
  45(35):355002, 2012.

\bibitem{tracy1993introduction}
C.~A. Tracy;~H. Widom.
\newblock Introduction to random matrices.
\newblock In {\em Geometric and quantum aspects of integrable systems}, pages
  103--130. Springer, 1993.

\bibitem{tracy1994fredholm}
C.~A. Tracy;~H. Widom.
\newblock Fredholm determinants, differential equations and matrix models.
\newblock {\em Communications in mathematical physics}, 163(1):33--72, 1994.

\bibitem{tracy1994level}
C.~A. Tracy;~H. Widom.
\newblock Level-spacing distributions and the airy kernel.
\newblock {\em Communications in Mathematical Physics}, 159(1):151--174, 1994.

\bibitem{tracy2007nonintersecting}
C.~A. Tracy;~H. Widom.
\newblock Nonintersecting brownian excursions.
\newblock {\em The Annals of Applied Probability}, 17(3):953--979, 2007.

\bibitem{tracy1994level2}
C.~Tracy;~H. Widom.
\newblock Level spacing distributions and the bessel kernel.
\newblock {\em Communications in mathematical physics}, 161(2):289--309, 1994.

\bibitem{sadhu2018generalized}
T.~Sadhu; M. Delorme; K.~J. Wiese.
\newblock Generalized arcsine laws for fractional brownian motion.
\newblock {\em Physical review letters}, 120(4):040603, 2018.

\bibitem{wigner1951statistical}
E.~P. Wigner.
\newblock On the statistical distribution of the widths and spacings of nuclear
  resonance levels.
\newblock In {\em Mathematical Proceedings of the Cambridge Philosophical
  Society}, volume~47, pages 790--798. Cambridge University Press, 1951.

\bibitem{wigner1958distribution}
E.~P. Wigner.
\newblock On the distribution of the roots of certain symmetric matrices.
\newblock {\em Ann. Math}, 67(2):325--327, 1958.

\bibitem{wigner1993characteristic}
E.~P. Wigner.
\newblock Characteristic vectors of bordered matrices with infinite dimensions
  i.
\newblock In {\em The Collected Works of Eugene Paul Wigner}, pages 524--540.
  Springer, 1993.

\bibitem{wigner1997quantum}
E.~P. Wigner.
\newblock On the quantum correction for thermodynamic equilibrium.
\newblock In {\em Part I: Physical Chemistry. Part II: Solid State Physics},
  pages 110--120. Springer, 1997.

\bibitem{wishart1928generalised}
J.~Wishart.
\newblock The generalised product moment distribution in samples from a normal
  multivariate population.
\newblock {\em Biometrika}, pages 32--52, 1928.

\bibitem{gross1993possible}
D.~J. Gross;~E. Witten.
\newblock Possible third-order phase transition in the large-n lattice gauge
  theory.
\newblock In {\em The Large N Expansion In Quantum Field Theory And Statistical
  Physics: From Spin Systems to 2-Dimensional Gravity}, pages 584--591. World
  Scientific, 1993.

\bibitem{lee1995universal}
H.~Lee; L.S. Levitov;~A.Y. Yakovets.
\newblock Universal statistics of transport in disordered conductors.
\newblock {\em Physical Review B}, 51(7):4079, 1995.

\bibitem{revuz2013continuous}
D.~Revuz;~M. Yor.
\newblock {\em Continuous martingales and Brownian motion}, volume 293.
\newblock Springer Science \& Business Media, 2013.

\bibitem{pitman2018guide}
J.~Pitman;~M. Yor.
\newblock A guide to brownian motion and related stochastic processes.
\newblock {\em arXiv preprint arXiv:1802.09679}, 2018.

\bibitem{yor1995distribution}
M.~Yor.
\newblock The distribution of brownian quantiles.
\newblock {\em Journal of Applied Probability}, 32(2):405--416, 1995.

\bibitem{anderson2010introduction}
G.~W. Anderson; A. Guionnet;~O. Zeitouni.
\newblock {\em An introduction to random matrices}, volume 118.
\newblock Cambridge university press, 2010.

\bibitem{kardar1986dynamic}
M.~Kardar; G. Parisi; Y.~C. Zhang.
\newblock Dynamic scaling of growing interfaces.
\newblock {\em Physical Review Letters}, 56(9):889, 1986.

\bibitem{majumdar_ziff}
S.~N. Majumdar; R.~M. Ziff.
\newblock Universal record statistics of random walks and l\'evy flights.
\newblock {\em Phys. Rev. Lett.}, 101:050601, Aug 2008.

\bibitem{bloch2008many}
I.~Bloch; J. Dalibard;~W. Zwerger.
\newblock Many-body physics with ultracold gases.
\newblock {\em Reviews of modern physics}, 80(3):885, 2008.

\bibitem{mukherjee2017homogeneous}
B.~Mukherjee; Z. Yan; P.~B. Patel; Z. Hadzibabic; T. Yefsah; J. Struck; M.~W.
  Zwierlein.
\newblock Homogeneous atomic fermi gases.
\newblock {\em Physical review letters}, 118(12):123401, 2017.

\bibitem{cheuk2015quantum}
L.~W. Cheuk; M.~A. Nichols; M. Okan; T. Gersdorf; V.~V. Ramasesh; W.~S. Bakr;
  T. Lompe; M.~W. Zwierlein.
\newblock Quantum-gas microscope for fermionic atoms.
\newblock {\em Physical review letters}, 114(19):193001, 2015.

\bibitem{ketterle2008making}
W.~Ketterle; M.~Z. Zwierlein.
\newblock Making, probing and understanding ultracold fermi gases.
\newblock {\em arXiv preprint arXiv:0801.2500}, 2008.

\end{thebibliography}

\chapter{Order statistics of random walks}
\label{ch:maxk}

\begin{figure}
\centering
\includegraphics[width=0.45\textwidth]{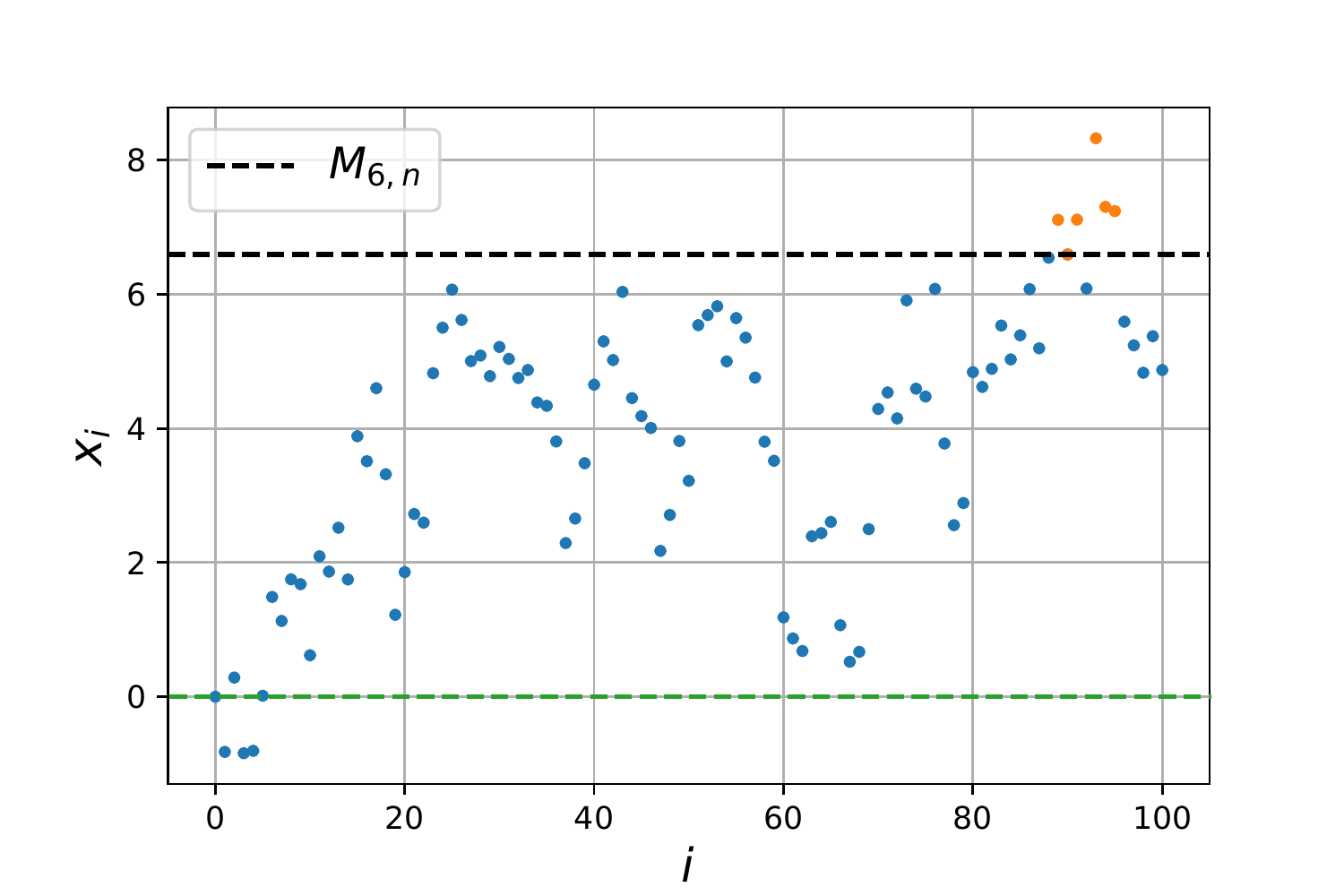}
\includegraphics[width=0.45\textwidth]{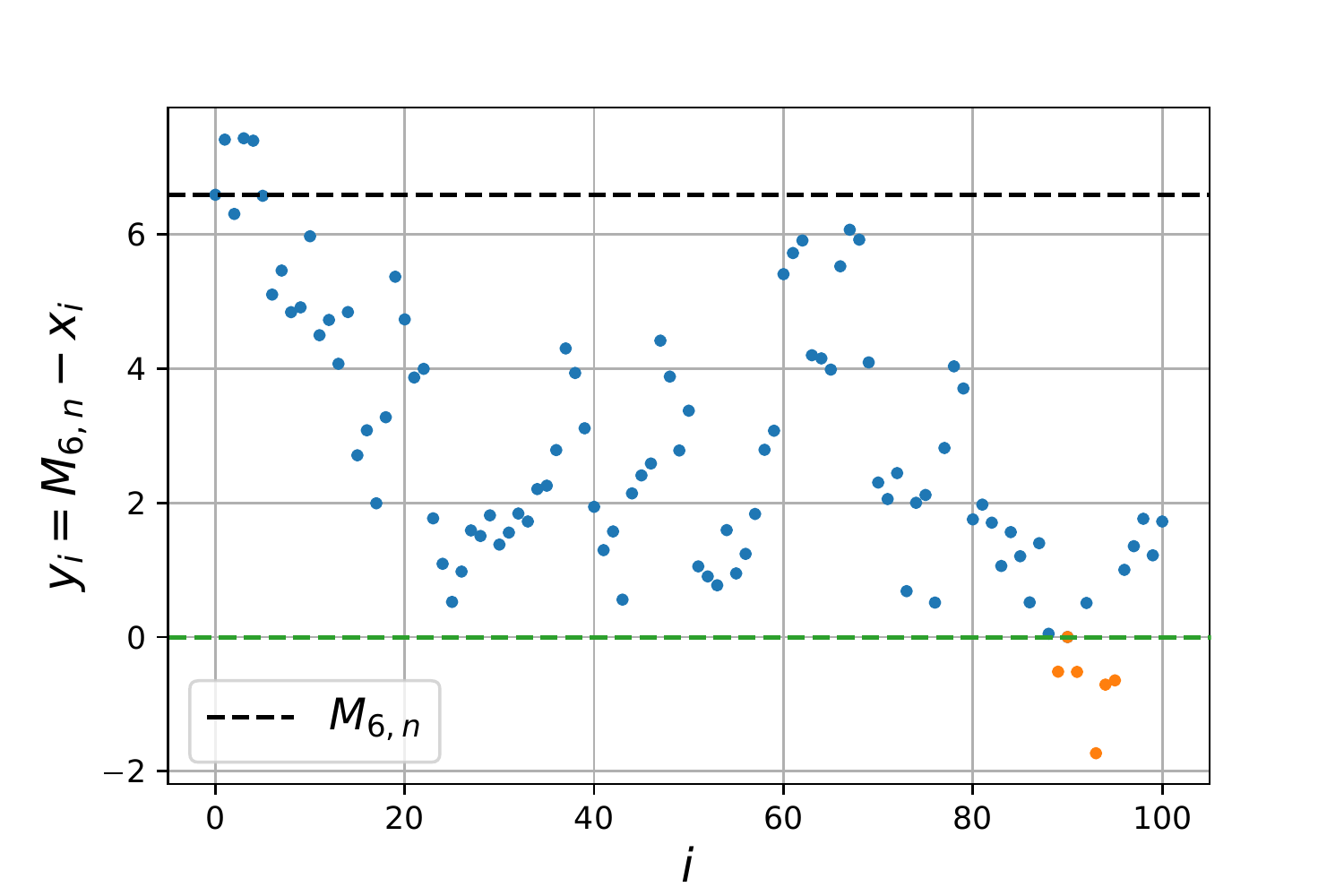}
\caption{The $6$ points of the random walk $x_i$ (in orange) lying above the level $x=M_{6,n}$ defined as its $6^{\rm th}$ maximum (represented by the dashed black line) on the left panel lie below the level $x=0$ (represented by the dashed green line) after the path transformation $y_i=M_{6,n}-x_i$ on the right panel.}\label{Fig_path_trans_max_occ}
\end{figure}

\begin{figure}
\centering
\includegraphics[width=0.45\textwidth]{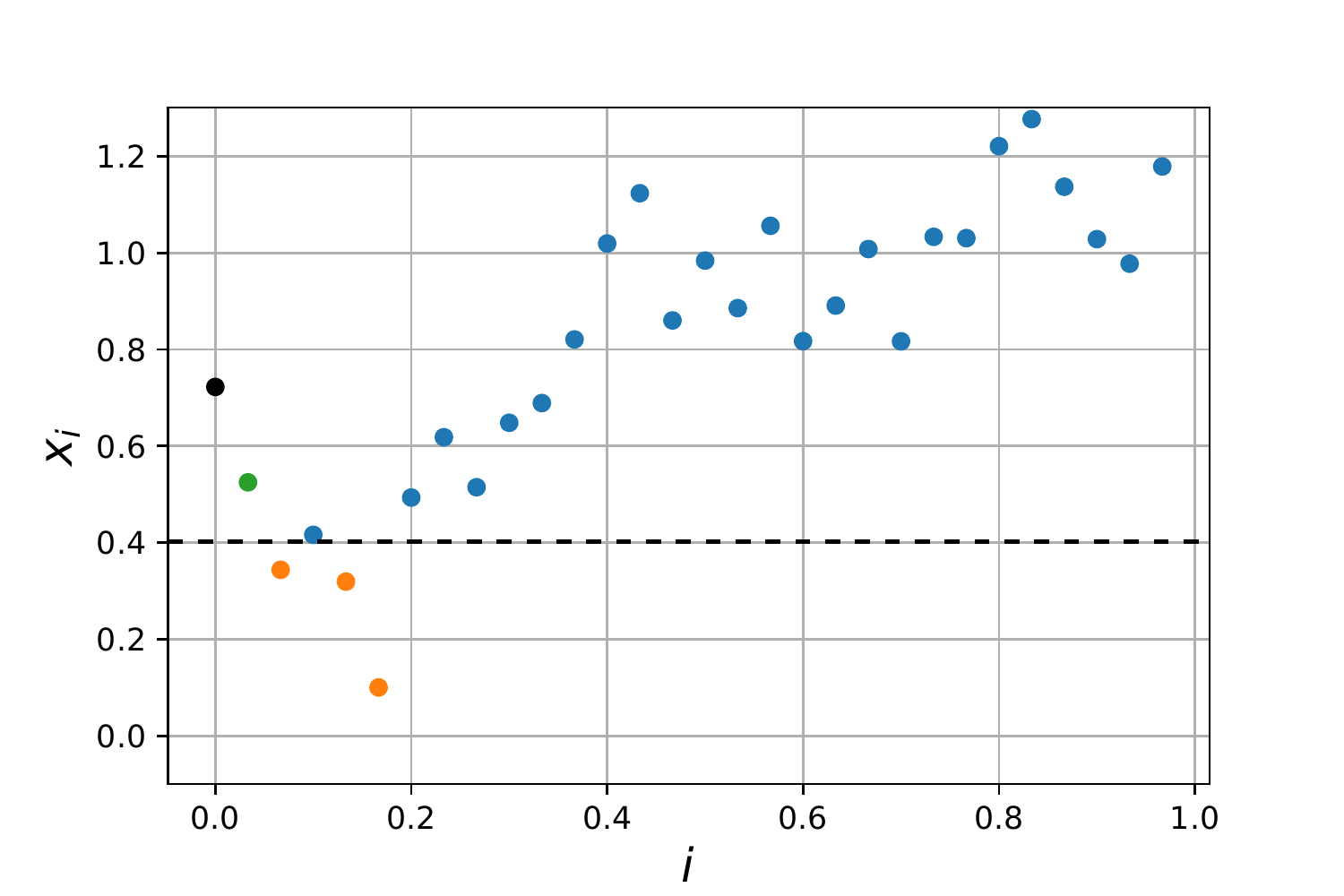}
\includegraphics[width=0.45\textwidth]{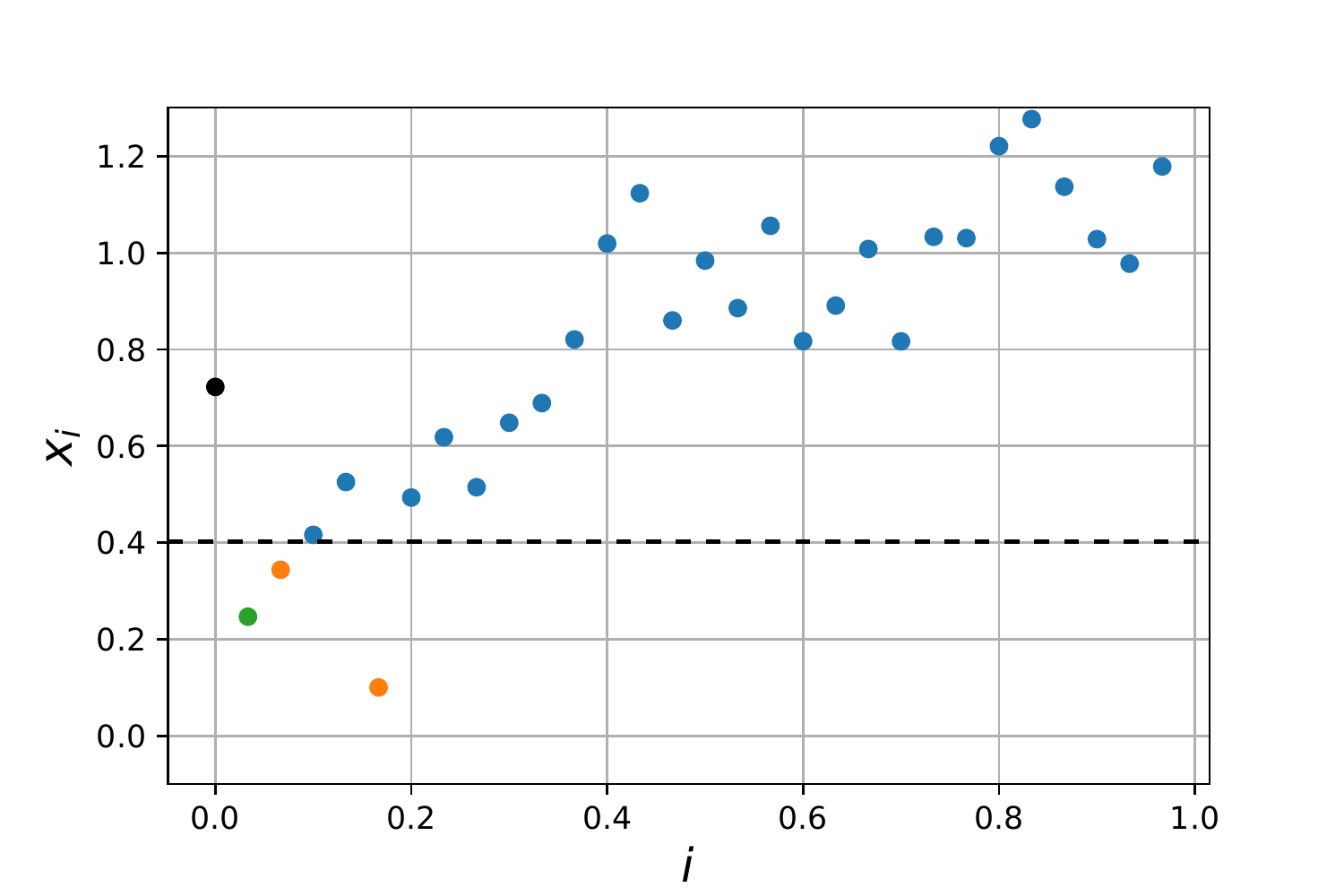}
\caption{On the left panel, the walk has an initial first step (in green) above level $x=0$ (represented by the black dashed line) and has $3$ positions below $x=0$ afterwards, for a total of $3$ positions below $x=0$. On the right panel, the walk has an initial first step (in green) below level $x=0$ (represented by the black dashed line) and has $2$ positions below $x=0$ afterwards, for a total of $3$ positions below $x=0$.}\label{Fig_prob_illus}
\end{figure}

In this chapter, we consider the symmetric random walks defined in Eq. \eqref{walk}. We order the positions of the walk by defining the $k^{\rm th}$ maximum $M_{k,n}$ such that
\be
M_{1,n}=x_{\max}\leq M_{2,n}\leq \cdots \leq M_{k,n}\leq M_{k+1,n}\leq \cdots\leq M_{n,n}\leq M_{n+1,n}=x_{\min}\;. 
\ee
Note that an alternative definition of the $k^{\rm th}$ maximum $M_{k,n}$ is the position $x_{l}=M_{k,n}$ such that there are exactly $k$ positions among the $n+1$ steps of the walk above $M_{k,n}$ (including $x_l=M_{k,n}$). It is then easy to deduce the identity 
\be
Q_{k,n}(x)=\Prob\left[M_{k,n}\leq x\right]=\Prob\left[N_{+}(x)< k\right]=\sum_{l=0}^{k-1} q_{l,n}(x)\;,\label{quant_occ_maxk}
\ee
where we introduced the probability of the occupation time $N_{+}(x)$ defined in Eq. \eqref{occ}
\be
q_{l,n}(x)=\Prob\left[N_{+}(x)=l\right]\;.
\ee
Note in particular that $q_{0,n}(x)=Q_n(x)$ is the CDF of the maximum. Introducing a path transformation by defining $y_i=x-x_i$, for $i=0,\cdots,n$, we get that $q_{l,n}(x)$ is also the probability that a walk of $n$ steps starting from position $y_0=x$ has exactly $k$ negative positions (c.f. Fig \ref{Fig_path_trans_max_occ}) 
One can then write a backward recursion relation for the probability $q_{l,n}(x)$. It reads \cite{schehr2012universal}
\be\label{WH_maxk}
q_{l,n+1}(x)=\int_{0}^{\infty}q_{l,n}(x')f(x'-x)dx'+\int_{-\infty}^{0}q_{l-1,n}(x')f(x'-x)dx'\;,\;\;x>0\;,
\ee
where $q_{0,0}(x)=1$ and $q_{l,n}(x)=0$ for $l>n$. Considering first the case where the walker arrives at a position $x'>0$ after the first jump, 
%
it must have $l$ negative positions in the remaining $n$ steps afterwards to have $l$ negative positions among its total $n+1$ positions (c.f. Fig. \ref{Fig_prob_illus}). On the contrary, if it arrives at a position $x'<0$ after the first jump, it must only have $l-1$ negative positions in the remaining $n$ steps afterwards to have a total of $l$ negative positions among its total $n+1$ positions (c.f. Fig. \ref{Fig_prob_illus}). Using $q_{l,n}(x)=\Prob\left[N_+(x)=l\right]=\Prob\left[N_-(x)=n+1-l\right]$, together with the symmetry of the walk, such that $q_{l,n}(-x)=\Prob\left[N_+(-x)=l\right]=\Prob\left[N_-(x)=l\right]$, we obtain the identity
\be
q_{l,n}(-x)=\Prob\left[N_-(-x)=n+1-l\right]=\Prob\left[N_+(x)=n+1-l\right]=q_{n+1-l,n}(x)\;.
\ee
It allows to simplify Eq. \eqref{WH_maxk} by only considering probabilities for $x>0$ as
\be
q_{l,n+1}(x)=\int_{0}^{\infty}q_{l,n}(x')f(x'-x)dx'+\int_{0}^{\infty}q_{n+2-l,n}(x')f(x'+x)dx'\;,\;\;x>0\;.\label{WH_q_n_l}
\ee
There is an exact representation for the Fourier transform in space and generating function in discrete time of the PDF $q_{k,n}'(x)$, similar to the Pollaczek-Spitzer formula in Eq. \eqref{PS}, known as the Pollaczek-Wendel formula \cite{pollaczek1952fonctions,pollaczek1975order,wendel1960order}.

This equation turns out to be rather difficult to analyse for a general jump distribution $f(\eta)$ and we will see in the following how to circumvent this difficulty. Let us first consider a few known results before stating the new results that were derived during this thesis.

\section{Distribution of the $k^{\rm th}$ maximum}

Quite remarkably, a general identity \cite{dassios1996sample,chaumont1999path} valid for any random walk $\lbrace x_i\rbrace$ (continuous or discrete, L\'evy or with finite variance jump PDF) allows to express the statistics of the $k^{\rm th}$ maximum only in terms of the statistics of the global maximum and global minimum
\be\label{k_max_id}
M_{k,n}{\buildrel d \over =} M_{1,n+1-k}+M_{k+1,k}\;,\;\;{\rm with}\;\;M_{1,n+1-k}=\max_{1\leq j\leq n+1-k} y_{j}\;\;{\rm and}\;\;M_{k+1,k}=\min_{1\leq l\leq k} z_l\;,
\ee
where the walks $y_j$ and $z_l$ are two independent realisations of random walks with the same initial position $x_0=y_0=z_0$ and jump PDF $f(\eta)$ as $x_i$. Note that for a symmetric random walk, this simplifies to
\be\label{k_max_id_sym}
M_{k,n}{\buildrel d \over =} M_{1,n+1-k}-M_{1,k-1}\;.
\ee
Therefore, if we know the PDF $F_{1,n}(x)=\partial_x Q_n(x)$ of the global maximum for any values of $n$ and $x$, then we may compute the PDF $F_{k,n}(x)=\partial_x Q_{k,n}(x)$ of $M_{k,n}$ for any value of $k$ as the convolution
\be\label{maxk}
F_{k,n}(x)=\int_{-\infty}^{\infty}F_{1,n+1-k}(z)F_{1,k-1}(z-x)dz\;.
\ee
This equation allows to circumvent the difficulty of solving the equation \eqref{WH_q_n_l}. Note that if one considers a random walk in discrete time and space, referred to in the following as {\it discrete random walk}, we expect in general that the value of the $k^{\rm th}$ maximum will be reached several times, i.e. at several steps. In this case, the identity in Eq. \eqref{k_max_id} remains valid if we order the positions that have same value in decreasing order of time step, i.e. if $x_l=x_m$, and $m>l$ then $x_m=M_{k,n}$ and $x_l=M_{k+1,n}$ \cite{chaumont1999path}. We now use this identity together with Eq. \eqref{mean_max_mu} to obtain the mean value of $M_{k,n}$. For general value of the L\'evy index $\mu$ (including $\mu=2$), we obtain for large $n$ and $k$,
\be\label{mean_max_k_mu}
\frac{\moy{M_{k,n}}}{a_\mu}\approx {\cal M}_{\mu}\left(\frac{k}{n}\right)\;\;{\rm with}\;\;{\cal M}_{\mu}\left(\alpha\right)=\frac{\mu}{\pi}\Gamma\left(1-\frac{1}{\mu}\right)\left[(1-\alpha)^{1/\mu}-\alpha^{1/\mu}\right]\;,\;\;\mu> 1\;.
\ee
For $\mu\leq 1$, the mean value of the jump PDF is not defined and $|\moy{M_{k,n}}|=\infty$ for all $k$ and $n>0$. More generally, in the case of a random walk with L\'evy index $0<\mu<2$, the variance is infinite $\sigma^2=\infty$. As seen in the previous section, the random walk does not converge towards a Brownian motion but towards a L\'evy flight. The PDF of the maximum of this process is not known explicitly for general values of $\mu$. Therefore, even though the identity \eqref{k_max_id_sym} remains valid, it cannot be used in practice to compute explicitly the PDF of the quantile of L\'evy flight. As there is no simple generalisation of the Feynman-Ka\v c formalism to L\'evy flights, we cannot either compute the PDF of the occupation time $N_+(x)$, which would allow us to obtain the distribution of the quantiles using Eq. \eqref{quant_occ_maxk}.

For random walks with finite variance jump PDF, we obtain
\be\label{mean_max_k_BM}
\frac{\moy{M_{k,n}}}{\sigma}\approx {\cal M}\left(\frac{k}{n}\right)\;\;{\rm with}\;\;{\cal M}\left(\alpha\right)=\frac{{\cal M}_{\mu=2}(\alpha)}{\sqrt{2}}=\sqrt{\frac{2}{\pi}}\left[\sqrt{1-\alpha}-\sqrt{\alpha}\right]\;.
\ee
As expected, the rescaled mean value is antisymmetric under the change $\alpha\to 1-\alpha$. In this case, the full PDF of $M_{k,n}$ can be obtained using the convergence to Brownian motion. We have seen in the previous section that the quantiles of Brownian motion are naturally defined as the analogous of the $k^{\rm th}$ maximum for a continuous process (see \eqref{def_quant} and the discussion afterwards). In the large $n$ limit, we therefore have
\be\label{q_a_m_k_n}
q(\alpha)=\lim_{n\to \infty} \frac{M_{\alpha n,n}}{\sigma \sqrt{n}}\;.
\ee
From this property, we expect the PDF of the $k^{\rm th}$ maximum for $n\to \infty$ with $\alpha=k/n$ to take the scaling form
\be
F_{k,n}(x)\approx \frac{1}{\sqrt{n}\sigma}P_{\alpha=k/n}\left(\frac{x}{\sqrt{n}\sigma}\right)\;,
\ee
where $P_{\alpha}(z)$ is the PDF of the alpha quantile of Brownian motion. We already obtained the CDF in Eq. \eqref{CDF_quantiles}, yielding
\be\label{PDF_quantile}
P_{\alpha}(z)=\begin{cases}
\displaystyle \sqrt{\frac{2}{\pi}}e^{-\frac{z^2}{2}}\erfc\left(z\sqrt{\frac{\alpha}{2(1-\alpha)}}\right)&\;,\;\;z\geq 0\\
&\\
\displaystyle \sqrt{\frac{2}{\pi}}e^{-\frac{z^2}{2}}\erfc\left(|z|\sqrt{\frac{1-\alpha}{2\alpha}}\right)&\;,\;\;z< 0\;.
\end{cases}
\ee
As the distribution of the global maximum $M_{1,n}$ is known explicitly for the Brownian motion, this PDF can alternatively be obtained using Eq. \eqref{maxk} \cite{dassios1995distribution}. This distribution has the symmetry $P_{\alpha}(-z)=P_{1-\alpha}(z)$. Note that taking the limit $\alpha\to 0$ (resp. $\alpha\to 1$), we recover the half Gaussian distribution of the maximum (resp. minimum). We now consider the distribution of the time to reach this $k^{\rm th}$ maximum.

\begin{figure}
\centering
\includegraphics[width=0.6\textwidth]{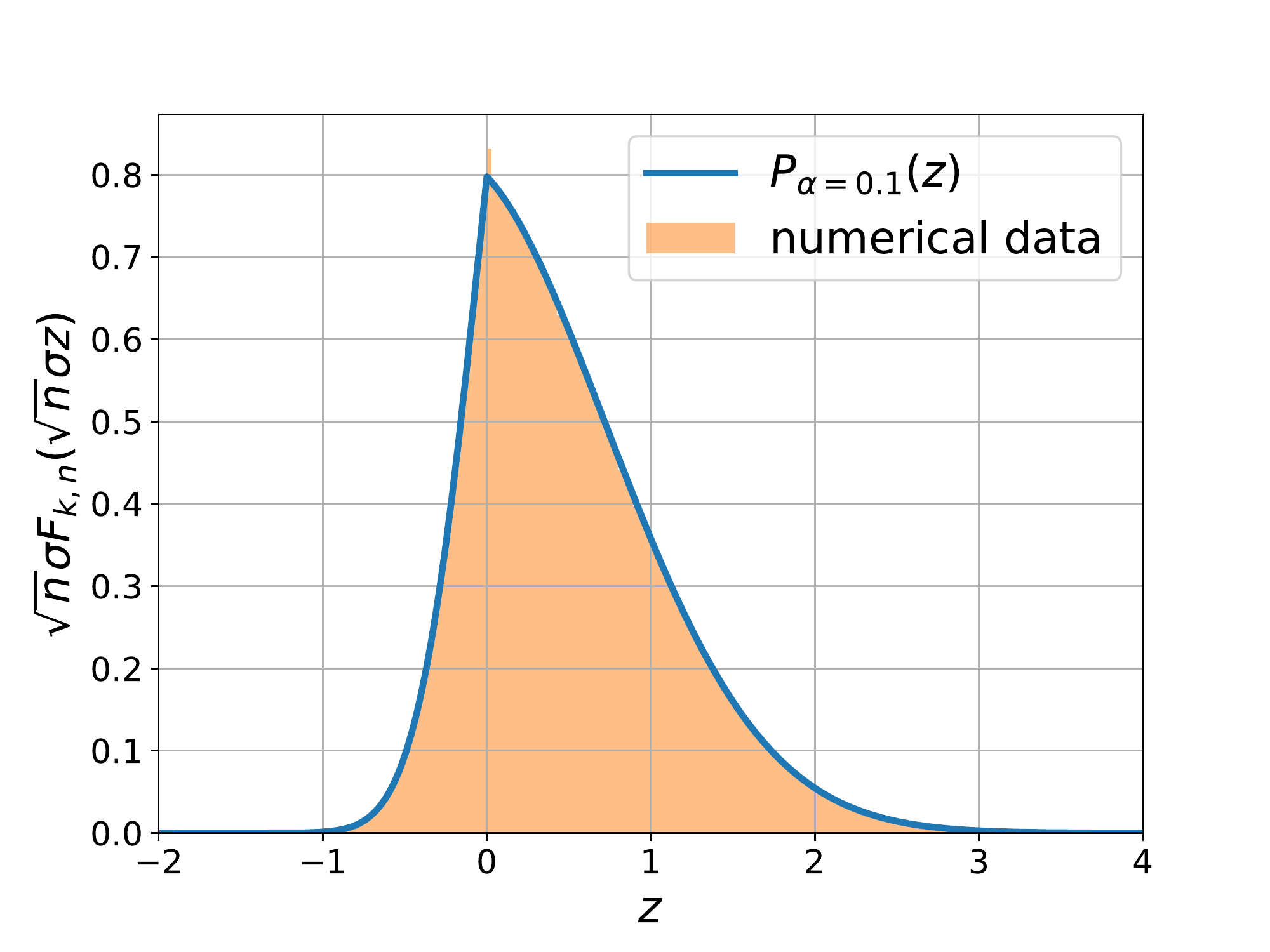}
\caption{Comparison between the rescaled PDF $\sqrt{n}\sigma F_{k,n}(\sqrt{n}\sigma z)$ of the maximum $M_{k,n}$ for $k=10^2$ and $n=10^3$ as a function of $z$ obtained from the simulation of $10^6$ random walks with Gaussian jump PDF and the scaling function $P_{\alpha}(z)$ in Eq. \eqref{PDF_quantile}. The numerical data shows a very good agreement with the analytical results.}\label{Fig_max_RW}
\end{figure}

\section{Time to reach the $k^{\rm th}$ maximum}

\begin{figure}
\centering
\includegraphics[width=0.6\textwidth]{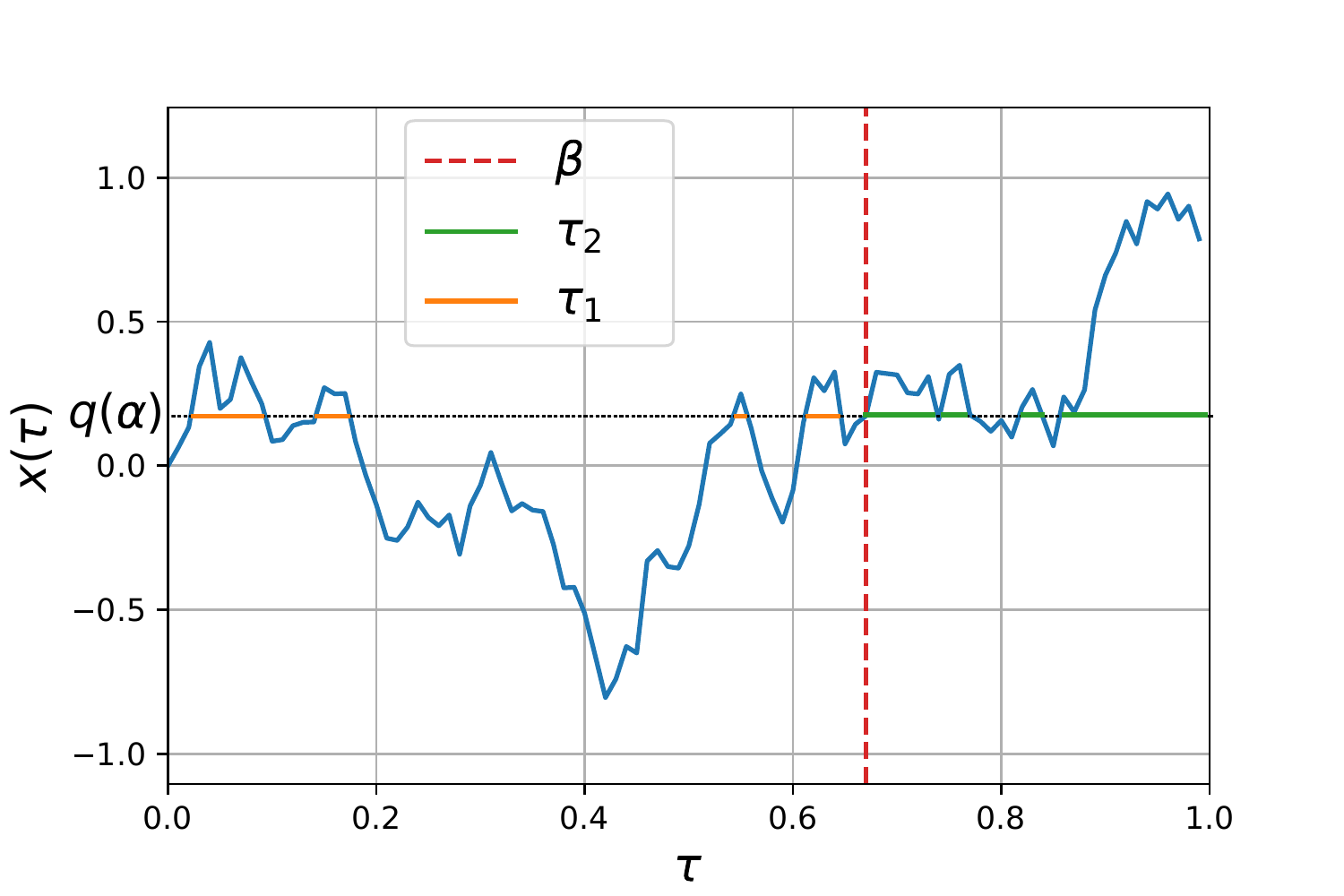}
\caption{Trajectory of a Brownian motion decomposed in two parts. In the first part $\tau\in[0,\beta]$, the Brownian spends a time $\tau_1$ above level $q(\alpha)$and reaches $q(\alpha)$ at $\tau=\beta$. In the second part $\tau\in[\beta,1]$, the Brownian starts from $q(\alpha)$ and spends a time $\tau_2$ above level $q(\alpha)$. As $q(\alpha)$ is the $\alpha$-quantile, $\tau_1+\tau_2=\alpha$.}\label{Fig_t_alphaq}
\end{figure}

\begin{figure}
\centering
\includegraphics[width=0.6\textwidth]{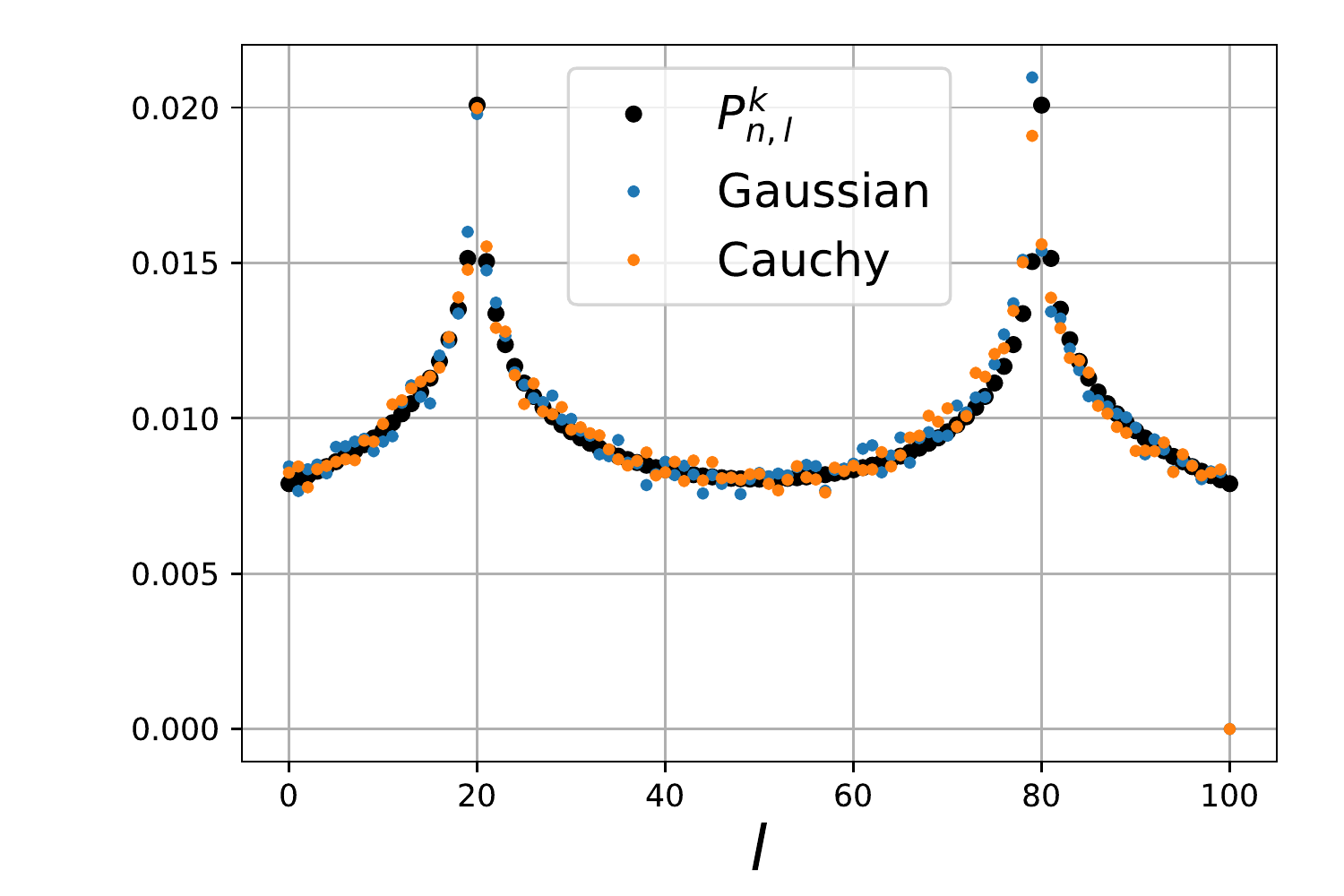}
\caption{Comparison between the analytical expression in Eq. \eqref{t_reach_max_k} (black) for $n=100$ and $k=20$ and simulations of random walks with Gaussian (blue) and Cauchy (orange) jump PDF, showing an excellent agreement.}\label{Fig_t_reach_max_k}
\end{figure}

It is rather easy to generalise the computation for the time to reach the global maximum to this more general setting of the $k^{\rm th}$ maximum. Considering that it is reached at the step $l$ such that $x_l=M_{k,n}$, we define the two independent random walks $y_{j}=x_{j}-x_{l}$ for $j=0,\cdots,l$ and $z_{m}=x_{m}-x_l$ for $m=l+1,\cdots,n$. The total occupation time of the walk $x_i$ above $x_l$ is $N_+(x_l)=k$ and must be splitted in each sub-walk $y_j$ and $z_m$. Therefore, there is respectively a number $i=N_+(0)$ for the walk $y_j$ and $k-i=N_+(0)$ for the walk $z_m$ of steps above $z_0=y_0=0$. Summing over all the possibilities for $i$, we obtain
\be
\boxed{\frame{\boxed{\begin{array}{rl}
\displaystyle P_{n,l}^{k}&\displaystyle =\sum_{i=\max(1,l+k-n)}^{\min(n+1,k)} P_{l,i-1}P_{n-l,k-i}\\
&\displaystyle=2^{-2n}\sum_{i=\max(1,l+k-n)}^{\min(n+1,k)}{{2(i-1)}\choose{i-1}}{{2(k-i)}\choose{k-i}}{{2(l+1-i)}\choose{l+1-i}}{{2(n-l-k+i)}\choose{n-l-k+i}}\;.
\end{array}}}}\label{t_reach_max_k}
\ee
To the best of our knowledge, this formula appears neither in the physics literature nor in the {\it fluctuation theory} studied in maths. Note that for $k=1$, as $x_l=x_{\max}$, all the positions $y_j=x_{\max}-x_j< 0$ for $j>0$ such that only the term for $i=1$ is non-zero. We thus recover the time to reach the global maximum $P_{n,l}^{1}=P_{n,l}=2^{-2n}{{2l}\choose{l}}{{2(n-l)}\choose{n-l}}$. Similarly, setting $k=n+1$, as $x_l=x_{\min}$, all the positions $y_j=x_{\min}-x_j\geq 0$ for $j\geq 0$ such that only the term for $i=l+1$ is non-zero. Therefore we also have $P_{n,l}^{n+1}=P_{n,l}$. For a fixed $l$, the distribution is symmetric under $k\to n-k$, i.e. $P_{n,l}^{k}=P_{n,l}^{n-k}$, while for fixed $k$ it is symmetric under $l\to n-l$, i.e. $P_{n,n-l}^{k}=P_{n,l}^{k}$. In the large $n$ limit, we recover a PDF similar to the arcsine law but with divergences for the values $l=k$ and $l=n-k$. This probability is plotted in Fig. \ref{Fig_t_reach_max_k} for $n=100$ and $k=20$ for a Gaussian and a Cauchy distribution of jumps (with  infinite variance), highlighting the universality of this result which is valid for {\it any} symmetric distribution of jumps. In the large $n$ limit, this formula converges to the scaling form
\be
\boxed{\frame{\boxed{P_{n,l}^{k}\approx \frac{1}{n}{\cal P}_{\rm reach}\left(\frac{k}{n},\frac{l}{n}\right)\;,\;\;{\rm with}\;\;{\cal P}_{\rm reach}(\alpha,\beta)=\int_0^{1} \frac{\Theta(\alpha-\tau)\Theta(\beta-\tau)\Theta(1+\tau-\alpha-\beta)d\tau}{\pi^2\sqrt{\tau(\alpha-\tau)(\beta-\tau)(1+\tau-\alpha-\beta)}}\;.}}}
\ee
Note that this formula can be obtained directly in the continuous case by separating the path in two independent parts, the interval $[0,\beta]$ and the interval $[\beta,1]$. For each independent part, the PDF of the occupation time is given by the arcsine law and we simply ensure that the sum of the two times is exactly $\alpha$ (see Fig. \ref{Fig_t_alphaq})
\be
{\cal P}_{\rm reach}(\alpha,\beta)=\int_0^1 d\tau_1 \int_0^1 d\tau_2 \frac{\Theta(\beta-\tau_1)\Theta(1-\beta-\tau_2)\delta(\tau_1+\tau_2-\alpha)}{\sqrt{\tau_1(\beta-\tau_1)\tau_2(1-\beta-\tau_2)}}\;.
\ee

\begin{figure}
\centering
\includegraphics[width=0.6\textwidth]{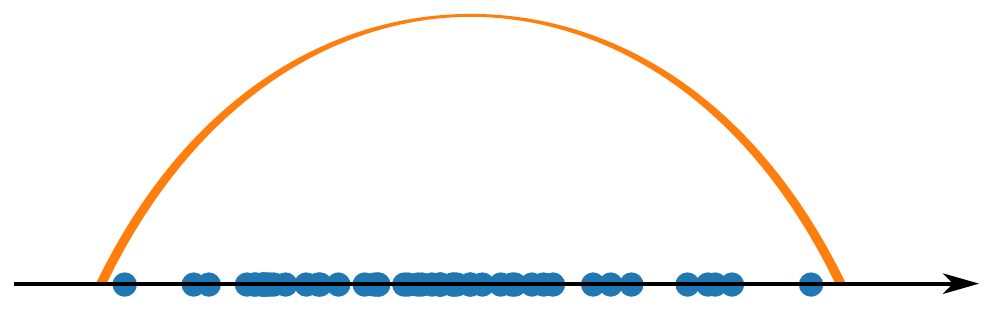}
\caption{Scheme of the repartition of maxima of the random walk for $n=50$ steps and of the associated density.}\label{Fig_quasi_cont}
\end{figure}

\section{Spatial annealed and quenched densities for the random walk}

In the limit of large $n$, the positions $M_{k,n}$ of the maxima form a quasi-continuum on the real axis (c.f. Fig. \ref{Fig_quasi_cont}). We may define a point process associated to these maxima. To first characterise this point process we will now compute its density. The most natural way to define the density of this point process is
\be
\rho_{n}^{\rm a}(x)=\frac{1}{n+1}\sum_{l=0}^{n}\moy{\delta(x-x_l)}=\frac{1}{n+1}\sum_{k=1}^{n+1}\moy{\delta(x-M_{k,n})}=\frac{1}{n+1}\sum_{k=1}^{n+1} F_{k,n}(x)\;.
\ee
Note that from the normalisation of $F_{k,n}(x)$, the density $\rho_n^{\rm a}(x)$ is also normalised to unity.
This average density can be obtained from the mean occupation time $N_+(x)$, as follows
\be
\rho_{n}^{\rm a}(x)=-\frac{1}{n+1}\partial_x\left[\sum_{k=0}^{n}\moy{\Theta(x_k-x)}\right]=-\frac{1}{n+1}\partial_x \moy{N_+(x)}\;.
\ee
This annealed density is naturally expressed in terms of the mean local time $\moy{\tau_{\rm loc}(x)}$ at position $x$, where this local time is defined as
\be
\tau_{\rm loc}(x)=\sum_{k=0}^n \delta(x_k-x)\;.
\ee

For a PDF $F_{k,n}(x)$ that is peaked about its mean value $\moy{M_{k,n}}$, we expect that it coincides with the 'typical' density defined as
\be
\rho_{n}^{\rm q}(x)=\frac{1}{n+1}\sum_{k=1}^{n+1}\delta(x-\moy{M_{k,n}})\;.
\ee 
However, we already saw from the PDF of the global maximum $M_{1,n}$, that the mean value $\moy{M_{1,n}}\approx \sqrt{2n/\pi}$ does not coincide with the maximum of $F_{1,n}(x)$ which is reached for $x=0$. Therefore we will use these two densities to characterise the point process. We distinguish the {\it annealed} density $\rho_{n}^{\rm a}(x)$ that will be obtained by averaging over several realisations of the process from the {\it quenched} density $\rho_{n}^{\rm q}(x)$ which is obtained for a typical realisation of the process. The terms of quenched and annealed densities appear more naturally when considering the Fourier transform of these PDFs, as
\be
\tilde \rho_{n}^{\rm a}(k)=\frac{1}{n+1}\sum_{k=1}^{n+1}\moy{e^{\I M_{k,n}}}\;,\;\;{\rm while}\;\;\tilde \rho_{n}^{\rm q}(k)=\frac{1}{n+1}\sum_{k=1}^{n+1}e^{\I \moy{M_{k,n}}}\;.
\ee
Note that in this case, the terms ``annealed'' and ``quenched'' do not refer to a disorder average but rather on an average with respect to a flat measure for all values of $k$.

We will now analyse in detail the behaviour of these maxima in the large $n$ limit.


%

\subsection{Spatial annealed and quenched densities for Brownian motion}

\begin{figure}
\centering
\includegraphics[width=0.6\textwidth]{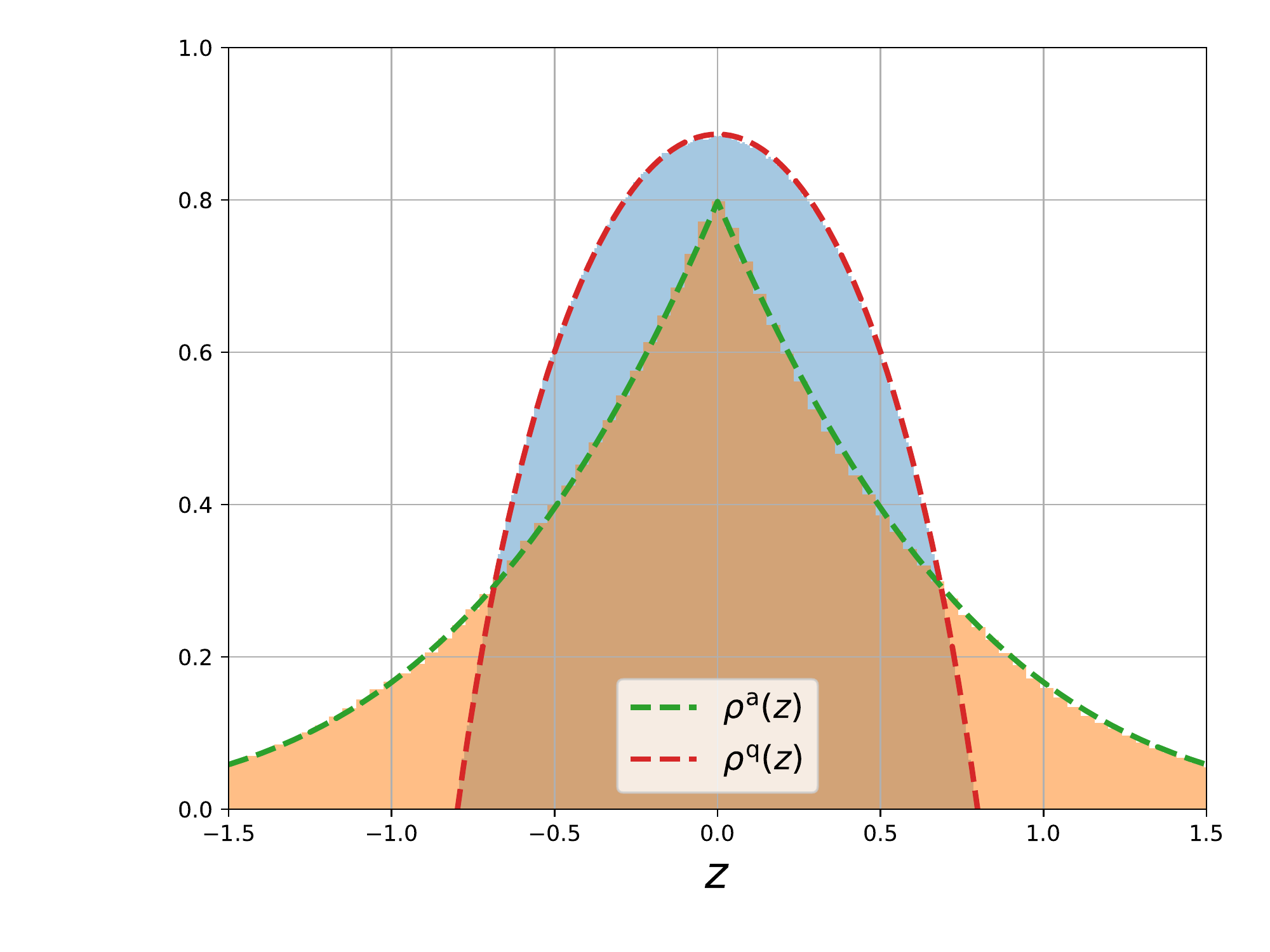}
\caption{Comparison between the rescaled annealed (orange) and quenched (blue) densities $\sqrt{n}\rho_n^{\rm a, q}(\sqrt{n}z)$ of maxima $M_{k,n}$ for $n=10^5$ as a function of $z$ obtained from the simulation of $10^5$ random walks with Gaussian jump PDF ($\mu=2$) and the scaling functions $\rho^{\rm a,q}(z)$ (respectively in dashed green and red)  in Eqs. \eqref{rho_a_BM} and \eqref{rho_q_BM}. The numerical data shows a very good agreement with the analytical results.}\label{Fig_quenched_bro}
\end{figure}

In the large $n$ limit, the full process of the random walk converges towards the Brownian motion. We will now try to characterize the process of the maxima by first computing their associated densities in the large $n$ limit. We expect from the Brownian scaling that the annealed and quenched density of the random walk with finite variance takes in the large $n$ limit the scaling form,
\be\label{scal_density}
\rho_n^{\rm a, q}(x)\approx \frac{1}{\sigma\sqrt{n}} \rho^{\rm a, q}\left(\frac{x}{\sigma\sqrt{n}}\right).
\ee
Using the Markov propagator of Brownian motion defined in Eq. \eqref{prop_BM}, we may compute explicitly the annealed density
\be
\boxed{\frame{\boxed{\rho^{\rm a}(z)=\int_0^{1}dt\moy{\delta(x(t)-z)}=\int_0^{1} G(z,\tau)d\tau=\sqrt{\frac{2}{\pi}}e^{-\frac{z^2}{2}}-|z|\erfc\left(\frac{|z|}{\sqrt{2}}\right)\;.}}}\label{rho_a_BM}
\ee
This density is symmetric and does not have a finite edge but vanishes at infinity as $\rho^{\rm a}(z)\approx \sqrt{2/\pi}z^{-2} e^{-z^2/2}$. For a single realisation, it is clear that the Brownian motion always has a finite maximum and minimum and the 'typical' density must also have finite edges. 
The quenched density for Brownian motion can be obtained from the result for the mean value of $M_{k,n}$ in Eq. \eqref{mean_max_k_BM}. Taking the large $n$ limit, we obtain
\be
\rho^{\rm q}(z)=\int_{0}^1 d\alpha \delta\left(z-{\cal M}(\alpha)\right)=\int_{0}^1 \frac{d\alpha}{|{\cal M}'(\alpha)|}\delta(\alpha-{\cal A}(z))=\frac{1}{|{\cal M}'({\cal A}(z))|}=|\partial_z {\cal A}(z)|\;,\label{quenched_BM_1}
\ee
where the function ${\cal A}(z)$ is the inverse function of ${\cal M}(\alpha)$ defined in Eq. \eqref{mean_max_k_BM}. In the case of the Brownian motion, this inverse function can be obtained explicitly ${\cal A}(z)=\frac{1}{2}-\frac{z}{4}\sqrt{\pi(4-\pi z^2)}$ and we may therefore obtain the quenched density of Brownian motion
\be
\boxed{\frame{\boxed{\rho^{\rm q}(z)=\int_{0}^1 d\alpha \delta\left(z-{\cal M}(\alpha)\right)=-\partial_z {\cal A}(z)=\frac{\sqrt{\pi}}{2}\frac{2-\pi z^2}{\sqrt{4-\pi z^2}}=\frac{1}{z_{\rm e}^2}\frac{z_{\rm e}^2-z^2}{\sqrt{2 z_{\rm e}^2-z^2}}\;,}}}\label{rho_q_BM}
\ee
where the edges of this symmetric density are at $\pm z_{\rm e}=\pm \sqrt{\frac{2}{\pi}}$. At these edges the density vanishes linearly $\rho^{\rm q}(z)\approx 4\pi(z_{\rm e}-z)$. 
Note that we have also computed these densities for Brownian bridges, Brownian excursions and reflected Brownian motion and obtained that the linear behaviour of the quenched density close to its edge is universal. The quenched and annealed densities are plotted in Fig. \ref{Fig_quenched_bro} together with numerical data, showing excellent agreement.  

 This concludes our study of $M_{k,n}$ in the case of finite variance jump PDF. We will now analyse the case of L\'evy flights, where the process does not converge to Brownian motion.

\subsection{Spatial annealed and quenched densities for L\'evy flights}

\begin{figure}
\centering
\includegraphics[width=0.6\textwidth]{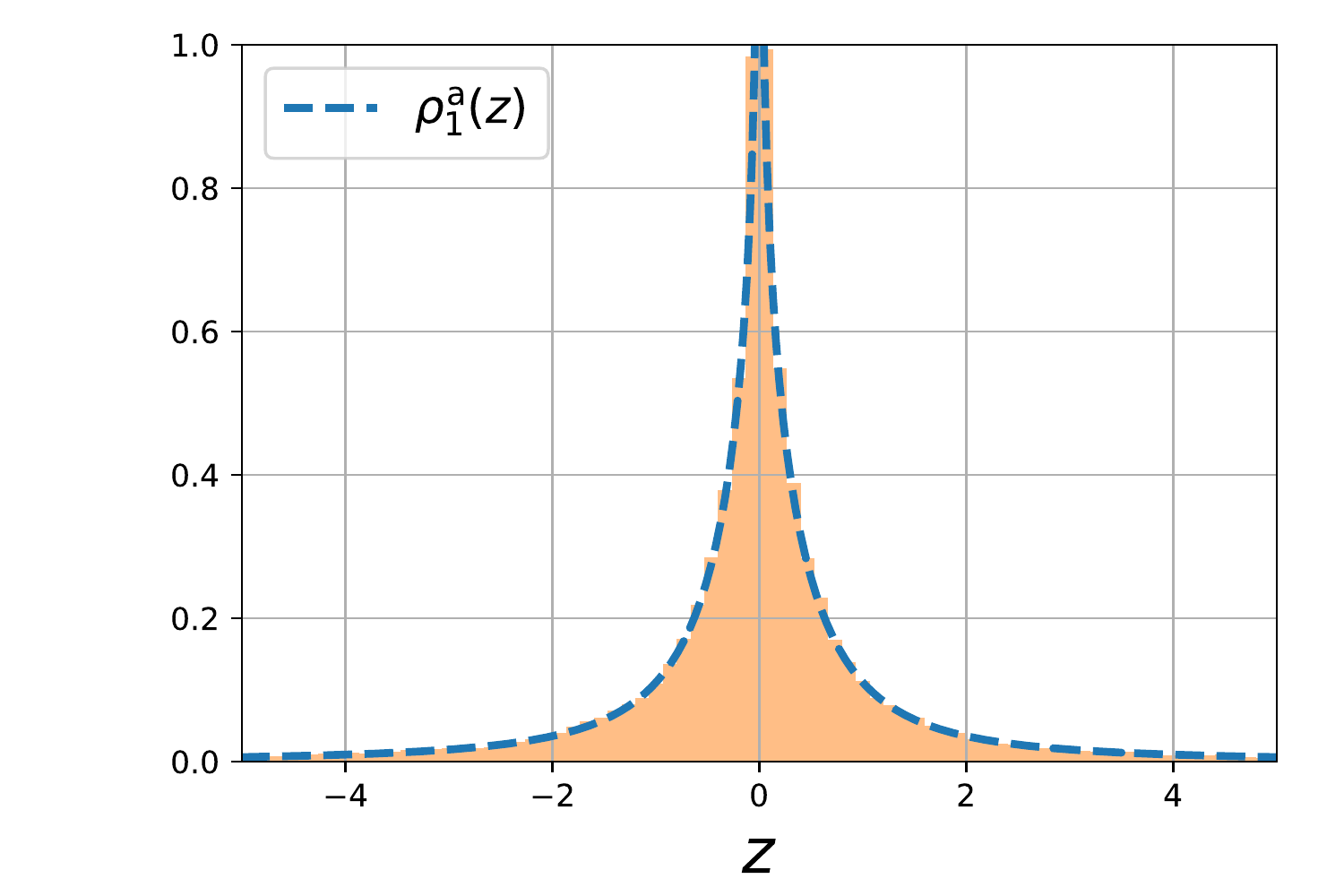}
\caption{Comparison between the rescaled annealed density $n \rho_n^{\rm a}(n z)$ of maxima $M_{k,n}$ for $n=10^4$ as a function of $z$ obtained from the simulation of $10^4$ random walks with Cauchy jump PDF ($\mu=1$) and the scaling function $\rho_1^{\rm a}(z)=\ln(1+z^{-2})/(2\pi)$ in Eq. \eqref{annealed_dens}. This density diverges logarithmically close to the origin. The numerical data shows a very good agreement with the analytical results.}\label{Fig_annealed cauchy}
\end{figure}

As seen previously, few information have been obtained analytically to characterised the quantiles of L\'evy flights. As a first step to characterise the process of these quantiles, we may compute the quenched and annealed densities of the process. From the scaling of the propagator in Eq. \eqref{levy_scal}, we expect the densities to take the scaling forms
\be
\rho_n^{\rm a,q}(x)\approx \frac{1}{a_\mu n^{1/\mu}} \rho^{\rm a,q}_{\mu}\left(\frac{x}{a_\mu n^{1/\mu}}\right)\;,
\ee
where $\rho^{\rm a,q}_{\mu}(z)$ is the annealed (resp. quenched) density of the L\'evy flight of index $\mu$.
The annealed density can be computed from the propagator ${\cal L}_{\mu}(z)=\int_{-\infty}^{\infty}\frac{dk}{2\pi}e^{-\I k z-|k|^{\mu}}$ of the L\'evy flight as
\be
\boxed{\frame{\boxed{\rho^{\rm a}_{\mu}(z)=\int_0^{1}d\tau \frac{1}{\tau^{1/\mu}}{\cal L}_{\mu}\left(\frac{z}{\tau^{1/\mu}}\right)=\int_0^{\infty}\frac{dk}{\pi k^{\mu}}\left(1-e^{-k^{\mu}}\right)\cos(k z)\;.}}}\label{annealed_dens}
\ee
Taking $\mu=2$ in this expression, we recover $\rho^{\rm a}_{\mu=2}(z)=2^{-1/2}\rho^{\rm a}\left(z/\sqrt{2}\right)$. For a value $0<\mu<2$, inserting the large $z$ behaviour of ${\cal L}_{\mu}(z)\sim z^{-\mu-1}$ in Eq. \eqref{annealed_dens}, we obtain the same algebraic decay of the density $\rho^{\rm a}_{\mu}(z)\sim z^{-\mu-1}$. For a value $0<\mu<1$, the density diverges close to the origin as $\rho^{\rm a}_{\mu}(z)\sim z^{\mu-1}$. This function is plotted in Fig. \ref{Fig_annealed cauchy} for $\mu=1$ together with a comparison with results from numerical simulations of Cauchy random walks. We now consider the quenched density of L\'evy flights. 

The quenched density is not defined for $\mu\leq 1$ as we have seen that the mean value $|\moy{M_{k,n}}|=\infty$. For general $\mu>1$, as seen in Eq. \eqref{quenched_BM_1} in the case of the Brownian motion, the quenched density can be expressed in terms of the inverse ${\cal A}_{\mu}(z)$ of the function ${\cal M}_{\mu}(\alpha)$, given in Eq. \eqref{mean_max_k_mu}, as
\be
\boxed{\frame{\boxed{\rho^{\rm q}_{\mu}(z)=\frac{\pi}{\Gamma\left(1-\frac{1}{\mu}\right)}\frac{\left[{\cal A}_{\mu}(z)(1-{\cal A}_{\mu}(z))\right]^{1-1/\mu}}{{\cal A}_{\mu}(z)^{1-1/\mu}+(1-{\cal A}_{\mu}(z))^{1-1/\mu}}\;.}}}\label{quenched_levy_flight}
\ee
The function ${\cal A}_{\mu}(z)$ does not have an analytical expression in general but can be obtained numerically. Close to the edges $\pm z_{{\rm e},\mu}=\frac{\mu}{\pi}\Gamma(1-\frac{1}{\mu})$, we can show that the density vanishes as a power law that depends on $\mu$,
\be
\rho^{\rm q}_{\mu}(z)\approx \frac{(z_{{\rm e},\mu}-z)^{\mu-1}}{\mu\,z_{{\rm e},\mu}^\mu}\;.
\ee
The quenched and annealed densities of L\'evy flight are plotted in Fig. \ref{Fig_annealed cauchy} for $\mu=3/2$ together with a comparison with numerical data, showing a good agreement. 

We now briefly summarised the results of this chapter before considering the gap statistics of random walks. 

\begin{figure}
\centering
\includegraphics[width=0.6\textwidth]{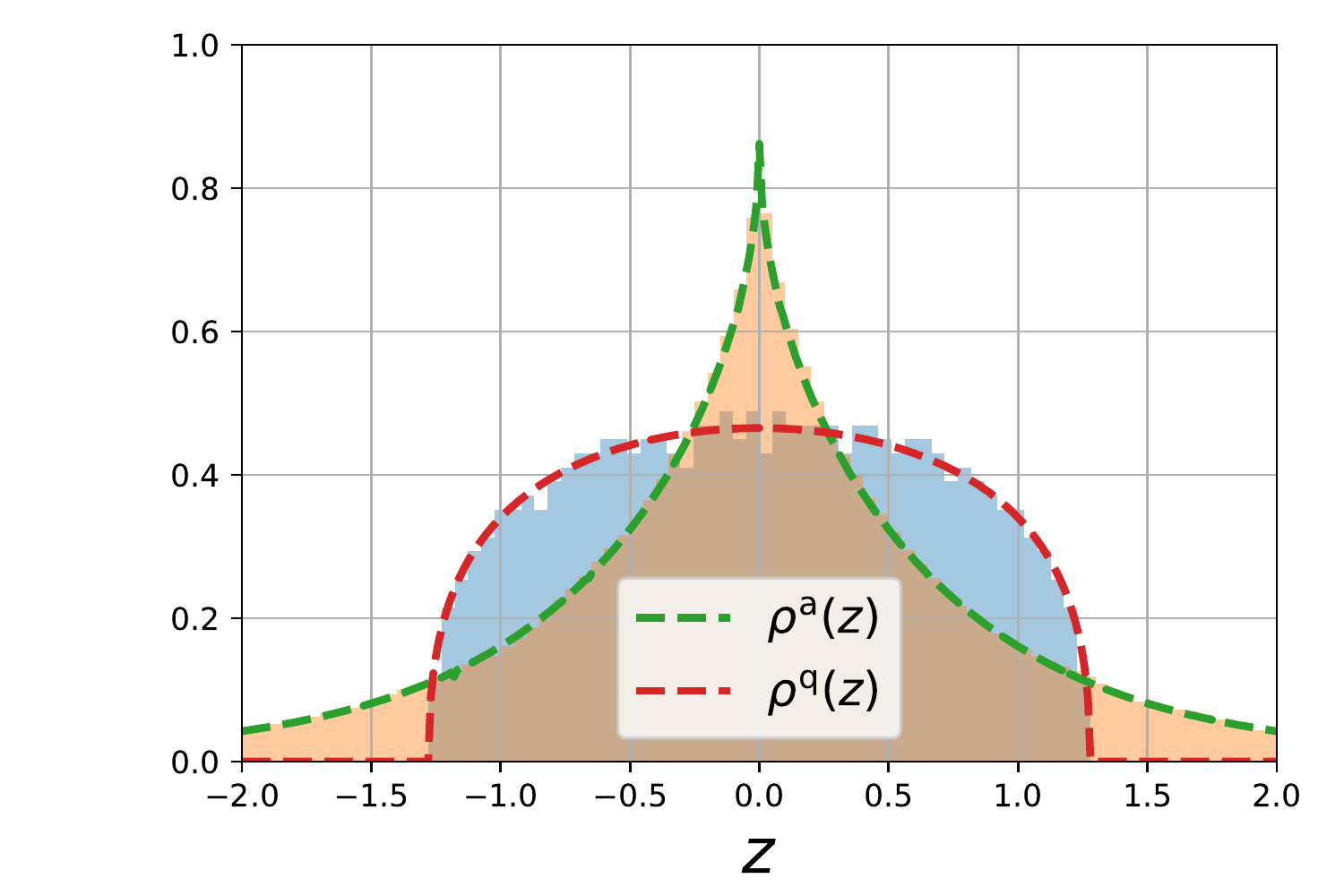}
\caption{Comparison between the rescaled annealed (orange) and quenched (blue) densities $\sqrt{n}\rho_n^{\rm a, q}(\sqrt{n}z)$ of maxima $M_{k,n}$ for $n=10^3$ as a function of $z$ obtained from the simulation of $10^6$ L\'evy flights of index $\mu=3/2$ and the scaling functions $\rho_{3/2}^{\rm a,q}(z)$ (respectively in dashed green and red)  in Eqs. \eqref{annealed_dens} and \eqref{quenched_levy_flight}. The numerical data shows a very good agreement with the analytical results.}\label{Fig_quenched_lev}
\end{figure}

\section{Summary of the results on ordered statistics of random walks}

We have considered in this chapter the order statistics of random walks, both with finite variance and L\'evy flights. We have obtained in Eq. \eqref{t_reach_max_k} an {\it exact} formula for the time to reach the $k^{\rm th}$ maxima of a random walk with {\it finite} $n$ steps which did not appear in the literature, to the best of our knowledge. This formula is {\it universal} and holds for any distribution of jumps of the random walk. We introduced the difference between the average density of positions/maxima of a random walk, referred to as {\it annealed density}, of the random walk and the {\it typical} distribution of maxima, referred to as {\it quenched density}. We have computed these densities in the large $n$ limit both for random walks with finite variance (converging to Brownian motion) in Eqs. \eqref{rho_a_BM} and \eqref{rho_q_BM} and for L\'evy flights in Eqs. \eqref{annealed_dens} and \eqref{quenched_levy_flight}. We have obtained, as for many properties of random walks, the emergence of universality classes for these densities depending on the L\'evy index $\mu$. Note finally that we have considered an extension of the results obtained in this section, for $M$ independent random walks, that should lead to future publication.

\chapter{Gap statistics of random walks}
\label{ch:gapk}

In this chapter, we consider the statistics of the gap $d_{k,n}$ between consecutive maxima of a random walk.
These variables are defined for $k=1,\cdots,n$ as (see also Fig. \ref{})
\be
d_{k,n}=M_{k,n}-M_{k+1,n}\geq 0\;.
\ee
For the probability of the gap and at variance with the maximum, we can only obtain analytical results for a few particular cases. We will first treat the singular case of a discrete simple random walk for which we can obtain the full distribution of $d_{k,n}$ explicitly before turning to continuous walks.

\begin{figure}
\centering
\includegraphics[width=0.6\textwidth]{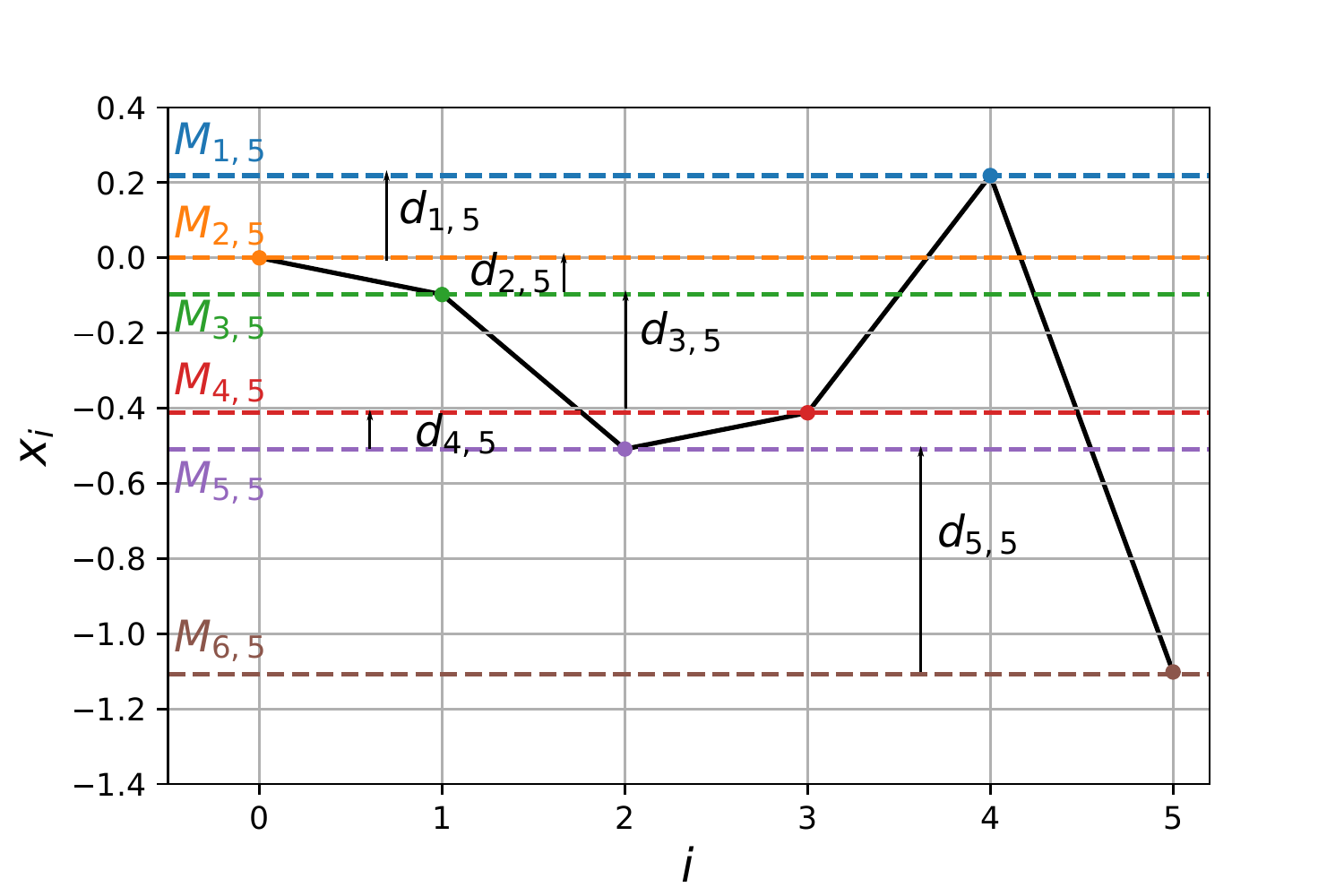}
\caption{Scheme of the maxima $M_{k,n}$ and gaps $d_{k,n}$ for a random walk.}\label{Fig_scheme_gap_max}
\end{figure}

\section{The singular discrete case: Simple random walk}

For the simple random walk characterised by the jump distribution
\be
f(\eta)=\frac{1}{2}\left[\delta(\eta+1)+\delta(\eta-1)\right]\;,
\ee
the walk only makes jump of length $1$, we must therefore have either $d_{k,n}=0$ or $d_{k,n}=1$. To characterise fully the statistics of $d_{k,n}$, we only need to compute the probability $\Pi_{k,n}$ that $d_{k,n}=1$. This probability can be obtained as the mean value of the gap
\be
\moy{d_{k,n}}=\Pi_{k,n}=\moy{M_{k,n}}-\moy{M_{k+1,n}}.
\ee
 Using additionally the identity \eqref{k_max_id_sym}, the probability can be expressed only in terms of the mean value of the global maximum
\be
\boxed{\frame{\boxed{\Pi_{k,n}=\moy{d_{k,n}}=\moy{M_{1,k+1}}-\moy{M_{1,k}}+\moy{M_{1,n+1-k}}-\moy{M_{1,n-k}}}}}\;.\label{pi_k_n}
\ee
The probability of the maximum of the simple random walk reads \cite{comtet2005precise,majumdar2012record}
\be
V_n(M)=\Prob\left[x_{\max,n}=M\right]=\frac{1}{2^{n}}{{n}\choose{\left\lfloor\frac{n+M+1}{2}\right\rfloor}}\;,
\ee
from which we deduce the mean value
\be
\moy{M_{1,k}}=\frac{1}{2^{k}}\sum_{M=1}^k M {{k}\choose{\left\lfloor\frac{k+M+1}{2}\right\rfloor}}\;.\label{moy_max_discrete}
\ee

\begin{figure}
\centering
\includegraphics[width=0.6\textwidth]{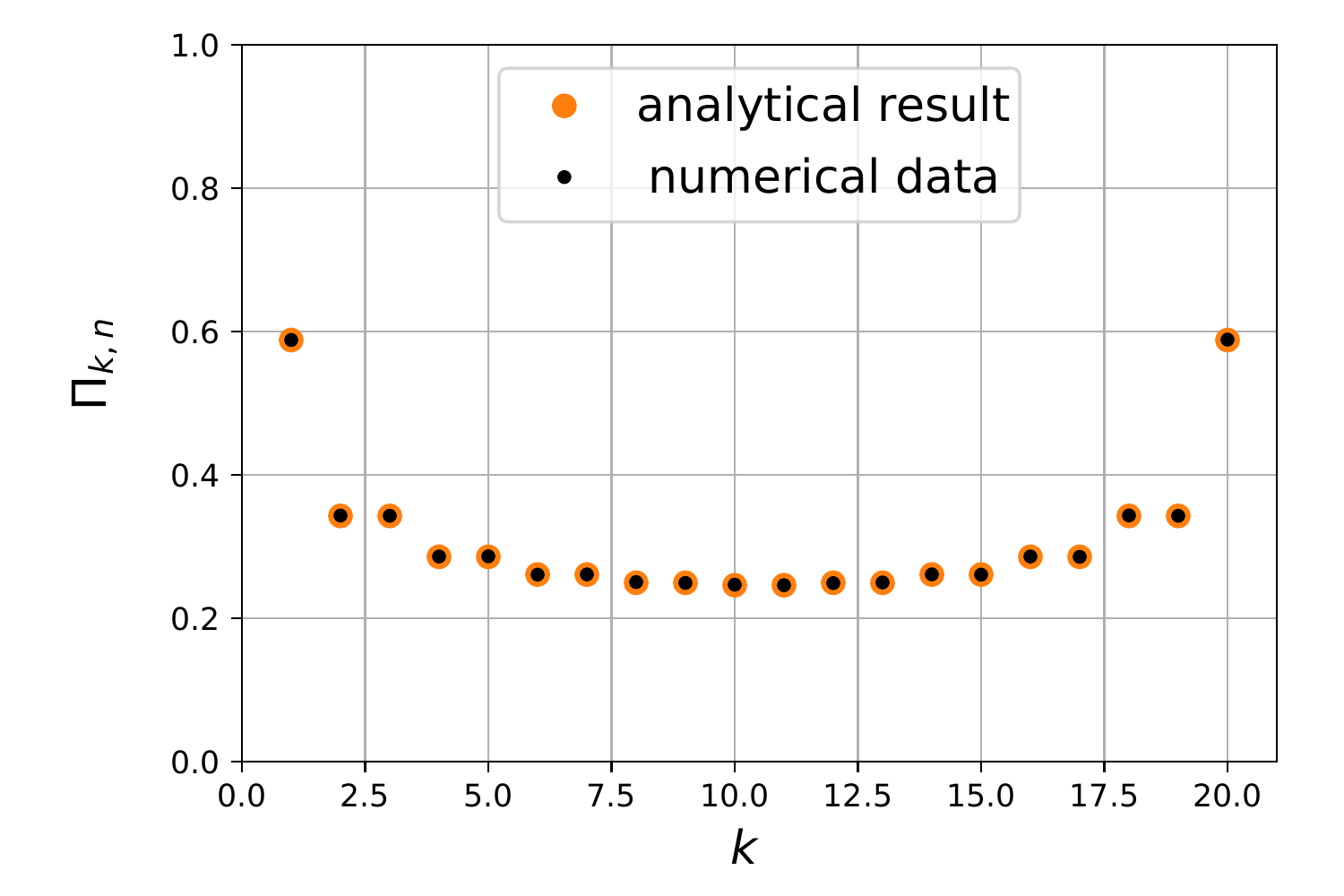}
\caption{Comparison between the probability $\Pi_{k,n}$ obtained by simulation of $10^6$ discrete random walks of $n=20$ steps (in black) and the exact analytical formula obtained by inserting Eq. \eqref{moy_max_discrete} in Eq. \eqref{pi_k_n} (in orange), showing an excellent agreement.}\label{Fig_prob_pi_k_n}
\end{figure}

In the limit of large $k$, this expression goes to $\moy{M_{1,k}}=\sqrt{2k/\pi}$, which coincides with the Brownian result in Eq. \eqref{x_max_mean_BM}. This is expected as the PDF $f(\eta)$ has a finite variance $\sigma^2=1$. Using this result, we obtain that for large $n$ and $k$ with $\alpha=k/n$ fixed, the probability $\Pi_{k,n}$ decays as
\be\label{mean_gap}
\boxed{\frame{\boxed{\Pi_{k,n}=\frac{1}{\sqrt{n}}m_1\left(\frac{k}{n}\right)\;,\;\;{\rm with}\;\;m_1(\alpha)=-\partial_{\alpha}{\cal M}(\alpha)=\frac{1}{\sqrt{2\pi\alpha}}+\frac{1}{\sqrt{2\pi(1-\alpha)}}}}}\;,
\ee
while it remains finite close to the global maximum, for $k=O(1)$,
\be
\Pi_{k,n}\approx \moy{M_{1,k+1}}-\moy{M_{1,k}}\;.
\ee
As we have seen, the discrete case is quite singular as the gap can only take two values. We will now consider the more general case of a continuous random walk.

\section{Continuous random walks}

\begin{figure}
\centering
\includegraphics[width=0.45\textwidth]{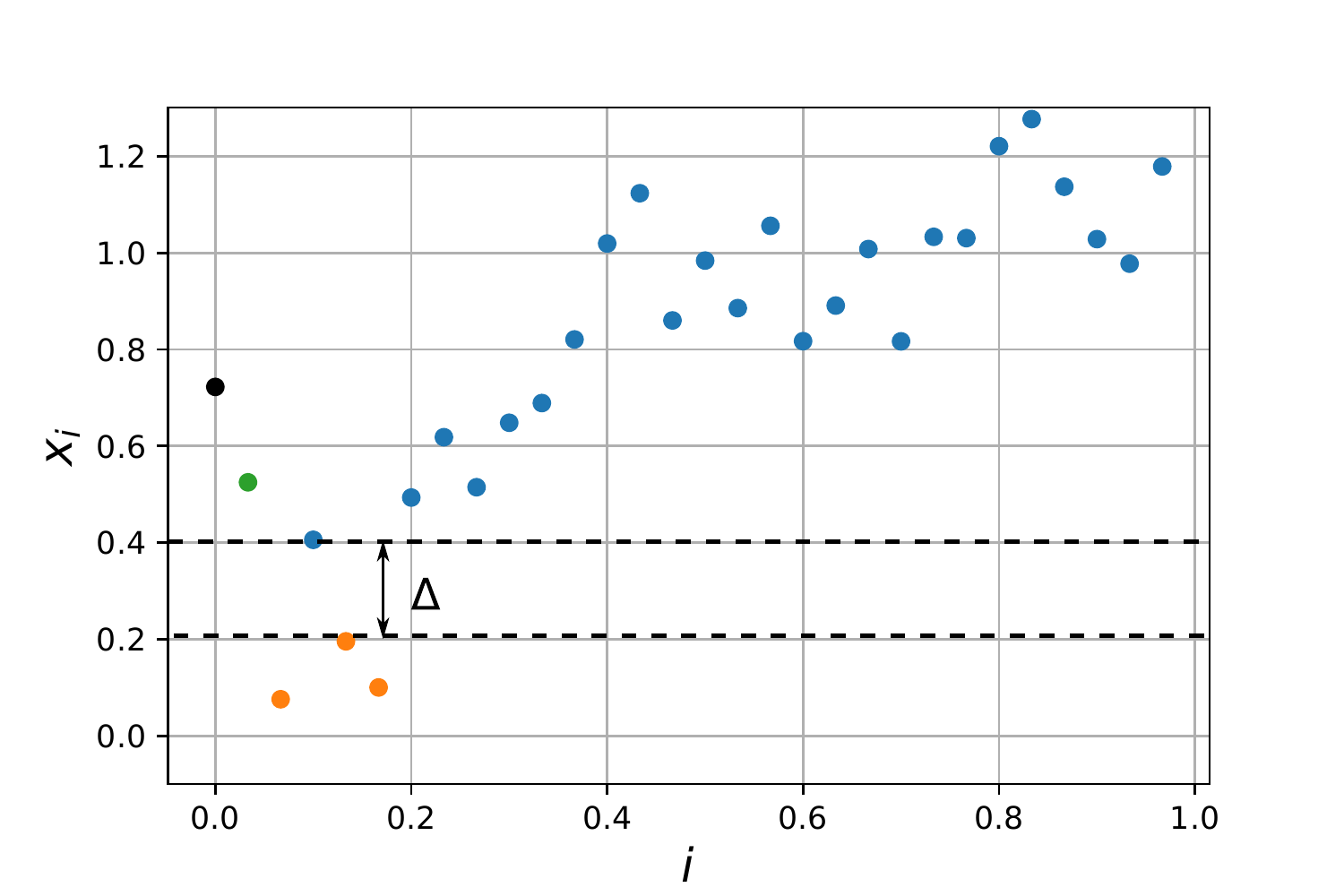}
\includegraphics[width=0.45\textwidth]{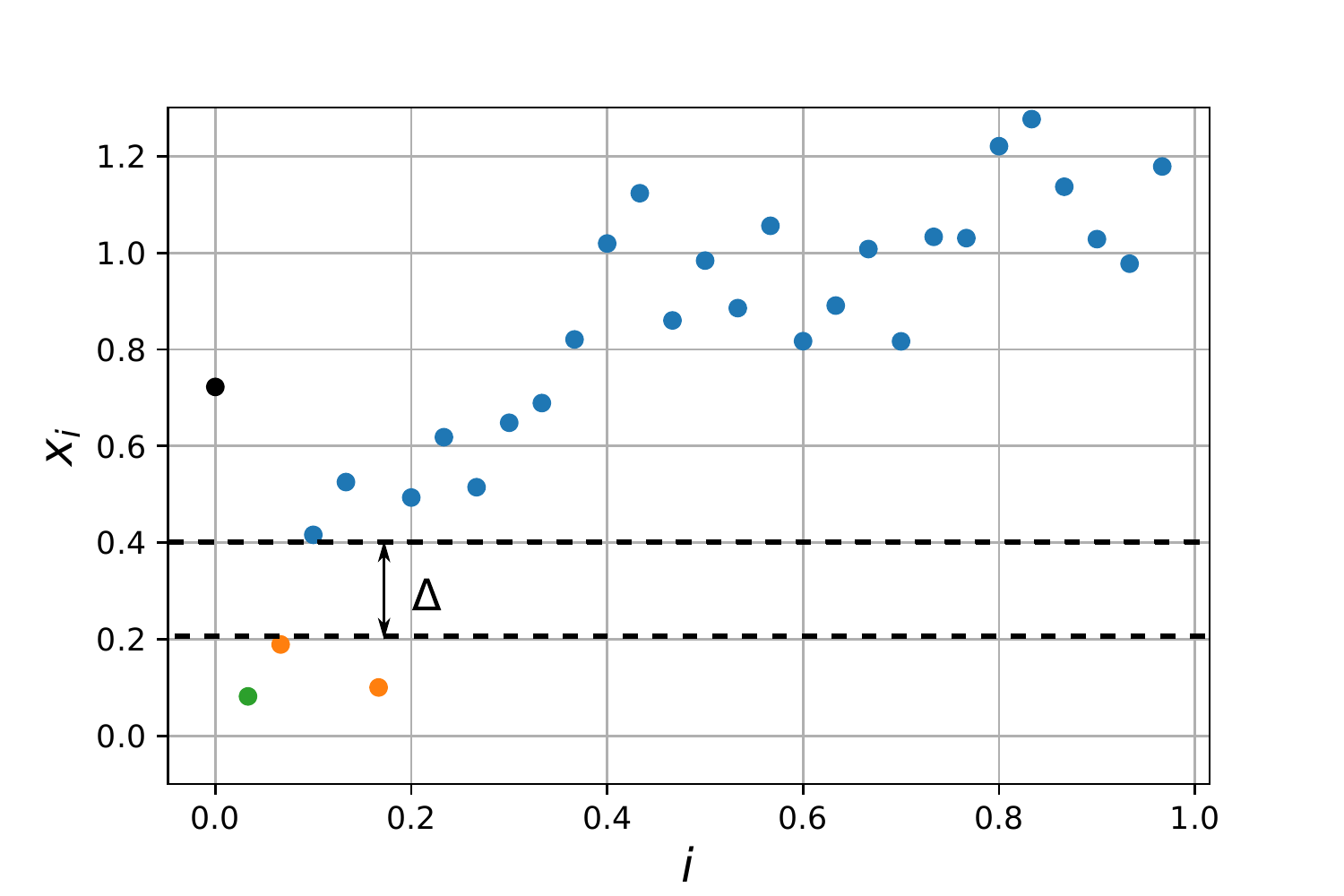}
\caption{On the left panel, the walk has an initial first step (in green) above level $x=0$ (represented by the black dashed line) has $3$ positions below $x=0$ afterwards, for a total of $3$ positions below $x=0$ and no position in the interval $(-\Delta,0)$. On the right panel, the walk has an initial first step (in green) below level $x=0$ (represented by the black dashed line) has $2$ positions below $x=0$ afterwards, for a total of $3$ positions below $x=0$ and no position in the interval $(-\Delta,0)$.}\label{Fig_prob_illus_gap}
\end{figure}

For a general continuous random walk, obtaining the PDF of the gap $d_{k,n}$ requires the joint PDF $-\partial^2_{xy} S_{k,n}(x,y)$ of the maxima $M_{k,n}$ and $M_{k+1,n}$, where 
\be\label{j_CDF}
S_{k,n}(x,y)=\Prob\left[M_{k,n}\leq x, M_{k+1,n}\geq y\right]\;.
\ee
This joint CDF encodes not only the values of the two consecutive maxima but also their correlations. From the joint PDF, it is rather simple to obtain the PDF of the gap $d_{k,n}$ as
\be
p_{k,n}(\Delta)=-\int_{-\infty}^{\infty}dx\int_{-\infty}^{\infty}dy \Theta(x-y)\partial^2_{xy} S_{k,n}(x,y)\delta(x-y-\Delta)\;.
\ee
One can express the joint CDF $S_{k,n}(x,y)$ in Eq. \eqref{j_CDF} as
\be
S_{k,n}(x,y)=\begin{cases}
q_{k,n}(x,x-y)&\;,\;\;x>0\;,\\
0&\;,\;\;x<0\;\&\;y>0\;,\\
q_{k,n}(-y,y-x)&\;,\;\;x<0\;\&\;y<0\;,
\end{cases}
\ee
where the probability $q_{k,n}(x,\Delta)$ is similar to the probability $q_{k,n}(x)=\Prob\left[N_+(x)=k\right]$ of the occupation time. It is defined as the probability that a random walk starting from position $x_0=x$ has exactly $k$ negative steps and no steps in the interval $(-\Delta,0)$ (c.f. Fig. \ref{Fig_prob_illus_gap}). Note that taking the limit $\Delta\to 0$, we recover $q_{k,n}(x,\Delta=0)=q_{k,n}(x)$. This probability $q_{k,n}(x,\Delta)$ is solution of a Wiener-Hopf type of recursion relation \cite{schehr2012universal}
\be\label{WH_2}
q_{l,n+1}(x,\Delta)=\int_{0}^{\infty}q_{l,n}(x',\Delta)f(x-x')dx'+\int_{0}^{\infty} q_{n+1-l,n}(x',\Delta)f(x'+x+\Delta)dx\;.
\ee

As a first step to obtain explicit results, we define the double generating functions of $q_{k,n}(x,\Delta)$ and $q_{n-k,n}(x,\Delta)$ respectively as
\begin{align}
\tilde q(z,s;x,\Delta)&=\sum_{n=0}^{\infty}\sum_{k=0}^{\infty}s^n z^k q_{k,n}(x,\Delta)\;,\\
\tilde r(z,s;x,\Delta)&=\sum_{n=0}^{\infty}\sum_{k=0}^{\infty}s^n z^k q_{n-k,n}(x,\Delta)\;,
\end{align}
which yields after multiplying Eq. \eqref{WH_2} by $z^k s^n$ and summing over $n$ and $k$ \cite{schehr2012universal}
\begin{align}
\tilde q(z,s;x,\Delta)&=1+s\int_0^{\infty}dx' f(x-x')\tilde q(z,s;x',\Delta) \nonumber \\
&+zs\int_0^{\infty}dx' f(x+x'+\Delta)\tilde r(z,s;x',\Delta) \;, \label{inteq_s}\\
\tilde r(z,s;x,\Delta)&=1+zs\int_0^{\infty}dx' f(x-x')\tilde r(z,s;x',\Delta) \nonumber \\
&+s\int_0^{\infty}dx' f(x+x'+\Delta)\tilde q(z,s;x',\Delta)\label{inteq_r}\;.
\end{align} 
This equation cannot be solved for a general jump PDF $f(\eta)$. Some information on the PDF can still be obtained using the results on $M_{k,n}$. For instance, we have seen that for a jump PDF with finite variance $\sigma^2$, the mean value $\moy{M_{k,n}}/\sigma$ takes a universal scaling form in the limit $n, k\to \infty$ with $\alpha=k/n=O(1)$ given in Eq. \eqref{mean_max_k_BM}. From this result, we obtain that in this limit, the mean value of the gap 
\be
\frac{\moy{d_{k,n}}}{\sigma}=\frac{\moy{M_{k,n}}}{\sigma}-\frac{\moy{M_{k+1,n}}}{\sigma}=\Pi_{k,n}
\ee
is also a universal quantity that we already computed for the simple random walk in Eq. \eqref{mean_gap}. We might therefore expect that the whole PDF $p_{k,n}(\Delta)$ of the gap $d_{k,n}$ is universal for finite variance jump PDF. We will therefore consider a case that is exactly solvable, obtaining an analytical formula for the scaling form of the PDF and investigate numerically whether it is universal.

\section{Special case of the Laplace jump PDF}

\begin{figure}
\centering
\includegraphics[width = 0.8\textwidth]{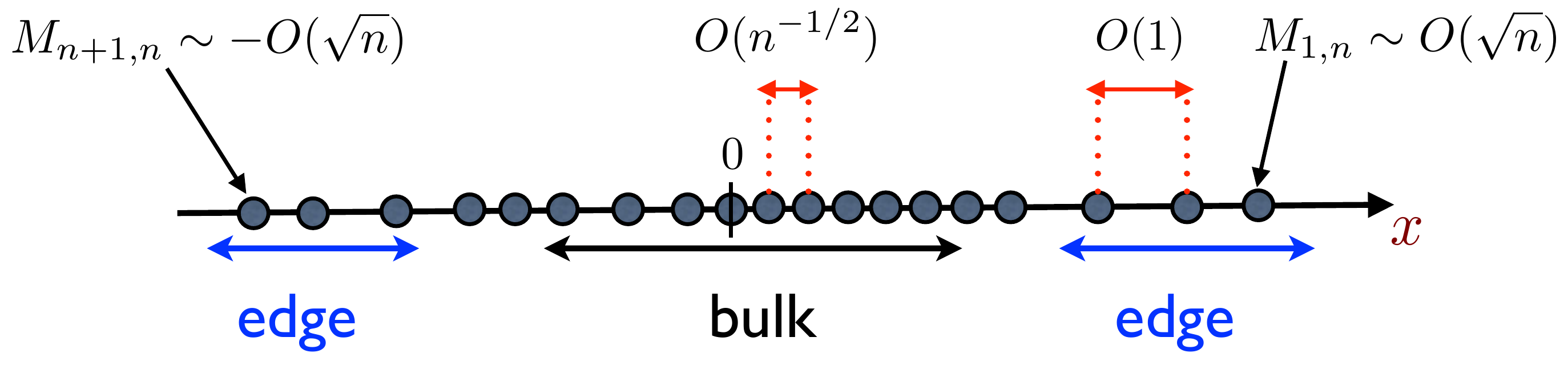}
\caption{Sketch of the point process constituted by the $k^{\rm th}$ maxima $M_{k,n}$ of the random walk (\ref{walk}) starting at $x_0=0$ after a large number of steps $n \gg 1$. At {\it the edges}, i.e. near the maximum $M_{1,n}$ and the minimum $M_{n+1,n}$ the gaps are of order $O(1)$, for large $n$, while {\it in the bulk}, i.e. far from the minimum and the maximum, the gaps are much smaller and of order $O(n^{-1/2})$.}\label{Fig_pp}
\end{figure}

We consider the special case of the Laplace jump PDF
\be
f(\eta)=\frac{e^{-\frac{\sqrt{2}|\eta|}{\sigma}}}{\sqrt{2}\sigma}\;.
\ee 
This Laplace distribution fulfils the simple identity
\be
f''(\eta)=\frac{2}{\sigma^2}\left[f(\eta)-\delta(\eta)\right]\;,\label{iden_Lap}
\ee
which will be the key property allowing us to simplify greatly the computations.
Deriving Eqs. \eqref{inteq_s} and \eqref{inteq_r} twice with respect to $x$ and using the identity in Eq. \eqref{iden_Lap}, we obtain a set of decoupled differential equations
\begin{align}
\frac{\sigma^2}{2}\partial_{x}^2\tilde q(z,s;x,\Delta)&=(1-s)\tilde q(z,s;x,\Delta)-1\;,\\
\frac{\sigma^2}{2}\partial_{x}^2\tilde r(z,s;x,\Delta)&=(1-z s)\tilde r(z,s;x,\Delta)-1\;.
\end{align}
These equations can be solved exactly, yielding
\begin{align}
&\tilde q(z,s;x,\Delta)=A(z,s;\Delta)e^{-\sqrt{2(1-s)}\frac{x}{\sigma}}+\frac{1}{1-s}\;,\\
&\tilde r(z,s;x,\Delta)=B(z,s;\Delta)e^{-\sqrt{2(1-zs)}\frac{x}{\sigma}}+\frac{1}{1-zs}\;,
\end{align}
where the coefficients $A$ and $B$ can be obtained explicitly by injecting this ansatz in Eq. \eqref{inteq_s} and \eqref{inteq_r} \cite{schehr2012universal}
\be\label{A}
A(z,s;\Delta)=\frac{\frac{zs}{\sqrt{1-zs}}-\frac{s}{1-s}\left[\sqrt{1-zs}\cosh\left(\frac{\sqrt{2}\Delta}{\sigma}\right)+\sinh\left(\frac{\sqrt{2}\Delta}{\sigma}\right)\right]}{(\sqrt{(1-zs)(1-s)}+1)\sinh\left(\frac{\sqrt{2}\Delta}{\sigma}\right)+(\sqrt{1-zs}+\sqrt{1-s})\cosh\left(\frac{\sqrt{2}\Delta}{\sigma}\right)}\;,
\ee
and $B(z,s;\Delta)=A(z^{-1},zs;\Delta)$. Note that these two coefficients $A$ and $B$ allow to reconstruct directly the double generating function of the PDF $p_{k,n}(\Delta)$ of the gap $d_{k,n}$ as \cite{schehr2012universal}
\begin{align}
\tilde p(z,s;\Delta)&=\sum_{n=0}^{\infty}\sum_{k=0}^{\infty}s^n z^k p_{k,n}(\Delta)\label{p_gap_GF}\\
&=\partial_{\Delta}A(z,s;\Delta)+\frac{\sigma}{\sqrt{2(1-s)}}\partial_{\Delta}^2 A(z,s;\Delta)\nn\\
&+z e^{\sqrt{2(1-zs)}\frac{\Delta}{\sigma}}\left(\partial_{\Delta}B(z,s;\Delta)+\frac{\sigma}{\sqrt{2(1-zs)}}\partial_{\Delta}^2 B(z,s;\Delta)\right)\;.\nn
\end{align}
These results for the generating function are exact. We will first analyse the large $n$ limit for $k=O(1)$, i.e. close to the global maximum, and then for $\alpha=k/n=O(1)$, i.e. in the bulk of the density of maxima (see Fig. \ref{Fig_pp}).

\subsection{Large $n$ limit for fixed $k$: gaps close to the global maximum}

In \cite{schehr2012universal}, Schehr and Majumdar studied the regime of fixed $k$ and large $n$. In this regime, the typical scaling of the gap is obtained by using the behaviour close to its edge of the quenched density of maxima $\rho_n^{\rm q}(x)$ computed in Eq. \eqref{rho_q_BM},
\be
\rho_n^{\rm q}(x)=\frac{1}{\sigma\sqrt{n}}\rho^{\rm q}\left(\frac{x}{\sigma\sqrt{n}}\right)\;,\;\;\rho^{\rm q}(z)\approx \sqrt{4\pi}(z_{\rm e}-z)\;.
\ee
Using a scaling argument, ensuring that there is $O(1)$ maxima in the interval $[z_{\rm e}-\delta,z_{\rm e}]$, we obtain the typical scaling of the density at the edge
\be
\int_{z_{\rm e}-\frac{\delta}{\sigma\sqrt{n}}}^{z_{\rm e}}\rho^{\rm q}(z)dz\sim \frac{1}{n}\Rightarrow \int_{0}^{\frac{\delta}{\sigma\sqrt{n}}}4\pi u du\sim \frac{1}{n}\Rightarrow \delta=O(1)\;.
\ee
We therefore expect that the PDF $p_{k,n}(x)$ should not depend on $n$ in this limit. To confirm this
we may compute the generating function in the limit $s\to 1$ in Eq. \eqref{A}, keeping $z$ and $\Delta$ fixed. This yields
\be
A(z,s;\Delta)\approx -\frac{1}{1-s}+\frac{A_1\left(z;\frac{\Delta}{\sigma}\right)}{\sqrt{1-s}}\;,\;\;A_1\left(z;\delta\right)=\frac{\cosh(\sqrt{2}\delta)+\sqrt{1-z}\sinh(\sqrt{2}\delta)}{\sqrt{1-z}\cosh(\sqrt{2}\delta)+\sinh(\sqrt{2}\delta)}\;.
\ee
In this limit of large $n$, keeping the leading order terms for $s\to 1$, we may therefore obtain the relation
\be
\tilde p(z,s;\Delta)\approx \frac{\sigma}{\sqrt{2}}\frac{\partial_{\Delta}^2 A_1\left(z;\frac{\Delta}{\sigma}\right)}{1-s}\;.
\ee
From this scaling form, we expect that the PDF of the gap in this regime does not depend on $n$, i.e. $\sum_{s\geq 0}s^n=(1-s)^{-1}$. For $n\gg k\gg 1$ (corresponding to the limit $z\to 1$), the authors were able to obtain the regime of typical fluctuations of the gaps close to the global maximum. The scaling form reads \cite{schehr2012universal}
\be
p_{k,n}(\Delta)\approx \frac{\sqrt{k}}{\sigma}P\left(\frac{\sqrt{k}\Delta}{\sigma}\right)\;,\;\;P(\delta)=4\left[\sqrt{\frac{2}{\pi}}(1+2\delta^2)-\delta(4\delta^2+3)e^{2\delta^2}\erfc(\sqrt{2}\delta)\right]\;.\label{P_gap_edge}
\ee
It was shown numerically that this regime of typical fluctuations is not limited to the Laplace jump PDF but holds more generally for finite variance jump PDF. An additional indication of the wideness of the universality class was obtained in \cite{battilana2017gap}, where the authors found the same gap probability $P(\delta)$ for symmetric gamma distributed jumps. Note that in the latter, an intriguing similarity with the study of thermodynamics of a classical Ising spin chain in a gamma distributed random magnetic field was noticed \cite{luck1991low}. The regime of atypical fluctuations can be computed for the Laplace jump PDF but it depends explicitly on the jump PDF and is therefore somewhat less interesting. Note finally that this PDF $P(\delta)$ has a non-trivial power law tail
\be\label{tail_P}
P(\delta)\approx \frac{3}{\sqrt{8\pi}}\frac{1}{\delta^4}\;,\;\;\delta\to\infty\;.
\ee
This result pushed us to investigate further and obtain the scaling form equivalent to Eq. \eqref{P_gap_edge} but in the bulk of the quenched density $\rho_n^{\rm q}(x)$ instead of the edge (see Fig. \ref{Fig_pp}). This is the main purpose of Article \ref{Art:gap}.

\subsection{Large $n$ limit for fixed $\alpha=k/n$: gaps in the bulk}

\begin{figure}
\centering
\includegraphics[width=0.7\textwidth]{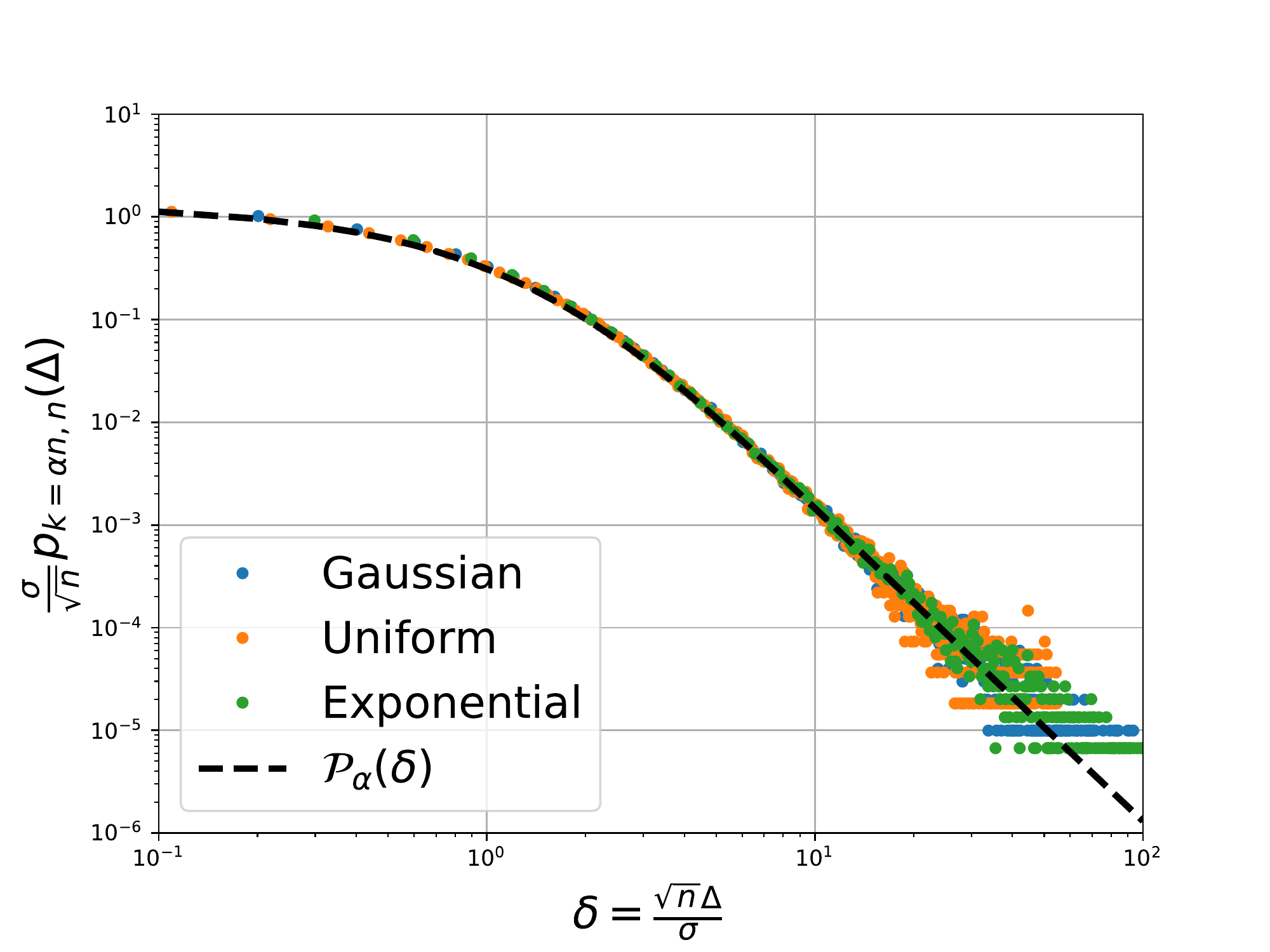}
\caption{Comparison between the rescaled PDF $\frac{\sigma}{\sqrt{n}}p_{k=\alpha n,n}(\Delta)$ of the gap $d_{k,n}$ obtained numerically for $10^6$ random walks of $n=10^3$ steps and $k=500$, hence $\alpha = k/n = 1/2$ with Gaussian (in blue), uniform (in orange) and exponential (in green) PDF of jumps $f(\eta)$ and the scaling function ${\cal P}_{\alpha=1/2}(\delta=\sqrt{n}\Delta/\sigma)$ (dashed line) given in Eq. \eqref{P_a_d}. The curves for different jumps PDF all collapse on the same master curve described by ${\cal P}_{1/2}(\delta)$, suggesting the universality of the scaling function (\ref{P_a_d}).}\label{Fig_PDF_gap}
\end{figure}

Applying the same scaling argument as we did for the edge, in this bulk limit, we use that the density appears locally flat close to a point $z$ in the bulk,
\be
\int_{z-\frac{\delta}{\sigma\sqrt{n}}}^{z}\rho^{\rm q}(z')dz'\sim \frac{1}{n}\Rightarrow \rho^{\rm q}(z)\int_{0}^{\frac{\delta}{\sigma\sqrt{n}}}du\sim \frac{1}{n}\Rightarrow \delta=O(n^{-1/2})\;.
\ee
We therefore expect a scaling regime of the type
\be\label{scal_form_p_k_n}
p_{k,n}(\Delta)\approx \frac{\sqrt{n}}{\sigma}{\cal P}_{k/n}\left(\frac{\sqrt{n}\Delta}{\sigma}\right)\;.
\ee
Note that this scaling form is fully consistent with the universal mean value of the gap $\Pi_{k,n}=\moy{d_{k,n}}/\sigma$, that we obtained in this regime \eqref{mean_gap}.
To analyse the generating function in this regime, we should take the limit $s,z\to 1$, with $(1-z)=q\sim(1-s)=p$ and $\Delta\sim \sqrt{p}$. In this case, we obtain
\be
A(z,s;\Delta)\approx a\left(p+q,p;\frac{\Delta}{\sigma}\right)=-\frac{(p+q)-p+\frac{\sqrt{2(p+q)}\Delta}{\sigma}}{p\sqrt{p+q}\left(\sqrt{p}+\sqrt{p+q}+\frac{\sqrt{2}\Delta}{\sigma}\right)}\;.
\ee
Note that this expression only depends on $q$ through $r=p+q$. Computing $B(z,s;\Delta)\approx a(p,p+q;\Delta/\sigma)$ leads to the symmetric result exchanging $p\to r=p+q$.
Inserting these expressions in Eq. \eqref{p_gap_GF} and keeping only the leading order term, the generating function of the PDF reads
\be
\tilde p(z=1-q,s=1-p;\Delta)\approx \frac{\sigma}{\sqrt{2}}\partial_{\Delta^2}\left[\frac{a\left(p+q,p;\frac{\Delta}{\sigma}\right)}{\sqrt{p}}+\frac{a\left(p,p+q;\frac{\Delta}{\sigma}\right)}{\sqrt{p+q}}\right]\;.
\ee
Inverting this generating function, we obtain our final result for the scaling function in Eq. \eqref{scal_form_p_k_n}, which corresponds to gaps $d_{k,n}$ in the regime $n\to \infty$ and $k\to \infty$, keeping $0<\alpha<1$, (see Article \ref{Art:gap} for further details)
\be\label{P_a_d}
\boxed{
\begin{array}{rl}
\displaystyle{\cal P}_\alpha(\delta)=&\displaystyle\int_{0}^{\infty} y^{2}e^{-\delta y}\left[\frac{e^{-\frac{y^{2}}{8\alpha(1-\alpha)}}}{\pi\sqrt{\alpha(1-\alpha)}}+\frac{ye^{-\frac{y^{2}}{8(1-\alpha)}}}{4\sqrt{2\pi}(1-\alpha)^{\frac{3}{2}}}\erfc\left(\frac{y}{2\sqrt{2\alpha}}\right)\right.\\
&\displaystyle\left.+\frac{y e^{-\frac{y^{2}}{8\alpha}}}{4\sqrt{2\pi}\alpha^{\frac{3}{2}}}\erfc\left(\frac{y}{2\sqrt{2(1-\alpha)}}\right)\right]dy\;.
\end{array}
}
\ee
This scaling function is plotted in Fig. \ref{Fig_PDF_gap} together with the PDF of the gap $d_{k,n}$ for $n=10^3$ and $k=500$ obtained  numerically for different jump PDF. The agreement is excellent for all the jump PDF up until the regime of atypical fluctuations $d_{k,n}\gg n^{-1/2}$ is reached, indicating the universality of this result. Note that the scaling function in Eq. \eqref{P_a_d} is clearly symmetric in $\alpha\to 1-\alpha$. 

In the limit $\alpha\to 0$, the distribution $P(\delta)$ in Eq. \eqref{P_gap_edge} describing the gaps close to the global maximum is recovered
\be
P(\delta)=\lim_{\alpha\to 0}\frac{1}{\sqrt{\alpha}}{\cal P}_{\alpha}\left(\frac{\delta}{\sqrt{\alpha}}\right)\;.
\ee
The asymptotic behaviours of Eq. \eqref{P_a_d} are obtained as
\be
{\cal P}_\alpha(\delta)\approx\begin{cases}  \displaystyle4\sqrt{\frac{2}{\pi}}(\sqrt{\alpha}+\sqrt{1-\alpha}-1)&\;,\;\;\delta\to 0\;,\\
&\\
\displaystyle \frac{2}{\pi\sqrt{\alpha(1-\alpha)}}\frac{1}{\delta^3}&\;,\;\;\delta\to\infty\;.
\end{cases}\label{tail_P_a}
\ee
Note in particular the tail ${\cal P}_\alpha(\delta)\propto\delta^{-3}$ in Eq. \eqref{tail_P_a} for large $\delta$, which is different from the tail of $P(\delta)\propto \delta^{-4}$ in Eq. \eqref{tail_P}. This indicates clearly that the limits $\alpha\to 0$ and $\delta\to \infty$ do not commute. Indeed, we find that in the limit $\delta\to \infty$ and $\alpha\to 0$, keeping $\xi=\alpha\delta$, there exists a non-trivial scaling form allowing to match both tails in Eqs. \eqref{tail_P} and \eqref{tail_P_a}
\be\label{F}
{\cal P}_\alpha(\delta) \approx \alpha^{5/2}{\cal F}(\alpha \delta)\;\;{\rm with}\;\;{\cal F}(\xi)=\frac{2}{\pi}\frac{1}{\xi^3}+\frac{3}{\sqrt{8\pi}}\frac{1}{\xi^4}\;.
\ee
The tail in Eq. \eqref{tail_P_a} also suggests a different scaling of the moments $\moy{d_{k,n}^p}/\sigma^p$ for $p>1$. These moments together with the regime of large fluctuations are obtained in Article \ref{Art:gap} for the Laplace jump PDF and depend on the specific distribution of jumps. 

\section{Summary of the results for the gap statistics of random walks}

In this chapter, we have considered the gap statistics of random walks. We have obtained an {\it exact} expression for the probability $\Pi_{k,n}$ that the $k^{\rm th}$ gap of a random walk {\it discrete in space and time} is equal to $d_{k,n}=+1$ in Eq. \eqref{pi_k_n}. We have also obtained an exact expression for the typical distribution of the $k^{\rm th}$ gap of a space continuous random walk with Laplace distribution of gaps in the ``bulk'' regime, i.e. for large $k$ and $n$ with fixed $\alpha=k/n$, in Eq. \eqref{P_a_d}. This distribution is non trivial and has been shown numerically to be universal for any distribution of jumps with finite variance.






 \AddArticle{Art:gap}{Gap statistics close to the quantile of a random walk}

 \begin{center}
   {\large \textbf{Gap statistics close to the quantile of a random walk}}
 \end{center}


 \vspace{2cm}

 \noindent B. Lacroix-A-Chez-Toine, S.~N. Majumdar, G. Schehr,\\
 arXiv preprint,  arXiv: \textbf{1812.08543}, (2018).\\

 \ding{43}
 \href{https://arxiv.org/abs/1812.08543}{https://arxiv.org/abs/1812.08543}

\begin{abstract}
We consider a random walk of $n$ steps starting at $x_0=0$ with a double exponential (Laplace) jump distribution. We compute exactly the distribution $p_{k,n}(\Delta)$ of the gap $d_{k,n}$ between the $k^{\rm th}$ and $(k+1)^{\rm th}$ maxima in the limit of large $n$ and large $k$, with $\alpha=k/n$ fixed. We show that the typical fluctuations of the gaps, which are of order $O( n^{-1/2})$, are described by a universal $\alpha$-dependent 
distribution, which we compute explicitly. Interestingly, this distribution has an inverse cubic tail, which implies a non-trivial $n$-dependence of the moments of the gaps. We also argue, based on numerical simulations, that this distribution is universal, i.e. it holds for more general jump distributions (not only the Laplace distribution), which are continuous, symmetric with a well defined second moment. Finally, we also compute the large deviation form of the gap distribution $p_{\alpha n,n}(\Delta)$ for $\Delta=O(1)$, which turns out to be non-universal. 
\end{abstract}





\counterwithout{table}{chapter}







\chapter*{Conclusion}
\addcontentsline{toc}{chapter}{Conclusion}

Extreme value statistics is a central theme in statistical mechanics, and in the applications of statistics in general, and as such, it has been studied over the last decades. However, obtaining exact results for strongly correlated random variables remains challenging. In this thesis, we have provided results that broaden the knowledge on this field of research, focusing on three different physical models:

\section*{Non interacting fermions}

In part \ref{Part:Fer}, we have considered the spatial statistics of non-interacting fermions. Even in the absence of interaction, the correlations at low temperature remain non-trivial in this system because of the Pauli exclusion principle. As the current experimental set-ups allow to obtain precise imaging of the individual positions of cold fermions \cite{parsons2015site , cheuk2015quantum, haller2015single} it becomes crucial to have an accurate description of the spatial statistics in this system. In the bulk of the gas, where the density is large, semi-classical approximations allow to simply describe these statistics \cite{castin2006basic}. However, the confining potential needed to conduct experiments creates edges to the density near which the statistics are non-trivial \cite{kohn1998edge}. The fluctuations for smooth confining potentials -- e.g. $V(x)\sim |x|^p$ with $p>0$ -- creating {\it soft edges}, were described and shown to be {\it universal} with respect to the potential in a recent series of papers (c.f. \cite{Dean_2019, dean2016noninteracting} for recent reviews) using the tools of {\it determinantal point process}. However, a proper understanding of the spatial statistics for potentials creating {\it hard edges}, where the density vanishes abruptly, was still lacking.\\

 \ding{42} In chapter \ref{ch: ferm_hard_edge}, we have extended the description to {\it hard edges} (e.g. for hard box potentials or $V(x)\sim x^{-\gamma}$ with $\gamma>1$), obtaining exact and {\it universal} results for the correlation kernel, which controls all the fluctuations in the gas. In dimension one, we have shown an exact mapping between the ground state of fermions in a hard box potential and the Jacobi Unitary Ensemble of random matrix theory. We have used these results to analyse the extreme value statistics in the gas, i.e. the position of the particle the closest to the hard edge. In particular we have shown the emergence in dimension $d>1$ of an {\it intermediate deviation regime} connecting smoothly the regime of typical fluctuations and large deviations. These results are described in Articles \ref{Art:fermions_lett} and \ref{Art:ferm_long}.\\

For a free gas of non-interacting fermions, the bipartite entanglement entropy can be obtained using field theory techniques \cite{calabrese2004entanglement, calabrese2011entanglement, calabrese2011entanglement2}. However, for {\it trapped} Fermi gases, the potential breaks explicitly the translation invariance, and these techniques cannot generally be used (see however \cite{dubail2017conformal}). Recent exact results have been obtained for a system with an explicit connection to random matrix theory \cite{calabrese2015random}. A connection between entropy and full counting statistics for non-interacting Fermi gases \cite{song2012bipartite, song2011entanglement, klich2006lower} was used to obtain the entropy in the bulk but even in this case, the edge behaviour of the entropy is rather hard to analyse.  \\

\ding{42} In chapter \ref{ch: rot_trap}, we have shown an exact mapping between the ground state of a system of $N$ non interacting fermions in a two-dimensional rotating harmonic trap and the complex Ginibre ensemble. For this model of fermions we obtained {\it exact} results for the full counting statistics -- i.e. the number of fermions in a disk of radius $r$ -- and the entanglement entropy, valid for any {\it finite number} of particles $N$. In particular, we have shown in the large $N$ limit that while the number variance and the entanglement entropy are proportional to each other in the bulk of the gas it is not the case close to the edge of the density.  These results are described in Article \ref{Art:rot}.\\

The analysis of these models of non-interacting fermions left a few open questions:

\begin{enumerate}[label=\roman*.]
\item While the computations of the large deviation functions of general observables is possible for models of fermions with an explicit mapping to ensembles of random matrix theory, a general theory valid for any potential is still lacking. Furthermore even in cases where there is a mapping, it only holds at zero temperature and there is no framework to compute the large deviations at finite temperatures.

\item As the current technology allows to measure experimentally the full counting statistics, it would be interesting to extend recent results \cite{grabsch2018fluctuations, marino2016number} and obtain the fluctuations of the number of fermions in a given interval at finite temperature in dimension $d$, allowing for an indirect measurement of the finite temperature entanglement entropy via the full counting statistics. 

\item For smooth confining potentials, the study of the equilibrium properties of the spatial statistics has led to the investigation of numerous properties of the Fermi gas: the statistics of the momenta \cite{le2018multicritical}, the study of the Wigner function \cite{dean2018wigner}, the equilibrium dynamics \cite{le2017periodic} or the non-equilibrium properties of quantum quenches \cite{dean2019nonequilibrium}. The non-equilibrium dynamics in hard edge potentials has recently been considered \cite{kulkarni2018quantum} but all these questions could be studied for hard edge potentials.



\end{enumerate}

\section*{Random matrix theory}

There exists several exact connections between random matrix theory and models of non interacting fermions (c.f. \cite{Dean_2019}). Despite an extensive literature in random matrix theory \cite{forrester2010log, akemann2011oxford, mehta2004random}, many problems are still open. In particular, while the matching between the regime of typical fluctuations and large deviations of observables in invariant ensembles is well-understood, it is not the case for non-Hermitian ensembles \cite{cunden2016large}. These non-Hermitian models, and in particular the complex Ginibre ensemble, are also relevant in Physics, with natural connections to problems such as the two-dimensional one component plasma at equilibrium at inverse temperature $\beta$ \cite{forrester1998exact} or the Laughlin states in the context of the quantum Hall effects \cite{ezawa}.\\

\ding{42} In chapter \ref{ch: rot_trap} we have shown for $\beta=2$ the presence of an {\it intermediate regime} of fluctuations connecting the typical fluctuations and the large deviations for both the full counting statistics and the extreme value statistics of the two-dimensional one component plasma ($2d$ OCP) in a rather general {\it rotationally symmetric} potential, which holds in particular for the complex Ginibre ensemble. We have argued that this intermediate deviation regime holds for a general class of observables of the gas restricted to a finite {\it rotationally symmetric} domain. These results are described in Articles \ref{Art:r_max} and \ref{Art:FCSGin}.\\

The analysis of non-Hermitian models and Coulomb gases conducted in this thesis raised open questions:

\begin{enumerate}[label=\roman*.]
\item The techniques used to obtain our results rely heavily on the rotational symmetry of the problem. A natural question is then to wonder if intermediate deviation regimes also emerge when the system is not symmetric, e.g. for the number $N_+$ of eigenvalues of the complex Ginibre ensemble with positive real part, and how to capture them.


\item These intermediate deviation regimes have been shown to emerge in several examples of determinantal point processes with rotational symmetry (fermions, $2d$ OCP). In these cases, the typical regime of fluctuation is given by the same statistics as for i.i.d. random variables, while the intermediate regime is non-trivial. However, there are also examples for which the typical fluctuations are non trivial and match smoothly with the large deviation regime, e.g. for the smallest radius $r_{\min}$ in the complex Ginibre ensemble. Is there a standard criterion to distinguish whether the fluctuations of an observable will be the former or the latter? 


\item The extreme value statistics for the $2d$ OCP have been explored in details both for the typical and large deviation regime at inverse temperature $\beta=2$, where the system is determinantal \cite{cunden2017universality, chafai2014note}. Recently, it has been considered at any temperature for the one-dimensional Coulomb gas \cite{dhar2017exact, dhar2018extreme}. A natural question is then the extension in dimension $d>2$ or at inverse temperature $\beta\neq 2$. While the large deviations were obtained for general $d$ and $\beta$ \cite{cunden2017universality} the regime of typical fluctuations is only conjectured to be given by a Gumbel distribution \cite{chafai2019simulating}. Even if that conjecture actually holds, the same problem of matching between typical and large deviations emerges as in dimension $d=2$ and inverse temperature $\beta=2$. The regime of intermediate fluctuations connecting these two regimes remains therefore to be characterised in these cases.


\end{enumerate}


\section*{Random walks and Brownian motions}

In part \ref{Part:gaps} we have considered the order and gap statistics associated to random walks and Brownian motion. The order statistics have been studied extensively for random walks \cite{dassios1996sample, yor1995distribution,  dassios1995distribution} where many results can be obtained using the convergence to Brownian motion. If one considers instead the gap statistics, which is characteristic of discrete processes, this property cannot be used and the problem is much harder to solve. The statistics of these gaps were obtained close to the global maximum \cite{schehr2012universal} for the Laplace distribution of jumps and were argued to hold for any distribution with finite variance. \\

\ding{42} In chapter \ref{ch:gapk} we have obtained a similar {\it exact} result for the gaps of random walks with a Laplace distribution of jumps in a different regime: deep in the bulk of maxima, i.e. far from the global maximum. We have also argued, on numerical ground, that this distribution holds for a random walk with any jump distribution provided it has a finite variance. These results are described in Article \ref{Art:gap}.\\

\ding{42} In chapter \ref{ch:maxk} we have defined and computed exactly the annealed and quenched density of Brownian motion and L\'evy flights, corresponding to the average and typical distribution of maxima of the process.\\

After the analysis of random walks and Brownian motions conducted in this thesis, some questions remain open:

\begin{enumerate}[label=\roman*.]
\item The results obtained in this thesis for the gap statistics are only shown exactly for the Laplace random walk. While numerical simulations suggest the universality of this result, an analytical proof is still missing (see however \cite{battilana2017gap}).

\item The analogous distribution of gaps for L\'evy flights remains to be characterised.

\item The framework developed in this thesis should allow to compute other observables such as the time between two consecutive maxima of a random walk.


\end{enumerate}

\label{conclu}

\counterwithin{table}{chapter}



\appendix

 \AddArticle{Art:BMcoi}{Distribution of Brownian coincidences }

 \begin{center}
   {\large \textbf{Distribution of Brownian coincidences }}
 \end{center}


 \vspace{2cm}

 \noindent A. Krajenbrink, B. Lacroix-A-Chez-Toine, P. Le Doussal,\\
 arXiv preprint,  arXiv: \textbf{1903.06511}, (2019).\\

 \ding{43}
 \href{https://arxiv.org/abs/1903.06511}{https://arxiv.org/abs/1903.06511}

\begin{abstract}
We study the probability distribution, $P_N(T)$, of the coincidence time $T$, i.e. the total local time of all pairwise coincidences
of $N$ independent Brownian walkers. We consider in details two geometries: Brownian motions 
all starting from $0$, and Brownian bridges. Using a Feynman-Ka\v c representation for the moment generating function of this coincidence time, we map this problem onto some observables in three related models
(i) the propagator of the Lieb Liniger model of quantum particles with pairwise delta function interactions (ii) the moments of the partition function of a directed polymer in a random medium (iii) the exponential moments of the solution of the Kardar-Parisi-Zhang equation. Using these mappings, we obtain closed formulae for the probability distribution of the coincidence time, its tails and some of its moments. Its asymptotics at large and small coincidence time are also obtained for arbitrary fixed endpoints. The universal large $T$ tail, $P_N(T) \sim \exp(- 3 T^2/(N^3-N))$ is obtained, and is independent of the geometry. 
We investigate the large deviations in the limit of a large number of walkers through a Coulomb gas approach.
Some of our analytical results are compared with numerical simulations.
\end{abstract}

\part*{Appendices}
\addcontentsline{toc}{part}{Appendices}

\renewcommand*{\printchaptername}{%
  \chapnamefont\centering Appendix}







\chapter{Properties and table of some useful Laplace transform}\label{LT}

Inversion formula
\be
{\cal L}_{s\to t}^{-1}(\tilde f(s))=\int_{\cal C}\frac{ds}{2\pi i}e^{s t}\tilde f(s)\;,
\ee
where ${\cal C}$ is the Bromwich contour which goes from $c-\I\infty$ to $c+\I \infty$ and $c\in \mathbb{R}$ is to the right of all the singularities of $\tilde f(s)$.

A few properties of the Laplace transform and its inverse:
\begin{itemize}
\item Linearity
\be
{\cal L}_{t\to s}(\lambda f(t)+g(t))=\lambda {\cal L}_{t\to s}(f(t))+{\cal L}_{t\to s}(g(t))=\lambda \tilde f(s)+\tilde g(s)\;.
\ee

\item Derivative
\be
{\cal L}_{t\to s}(f'(t))=s{\cal L}_{t\to s}(f(t))-f(0)=\tilde f(s)-f(0)\;.
\ee
\item Translation
\be
{\cal L}^{-1}_{s\to t}(\tilde f(s+r))=\Theta(s+r)e^{-r t}{\cal L}^{-1}_{s\to t}(\tilde f(s))=\Theta(s+r)e^{-r t}{\cal L}^{-1}_{s\to t} f(t)\;.
\ee
\item Multiplication
\be
{\cal L}^{-1}_{s\to t}(\tilde f(s)\tilde g(s))=(f*g)(t)=\int_0^{t}f(t-\tau)g(\tau)d\tau\;.
\ee
\end{itemize}

We list a few Laplace transform that were used in the manuscript:

\begin{itemize}

\item Dirac delta function  $\delta(t-\tau)$
\be
{\cal L}_{t\to s}(\delta(t-\tau))=\int_0^{\infty}e^{-s t}\delta(t-\tau) dt = e^{-s \tau}\;.   \label{LT_dirac}
\ee
\item Heaviside step function $\Theta(t-\tau)$
\be
{\cal L}_{t\to s}(\Theta(t-\tau))=\int_0^{\infty}e^{-s t}\Theta(t-\tau) dt = \frac{e^{-s \tau}}{s}\;.   \label{LT_step}
\ee
\item Power laws $t^{a}$
\be
{\cal L}_{t\to s}(t^a)=\int_0^{\infty}e^{-s t}t^{a} dt = \frac{\Gamma(a+1)}{s^{a+1}}\;,\;\;a>-1\;.  \label{LT_power_law}
\ee
\item Bessel function
\be
{\cal L}_{t\to s}\left(\left(\frac{t}{a}\right)^{\frac{d}{4}}\J_{\frac{d}{2}}\left(2\sqrt{a t}\right)\right)=\int_0^{\infty}e^{-s t}\left(\frac{t}{a}\right)^{\frac{d}{4}}\J_{\frac{d}{2}}\left(2\sqrt{a t}\right) dt = \frac{e^{-\frac{a}{s}}}{s^{\frac{d}{2}+1}}\;,\;\;a,d>0\;.   \label{LT_bessel}
\ee
\item Free diffusion propagator in time
\be
{\cal L}_{t\to s}\left(\frac{e^{-\frac{a^2}{2t}}}{\sqrt{2\pi t}} \right)=\int_0^{\infty}e^{-s t}\frac{e^{-\frac{a^2}{2t}}}{\sqrt{2\pi t}} dt = \frac{e^{-\sqrt{2s}|a|}}{\sqrt{2s}}\;.   \label{LT_diff_prop}
\ee
\item Integrated free diffusion propagator in time
\be
{\cal L}_{t\to s}\left(\erf\left(\frac{a}{\sqrt{2t}}\right) \right)=\int_0^{\infty}e^{-s t}\erf\left(\frac{a}{\sqrt{2t}}\right) dt = \left|\frac{1-e^{-\sqrt{2s}|a|}}{s}\right|\;.   \label{LT_diff_prop_int}
\ee
\end{itemize}

\chapter{A few properties of random matrices}\label{RMT_app}

In this chapter, we review some aspects of random matrix theory that were not treated in the main chapters. For further details, we refer to the extensive literature on random matrix theory \cite{mehta2004random, anderson2010introduction, tracy1993introduction, livan2018introduction, akemann2011oxford, forrester2010log}.

\subsubsection{Dyson's three fold way}

A natural application of random matrices in physics is to model an Hamiltonian (acting on a finite space of dimension $N$) as a random matrix. In quantum mechanics, it is natural to ensure that this Hamiltonian $H$ is hermitian $H=H^{\dag}$. This symmetry can be imposed on the random matrix by ensuring that the statistical properties are invariant under unitary transformation
\be
H\to U^{\dag} H U\;,\;\;{\rm with}\;\;U\in U(N)\;.
\ee
Imposing further symmetries of the system will impose an invariance under a different group of transformation. Dyson introduced a classification  of these symmetries with an associated index $\beta=1,2,4$ \cite{dyson1962threefold}. The case $\beta=2$ refers to the unitary invariance that we have seen.
\begin{itemize}
\item For $\beta=1$, the system is invariant under time-reversal symmetry. The Hamiltonian is invariant under orthogonal transformations
\be
H\to {}^{T}O H O\;,\;\;{\rm with}\;\;O\in O(N)\;.
\ee
\item For $\beta=4$, the Hamiltonian is invariant under symplectic transformations
\be
H\to S^{-1} H S\;,\;\;{\rm with}\;\;S\in Sp(N)\;.
\ee
\end{itemize}
In this thesis, we mainly focused the discussion on the case $\beta=2$ where the statistical properties of the matrices are invariant under unitary transformations.

\subsubsection{Joint PDF of the eigenvalues and Vandermonde}

We consider a matrix $M$ of size $N$ which can be diagonalised by orthogonal ($\beta=1$), unitary ($\beta=2$) or symplectic ($\beta=4$) transformations
\be
M=U^{\dag}\Lambda U\;.
\ee
where $U$ is a matrix in $O(N),U(N),Sp(N)$ respectively for $\beta=1,2,4$ that encompasses all eigenvectors degrees of freedom while $\Lambda={\rm diag}(\lambda_1,\cdots,\lambda_N)$ is the diagonal matrix of the eigenvalues. To define a probability weight $P(M)$ associated to a realisation of the matrix $M$, there are two possibilities
\begin{itemize}
\item We may either define independent weight for the entries (ensuring symmetric, hermitian or symplectic symmetries) 
\be\label{P_entries}
P(M)dM=\prod_{i,j}p_{i,j}(m_{ij})dm_{ij}\;.
\ee
\item Or we may define a weight on the eigenvalues and eigenvectors
\be\label{P_U_l}
P(M)dM=P(U^{\dag}\Lambda U)|\Delta(\Lambda)|^{\beta} d\mu(U)d\Lambda =P(U^{\dag}\Lambda U)\prod_{i<j}|\lambda_i-\lambda_j|^{\beta} d\mu(U) \prod_{i=1}^N d\lambda_i\;,
\ee
where $d\mu(U)$ is the Haar measure on the orthogonal, unitary or symplectic matrices. The Vandermonde term $|\Delta(\Lambda)|^{\beta}$ comes from the Jacobian of the transformation $\{m_{ij}\}\to (U,\Lambda)$. The number $\beta$ corresponds to the number of real independent degrees of freedom used to define an entry in the matrix (one for real numbers, two for complex, four for quaternions).
\end{itemize}
We want to impose that the statistical properties are invariant under unitary transformation, which can be obtained by considering a  measure invariant by unitary transformations
\be
P(M=U^{\dag}\Lambda U)=\frac{e^{-\frac{\beta N}{2}\Tr[v(M)]}}{z_N}=\frac{e^{-\frac{\beta N}{2} \Tr[v(\Lambda)]}}{z_N}\;.
\ee
Integrating Eq. \eqref{P_U_l} with this weight over the Haar measure, i.e. the eigenvectors degrees of freedom, one obtains an explicit formula for joint PDF of the eigenvalues
\be\label{P_joint_v}
P_{\rm joint}(\lambda_1,\cdots,\lambda_N)=\int d\mu(U) P(\Lambda)\prod_{i<j}|\lambda_i-\lambda_j|^{\beta}=\frac{1}{Z_N} \prod_{i<j}|\lambda_i-\lambda_j|^{\beta} \prod_{i=1}^N e^{-\frac{\beta N}{2}v(\lambda_i)}\;,
\ee 
where $Z_N$ is a normalisation factor. It is clear from this equation that the eigenvalues are strongly correlated random variables. 
Rewriting the probability weight as
\begin{align}
P_{\rm joint}(\lambda_1,\cdots,\lambda_N)&=\frac{\displaystyle e^{-\frac{\beta N^2}{2} E_N(\lambda_1,\cdots,\lambda_N)}}{Z_N}\;,\nn\\E_N(\lambda_1,\cdots,\lambda_N)&=\frac{1}{N}\sum_{i=1}^N v(\lambda_i) -\frac{2}{N^2}\sum_{i<j}\ln|\lambda_i-\lambda_j|\;,\label{E_coulomb_gas}
\end{align}
we can reinterpret these eigenvalues as a gas of ``charges'' (with logarithmic $2d$ Coulomb repulsion) and confined on the real line by a one-dimensional potential $v(r)$ \cite{dyson1962statistical,dyson1962statistical2,dyson1962statistical3}.

Note finally that if one gives a weight associated to the entries of the random matrix in Eq. \eqref{P_entries}, this weight will not in general be invariant under orthogonal, unitary or symplectic transformations and computing the joint PDF of the eigenvalues becomes highly non-trivial. In this sense, the Gaussian $\beta$ Ensembles play a singular role as they do match both the criteria of independent (Gaussian) entries and invariance.

\chapter{Basic definition of Fredholm determinant}\label{Fred_det_app}

In this chapter, we define a few concepts on Fredholm determinants. We define the integral operator acting on a function $\Psi(x)$ as
\be
\tilde \Psi(x)=\int dy K(x,y)\Psi(y)\;,
\ee
where $K(x,y)$ is called the kernel of the operator. The Fredholm determinant $\Det(\mathbb{I}-s K)$ is defined as a product over the (infinitely many) eigenvalues $\lambda_k$'s of this operator
\be
\Det(\mathbb{I}-s K)=\prod_{k=1}^{\infty}(1-s\lambda_k)\;.
\ee
Defining the trace of the kernel as
\be
\Tr(K^p)=\int dx_1\cdots dx_p K(x_1,x_2)K(x_2,x_3)\cdots K(x_p,x_1)\;,
\ee
one can expand the Fredholm determinant in terms of these traces
\be
\Det(\mathbb{I}-s K)=\exp\left(\Tr\left[\ln(\mathbb{I}-s K)\right]\right)=\exp\left(-\sum_{p=1}^{\infty}\frac{s^p}{p}\Tr\left[K^p\right]\right)\;.
\ee


\bibliographystyle{plain}





\cleartoevenpage

 \includepdf[pages={1}]{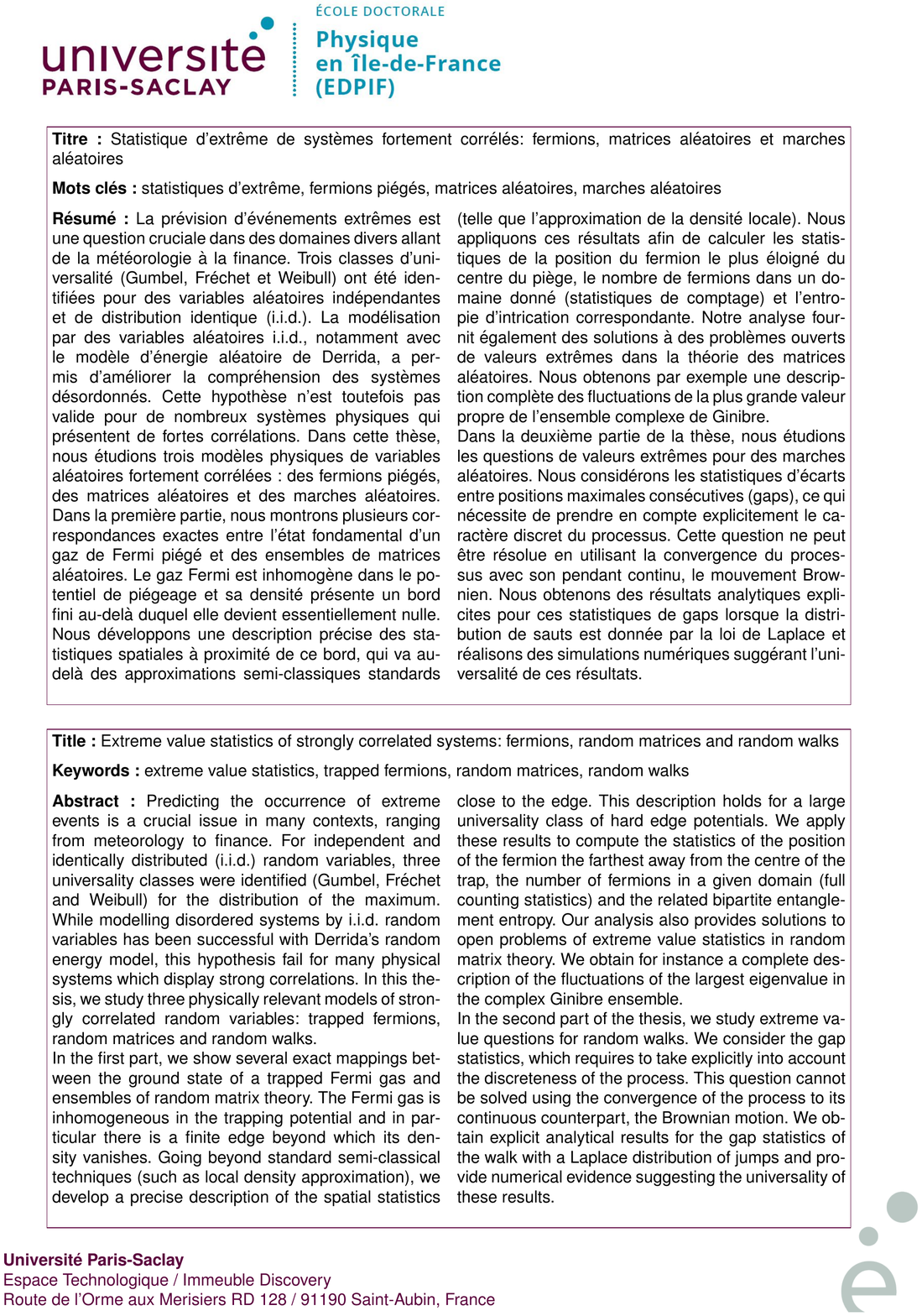}

\end{document}